\documentclass[letter,seceqn,secthm]{elsarticle}
\usepackage{latexsym}
\usepackage{amssymb}
\usepackage{amsmath}
\usepackage{stmaryrd}
\usepackage{graphicx}
\usepackage{pst-all}
\usepackage{times}
\usepackage{comment}
\usepackage{bbold}

\psset{arrowscale=1.4,arrowinset=0.15}

\newcommand{\aTr}{\widetilde{\text{Tr}}}
\newcommand{\arho}{\tilde{\rho}}
\newcommand{\aS}{\tilde{S}}
\newcommand{\bra}[1]{\left\langle #1 \right|}
\newcommand{\ket}[1]{\left| #1 \right\rangle}
\newcommand{\rhoH}{\tilde{\rho}_{\mathcal{A}}}
\newcommand{\rhoD}{\tilde{\rho}_{\mathbb{A}}}

\newcommand{\AD}{\mathbb{A}}

\journal{arXiv}

\begin{document}

\begin{frontmatter}
\title{Anyonic Entanglement and Topological Entanglement Entropy}

\author[stq]{Parsa Bonderson}
\author[ucsb]{Christina Knapp\corref{cor1}}
\ead{cknapp@physics.ucsb.edu}
\author[ucsb]{Kaushal Patel}

\cortext[cor1]{Corresponding author}
\address[stq]{Station Q, Microsoft Research, Santa Barbara, California 93106-6105, USA}
\address[ucsb]{Department of Physics, University of California, Santa Barbara, California 93106, USA}

\begin{abstract}
We study the properties of entanglement in two-dimensional topologically ordered phases of matter. Such phases support anyons, quasiparticles with exotic exchange statistics. The emergent nonlocal state spaces of anyonic systems admit a particular form of entanglement that does not exist in conventional quantum mechanical systems.  We study this entanglement by adapting standard notions of entropy to anyonic systems.  We use the algebraic theory of anyon models (modular tensor categories) to illustrate the nonlocal entanglement structure of anyonic systems. Using this formalism, we present a general method of deriving the universal topological contributions to the entanglement entropy for general system configurations of a topological phase, including surfaces of arbitrary genus, punctures, and quasiparticle content. We analyze a number of examples in detail. Our results recover and extend prior results for anyonic entanglement and the topological entanglement entropy.
\end{abstract}


\end{frontmatter}

\section{Introduction}

Entanglement, ``the characteristic trait of quantum mechanics"~\cite{Schroedinger35}, underlies some of the most exotic phenomena in condensed matter physics, including quantum critical points~\cite{Calabrese04,Calabrese09}, quantum spin liquids~\cite{Savary16}, and topologically ordered phases of matter~\cite{Wen90,Nayak08}. Topological order occurs in gapped, many-body systems whose microscopic degrees of freedom possess daedal entanglement in their ground states. In particular, topological phases exhibit emergent universal phenomena that depend only on the global (topological) properties of the system, making them robust to local perturbations and incapable of being identified by any local probe of the system. Among the most intriguing of such emergent phenomena is the ability to support anyons -- quasiparticle excitations with a topological (nonlocal) state space and exotic exchange statistics characterized by braiding~\cite{Leinaas77,Wilczek82b,Goldin85,Fredenhagen89,Froehlich90}.

Beyond their fundamental interest as exemplars of the ways nature can give rise to emergent properties that are not intrinsic to the microscopic degrees of freedom, anyons provide a technologically promising platform for quantum information processing. Topological quantum computing~\cite{Kitaev03,Freedman98,Nayak08}, the nonlocal storage and manipulation of quantum information in an anyonic system, is robust against errors due to local perturbations and noise from the environment.

The topological entanglement entropy (TEE)~\cite{Kitaev06b,Levin06a} is a signature of topological order that has been the focus of numerous theoretical~\cite{Hamma05,Dong08,Castelnovo08,Rodriguez08,Grover11,Brown11,Kim13,Kim15,Wen16a,Bullivant16} and numerical studies~\cite{Furukawa07,Haque07,Zozulya07,Zozulya08,Yao10,Isakov11,Sterdyniak11,Zhang12,Jiang12,Zaletel12,Zaletel13,Cincio13,Estienne14,Grushin15}. Despite these efforts, an intuitive understanding of the origin and form of the TEE has remained elusive and only an inchoate connection between the TEE and the anyonic excitations of the system has been established.

In this work, we examine entanglement and entropy of anyonic systems.  In doing so, we demonstrate that TEE is a natural consequence of the conservation of topological charge.  We obtain our results using anyon models, which are the algebraic description of the long-ranged, low-energy effective theories of quasiparticles.  Mathematically, anyon models are known as unitary modular tensor categories (UMTCs) and apply beyond the context of anyons~\cite{Fredenhagen89,Moore89b,Froehlich90,Turaev94,Bakalov01,Preskill-lectures,Kitaev06a}.  We use the formalism for anyonic density matrices developed in Refs.~\cite{Bonderson07b,Bonderson07c}. Our analysis applies to bosonic topological phases of matter on compact, orientable surfaces in two spatial dimensions.

This paper is organized as follows.  In Section~\ref{sec:entropy}, we briefly review classical and quantum entropy.  In Section~\ref{sec:anyonicEntropy}, we discuss anyonic entanglement, introducing the anyonic entanglement entropy (AEE) and entropy of anyonic charge entanglement, as well as presenting a new derivation of the TEE for a disk in the plane.  In Section~\ref{sec:anyonmodelshigher}, we discuss the state space of anyon models on higher genus surfaces.  In Section~\ref{sec:ATEE}, we apply this formalism to derive the TEE on higher genus surfaces.  In Section~\ref{sec:discussion}, we conclude and place our results in the broader context of lattice models, topological defects, fermionic topological phases, non-orientable surfaces, and three-dimensional topological phases.

\section{Entropy}
\label{sec:entropy}

\subsection{Classical and Quantum Entropies}

Entropy is the measure of uncertainty in a state of a physical system.  Classically, if an unknown variable $X$ has value $x$ with probability $p_x$, the Shannon entropy is
 \begin{equation}
 H(\{p_x\})\equiv -\sum_x p_x \log p_x.
 \end{equation}
 The Shannon entropy quantifies our uncertainty in the value of $X$, or equivalently, how much information we gain by learning the value of $X$.

The classical R\'enyi entropy of order $\alpha$ is defined by
\begin{equation}
H_\alpha (\{p_x\})\equiv \frac{1}{1-\alpha}\log\left(\sum_x p_x^\alpha\right)
\end{equation}
for $\alpha>0$.  Note that $\lim_{\alpha\to 1}H_\alpha(\{p_x\})=H(\{p_x\})$, thus the R\'enyi entropies may be understood as a generalization of the Shannon entropies. The R\'enyi entropies are normalized to vanish for a pure state ($\{p_x\}=\{\delta_{xy}\}$ for some $y$) and to be maximized for a uniform distribution ($\{p_x\}=\{1/N\}$).

Classical entropies can be easily extended to describe quantum states by replacing probability distributions with density matrices and sums with traces over the degrees of freedom in the system.  The quantum analogue of the Shannon entropy for a quantum state $\rho$ is the von Neumann entropy,
\begin{equation}
\label{eq:vonNeumann}
S(\rho)\equiv -\text{Tr}(\rho\log\rho),
\end{equation}
which can be re-expressed as the Shannon entropy of the eigenvalues $\lambda_x$ of $\rho$,
\begin{equation}
S(\rho)=H(\{\lambda_x\})=-\sum_x \lambda_x \log \lambda_x.
\end{equation}
The quantum R\'enyi entropy of order $\alpha$ is similarly generalized as
\begin{equation}
\label{eq:qRenyi}
S^{(\alpha)}(\rho) \equiv \frac{1}{1-\alpha}\log[\text{Tr}(\rho^\alpha)].
\end{equation}
There exist many other entropy-related quantities.  The relative entropy measures the closeness of two quantum states $\rho$ and $\sigma$:
\begin{equation}
\label{eq:relEntropy}
 S(\rho||\sigma) \equiv \text{Tr}\left(\rho\log\rho\right)-\text{Tr}\left(\rho\log\sigma\right).
\end{equation}
The mutual information measures how much information is shared between two subsystems.  That is, if a system with state $\rho$ has two subsystems $A$ and $B$, then the mutual information is
\begin{equation}
\label{eq:mutInformation}
I(A:B)\equiv S(\rho_A)+S(\rho_B)-S(\rho),
\end{equation}
where $\rho_A=\text{Tr}_{B}\rho$ and $\rho_B=\text{Tr}_A \rho$.  Both the relative entropy and the mutual information can be defined for classical probability distributions in the natural way.

\subsection{Entanglement Entropy}\label{sec:EE}

Consider partitioning a system into a region $A$ and its complement $\bar{A}$. If we are interested only in $A$, then we would like to describe the state with degrees of freedom local to $A$, rather than the state of the full system, $\rho$.  When the Hilbert space of the system admits a factorization
 \begin{equation}
\mathcal{H}=\mathcal{H}_A\otimes\mathcal{H}_{\bar{A}},
\end{equation}
where $\mathcal{H}_{A}$ has support in $A$, then we can define the reduced density matrix $\rho_A$ by
\begin{equation}
\rho_A=\text{Tr}_{\bar{A}} \rho.
\end{equation}
The partial trace $\text{Tr}_{\bar{A}}$ means we sum over all degrees of freedom local to $\bar{A}$, essentially retaining only the information associated with $A$.  For any operator ${O=O_A\otimes O_{\bar{A}}}$, where $O_{A}$ has support in $A$, the partial trace is the unique operator satisfying ${\text{Tr}\left(\rho O\right)}={\text{Tr}_A\left( \rho_A O_A\right)}$~\cite{Nielsen11}.

Note that $\rho_A$ is a pure state only when $\rho=\rho_A\otimes\rho_B$ is separable and ${\rho_A=\ket{\psi_A}\bra{\psi_A}}$.  In general, if there is some entanglement between $A$ and $\bar{A}$, $\rho_A$ will be a mixed state.  The von Neumann entropy of the reduced density matrix,
 \begin{equation}
 \label{eq:EE}
 S(\rho_A)\equiv -\text{Tr}_A[\rho_A\log\rho_A],
 \end{equation}
 is a measure of this entanglement; it can only decrease when acted upon by operators local to $A$.  We call $S(\rho_A)$ the entanglement entropy.  If $\rho$ for the full system is a pure state, then $S(\rho_A)$ is the unique entanglement measure that is (1) invariant under operators acting only on $A$, (2) continuous, and (3) additive when there are several copies of the system.

\subsection{Topological Entanglement Entropy}\label{sec:TEE}

In a gapped two dimensional system partitioned into regions $A$ and $\bar{A}$ with smooth boundaries, the ground state of the $A\cup\bar{A}$ is expected to have entanglement entropy that scales linearly with the boundary separating $A$ and $\bar{A}$.  If the state is topologically ordered, the entanglement entropy will have a universal constant correction to this ``boundary law" that is completely determined by topological invariants~\cite{Hamma05,Kitaev06b,Levin06a}.  The ground state wavefunction of a topological phase on the plane, partitioned into a disk $A$ and its complement $\bar{A}$, has entanglement entropy
\begin{equation}\label{eq:ententropy}
S_A = \alpha L +S_{\text{topo}} +\mathcal{O}(L^{-1})
\end{equation}
where $L$ is the linear size of $A$, $\alpha$ is a non-universal constant dependent upon the short distance physics of the system, and
\begin{equation}\label{eq:Stopo}
  S_{\text{topo}}\equiv-\log \mathcal{D}
\end{equation}
is the topological entanglement entropy (TEE)~\cite{Kitaev06b}.  The quantity $\mathcal{D}$ is the total quantum dimension of the system.  For a topological phase whose corresponding TQFT is described by the UMTC $\mathcal{C}$, the total quantum dimension is defined by
\begin{equation}
\mathcal{D}=\sqrt{\sum_{a\in \mathcal{C}} d_a^2},
\end{equation}
where $d_a$ is the quantum dimension of the anyon with topological charge $a$ (see \ref{sec:anyonmodelssphere} for a review). Eq.~(\ref{eq:ententropy}) also holds in the context of string-nets~\cite{Levin06a,Turaev92,Levin05a}, see Section~\ref{sec:stringnets} for further discussion.

At first consideration, $S_{\text{topo}}$ might seem like a rather crude quantity to use for characterizing a topological phase, as it is a single number. Indeed, other entanglement-based probes of the system, such as the ``entanglement spectrum''~\cite{Li08}, will generally provide more information about the system. However, topological order is highly constrained, so the information contained in the single number $S_{\text{topo}}$ can be used, with a bit of algebraic effort, to significantly narrow the field of possibilities when trying to identify a topological phase. Indeed, for many cases, knowing $S_{\text{topo}}$ is sufficient to completely determine the topological order (up to chirality). To be more specific, in the context of anyon models, if $N$ refers to the number of anyon types in a theory, one can easily show (from the fusion rules) that $N \leq \mathcal{D}^{2}$. It was shown in Ref.~\cite{Bruillard13} that, for a given rank $N$, there are only a finite number of possible UMTCs. It follows that there are only a finite number of possible UMTCs for a particular value of $\mathcal{D}$. Moreover, the UMTCs with a given value of $\mathcal{D}$ are usually very closely related. In Table~\ref{Table:D}, we list all UMTCs for $\mathcal{D}^2<8$.

\begin{table}
\[
\begin{array}{c|l}
\mathcal{D}^{2} & \text{UMTCs}  \\
\hline \hline
1 &  \mathbb{Z}_1 \quad \text{(Trivial)}  \\
\hline
2 &\mathbb{Z}_2^{(p)},  ~p=\frac{1}{2},\frac{3}{2}  \\
\hline
3 &  \mathbb{Z}_3^{(p)}, ~p=1,2 \\
\hline
\phi+2 &  \text{Fib}^{\pm 1}\\
\hline
4 & \mathbb{Z}_2^{(1/2)} \times \mathbb{Z}_2^{(3/2)} ; \,\, \mathcal{K}_{\nu}, ~\nu=0,1, \ldots , 15 \\
\hline
5 & \mathbb{Z}_5^{(p)}, ~p=1,2 \\
\hline
6 &  \mathbb{Z}_6^{(p)},  ~p=\frac{1}{2},\frac{5}{2},\frac{7}{2},\frac{11}{2} \\
\hline
7 &  \mathbb{Z}_7^{(p)}, ~p=1,3\\
\hline
2\left(\phi+2\right)  & \text{Fib}^{\pm 1} \times \mathbb{Z}_2^{(p)},  ~p=\frac{1}{2},\frac{3}{2}\\
\hline
\end{array}
\]
\caption{A TQFT in ($2+1$)D is described by a UMTC, which can be classified according to its value of the total quantum dimension $\mathcal{D}$. This table lists all distinct UMTCs with $\mathcal{D}^{2} < 8$, as determined from Refs.~\cite{Bonderson07b,Rowell07,BondersonWIP}. ($\phi = \frac{1+\sqrt{5}}{2} \approx 1.6$ is the Golden ratio.) For most values of $\mathcal{D}$, there are very few possible UMTCs. Moreover, the UMTCs with a given value of $\mathcal{D}$ are usually very closely related. Additional details may be found in \ref{sec:BTCs}.}
\label{Table:D}
\end{table}

Since the seminal works of Refs.~\cite{Kitaev06b, Levin06a}, TEE has received a significant amount of attention.
Theoretical studies have investigated the connections between TEE and ground state degeneracy~\cite{Kim13}, derived the TEE for Chern-Simons theories on higher genus surfaces~\cite{Dong08,Wen16a}, derived TEE for certain systems with topological defects~\cite{Brown11}, and explored the TEE in the context of $(3+1)$-dimensional topological phases~\cite{Castelnovo08,Grover11,Kim15,Bullivant16}.
In numerical studies, TEE has become a useful quantity for identifying topological phases~\cite{Furukawa07,Haque07,Zozulya07,Zozulya08,Yao10,Isakov11,Zhang12,Jiang12,Zaletel12,Zaletel13,Cincio13,Estienne14,Grushin15}  (though it has been demonstrated that the accuracy of numerical extractions of TEE requires some caution~\cite{Zozulya08,Sterdyniak11}).  Nonetheless, the meaning and origin of $S_{\text{topo}}$ has remained somewhat nebulous.  In this paper, we attempt to demystify these concepts by analyzing entanglement entropy and TEE in the context of anyon models.

Our calculations of the entanglement entropy and the TEE only take into account the long-range physics encoded in the TQFT describing the topological phase.  If the system is away from the purely topological, zero correlation length limit, microscopic details of the system will modify the length-dependent terms in Eq.~(\ref{eq:ententropy}). However, the universal contribution to the entanglement entropy, $S_{\text{topo}}$, will be the same.

For an arbitrary compact, orientable surface (possibly including genus, punctures, and quasiparticles) partitioned into regions $A$ and $\bar{A}$, the entanglement entropy between $A$ and $\bar{A}$ takes the form
\begin{equation}
\label{eq:general_result}
S_{A} = \sum_{k=1}^{N} \left( \alpha L_{k} - \log \mathcal{D} + \sum_{c} p_{c}^{(k)} \log d_{c} \right) + \aS(\tilde{\rho}_{A}) + \mathcal{O}(L_{k}^{-1})
,
\end{equation}
where $k=1,\ldots,N$ labels the connected components of the partition boundary between $A$ and $\bar{A}$; $L_k$ is the length of the $k$th connected component of the partition boundary; $\tilde{\rho}_{A}$ is the anyonic reduced density matrix for region $A$ (including boundaries); $p_{c}^{(k)}$ the probability of the state $\tilde{\rho}_{A}$ being in a configuration wherein the $k$th joint boundary component carries topological charge $c$; $d_{c}$ is the quantum dimension of topological charge $c$; and $\aS(\tilde{\rho})$ is the anyonic entropy of the anyonic state $\tilde{\rho}$. These quantities will be defined and explained in detail in this paper.

\section{Anyonic Entropy and Entanglement}
\label{sec:anyonicEntropy}

We proceed by applying the standard notions of entropy, discussed in Section~\ref{sec:entropy}, to anyon models, reviewed in \ref{sec:anyonmodelssphere}.  In doing so, we elucidate the unique ways in which entanglement arises in a topologically ordered system.  For clarity, we denote an anyonic state (density matrix) and its associated entropy with a tilde; $\arho$ and $\aS(\arho)$ respectively.

The anyonic von Neumann entropy is
\begin{equation}
    \aS(\arho) =-\aTr(\arho \log \arho),
\end{equation}
where $\aTr$ denotes the quantum trace, see \ref{sec:anyonmodelssphere}.
In \ref{app:proofs}, we prove that the anyonic von Neumann entropy has many of the important properties that the conventional von Neumann entropy has.  Moreover, when the state has Abelian total charge, the quantum trace is equivalent to the conventional trace, in which case the anyonic density matrix $\arho$ is a properly normalized conventional density matrix and $\aS(\arho)=S(\arho)$.

The anyonic R\'enyi entropy is
\begin{equation}
    \aS^{(\alpha)}(\arho) =\frac{1}{1-\alpha}\log\aTr(\arho^\alpha).
\end{equation}
The relation between the conventional von Neumann and R\'enyi entropies holds for the anyonic counterparts:
\begin{equation}
\label{eq:replica}
\lim_{\alpha\to 1} \aS^{(\alpha)}(\arho) = \aS(\arho).
\end{equation}

\subsection{Pure States and Mixed States}

An anyonic state on the sphere (or plane with no topological charge on the boundary) must have trivial total fusion channel.  This constraint derives from the conservation of topological charge; a single anyon with nontrivial charge cannot be created from the vacuum.  This simple statement has important consequences for anyonic entanglement, which we now explore.

Similar to the conventional quantum states, we define an {\it anyonic pure state} to be the ones whose anyonic density matrix $\arho$ has vanishing anyonic von Neumann entropy, or equivalently, $\aTr \left(\arho^2\right)=1$.  When $\aTr \left(\arho^2\right)<1$, the anyonic state is mixed.

Our intuition from conventional quantum mechanics can be misleading when applied to anyonic states.  As an illustrative example, consider the density matrix of a single anyon with definite charge $a$:
\begin{equation}
\label{eq:rho_a}
\arho_{a}=\frac{1}{d_{a}}\left| a\right\rangle \left\langle a\right| = \frac{1}{d_{a}} \mathbb{1}_{a}
= \frac{1}{d_{a}}
\pspicture[shift=-0.7](0.4,-0.1)(1.05,1.5)
  \scriptsize
  \psline[ArrowInside=->](0.6,0.1)(0.6,1.3) \rput[bl]{0}(0.75,0.41){$a$}
  \endpspicture
.
\end{equation}%
One can write
$\left| a\right\rangle \left\langle a\right|$ as
$\left| a,0;a\right\rangle \left\langle a,0;a\right|$ to maintain the proper association of bras and kets with trivalent vertices. At first glance, Eq.~(\ref{eq:rho_a}) may appear to be a pure state, as there is no degeneracy in the local state space associated with a single anyon.   However, it must be kept in mind that, due to conservation of topological charge, a single anyon cannot truly exist by itself.  Such a nontrivial state must be obtained from the state of multiple anyons by tracing out all but one, e.g.,
\begin{equation}
\arho_a = \aTr_{\bar{a}} \left( \frac{1}{d_a} \begin{pspicture}[shift=-0.6](-0.2,0)(1.2,1.5)
        \scriptsize
        \psline[ArrowInside=->](0,0)(0.5,0.5)\rput(0,0.25){$a$}
        \psline[ArrowInside=->](1,0)(0.5,0.5)\rput(1,0.25){$\bar{a}$}
        \psline[ArrowInside=->](0.5,1)(0,1.5)\rput(0,1.25){$a$}
        \psline[ArrowInside=->,](0.5,1)(1,1.5)\rput(1,1.25){$\bar{a}$}
\end{pspicture} \right)
     = \frac{1}{d_a}
\begin{pspicture}[shift=-0.6](-0.4,0)(1.6,1.5)
        \scriptsize
        \psline(0,0)(0.5,0.5)(1,0)(1.5,.5)(1.5,1)(1,1.5)(0.5,1)
        \psline[ArrowInside=->](0.5,1)(0,1.5)\rput(0,1.25){$a$}
        \psline[ArrowInside=->](0.5,1)(1,1.5)\rput(1,1.25){$\bar{a}$}
        \psline[ArrowInside=->](0,0)(0.5,.5)\rput(0,.25){$a$}
        \psline[ArrowInside=->](1,0)(0.5,.5)\rput(1,.25){$\bar{a}$}
\end{pspicture}
     = \frac{1}{d_a}
\begin{pspicture}[shift=-0.6](-0.35,0)(.2,1.5)
        \scriptsize
        \psline[ArrowInside=->](0,0)(0,1.5)\rput(-.2,0.8){$a$}
\end{pspicture} .
\end{equation}
If the charge $a$ of the remaining anyon is non-Abelian, and hence $d_{a} > 1$, this state is not pure, as can be seen from
\begin{equation}
\aTr \left[ \arho_a^{2} \right] = 1/d_{a} < 1.
\end{equation}
The remaining single anyon is in an anyonic mixed state as a consequence of the anyonic entanglement it had with the other anyons from the traced out subsystem.
This simple example highlights the type entanglement we wish to quantify.

One can check that $\aS \left(\arho_a \right)$ is nonzero.  The anyonic R\'enyi entropy of $\arho$ is
\begin{equation}
\begin{split}
\aS^{(\alpha)}\left(\arho_a\right)&= \frac{1}{1-\alpha}\log \aTr\left(\arho_a^{\alpha}\right)
= \frac{1}{1-\alpha} \log \aTr \left( \frac{1}{d_a^\alpha}
\psscalebox{.9}{
\pspicture[shift=-0.7](-.1,-0.1)(.3,1.4)
\scriptsize
\psline[ArrowInside=->] (.2,.1)(.2,1.3)\rput(-0,.75){$a$}
\endpspicture}\right)
= \frac{1}{1-\alpha}\log \left( \frac{1}{d_a^{\alpha}}
\psscalebox{.9}{
\pspicture[shift=-0.7](0,-0.1)(1.,1.4)
 \scriptsize
 \psline[ArrowInside=->](.3,0.5)(.3,0.9)
  \psline(0.6,.2)(0.3,0.5)(0.3,0.9)(0.6,1.2) (.9,0.9)(.9,0.5)(0.6,0.2)\rput(.13,.75){$a$}
 \endpspicture}
 \right)
\\ &= \frac{1}{1-\alpha}\log d_a^{1-\alpha}
=\log d_a.
\end{split}
\end{equation}
Taking the (trivial in this example) limit $\alpha\to 1$, we see
\begin{equation}
\label{eq:Sa}
\aS(\arho_a)=\log d_a \equiv \aS_a,
\end{equation}
which is nonzero when $a$ is non-Abelian. Eq.~(\ref{eq:Sa}) is the anyonic entropy associated with the topological charge $a$, due solely to the topological nature of the system.  Recall from regular quantum mechanics that a quantum system with a $d$-dimensional Hilbert space has $\log d$ as its maximal von Neumann entropy.  From this perspective, one may think of this anyonic entropy as arising from some locally inaccessible internal degrees of freedom of anyons. This is precisely what gives rise to the nonlocal topological state space associated with non-Abelian anyons.

Let $\ket{\psi_c}$ denote a state with overall topological charge $c$.  From the above example, we see that an anyonic pure state has anyonic density matrix $\arho$ that can be written as $\arho = \left| \psi_c \right\rangle \left\langle \psi_c \right|$, such that $c$ is Abelian. The term ``anyonic pure state" is sometimes defined to only include states with trivial overall topological charge $0$, but here we expand the definition to include states with overall Abelian charge, because from the entropic perspective they have all the same properties.

A general state of a system of two anyons can be diagonalized into sectors of distinct charge.
Let $\arho_{AB}$ be the state of a system of two anyons $A$ and $B$, where the capital letters denote that there can be sums over external fusion trees.  We can write
\begin{equation}
\begin{split}
\arho_{AB} &=\sum\limits_{c,\mu _{c}}\frac{p_{\mu _{c}}^{AB}}{d_{c}}\left| \mu
_{c}\right\rangle \left\langle \mu _{c}\right|
=\sum_{c,\mu_c} \frac{p_{\mu_c}^{AB}}{d_c} \sum_{\substack{a,b,\alpha,\\ a',b',\alpha'}} \frac{\psi_{a,b,c,\alpha}^{(\mu)} \left(\psi_{a',b',c,\alpha'}^{(\mu)}\right)^*\sqrt{d_c}}{\left(d_a d_b d_{a'} d_{b'} \right)^{1/4}}
\pspicture[shift=-0.6](-0.1,-0.45)(1.5,1)
 \scriptsize
  \psline[ArrowInside=->](0.7,0)(0.7,0.55)
  \psline[ArrowInside=->](0.7,0.55) (0.25,1)
  \psline[ArrowInside=->](0.7,0.55) (1.15,1)
  \rput[bl]{0}(0.38,0.2){$c$}
  \rput[bl]{0}(1.3,0.8){$b$}
  \rput[bl]{0}(0,0.8){$a$}
  \psline[ArrowInside=->](0.25,-0.45)(0.7,0) \rput[bl]{0}(0,-0.4){$a'$}
  \psline[ArrowInside=->](1.15,-0.45)(0.7,0) \rput[bl]{0}(1.3,-0.4){$b'$}
  \rput[bl]{0}(0.85,0.38){$\alpha$}
  \rput[bl]{0}(0.85,-0.05){$\alpha'$}
  \endpspicture,
\end{split}
\end{equation}%
where the state vectors
\begin{equation}
\ket{\mu_c}=\sum_{a,b,\alpha} \psi_{a,b,c,\alpha}^{(\mu)}\left( \frac{d_c}{d_ad_b}\right)^{1/4}
\pspicture[shift=-0.6](-0.1,-0.2)(1.5,1)
 \scriptsize
  \psline[ArrowInside=->](0.7,0)(0.7,0.55)
  \psline[ArrowInside=->](0.7,0.55) (0.25,1)
  \psline[ArrowInside=->](0.7,0.55) (1.15,1)
  \rput[bl]{0}(0.38,0.2){$c$}
  \rput[bl]{0}(1.3,0.8){$b$}
  \rput[bl]{0}(0,0.8){$a$}
  \rput[bl]{0}(0.85,0.38){$\alpha$}
  \endpspicture
\end{equation}
have coefficients $\psi_{a,b,c,\alpha}^{(\mu)}$ chosen such that
\begin{equation}
\bra{\nu_c}\mu_c\rangle = \sum_{a,b,\alpha} \psi^{(\mu)}_{a,b,c,\alpha}\left(\psi^{(\nu)}_{a,b,c,\alpha}\right)^*
\begin{pspicture}[shift=-0.6](-0.1,-0.2)(.4,1)
\scriptsize
\psline[ArrowInside=->](0,0)(0,1)
\rput(.2,.6){$c$}
\end{pspicture}
=\delta_{\mu,\nu} \mathbb{1}_c.
\end{equation}
The decomposition can always be done in terms of vectors $\ket{\mu_c}$ with definite overall charge $c$ because superpositions of different values of overall topological charge are always incoherent, i.e. the density matrix is always block diagonal in sectors of distinct overall topological charge $c$.

The anyonic von Neumann entropy of $\arho_{AB}$ is
\begin{equation}
\begin{split}
\aS\left( \arho_{AB}\right)  &=-\partial_{\alpha} \left( \aTr\left( \arho_{AB}\right)^\alpha\right)_{\alpha=1}
= -\partial_{\alpha}\left( \sum_{\mu_c,c} d_c \left( \frac{p_{\mu_c}^{AB}}{d_c}\right)^\alpha \right)_{\alpha=1}
\\ &=-\sum\limits_{c,\mu _{c}}p_{\mu
_{c}}^{AB}\log \left( \frac{p_{\mu _{c}}^{AB}}{d_{c}}\right)
= \sum_c H\left(\{p_{\mu_c}^{AB}\}\right) +\sum_{c} p_{_c}^{AB}\aS_c,
\end{split}
\end{equation}
where
\begin{equation}
p_c^{AB}=\sum_{\mu_c} p_{\mu_c}^{AB}
\end{equation}
is the probability of the state having overall topological charge $c$.
In particular, the only way for Abelian anyonic states to have nonzero entropy is through incoherent superpositions of the charges of localized anyons, which is just the Shannon entropy of classical origin. This represents the fact that there are no fusion degeneracies to evoke a multidimensional state space for Abelian anyons.

One might be tempted to think of the term
\begin{equation}
 \sum\limits_{c,\mu _{c}}p_{\mu _{c}}^{AB}\aS_c
\end{equation}
as the ``topological'' contribution to the entropy of this system, since it results from the overall charge of the system, and it appears to be the difference between the anyonic entropy and the entropy of a non-anyonic system with orthonormal decomposition coefficients $p_{\mu _{c}}^{AB}$. However, this is a misleading superficiality and one cannot partition the provenance of entropy in this manner.  The fusion category structure of anyon models is not a simple tensor product and the topological effects and qualities of the system are subtly encoded throughout the fusion channel description of an anyonic state.

\subsection{Anyonic Entanglement}

Having gained some insight from the examples of the previous section, we turn now to characterizations of anyonic entanglement.
In ordinary quantum mechanics, entanglement arises from correlations between local degrees of freedom.  For example, in the Bell state
\begin{equation}
\ket{\Phi^+}=\frac{1}{\sqrt{2}}\left( \ket{0}_A\otimes \ket{0}_B+\ket{1}_A\otimes\ket{1}_B\right)
\end{equation}
all degrees of freedom of the system are local to either qubit $A$ or qubit $B$, and the state of qubit $A$ is correlated with that of qubit $B$.
In a topological phase, the anyonic Hilbert space generally does not admit a tensor product structure.  Thus, there exist nonlocal emergent degrees of freedom which cannot be assigned to a particular region, e.g. the total topological charge of a collection of anyons.  These nonlocal degrees of freedom arise from topological correlations in the system and imprint signatures in the entanglement of the state.

One probe of the system's topological correlations is the \emph{entropy of anyonic charge entanglement}
\begin{equation}\label{eq:Sace}
\aS_{\text{ace}} (A:B) = \aS\left( D_{A:B}[\arho]\right) - \aS\left( \arho\right)
,
\end{equation}
where $D_{A:B}$ is the charge line decoherence superoperator that severs charge lines in the density matrix that connect the subsystems $A$ and $B$. $D_{A:B}$ may be enacted by a vertical $\omega_0$-loop applied to the diagrammatic density matrix that encloses topological charge lines connecting the two regions. This definition of $\aS_{\text{ace}} (A:B)$ is intended to extract only the entropy associated directly with the anyonic charge lines that connect the two subsystems $A$ and $B$ (as will be made more clear).

More explicitly, if subsystems $A$ and $B$ are connected by the diagram (suppressing vertex labels and the fusion trees of anyons within subsystems $A$ and $B$)
\begin{equation}
\begin{split}
 & \begin{pspicture}[shift=-1.05](1.1,-3.3)(2.8,-1.2)
        \scriptsize
        \psline[border=2pt](2.6,-1.2)(1.9,-1.9)
        \psline(1.2,-1.2)(1.9,-1.9)
        \psline[ArrowInside=->](1.7,-1.7)(1.4,-1.4)\rput(1.6,-1.4){$a$}
        \psline[ArrowInside=->](2.1,-1.7)(2.4,-1.4)\rput(2.2,-1.4){$b$}
        \psline[ArrowInside=->](1.9,-2.5)(1.9,-1.9)\rput(1.98,-2.1){$c$}
        \psline[border=2pt](2.6,-3.2)(1.9,-2.5)
        \psline(1.2,-3.2)(1.9,-2.5)
        \psline[ArrowInside=->](1.4,-3)(1.7,-2.7)\rput(1.6,-3){$a'$}
        \psline[ArrowInside=->](2.4,-3)(2.1,-2.7)\rput(2.15,-3){$b'$}
    \end{pspicture},
\end{split}
\end{equation}
then $D_{A:B}$ acts on the system by applying the $\omega_0$-loop as shown below~\cite{Bonderson09}:
\begin{equation}
\begin{split}
 & \begin{pspicture}[shift=-1.05](1.1,-3.3)(2.8,-1.2)
        \scriptsize
        \psellipse[linecolor=black,border=0](1.9,-2.2)(0.2,.6) \rput(1.45,-2.35){$\omega_0$}
        \psline{>-}(1.71,-2.25)(1.71,-2.15)
        \psline[border=2pt](2.6,-1.2)(1.9,-1.9)
        \psline(1.2,-1.2)(1.73,-1.73)
        \psline(1.83,-1.83)(1.9,-1.9)
        \psline[ArrowInside=->](1.7,-1.7)(1.4,-1.4)\rput(1.6,-1.4){$a$}
        \psline[ArrowInside=->](2.1,-1.7)(2.4,-1.4)\rput(2.2,-1.4){$b$}
        \psline[ArrowInside=->](1.9,-2.5)(1.9,-1.9)\rput(1.98,-2.1){$c$}
        \psline[border=2pt](2.6,-3.2)(1.9,-2.5)
        \psline(1.2,-3.2)(1.73,-2.67)
        \psline(1.83,-2.57)(1.9,-2.5)
        \psline[ArrowInside=->](1.4,-3)(1.7,-2.7)\rput(1.6,-3){$a'$}
        \psline[ArrowInside=->](2.4,-3)(2.1,-2.7)\rput(2.15,-3){$b'$}
    \end{pspicture}
 = \sum_{e} \Big[ F^{ab}_{a'b'}\Big]_{ce}
 \begin{pspicture}[shift=-1.05](1.1,-3.3)(2.5,-1.2)
        \scriptsize
        \psellipse[linecolor=black,border=0](1.8,-2.2)(0.2,.6) \rput(1.8,-1.5){$\omega_0$}
        \psline{>-}(1.61,-2.25)(1.61,-2.15)
        \psline[border=1.5pt](1.7,-2.1)(2.3,-2.4)
        \psline[ArrowInside=->](2.3,-2.4)(1.9,-2.2)
        \psline(1.55,-2.025)(1.3,-1.9)
        \psline[ArrowInside=->](1.3,-1.8)(1.3,-1.5)
        \psline[ArrowInside=->](1.3,-3)(1.3,-2.5)
        \psline(1.3,-3)(1.3,-1.5)
        \psline[ArrowInside=->](2.3,-1.8)(2.3,-1.5)
        \psline[ArrowInside=->](2.3,-3)(2.3,-2.5)
        \psline(2.3,-3)(2.3,-1.5)
        \rput(2.3,-1.3){$b$}
        \rput(1.3,-1.3){$a$}
        \rput(2.3,-3.2){$b'$}
        \rput(1.3,-3.2){$a'$}
        \rput(2.15,-2.125){$e$}
 \end{pspicture}
 = \sqrt{\frac{d_c}{d_a d_b}} \delta_{a,a'}\delta_{b,b'}
 \begin{pspicture}[shift=-1.05](1.1,-3.3)(2.5,-1.2)
        \scriptsize
        \psline[ArrowInside=->](1.3,-3)(1.3,-1.5)
        \psline[ArrowInside=->](2.3,-3)(2.3,-1.5)
        \rput(2.1,-2.2){$b$}
        \rput(1.5,-2.22){$a$}
 \end{pspicture}
.
\end{split}
\end{equation}

The state $\arho_{AB}$ has no \emph{anyonic charge entanglement} between subsystems $A$ and $B$ if
\begin{equation}
\arho_{AB}\in V^{A_1,...,A_m}_{A_1',...,A_m'}\otimes V^{B_1,...,B_n}_{B_1',...,B_n'},
\end{equation}
which implies that $\arho_{AB}=D_{A:B}[\arho_{AB}]$.  Again, the capital letters imply that there can be sums over external fusion trees.
Diagrammatically, $\arho_{AB}$ can be written such that no nontrivial charge lines connect the anyons of subsystem $A$ with those of subsystem $B$~\cite{Bonderson07c}.

Alternatively, we can investigate the entanglement using the anyonic analogue to Eq.~(\ref{eq:EE}).  For a state $\arho_{AB}$ in region $A\cup B$, the \emph{anyonic entanglement entropy} (AEE) of $A$ with $B$ is
\begin{equation}
\label{eq:AEE}
    \aS(\arho_A) \equiv - \aTr(\arho_A \log \arho_A),
\end{equation}
where $\arho_A =\aTr_B(\arho_{AB})$ is the reduced density matrix of subregion $A$.

The AEE captures all correlations between the two subsystems, while the entropy of anyonic charge entanglement extracts the correlations due to nontrivial dimension of the charge line connecting the two subsystems.  This distinction becomes more apparent when comparing the following three states:
\begin{align}
\arho_1 &\equiv \sum_a p_a \arho_a \otimes \arho_{\bar{a}}= \sum_a \frac{p_a}{d_a^2}
  \pspicture[shift=-0.7](0.15,-0.1)(1.35,1.5)
  \scriptsize
  \psline[ArrowInside=->](0.6,0.1)(0.6,1.3)\rput(0.3,0.41){$a$}
  \psline[ArrowInside=->](1,0.1)(1.0,1.3)\rput(1.15,0.41){$\bar{a}$}
  \endpspicture \label{eq:rhoArhoA}
\\ \arho_2 &\equiv \sum_{a} \frac{p_a}{d_a} \begin{pspicture}[shift=-0.6](-0.2,0)(1.2,1.75)
        \scriptsize
        \psline[ArrowInside=->](0,0)(0.5,0.5)\rput(0,0.25){$a$}
        \psline[ArrowInside=->](1,0)(0.5,0.5)\rput(1,0.25){$\bar{a}$}
        \psline[ArrowInside=->](0.5,1)(0,1.5)\rput(0,1.25){$a$}
        \psline[ArrowInside=->,](0.5,1)(1,1.5)\rput(1,1.25){$\bar{a}$}
    \end{pspicture} \label{eq:rhoAA}
\\ \arho_3 &\equiv \sum_{a,a'}\sqrt{ \frac{p_a p_{a'}}{d_ad_{a'}}} \begin{pspicture}[shift=-0.6](-0.4,0)(1.2,1.75)
        \scriptsize
        \psline[ArrowInside=->](0,0)(0.5,0.5)\rput(0,0.25){$a'$}
        \psline[ArrowInside=->](1,0)(0.5,0.5)\rput(1.05,0.25){$\bar{a}'$}
        \psline[ArrowInside=->](0.5,1)(0,1.5)\rput(0,1.25){$a$}
        \psline[ArrowInside=->,](0.5,1)(1,1.5)\rput(1,1.25){$\bar{a}$}
    \end{pspicture} \label{eq:rho3}.
  \end{align}
By comparing $\arho_j$ with $\arho_j^2$, one can easily check that $\arho_1$ is a mixed state, $\arho_2$ is a mixed state unless $p_a=\delta_{a,b}$ for a particular charge $b$, and $\arho_3$ is a pure state.
(We note that when $p_a=\delta_{a,b}$ for a particular charge $b$, the states $\arho_1$ and $\arho_2$ can be obtained from each other through the use of an interferometric ``forced measurement" procedure~\cite{Bonderson08b}. With these operational resources, either of these states may be used as entanglement resources for an anyonic analogue of quantum state teleportation~\cite{Bonderson08a,Bonderson08b}.)

The states $\arho_1$, $\arho_2$, and $\arho_3$ have exactly the same reduced density matrix
\begin{equation}
\arho_A = \aTr_A\left( \arho_1\right)= \aTr_A \left(\arho_2\right)=\aTr_A\left(\arho_3\right)= \sum_a \frac{p_a}{d_a} \pspicture[shift=-0.7](0.15,-0.1)(.68,1.5)
  \scriptsize
  \psline[ArrowInside=->](0.6,0.1)(0.6,1.3)\rput[bl]{0}(0.3,0.41){$a$}
  \endpspicture
,
\end{equation}
and, therefore, the same AEE
\begin{equation}
\aS \left( \arho_A \right) = H \left(\{ p_a \}\right) +\sum_a p_a \aS_a
.
\end{equation}
However, the states have distinct entropy of anyonic charge entanglement:
\begin{align}
&\aS_{\text{ace}}\left( \arho_1\right) =0 \label{eq:ace-rho1}
\\ & \aS_{\text{ace}}\left(\arho_2\right)=2\sum_a p_a\aS_a \label{eq:ace-rho2}
\\ &\aS_{\text{ace}}\left(\arho_3\right)=H(\{p_a\})+2\sum_ap_a \aS_a \label{eq:ace-rho3}.
\end{align}
Eq.~(\ref{eq:ace-rho1}) is easily seen from the fact that no charge lines connect $A$ with $\bar{A}$ in $\arho_1$.  Eq.~(\ref{eq:ace-rho2}) differs from Eq.~(\ref{eq:ace-rho3}) because, even though $D_{A:\bar{A}}[\arho_2]=D_{A:\bar{A}}[\arho_3]= \arho_1$, $\aS(\arho_2)\neq \aS(\arho_3)$.

For a slightly more in-depth example of how to calculate $\aS(\arho)$ and $\aS_{\text{ace}}(\arho)$, consider the pure state
\begin{equation}\label{eq:purepsi}
\begin{split}
\ket{\psi} = \sum_{\substack{\vec{a},\vec{e},\vec{\mu},\\
\vec{b},\vec{f},\vec{\nu},c}}& \frac{\psi_{\vec{a},\vec{e},\vec{\mu},\vec{b},\vec{f},\vec{\nu},c}}{\left( d_{\vec{a}}d_{\vec{b}} \right)^{1/4}}
\begin{pspicture}[shift=-1](-.5,-2)(4,.3)
        \scriptsize
        \psline[ArrowInside=->](.3,-.3)(0,0)\rput(0,.2){$a_1$}
        \psline[ArrowInside=->](.3,-.3)(.6,0)\rput(.6,.2){$a_2$}
        \psline(.4,-.4)(.3,-.3)
        \psline[linestyle=dotted](.7,-.7)(.4,-.4)
        \rput(.1,-.45){$\mu_2$}\rput(.6,-.4){$e_2$}
        \psline(.8,-.8)(.7,-.7)
        \psline(.8,-.8)(1.6,0)\psline[ArrowInside=->](1.3,-.3)(1.6,0)\rput(1.6,.2){$a_n$}
        \rput(.6,-.85){$\mu_n$}
        \psline(.8,-.8)(1.9,-1.9)
        \psline[ArrowInside=->](3.5,-.3)(3.8,0)\rput(3.8,.2){$b_1$}
        \psline[ArrowInside=->](3.5,-.3)(3.2,0)\rput(3.2,.2){$b_2$}
        \psline(3.5,-.3)(3.4,-.4)
        \rput(3.7,-.45){$\nu_2$}\rput(3.15,-.35){$f_2$}
        \psline[linestyle=dotted](3.4,-.4)(3.1,-.7)
        \psline(3,-.8)(2.2,0)\psline[ArrowInside=->](2.5,-.3)(2.2,0)\rput(2.2,.2){$b_n$}
        \rput(3.3,-.85){$\nu_n$}
        \psline(3.1,-.7)(1.9,-1.9)
        \psline[ArrowInside=->](1.8,-1.8)(1.5,-1.5)\rput(1.7,-1.5){$c$}
        \psline[ArrowInside=->](2,-1.8)(2.3,-1.5)\rput(2.1,-1.5){$\bar{c}$}
    \end{pspicture}.
    \end{split}
\end{equation}
For brevity, we write the product of quantum dimension factors as $d_{\vec{a}}=d_{a_1}d_{a_2}\dots d_{a_n}$ and the index $\vec{a}$ to mean $a_1, a_2, \dots, a_n$; and use similar abbreviations $\vec{b},\vec{e},\vec{f},\vec{\mu},$ and $\vec{\nu}$.  We calculate the entropy of anyonic charge entanglement between the left charges $a_i$ and the right charges $b_i$.

The decohered state $D_{A:B}\Big[\ket{\psi}\bra{\psi}\Big]$ is
\begin{equation}
\begin{split}
 D_{A:B} \Big[\ket{\psi}\bra{\psi}\Big]
 &=\sum_{\substack{\vec{a},\vec{e},\vec{\mu}, \\
\vec{b},\vec{f},\vec{\nu},c\\
\vec{a}',\vec{e}',\vec{\mu}',\\
\vec{b}',\vec{f}',\vec{\nu}',c'}} \frac{\psi_{\vec{a},\vec{e},\vec{\mu},\vec{b},\vec{f},\vec{\nu},c} \psi^*_{\vec{a}',\vec{e}',\vec{\mu}',\vec{b}',\vec{f}',\vec{\nu}',c'}}{\left( d_{\vec{a}}d_{\vec{b}}d_{\vec{a}'}d_{\vec{b}'} \right)^{1/4} }
 \psscalebox{.8}{
 \begin{pspicture}[shift=-2.25](0,-5)(4,.3)
        \scriptsize
        \psellipse[linecolor=black,border=0](1.9,-2.2)(0.2,.6)\rput(1.55,-2.35){$\omega_0$}
        \psline{>-}(1.7,-2.25)(1.7,-2.15)
        \psline[ArrowInside=->](.3,-.3)(0,0)\rput(0,.2){$a_1$}
        \psline[ArrowInside=->](.3,-.3)(.6,0)\rput(.6,.2){$a_2$}
        \psline(.4,-.4)(.3,-.3)
        \psline[linestyle=dotted](.7,-.7)(.4,-.4)
        \rput(.1,-.45){$\mu_2$}\rput(.6,-.4){$e_2$}
        \psline(.8,-.8)(.7,-.7)
        \psline(.8,-.8)(1.6,0)\psline[ArrowInside=->](1.3,-.3)(1.6,0)\rput(1.6,.2){$a_n$}
        \rput(.6,-.85){$\mu_n$}
        \psline[ArrowInside=->](3.5,-.3)(3.8,0)\rput(3.8,.2){$b_1$}
        \psline[ArrowInside=->](3.5,-.3)(3.2,0)\rput(3.2,.2){$b_2$}
        \psline(3.5,-.3)(3.4,-.4)
        \rput(3.7,-.45){$\nu_2$}\rput(3.15,-.35){$f_2$}
        \psline[linestyle=dotted](3.4,-.4)(3.1,-.7)
        \rput(3.3,-.85){$\nu_n$}
        \psline[border=2pt](3.1,-.7)(1.9,-1.9)
        \psline(.8,-.8)(1.73,-1.73)
        \psline(1.83,-1.83)(1.9,-1.9)
        \psline(3,-.8)(2.2,0)\psline[ArrowInside=->](2.5,-.3)(2.2,0)\rput(2.2,.2){$b_n$}
        \psline[ArrowInside=->](1.7,-1.7)(1.4,-1.4)\rput(1.6,-1.4){$c$}
        \psline[ArrowInside=->](2.1,-1.7)(2.4,-1.4)\rput(2.2,-1.4){$\bar{c}$}
        \psline[border=2pt](3.1,-3.7)(1.9,-2.5)
        \psline[ArrowInside=->](0,-4.4)(.3,-4.1)\rput(0,-4.6){$a_1'$}
        \psline[ArrowInside=->](.6,-4.4)(.3,-4.1)\rput(.6,-4.6){$a_2'$}
        \psline(.4,-4)(.3,-4.1)
        \psline[linestyle=dotted](.7,-3.7)(.4,-4)
        \rput(.0,-4){$\mu_2'$}\rput(.6,-4){$e_2'$}
        \psline(.8,-3.6)(.7,-3.7)
        \psline(.8,-3.6)(1.6,-4.4)\psline[ArrowInside=->](1.6,-4.4)(1.3,-4.1)\rput(1.6,-4.6){$a_n'$}
        \rput(.55,-3.5){$\mu_n'$}
        \psline(.8,-3.6)(1.73,-2.67)
        \psline(1.83,-2.57)(1.9,-2.5)
        \psline[ArrowInside=->](3.8,-4.4)(3.5,-4.1)\rput(3.8,-4.6){$b_1'$}
        \psline[ArrowInside=->](3.2,-4.4)(3.5,-4.1)\rput(3.2,-4.6){$b_2'$}
        \psline(3.5,-4.1)(3.4,-4)
        \rput(3.7,-4){$\nu_2'$}\rput(3.15,-4.05){$f_2'$}
        \psline[linestyle=dotted](3.4,-4)(3.1,-3.7)
        \psline(3,-3.6)(2.2,-4.4)\psline[ArrowInside=->](2.2,-4.4)(2.5,-4.1)\rput(2.2,-4.6){$b_n'$}
        \rput(3.3,-3.6){$\nu_n'$}
        \psline[ArrowInside=->](1.4,-3)(1.7,-2.7)\rput(1.6,-3){$c'$}
        \psline[ArrowInside=->](2.4,-3)(2.1,-2.7)\rput(2.2,-3){$\bar{c}'$}
    \end{pspicture}}
\\ =\sum_{\substack{\vec{a},\vec{e},\vec{\mu}, \\
\vec{b},\vec{f},\vec{\nu},c\\
\vec{a}',\vec{e}',\vec{\mu}',\\
\vec{b}',\vec{f}',\vec{\nu}'}}&
\frac{\psi_{\vec{a},\vec{e},\vec{\mu},\vec{b},\vec{f},\vec{\nu},c} \psi^*_{\vec{a}',\vec{e}',\vec{\mu}',\vec{b}',\vec{f}',\vec{\nu}',c}}{\left( d_{\vec{a}}d_{\vec{b}}d_{\vec{a}'}d_{\vec{b}'} \right)^{1/4} } \frac{1}{d_c}
\psscalebox{.9}{
\begin{pspicture}[shift=-2.25](-.1,-3.5)(4,.3)
\scriptsize
        \psline[ArrowInside=->](.3,-.3)(0,0)\rput(0,.2){$a_1$}
        \psline[ArrowInside=->](.3,-.3)(.6,0)\rput(.6,.2){$a_2$}
        \psline(.4,-.4)(.3,-.3)
        \psline[linestyle=dotted](.7,-.7)(.4,-.4)
        \rput(.1,-.45){$\mu_2$}\rput(.6,-.4){$e_2$}
        \psline(.8,-.8)(.7,-.7)
        \psline(.8,-.8)(1.6,0)\psline[ArrowInside=->](1.3,-.3)(1.6,0)\rput(1.6,.2){$a_n$}
        \rput(.5,-.85){$\mu_n$}
        \psline(.8,-.8)(.8,-2.1)\rput(.6,-1.45){$c$}
        \rput(.3,0){
        \psline[ArrowInside=->](3.5,-.3)(3.8,0)\rput(3.8,.2){$b_1$}
        \psline[ArrowInside=->](3.5,-.3)(3.2,0)\rput(3.2,.2){$b_2$}
        \psline(3.5,-.3)(3.4,-.4)
        \rput(3.7,-.45){$\nu_2$}\rput(3.15,-.35){$f_2$}
        \psline[linestyle=dotted](3.4,-.4)(3.1,-.7)
        \rput(3.3,-.85){$\nu_n$}
        \psline(3,-.8)(2.2,0)\psline[ArrowInside=->](2.5,-.3)(2.2,0)\rput(2.2,.2){$b_n$}
        \psline(3,-.8)(3.1,-.7)
        \psline(3,-.8)(3,-2.1)\rput(3.2,-1.45){$\bar{c}$}}
        \rput(0,1.5){\psline[ArrowInside=->](0,-4.4)(.3,-4.1)\rput(0,-4.6){$a_1'$}
        \psline[ArrowInside=->](.6,-4.4)(.3,-4.1)\rput(.6,-4.6){$a_2'$}
        \psline(.4,-4)(.3,-4.1)
        \psline[linestyle=dotted](.7,-3.7)(.4,-4)
        \rput(.0,-4){$\mu_2'$}\rput(.6,-4){$e_2'$}
        \psline(.8,-3.6)(.7,-3.7)
        \psline(.8,-3.6)(1.6,-4.4)\psline[ArrowInside=->](1.6,-4.4)(1.3,-4.1)\rput(1.6,-4.6){$a_n'$}
        \rput(.55,-3.4){$\mu_n'$}
        \rput(.3,0){
        \psline[ArrowInside=->](3.8,-4.4)(3.5,-4.1)\rput(3.8,-4.6){$b_1'$}
        \psline[ArrowInside=->](3.2,-4.4)(3.5,-4.1)\rput(3.2,-4.6){$b_2'$}
        \psline(3.5,-4.1)(3.4,-4)
        \rput(3.7,-4){$\nu_2'$}\rput(3.05,-4.05){$f_2'$}
        \psline[linestyle=dotted](3.4,-4)(3.1,-3.7)
        \psline(3,-3.6)(2.2,-4.4)\psline[ArrowInside=->](2.2,-4.4)(2.5,-4.1)\rput(2.2,-4.6){$b_n'$}
        \psline(3.1,-3.7)(3,-3.6)
        \rput(3.3,-3.5){$\nu_n'$}}}
    \end{pspicture}}.
\end{split}
\end{equation}
The second equality follows from
\begin{equation}
\begin{split}
 & \begin{pspicture}[shift=-1.05](1.1,-3.3)(2.5,-1.2)
        \scriptsize
        \psellipse[linecolor=black,border=0](1.9,-2.2)(0.2,.6) \rput(1.45,-2.35){$\omega_0$}
        \psline{>-}(1.71,-2.25)(1.71,-2.15)
        \psline[border=2pt](2.6,-1.2)(1.9,-1.9)
        \psline(1.2,-1.2)(1.73,-1.73)
        \psline(1.83,-1.83)(1.9,-1.9)
        \psline[ArrowInside=->](1.7,-1.7)(1.4,-1.4)\rput(1.6,-1.4){$c$}
        \psline[ArrowInside=->](2.1,-1.7)(2.4,-1.4)\rput(2.2,-1.4){$\bar{c}$}
        \psline[border=2pt](2.6,-3.2)(1.9,-2.5)
        \psline(1.2,-3.2)(1.73,-2.67)
        \psline(1.83,-2.57)(1.9,-2.5)
        \psline[ArrowInside=->](1.4,-3)(1.7,-2.7)\rput(1.6,-3){$c'$}
        \psline[ArrowInside=->](2.4,-3)(2.1,-2.7)\rput(2.15,-3){$\bar{c}'$}
    \end{pspicture}
 = \sum_{e} \Big[ \left(F^{c\bar{c}}_{c'\bar{c}'}\right)^{-1}\Big]_{0e}
 \begin{pspicture}[shift=-1.05](1.1,-3.3)(2.5,-1.2)
        \scriptsize
        \psellipse[linecolor=black,border=0](1.8,-2.2)(0.2,.6) \rput(1.8,-1.5){$\omega_0$}
        \psline{>-}(1.61,-2.25)(1.61,-2.15)
        \psline[border=1.5pt](1.7,-2.1)(2.3,-2.4)
        \psline[ArrowInside=->](2.3,-2.4)(1.9,-2.2)
        \psline(1.55,-2.025)(1.3,-1.9)
        \psline[ArrowInside=->](1.3,-1.8)(1.3,-1.5)
        \psline[ArrowInside=->](1.3,-3)(1.3,-2.5)
        \psline(1.3,-3)(1.3,-1.5)
        \psline[ArrowInside=->](2.3,-1.8)(2.3,-1.5)
        \psline[ArrowInside=->](2.3,-3)(2.3,-2.5)
        \psline(2.3,-3)(2.3,-1.5)
        \rput(2.3,-1.3){$\bar{c}$}
        \rput(1.3,-1.3){$c$}
        \rput(2.3,-3.2){$\bar{c}'$}
        \rput(1.3,-3.2){$c'$}
        \rput(2.15,-2.125){$e$}
 \end{pspicture}
 = \Big[ \left(F^{c\bar{c}}_{c\bar{c}}\right)^{-1}\Big]_{00}\delta_{c,c'}
 \begin{pspicture}[shift=-1.05](1.1,-3.3)(2.1,-1.2)
        \scriptsize
        \psline[ArrowInside=->](1.3,-3)(1.3,-1.5)
        \psline[ArrowInside=->](1.8,-3)(1.8,-1.5)
        \rput(1.95,-2.2){$\bar{c}$}
        \rput(1.45,-2.22){$c$}
 \end{pspicture}
  = \frac{\delta_{c,c'}}{d_c}
 \begin{pspicture}[shift=-1.05](1.1,-3.3)(2.1,-1.2)
        \scriptsize
        \psline[ArrowInside=->](1.3,-3)(1.3,-1.5)
        \psline[ArrowInside=->](1.8,-3)(1.8,-1.5)
        \rput(1.95,-2.2){$\bar{c}$}
        \rput(1.45,-2.22){$c$}
 \end{pspicture}
.
\end{split}
\end{equation}

The entropy of anyonic charge entanglement is
\begin{equation}
\begin{split}
\label{eq:Dpsi}
\aS_{\text{ace}}\left( \ket{\psi}\bra{\psi} \right) &=\aS\left(\Big[ D_{A:B}\ket{\psi}\bra{\psi} \Big]\right) -\aS \left( \ket{\psi}\bra{\psi} \right)
= - \sum_c p_c \log \left( \frac{p_c}{d_c^2} \right) -0
\\ &=H \left(\{ p_c \}\right) +2\sum_c p_c \aS_c,
\end{split}
\end{equation}
where we have defined the probability of the anyons in subsystem $A$ fusing to $c$ (or the anyons in subsystem $B$ fusing to $\bar{c}$) to be
\begin{equation}\label{eq:pc-psi}
p_c=\sum_{\substack{\vec{a},\vec{e},\vec{\mu},\\ \vec{b},\vec{f},\vec{\nu}}}\psi_{\vec{a},\vec{e},\vec{\mu},\vec{b},\vec{f},\vec{\nu},c}\psi^*_{\vec{a},\vec{e},\vec{\mu},\vec{b},\vec{f},\vec{\nu},c}.
\end{equation}
We emphasize that the $\aS_{\text{ace}}$ has isolated entropic quantities that are solely associated with the anyonic charge lines connecting the subsystems $A$ and $B$:
the details of the state within the two subsystems are unimportant, as only the probability of the overall topological charge of each subsystem contributes to $\aS_{\text{ace}}$. Notice that Eqs.~(\ref{eq:Dpsi}) and (\ref{eq:ace-rho3}) are identical. The first term in Eq.~(\ref{eq:Dpsi}) is  the  classical Shannon entropy of the probability distribution $\{p_c\}$ associated with the charge $c$ lines connecting the subsystems $A$ and $B$. The second term, which is nonzero only if at least one of the charge lines connecting subsystems $A$ and $B$ is non-Abelian, is the anyonic entropy associated with the charge $c$ lines themselves.

We can check (e.g. using the method of Lagrange multipliers) that Eq.~(\ref{eq:Dpsi}) is maximized by $p_a=d_a^2/\mathcal{D}^2$, and the corresponding maximum value is
\begin{equation}
\label{eq:max_ace}
\max_{\ket{\psi}} \left[ \aS_{\text{ace}}\left( \ket{\psi}\bra{\psi} \right) \right] =  2\log \mathcal{D}
.
\end{equation}
In fact, this is the maximum value of $\aS_{\text{ace}}$ for a general (possibly mixed) state whose overall topological charge is trivial.
We return to this point in the next section when discussing anyon pair-production.

We now calculate the AEE for the pure state given in Eq.~(\ref{eq:purepsi}).  Tracing over the $b$ charges gives the reduced density matrix for the $a$ charges
\begin{equation}
\arho_A = \sum_{\substack{\vec{a},\vec{e},\vec{\mu}, \\ \vec{a}',\vec{e}',\vec{\mu}',\\\vec{b},\vec{f},\vec{\nu},c}} \frac{\psi_{\vec{a},\vec{e},\vec{\mu},\vec{b},\vec{f},\vec{\nu},c}\psi^*_{\vec{a}',\vec{e}',\vec{\mu}',\vec{b},\vec{f},\vec{\nu},c}}{\left( d_{\vec{a}} d_{\vec{a}'} \right)^{1/4} \sqrt{d_c}}
\begin{pspicture}[shift=-2](-.5,-3.5)(2,.3)
\scriptsize
        \psline[ArrowInside=->](.3,-.3)(0,0)\rput(0,.2){$a_1$}
        \psline[ArrowInside=->](.3,-.3)(.6,0)\rput(.6,.2){$a_2$}
        \psline(.4,-.4)(.3,-.3)
        \psline[linestyle=dotted](.7,-.7)(.4,-.4)
        \rput(.1,-.45){$\mu_2$}\rput(.6,-.4){$e_2$}
        \psline(.8,-.8)(.7,-.7)
        \psline(.8,-.8)(1.6,0)\psline[ArrowInside=->](1.3,-.3)(1.6,0)\rput(1.6,.2){$a_n$}
        \rput(.5,-.85){$\mu_n$}
        \psline(.8,-.8)(.8,-2.1)\rput(.6,-1.45){$c$}
        \rput(0,1.5){\psline[ArrowInside=->](.3,-4.1)(0,-4.4)\rput(0,-4.6){$a_1'$}
        \psline[ArrowInside=->](.3,-4.1)(.6,-4.4)\rput(.6,-4.6){$a_2'$}
        \psline(.4,-4)(.3,-4.1)
        \psline[linestyle=dotted](.7,-3.7)(.4,-4)
        \rput(.0,-4){$\mu_2'$}\rput(.6,-4){$e_2'$}
        \psline(.8,-3.6)(.7,-3.7)
        \psline(.8,-3.6)(1.6,-4.4)\psline[ArrowInside=->](1.3,-4.1)(1.6,-4.4)\rput(1.6,-4.6){$a_n'$}
        \rput(.55,-3.4){$\mu_n'$}
        }
    \end{pspicture}.
\end{equation}
We can define a matrix $M_c$ whose components are given by
\begin{equation}
[M_c]_{(\vec{a},\vec{e},\vec{\mu}),(\vec{a}',\vec{e}',\vec{\mu}')} = \sum_{\vec{b},\vec{f},\vec{\nu}} \psi_{\vec{a},\vec{e},\vec{\mu},\vec{b},\vec{f},\vec{\nu},c}\psi^*_{\vec{a}',\vec{e}',\vec{\mu}',\vec{b},\vec{f},\vec{\nu},c}.
\end{equation}
Then, one can easily check that
\begin{equation}
\arho_A^\alpha = \sum_{\substack{\vec{a},\vec{e},\vec{\mu}, \\ \vec{a}',\vec{e}',\vec{\mu}',\\c}}  \frac{[M_c^\alpha]_{(\vec{a},\vec{e},\vec{\mu}),(\vec{a}',\vec{e}',\vec{\mu}')}}{ d_c^{\alpha-1}\left( d_{\vec{a}} d_{\vec{a}'} \right)^{1/4} \sqrt{d_c}}
\begin{pspicture}[shift=-2](-.5,-3.5)(2,.3)
\scriptsize
        \psline[ArrowInside=->](.3,-.3)(0,0)\rput(0,.2){$a_1$}
        \psline[ArrowInside=->](.3,-.3)(.6,0)\rput(.6,.2){$a_2$}
        \psline(.4,-.4)(.3,-.3)
        \psline[linestyle=dotted](.7,-.7)(.4,-.4)
        \rput(.1,-.45){$\mu_2$}\rput(.6,-.4){$e_2$}
        \psline(.8,-.8)(.7,-.7)
        \psline(.8,-.8)(1.6,0)\psline[ArrowInside=->](1.3,-.3)(1.6,0)\rput(1.6,.2){$a_n$}
        \rput(.5,-.85){$\mu_n$}
        \psline(.8,-.8)(.8,-2.1)\rput(.6,-1.45){$c$}
        \rput(0,1.5){\psline[ArrowInside=->](.3,-4.1)(0,-4.4)\rput(0,-4.6){$a_1'$}
        \psline[ArrowInside=->](.3,-4.1)(.6,-4.4)\rput(.6,-4.6){$a_2'$}
        \psline(.4,-4)(.3,-4.1)
        \psline[linestyle=dotted](.7,-3.7)(.4,-4)
        \rput(.0,-4){$\mu_2'$}\rput(.6,-4){$e_2'$}
        \psline(.8,-3.6)(.7,-3.7)
        \psline(.8,-3.6)(1.6,-4.4)\psline[ArrowInside=->](1.3,-4.1)(1.6,-4.4)\rput(1.6,-4.6){$a_n'$}
        \rput(.55,-3.4){$\mu_n'$}
        }
    \end{pspicture},
\end{equation}
from which it follows that
\begin{equation}
\aTr[ \arho_A^\alpha] = \sum_{c} \frac{\text{Tr}[M_c^\alpha]}{d_c^\alpha} d_c = \sum_{c,j} \left( \frac{\lambda_c^{(j)}}{d_c}\right)^\alpha d_c.
\end{equation}
In the last equality, we have defined $\lambda_c^{(j)}$ to be the $j$th eigenvalue of $M_c$.  Therefore, the AEE is
\begin{equation}\label{eq:AEEpure}
\aS[\arho_A] = -\sum_{c,j} \lambda_c^{(j)}\log\left( \frac{\lambda_c^{(j)}}{d_c}\right) = \sum_c H(\{\lambda_c^{(j)}\}) +\sum_{c} p_c \aS_c,
\end{equation}
where in the last equality we have noted that $\sum_j\lambda_c^{(j)}=p_c$ from Eq.~(\ref{eq:pc-psi}).
The above result could have equivalently been achieved by first performing a Schmidt decomposition on the state $\ket{\psi}$.

Several previous works have investigated anyonic entanglement through the entanglement entropy.  Ref.~\cite{Hikami08} used a skein theory approach to evaluate the bipartite entanglement entropy of a pure state in the context of SU$(2)_k$ Chern-Simons theory.  Ref.~\cite{Kato14} defined an operational entanglement measure, based on Eq.~(\ref{eq:EE}), for bipartite anyonic pure states with vacuum total charge.  More generally, Ref.~\cite{Pfeifer14} used anyon models to evaluate the AEE on surfaces of arbitrary genus, constructing the reduced density matrix from a given partitioning of a surface.  We give an alternative construction in Section~\ref{sec:DSg}.  All three of the above-mentioned works identify the second term of Eq.~(\ref{eq:AEEpure}) as the TEE for an anyonic system. In this paper, we reserve the term TEE for $S_{\text{topo}}=-\log \mathcal{D}$ of Refs.~\cite{Kitaev06b,Levin06a}, which cannot be derived used the methods of Refs.~\cite{Hikami08,Kato14,Pfeifer14}. In the next section, we explain how $S_{\text{topo}}$ may be wheedled out of the anyonic state description.

\subsection{Topological Entanglement Entropy in Anyon Models I}\label{sec:ATEEI}

The extraction of the TEE in the context of anyonic states is subtle.  Consider a sphere partitioned into two disks: region $A$ and its complement, region $\bar{A}$.
In order to obtain the (microscopic) density matrix for the subsystem $A$, we trace out the subsystem $\bar{A}$. Topologically, we view this as first cutting the system along the partition boundary $\partial A = \partial \bar{A}$ to yield two disjoint compact systems (disks) $A$ and $\bar{A}$, for which $\partial A \neq \partial \bar{A}$, and then tracing out $\bar{A}$. When the system is cut into disjoint compact subsystems, each resulting connected genus zero surface must individually have trivial total topological charge. Thus, if the interior of a resulting disk contains topological charge $c$, e.g. from a collection of quasiparticles in that region, then its boundary must carry a total topological charge of $\bar{c}$.

In the case of the ground state on the sphere, there are no topological excitations in the system, so $\text{int}(A)$ and $\text{int}(\bar{A})$ have trivial topological charge ($c=\bar{c}=0$). Before cutting the surface, the anyonic state representing this configuration is the trivial (vacuum) state, i.e. the empty diagram. If we use the trivial anyonic state $|0\rangle$, the corresponding AEE obviously vanishes, so one might na\"ively expect the TEE between regions $A$ and $\bar{A}$ to also vanish. This deduction is clearly invalid~\cite{Kitaev06b,Levin06a}.

The resolution to this apparent discrepancy is that a spatial cut of the system is an operation that is both topological and microscopic. That is, a cut has effects on length scales that are large compared to the topological correlation length $\xi$ and length scales that are small compared to the regularization length $\ell$, i.e. the lattice spacing or magnetic length (roughly the correlation length). In particular, degrees of freedom along either side of the spatial partition boundary effectively change from being adjacent to being infinitely separated as a result of a spatial cut (i.e. from strongly-interacting to non-interacting). This process evinces anyonic correlations across the partition boundary that could not be resolved within the uncut system, because they exist below the regularization length, which is why they were not captured by the anyonic state describing the system before cutting. That is, one can think of cutting as locally creating many anyons along the newly created boundaries, but since the total topological charge of each boundary is trivial, there is a projection of the total charge of these anyons along each boundary. In this section, we provide a heuristic description of these subtle anyonic correlations that exist across a spatial partition and explain how the topological charge projection imposed on the partition boundaries by the cutting operation generates the decrease in entropy (increase in order) characterized by the TEE. We will return to a more rigorous derivation of these in Section~\ref{sec:ATEE}.

Since we are interested in the correlations across the partition boundary, let us begin by focusing on the local correlations across the boundary between degrees of freedom in a small disk-like region $\mathcal{B}_1$ straddling the partition boundary between $A$ and $\bar{A}$, whose linear size is on the order of the regularization length $\ell$. (As a notational note, we will denote regions that do not strictly belong to $A$, or which result from a discretization of $A$, with calligraphic letters.)  We choose $\mathcal{B}_1$ in this way to represent a short segment of the partition boundary. However, if we cut the system along the partition boundary, then we must similarly partition the region $\mathcal{B}_1$ along the same partition boundary. For this, we define $\partial \mathcal{A}_1 = A \cap \mathcal{B}_1$ and $\partial \bar{\mathcal{A}}_1 = \bar{A} \cap \mathcal{B}_1$, and wish to consider the correlations between degrees of freedom in regions $\partial \mathcal{A}_1$ and $\partial \bar{\mathcal{A}}_1$. In general, there will be non-universal contributions to the entanglement entropy from the microscopic details of the local correlations. We are, however, interested in extracting the universal contributions to the entanglement entropy, so we focus on the anyonic correlations captured by the anyonic state formalism.

Since we are now considering a region $\mathcal{B}_1$ whose size is smaller than the resolution length scale, we can heuristically think of the region as being microscopically populated with pair-created anyons; the separation of these anyons is too small to resolve their individual existence, and since they are pair-created from vacuum, the total topological charge within region $\mathcal{B}_1$ is trivial, as it should be. In this picture, the region $\partial \mathcal{A}_1$ will contain topological charge $a_1$ and $\partial \bar{\mathcal{A}}_1$ will necessarily contain the (pair-created partner) topological charge $\bar{a}_1$, with some probability $p_{a_1}$. When the entire uncut system is in the ground state, we expect that pair-produced anyons of region $\mathcal{B}_1$ will be unentangled with regions that are disjoint from $\mathcal{B}_1$, so the anyonic correlations between regions $\partial \mathcal{A}_1$ and $\partial \bar{\mathcal{A}}_1$ can be represented by a two-anyon pure state. Moreover, we expect the anyonic state representing the local anyonic correlations at the regularization scale to have maximal anyonic charge line entanglement between the two subsystems.  Therefore, the density matrix describing quasiparticle pair production is the pure state of Eq.~(\ref{eq:rho3}) with $p_{a_1}=d_{a_1}^2/\mathcal{D}^2$ [see discussion around Eq.~(\ref{eq:max_ace})]:
\begin{equation}
\arho_{\mathcal{B}_1} = \sum_{a_1,a'_1} \frac{d_{a_1} d_{a_1}'}{\mathcal{D}^2} \frac{1}{\sqrt{d_{a_1} d_{a_1}'}}
  \pspicture[shift=-1](-.7,-1.3)(.7,.7)
  \scriptsize
  \psline[ArrowInside=->](0,0)(-.5,.5)\rput(-.5,.6){$a_1$}
  \psline[ArrowInside=->](0,0)(.5,.5)\rput(.5,.6){$\bar{a}_1$}
  \psline[ArrowInside=->](-.5,-1)(0,-.5)\rput(-.5,-1.2){$a_1'$}
  \psline[ArrowInside=->](.5,-1)(0,-.5)\rput(.5,-1.2){$\bar{a}_1'$}.
  \endpspicture.
\end{equation}
If we trace out the anyon in region $\partial \bar{\mathcal{A}}_1$, the density matrix for region $\partial \mathcal{A}_1$ is given by
\begin{equation}
 \arho_{\partial \mathcal{A}_1} = \aTr_{\partial \bar{\mathcal{A}}_1} [ \arho_{\mathcal{B}_1} ]=\sum_{a_1} \frac{d_{a_1}}{\mathcal{D}^2}
  \pspicture[shift=-.65](-.2,-.2)(.2,1.4)
  \scriptsize
  \psline[ArrowInside=->](0,0)(0,1)\rput(0,1.2){$a_1$}
  \endpspicture.
\end{equation}

\begin{figure}
\begin{centering}
\includegraphics[width=0.75\linewidth]{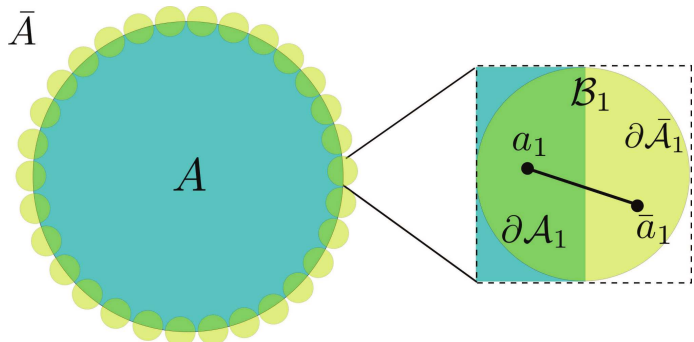}
\caption{The boundary between subsystems $A$ (blue) and $\bar{A}$ (white) is covered by a disjoint set of small disk-like regions $\mathcal{B}_j$ (yellow), each of which is partitioned into subregions $\partial \mathcal{A}_j$ and $\partial \bar{\mathcal{A}}_j$, which are contained in regions $A$ and $\bar{A}$, respectively. Cutting along the partition evinces local anyonic correlations between regions $\partial \mathcal{A}_j$ and $\partial \bar{\mathcal{A}}_j$ that can heuristically be thought of in terms of pair-created anyons in a maximally anyonic charge entangled state, which could not be resolved as separate anyons for the uncut system in the ground state. Cutting the system into disconnected, compact regions $A$ and $\bar{A}$ imposes a topological constraint that the total topological charge of regions $A$ is trivial (after the cut). This yields a topological correlation of the boundary anyons of region $A$, which is the origin of the TEE $S_{\text{topo}} \equiv  - \log \mathcal{D}$.
}
\label{fig:heuristic}
\end{centering}
\end{figure}

We now envision covering the partition boundary $\partial A$ with similar small disk-like regions $\mathcal{B}_1,\ldots,\mathcal{B}_n$ that are all disjoint from each other, as shown in Figure~\ref{fig:heuristic}. These divide the boundary $\partial A$ into $n$ segments $\partial \mathcal{A}_j = \partial A \cap \mathcal{B}_j$ associated with the local boundary regions. In this way, the boundary length is roughly $L \sim n \ell$. The same description of $\mathcal{B}_1$ above applies to each region $\mathcal{B}_j$. Thus, if we start with the ground state of the uncut system and trace out $\bar{A}$, we expect the state of subsystem $A$ after cutting to have an anyon corresponding to each segment of the discretized boundary, which is similarly described by the reduced density matrix
\begin{equation}
 \arho_{\partial \mathcal{A}_j} = \sum_{a_j} \frac{d_{a_j}}{\mathcal{D}^2}
  \pspicture[shift=-.65](-.2,-.2)(.2,1.4)
  \scriptsize
  \psline[ArrowInside=->](0,0)(0,1)\rput(0,1.2){$a_j$}
  \endpspicture
.
\end{equation}
However, the anyonic reduced density matrix for subsystem $A$ is not simply given by the tensor product
\begin{equation}
\label{eq:rho_partialA}
\arho_{\partial \mathcal{A}} \equiv \arho_{\partial \mathcal{A}_1} \otimes \dots \otimes \arho_{\partial \mathcal{A}_n}
\end{equation}
of those of the local boundary regions $\partial \mathcal{A}_j$. The compact region $A$ must have trivial total topological charge, so it is necessary to apply a projection of the overall topological charge onto the trivial charge. Denoting the anyonic reduced density matrix that takes into account the localized boundary charges as $\rhoH$, we have
\begin{equation}
\label{eq:rhoheuristic}
\begin{split}
\rhoH &\equiv \frac{\Pi_0 \arho_{\partial \mathcal{A}} \Pi_0}{\aTr\Big[\Pi_0 \arho_{\partial \mathcal{A}} \Pi_0 \Big]}
=\frac{\Pi_0 \left(  \arho_{\partial \mathcal{A}_1} \otimes \dots \otimes \arho_{\partial \mathcal{A}_n}\right)\Pi_0}{\aTr\Big[\Pi_0 \left(\arho_{\partial \mathcal{A}_1} \otimes \dots \otimes \arho_{\partial \mathcal{A}_n} \right)\Pi_0 \Big]}
\\ &= \sum_{\vec{a}} \frac{d_{\vec{a}}}{\mathcal{D}^{2n}}
\pspicture[shift=-0.9](-.75,-0.2)(2,1.75)
  \scriptsize
  \psellipse[linecolor=black,border=0](.8,.5)(1,.2)\rput(-.3,.75){$\omega_{0}$}
  \psline{>-}(0.7,.31)(.9,.315)
  \psline[border=2pt](0,0)(0,.3)
  \psline[border=2pt](0,.45)(0,1.5)\rput(0,1.6){$a_1$}
  \psline[border=2pt](0.4,0)(0.4,.26)
  \psline[border=2pt](0.4,.41)(0.4,1.5)\rput(0.4,1.6){$a_2$}
  \psline[border=2pt](0.8,0)(0.8,.22)
  \psline[border=2pt](0.8,.39)(0.8,1.5)
  \psline[border=2pt](1.2,0)(1.2,.24)
  \psline[border=2pt](1.2,.39)(1.2,1.5)\rput(1,1.6){\dots}
  \psline[border=2pt](1.6,0)(1.6,.26)
  \psline[border=2pt](1.6,.41)(1.6,1.5)\rput(1.6,1.6){$a_{n}$}
  \psline[ArrowInside=->](0,1.2)(0,1.5)
  \psline[ArrowInside=->](.4,1.2)(.4,1.5)
  \psline[ArrowInside=->](.8,1.2)(.8,1.5)
  \psline[ArrowInside=->](1.2,1.2)(1.2,1.5)
  \psline[ArrowInside=->](1.6,1.2)(1.6,1.5)
\endpspicture
= \sum_{\vec{a},\vec{e},\vec{\mu}} \frac{\sqrt{d_{\vec{a}}}}{\mathcal{D}^{2n-2} }
\begin{pspicture}[shift=-2.25](0,-4.7)(4,.3)
        \scriptsize
        \psline[ArrowInside=->](.3,-.3)(0,0)\rput(0,.2){$a_1$}
        \psline[ArrowInside=->](.6,-.6)(.3,-.3)\rput(.6,-.35){$e_2$}
        \rput(.1,-.45){$\mu_2$}
        \psline[ArrowInside=->](.3,-.3)(.6,0)\rput(.6,.2){$a_2$}
        \psline[ArrowInside=->](.9,-.9)(.6,-.6)\rput(.9,-.65){$e_3$}
        \rput(.4,-.75){$\mu_3$}
        \psline(.6,-.6)(1.2,0)\psline[ArrowInside=->](.9,-.3)(1.2,0)\rput(1.2,.2){$a_3$}
        \psline[linestyle=dotted](.9,-.9)(1.3,-1.3)
        \psline[ArrowInside=->](1.6,-1.6)(1.3,-1.3)\rput(1.55,-1.2){$e_{n-2}$}
        \rput(1.2,-1.75){$\mu_{n-1}$}
        \psline(1.6,-1.6)(3.2,0)\psline[ArrowInside=->](2.9,-.3)(3.2,0)\rput(3.2,.2){$a_{n-1}$}
        \psline[ArrowInside=->](1.9,-1.9)(1.6,-1.6)\rput(1.9,-1.65){$\bar{a}_n$}
        \psline(1.9,-1.9)(3.8,0)\psline[ArrowInside=->](3.5,-.3)(3.8,0)\rput(3.8,.2){$a_n$}
        \psline[ArrowInside=->](0,-4.4)(.3,-4.1)\rput(0,-4.6){$a_1$}
        \psline[ArrowInside=->](.3,-4.1)(.6,-3.8)\rput(.6,-4.1){$e_2$}
        \rput(.1,-4){$\mu_2$}
        \psline[ArrowInside=->](.6,-4.4)(.3,-4.1)\rput(.6,-4.6){$a_2$}
        \psline[ArrowInside=->](.6,-3.8)(.9,-3.5)\rput(.9,-3.8){$e_3$}
        \rput(.4,-3.7){$\mu_3$}
        \psline(.6,-3.8)(1.2,-4.4)\psline[ArrowInside=->](1.2,-4.4)(.9,-4.1)\rput(1.2,-4.6){$a_3$}
        \psline[linestyle=dotted](.9,-3.5)(1.3,-3.1)
        \psline[ArrowInside=->](1.3,-3.1)(1.6,-2.8)\rput(1.55,-3.3){$e_{n-2}$}
        \rput(1.2,-2.65){$\mu_{n-1}$}
        \psline(3.2,-4.4)(1.6,-2.8)\psline[ArrowInside=->](3.2,-4.4)(2.9,-4.1)\rput(3.2,-4.6){$a_{n-1}$}
        \psline[ArrowInside=->](1.6,-2.8)(1.9,-2.5)\rput(1.9,-2.8){$\bar{a}_n$}
        \psline(1.9,-2.5)(3.8,-4.4)\psline[ArrowInside=->](3.8,-4.4)(3.5,-4.1)\rput(3.8,-4.6){$a_n$}
    \end{pspicture} .
\end{split}
\end{equation}
The last equality of Eq.~(\ref{eq:rhoheuristic}) is obtained by performing a series of $F$-moves to write the state in a tree-like form, so that the $\omega_0$-loop is applied to a single charge line.

It follows that, when taking into account the anyonic correlations along the partition boundary, the anyonic entanglement entropy for the ground state is given by
\begin{equation}
\label{eq:Srho_A}
\aS \left( \rhoH \right)=n \aS \left(\arho_{\partial \mathcal{A}_j}\right) - 2 \log \mathcal{D}
,
\end{equation}
where we have written the anyonic entropy of a single ``boundary anyon'' as
\begin{equation}\label{eq:random-anyon-entropy}
\aS \left( \arho_{\partial \mathcal{A}_j}\right) = -\sum_{a_j} \frac{d_{a_j}^2}{\mathcal{D}^2} \log \left( \frac{d_{a_j}}{\mathcal{D}^2}\right).
\end{equation}
The explicit derivation of $\aS\left(\rhoH\right)$ from $\rhoH$ will be given in Section~\ref{sec:ATEE}.

A few comments are in order:
\begin{enumerate}

\item There is a subtle over-counting in this heuristic description of the anyonic correlations across the boundary that produces twice the actual amount of entanglement entropy between $A$ and $\bar{A}$. After correcting this inadvertent doubling found in Eq.~(\ref{eq:Srho_A}), the contribution to the entanglement entropy between regions $A$ and $\bar{A}$ is given by
    \begin{equation}\label{eq:Sactual}
    \aS_{A} =\frac{1}{2} \aS \left( \rhoH \right) = \frac{n}{2}\aS\left(\arho_{\partial \mathcal{A}_j} \right) - \log \mathcal{D} .
    \end{equation}
We address this point at the end of this section.

\item The first term of Eq.~(\ref{eq:Sactual}) describes a linear dependence of the anyonic entanglement entropy on the length $L$ of the boundary, since $n\sim L / \ell$.  The boundary length-dependent term $\alpha L$ of the entanglement entropy Eq.~(\ref{eq:ententropy}), in general, will have non-universal contributions from the microscopic details of the physical system. The term $\frac{n}{2}\aS\left(\arho_{\partial \mathcal{A}_j}\right)$ reflects a contribution to this from the topological sector of the theory, for which the non-universal aspect is determined by the short-distance regularization of the theory, i.e. giving $\alpha_{\text{topo}}= \frac{1}{2} \aS\left(\arho_{\partial \mathcal{A}_j} \right) \ell^{-1}$.

\item The second term is the universal $O(1)$ topological contributions to the entanglement entropy $S_{\text{topo}} \equiv  - \log \mathcal{D}$, i.e. the term that is independent of the size or shape of the boundary.  The origin of this term is the topological constraint that boundary anyons collectively have total topological charge $0$. This can be understood from considering the difference between the entropy of the boundary anyons before and after application of the topological charge projection, that is
    \begin{equation}
    \begin{split}
    \label{eq:DiffaS}
    \aS \left( \arho_{\partial \mathcal{A}}\right) -\aS \left( \rhoH \right)& = -2 S_{\text{topo}} =2 \log\mathcal{D}.
    \end{split}
    \end{equation}
    Thus, we view $S_{\text{topo}}$ the reduction in the entanglement entropy due to the topological constraint that the total topological charge of the compact subsystem $A$ must be trivial (after cutting the original system), which imposes a correlation of the boundary anyons charges. Notice that Eq.~(\ref{eq:DiffaS}) is the multipartite mutual information between the boundary anyons of regions $\partial \mathcal{A}_1,\dots, \partial \mathcal{A}_n$, which is a measure of the correlation between them, or the amount of information that is shared by them.  This information is only accessible by considering the boundary regions collectively. From this perspective, $\mathcal{D}$ can be thought of as the ``dimension" of the state space associated with a group of random anyons whose collective topological charge is $0$.

\end{enumerate}

When the system is not in the ground state, but has quasiparticle excitations, we can use this argument by including the anyonic state of the quasiparticles. We denote the reduced density matrix describing the quasiparticles in the interior of region $A$ as $\arho_{\text{int}(A)}$. In the case where there is a single quasiparticle of topological charge $c$ in region $A$, we have $\arho_{\text{int}(A)} = \arho_c$. Following the same arguments for this case, the anyonic reduced density matrix (including the localized boundary anyons) for the compact region $A$ after the cut is
\begin{equation}
\label{eq:rho_c_heuristic}
\begin{split}
    \rhoH &\equiv \frac{\Pi_0 \left( \arho_c \otimes \arho_{\partial \mathcal{A}}\right)\Pi_0}{\aTr\Big[\Pi_0 \left(\arho_c \otimes \arho_{\partial \mathcal{A}} \right)\Pi_0 \Big]}
= \sum_{\vec{a}} \frac{d_{\vec{a}}}{\mathcal{D}^{2n}}\frac{1}{d_c}
\pspicture[shift=-0.9](-.75,-0.2)(2.25,1.75)
  \scriptsize
  \psellipse[linecolor=black,border=0](1,.5)(1.4,.2)\rput(-.3,.75){$\omega_{0}$}
  \psline{>-}(0.7,.31)(.9,.31)
  \psline[border=2pt](2.0,0)(2.0,.3)
  \psline[border=2pt](2.0,.45)(2.0,1.5) \rput(2.0,1.6){$c$}
  \psline[border=2pt](0,0)(0,.3)
  \psline[border=2pt](0,.45)(0,1.5)\rput(0,1.6){$a_1$}
  \psline[border=2pt](0.4,0)(0.4,.26)
  \psline[border=2pt](0.4,.41)(0.4,1.5)\rput(0.4,1.6){$a_2$}
  \psline[border=2pt](0.8,0)(0.8,.22)
  \psline[border=2pt](0.8,.39)(0.8,1.5)
  \psline[border=2pt](1.2,0)(1.2,.24)
  \psline[border=2pt](1.2,.39)(1.2,1.5)\rput(1,1.6){\dots}
  \psline[border=2pt](1.6,0)(1.6,.26)
  \psline[border=2pt](1.6,.41)(1.6,1.5)\rput(1.6,1.6){$a_n$}
  \psline[ArrowInside=->](0,1.2)(0,1.5)
  \psline[ArrowInside=->](.4,1.2)(.4,1.5)
  \psline[ArrowInside=->](.8,1.2)(.8,1.5)
  \psline[ArrowInside=->](1.2,1.2)(1.2,1.5)
  \psline[ArrowInside=->](1.6,1.2)(1.6,1.5)
  \psline[ArrowInside=->](2,1.2)(2,1.5)
\endpspicture
\\ &= \sum_{\vec{a},\vec{e},\vec{\mu}} \frac{\sqrt{d_{\vec{a}}}}{\mathcal{D}^{2n-2} d_c^{3/2}}
\begin{pspicture}[shift=-2.25](0,-4.7)(4,.3)
        \scriptsize
        \psline[ArrowInside=->](.3,-.3)(0,0)\rput(0,.2){$a_1$}
        \psline[ArrowInside=->](.6,-.6)(.3,-.3)\rput(.6,-.35){$e_2$}
        \rput(.1,-.45){$\mu_2$}
        \psline[ArrowInside=->](.3,-.3)(.6,0)\rput(.6,.2){$a_2$}
        \psline[ArrowInside=->](.9,-.9)(.6,-.6)\rput(.9,-.65){$e_3$}
        \rput(.4,-.75){$\mu_3$}
        \psline(.6,-.6)(1.2,0)\psline[ArrowInside=->](.9,-.3)(1.2,0)\rput(1.2,.2){$a_3$}
        \psline[linestyle=dotted](.9,-.9)(1.3,-1.3)
        \psline[ArrowInside=->](1.6,-1.6)(1.3,-1.3)\rput(1.55,-1.2){$e_{n-2}$}
        \rput(1.2,-1.75){$\mu_{n-1}$}
        \psline(1.6,-1.6)(3.2,0)\psline[ArrowInside=->](2.9,-.3)(3.2,0)\rput(3.2,.2){$a_n$}
        \psline[ArrowInside=->](1.9,-1.9)(1.6,-1.6)\rput(1.9,-1.65){$\bar{c}$}
        \psline(1.9,-1.9)(3.8,0)\psline[ArrowInside=->](3.5,-.3)(3.8,0)\rput(3.8,.2){$c$}
        \psline[ArrowInside=->](0,-4.4)(.3,-4.1)\rput(0,-4.6){$a_1$}
        \psline[ArrowInside=->](.3,-4.1)(.6,-3.8)\rput(.6,-4.1){$e_2$}
        \rput(.1,-4){$\mu_2$}
        \psline[ArrowInside=->](.6,-4.4)(.3,-4.1)\rput(.6,-4.6){$a_2$}
        \psline[ArrowInside=->](.6,-3.8)(.9,-3.5)\rput(.9,-3.8){$e_3$}
        \rput(.4,-3.7){$\mu_3$}
        \psline(.6,-3.8)(1.2,-4.4)\psline[ArrowInside=->](1.2,-4.4)(.9,-4.1)\rput(1.2,-4.6){$a_3$}
        \psline[linestyle=dotted](.9,-3.5)(1.3,-3.1)
        \psline[ArrowInside=->](1.3,-3.1)(1.6,-2.8)\rput(1.55,-3.3){$e_{n-2}$}
        \rput(1.2,-2.65){$\mu_{n-1}$}
        \psline(3.2,-4.4)(1.6,-2.8)\psline[ArrowInside=->](3.2,-4.4)(2.9,-4.1)\rput(3.2,-4.6){$a_n$}
        \psline[ArrowInside=->](1.6,-2.8)(1.9,-2.5)\rput(1.9,-2.8){$\bar{c}$}
        \psline(1.9,-2.5)(3.8,-4.4)\psline[ArrowInside=->](3.8,-4.4)(3.5,-4.1)\rput(3.8,-4.6){$c$}
    \end{pspicture} .
\end{split}
\end{equation}
The corresponding anyonic entanglement entropy is given by
\begin{equation}
\label{eq:Srhonc}
\aS \left( \rhoH \right)=n \aS \left(\arho_{\partial \mathcal{A}_j}\right)+2S_{\text{topo}}+\aS_c
,
\end{equation}
where $\aS_c=\log d_c$ is the anyonic entropy associated with the topological charge $c$, as in Eq.~(\ref{eq:Sa}). For anyonic states, $\aS_c$ was associated with the system having overall topological charge $c$. Here, $\aS_c$ is associated with the the topological charge $\bar{c}$ on the boundary formed by the partition, which is the same thing as the interior of $A$ having overall topological charge $c$.

In the case of a more general configuration of quasiparticles, it is straightforward to see that the reduced density matrix
\begin{equation}
\label{eq:rhoheuristic_general}
\rhoH  \equiv \frac{\Pi_0 \left( \arho_{\text{int}(A)} \otimes \arho_{\partial \mathcal{A}} \right) \Pi_0}{\aTr\Big[\Pi_0 \left( \arho_{\text{int}(A)} \otimes \arho_{\partial \mathcal{A}} \right) \Pi_0 \Big]}
=\frac{\Pi_0 \left( \arho_{\text{int}(A)} \otimes \arho_{\partial \mathcal{A}_1} \otimes \dots \otimes \arho_{\partial \mathcal{A}_n}\right)\Pi_0}{\aTr\Big[\Pi_0 \left(\arho_{\text{int}(A)}\otimes \arho_{\partial \mathcal{A}_1} \otimes \dots \otimes \arho_{\partial \mathcal{A}_n} \right)\Pi_0 \Big]}
\end{equation}
yields
\begin{equation}
\label{eq:Srho_general}
\aS \left( \rhoH\right)= n \aS \left(\arho_{\partial \mathcal{A}_j}\right)+2S_{\text{topo}}+ \aS \left( \arho_{\text{int}(A)} \right)
,
\end{equation}
where $\aS \left( \arho_{\text{int}(A)} \right)$ is the anyonic entanglement entropy of the quasiparticles contained within region $A$ (before the cut), as defined in Eq.~(\ref{eq:AEEpure}). For the purposes of separating the contributions of the quasiparticles and the partition boundary to the entanglement entropy, it is useful to write this last term as
\begin{equation}
\aS \left( \arho_{\text{int}(A)} \right) =  \sum_{c} p_c \aS_c + \aS \left( \arho_{A} \right)
,
\end{equation}
where $p_c$ is the probability of the anyonic state $\arho_{A}$ being in a configuration with topological charge $c$ on the partition boundary.

This leads us to one additional comment:

\begin{enumerate}
\setcounter{enumi}{3}
\item The contribution to the entanglement entropy coming from the quasiparticle content and total topological charge on the partition boundary for region $A$ is not inadvertently doubled in this heuristic argument, so the total contribution of the anyonic correlations to the entanglement entropy between regions $A$ and $\bar{A}$ is given by
    \begin{equation}\label{eq:Sactual_general}
    \aS_{A} = \frac{n}{2}\aS\left(\arho_{\partial \mathcal{A}_j} \right) + S_{\text{topo}} + \sum_{c} p_c \aS_c + \aS \left( \arho_{A} \right) .
\end{equation}

\end{enumerate}

The fallacious doubling of the boundary contribution to the entanglement entropy discussed above resulted from the improper assumption that the local anyonic correlations across the boundary could be represented by localized anyons at fixed locations along the partition boundary in the manner described above. For example, a system in a chiral topological phase on a surface with boundary (e.g. a disk) will have a chiral, gapless CFT on the edge. Unlike in the (gapped) bulk, anyonic excitations on such an edge cannot be localized at a fixed point in space. While the heuristic picture described in this section is, strictly speaking, incorrect, the concept contains some truth and can be salvaged to represent a doubling of the degrees of freedom. This may be understood from a number of related perspectives.

One of these perspectives, which we will detail and utilize in Section~\ref{sec:ATEE}, stems from the method used in Ref.~\cite{Kitaev06b} to derive the TEE. In particular, the Kitaev-Preskill derivation involves (conceptually) introducing a time-reversal conjugate copy of the system and connecting the two systems at various locations by wormholes threaded by trivial topological flux. By locating such wormholes along the partition boundary (which is mirrored on the conjugate copy of the surface), the partition boundary will pass through the wormholes. In the doubled system with wormholes, the partition cut will cut the tubes connecting the (now doubled) regions $\AD$ and $\bar{\AD}$ (respectively corresponding to the un-doubled regions $A$ and $\bar{A}$ of the original surface), giving rise to boundaries (the circles along which the tubes are cut) which carry topological charge values.  The anyonic state $\rhoD$ turns out to be equivalent to the anyonic state $\rhoH$ described above (see Section~\ref{sec:ATEE} for details). The doubling of the boundary contribution to the entanglement entropy arises in this picture because the system itself was doubled.

This doubling can also be understood in the context of state-sum and string-net models. From this perspective, the Kitaev-Preskill surface doubling is interpreted as representing the two chiral sectors of the emergent TQFT. More specifically, for a (spherical) fusion tensor category $\mathcal{F}$ that describes the fusion structure of a MTC $\mathcal{C}$, the emergent TQFT associated with a state-sum or string-net model based on $\mathcal{F}$ is the Drinfeld quantum double $\text{D}(\mathcal{F})=\mathcal{C} \times \overline{\mathcal{C}}$. One can think of $\mathcal{C}$ as living on one surface and its time-reversal conjugate $\overline{\mathcal{C}}$ on another, and the wormholes connecting these surfaces represent the plaquette centers of the string-net lattice model (which is the lattice dual of the state-sum triangulation). In this way, the lattice degrees of freedom on the links, which are described by $\mathcal{F}$, are what is captured by the anyonic state $\rhoH$ at the partition boundary. As such, the lattice model with degrees of freedom in $\mathcal{F}$ provides a microscopic regularization and correct accounting of the entropy for the TQFT $\text{D}(\mathcal{F})$, which is double that of $\mathcal{C}$; for example, $\mathcal{D}_{\mathcal{F}} = \mathcal{D}_{\mathcal{C}} = \sqrt{\mathcal{D}_{\text{D}(\mathcal{F})}}$. (See Section~\ref{sec:stringnets} for more details.)

Another perspective on the boundary entropy doubling comes from considering the boundary degrees of freedom as an edge CFT, e.g. for a chiral topological phase. As mentioned, such an edge cannot localize topological charge at specific locations along the edge. Moreover, one cannot simply break such an edge into segments, as the chiral CFT cannot terminate at the segment endpoints. In order to break the edge into segments in a manner that is well-defined for the CFT, one can use a boundary CFT~\cite{Cardy04} (``boundary'' here refers to the endpoints of a 1D spatial segment on which the $(1+1)$D CFT lives, not the 1D boundary of the 2D bulk region). Such boundary CFTs always have both holomorphic and anti-holomorphic modes that are coupled to each other by the boundary conditions, so the edge CFT degrees of freedom are necessarily doubled. This can also be understood as another perspective on the Kitaev-Preskill derivation, wherein doubling the surface and introducing wormholes creates boundary segments on the conjugate surface carrying CFT modes that propagate in the opposite direction as that of the original boundary. In other words, the boundary edge is split up into boundary circles of the tubes connecting regions $\AD$ and $\bar{\AD}$ and the edge segment on the original surface can be viewed as carrying the holomorphic modes while the edge segment on the conjugate surface carries the anti-holomorphic modes.

In Section~\ref{sec:ATEE}, we provide the more rigorous derivation of Eqs.~(\ref{eq:rhoheuristic}) and (\ref{eq:Sactual}) using a generalization of the Kitaev-Preskill arguments. This approach requires TQFT methods in which we evaluate anyon diagrams associated with the topological state space of higher genus surfaces.  To aid our discussion, we develop the formalism of anyon models for higher genus surfaces in the next section.

\section{Anyon Models for Higher Genus Surfaces}\label{sec:anyonmodelshigher}

We now generalize the anyon model formalism, reviewed in \ref{sec:anyonmodelssphere} for a surface of genus zero, to higher genus, orientable, compact surfaces (possibly with boundary). The genus $g$ of a surface is the number of handles on it. The topology of an orientable, compact surface is classified by its genus $g$ and the number $n$ of punctures, i.e., connected boundary components.

The state space of anyon models on higher genus surfaces has previously been discussed by Ref.~\cite{Pfeifer10} and applied to anyonic entanglement in Ref.~\cite{Pfeifer14}.  Our presentation differs from that of Ref.~\cite{Pfeifer10} in notation and normalization conventions, but the fundamental understanding is the same.  Our discussion of anyonic entanglement, particularly our derivation of the reduced density matrix, differs from that of Ref.~\cite{Pfeifer14}.

Ref.~\cite{Pfeifer14} focuses on the entanglement of anyonic states associated with the quasiparticles in a subregion of the higher genus surface, rather than the entanglement between different regions of the surfaces.  Thus, when partitioning the surface into regions $A$ and $\bar{A}$, Ref.~\cite{Pfeifer14} traces over the topological charge lines threading the boundary between $A$ and $\bar{A}$.  In our treatment, we wish to examine both the entanglement associated with the anyonic states as well as the entanglement between $A$ and $\bar{A}$.  We therefore include the charge lines threading the boundary between $A$ and $\bar{A}$ in our reduced density matrix $\arho_A$, which is what allows us to calculate $S_{\text{topo}}$ in Section~\ref{sec:ATEE}.

\subsection{Topological State Space of a Higher Genus Surface}\label{sec:inside-outside-basis}

The topological Hilbert space of a compact surface with genus $g$ and $n$ punctures can be constructed from that of the $(2g+n)$-punctured sphere with puncture labels $a_1, \bar{a}_1,\dots, a_g, \bar{a}_g$ and $c_1, \dots, c_n$. The Hilbert space can be spanned by two canonical bases: the ``inside'' basis and the ``outside'' basis.

The inside basis is formed by expressing the fusion tree for the punctures inside the sphere and gluing the punctures labeled $a_1, \dots, a_g$ to their respective punctures labeled $\bar{a}_1, \dots, \bar{a}_g$  outside the sphere. This leaves all the anyonic charge lines enclosed in the interior of the resulting surface or ending at a remaining puncture.

\begin{center}
\includegraphics[width=0.7\linewidth]{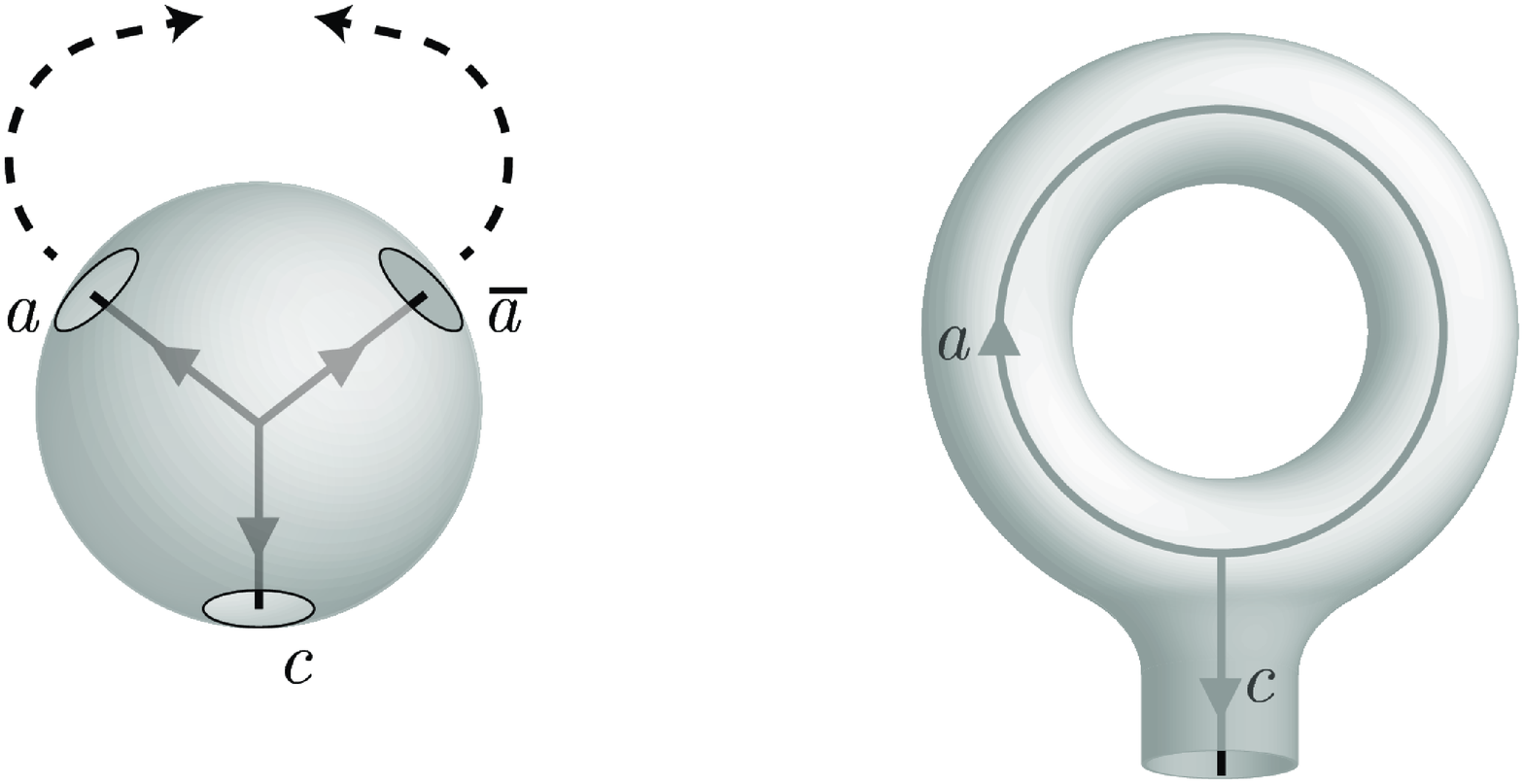}
\end{center}

The outside basis is formed by expressing the fusion tree for the punctures outside the sphere and gluing the punctures labeled $a_1, \dots, a_g$ to their respective punctures labeled $\bar{a}_1, \dots, \bar{a}_g$  inside (through) the sphere. This leaves all the anyonic charge lines in the region exterior to the resulting surface or ending at a remaining puncture.

\begin{center}
\includegraphics[width=0.7\linewidth]{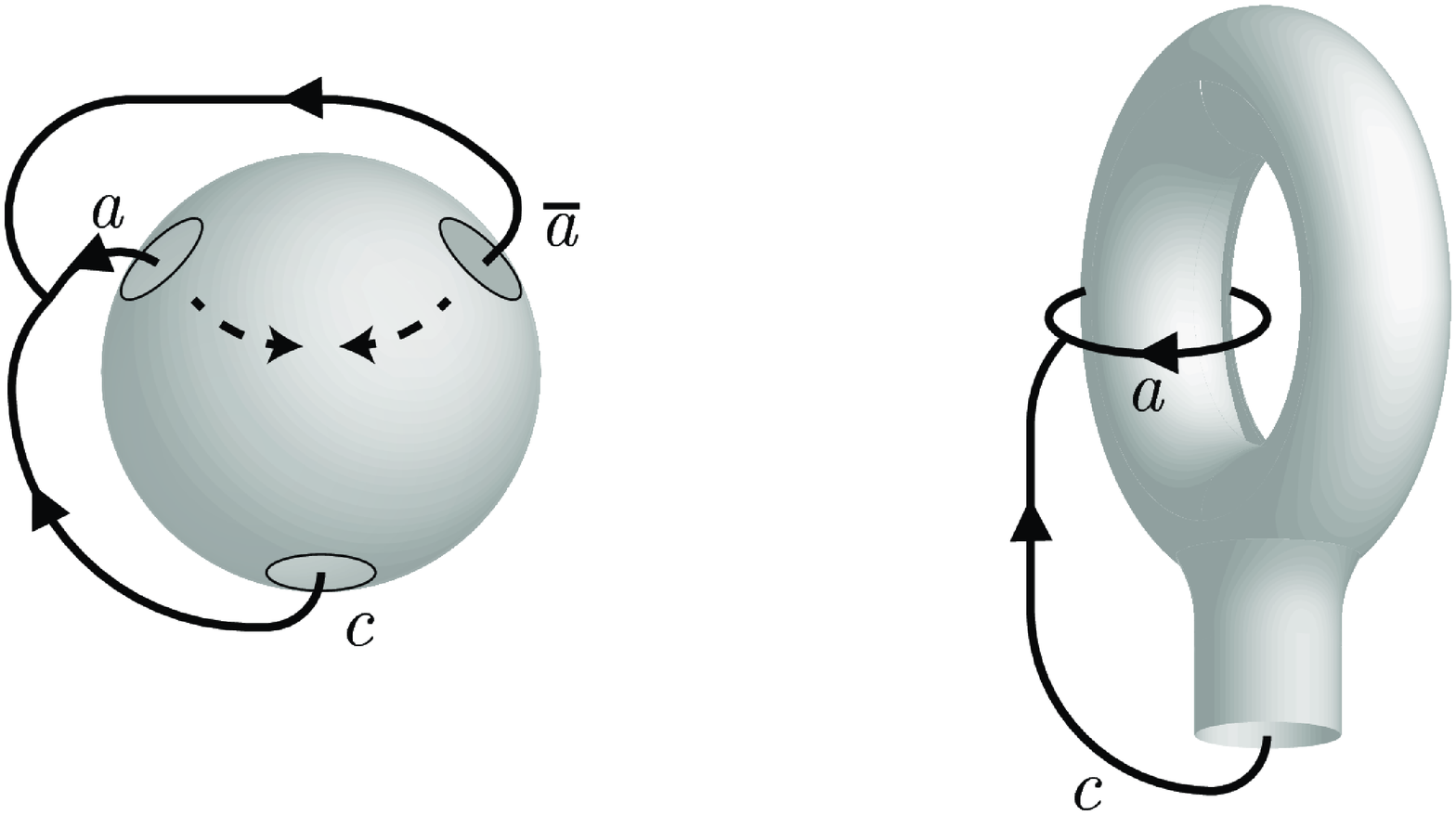}
\end{center}

The modular $\mathcal{S}$-transformations interchange the two complementary cycles associated with a given handle and, thus, provides a basis change between the inside and outside bases.

\begin{center}
\includegraphics[width=0.7\linewidth]{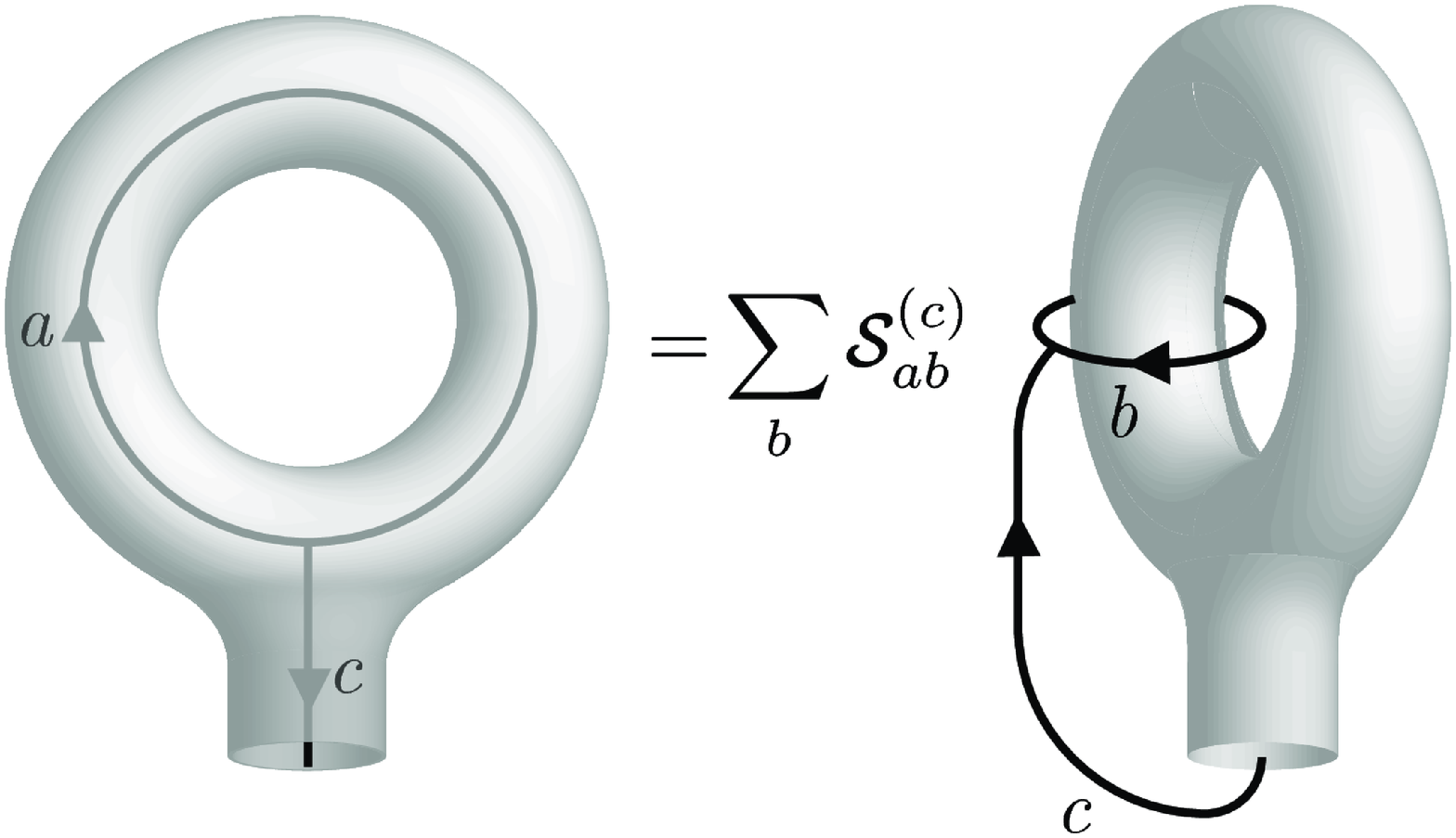}
\end{center}
In the following, we primarily work with the inside basis.

\subsubsection{Basis}

The topological Hilbert space on a sphere is constructed from the fusion and splitting spaces $V_{ab}^e$ and $V_{e}^{ab}$, see \ref{sec:anyonmodelssphere} for a review.  These vector spaces are supplemented on a higher genus surface by spaces involving topological charge lines circling non-contractible cycles, which we denote as $V_{e}^{(a)}$ and $V_{(a)}^e$.  The space $V^{(a)}_e$ is spanned by the vectors
\begin{equation}
    \ket{(a);e,\mu}=d_e^{1/4}
    \begin{pspicture}[shift=-0.7](-0.2,-0.5)(1.2,1.2)
        \scriptsize
        \rput(0.5,0.5){$\otimes$}
        \psline[border=1.5pt](0.5,0)(1,0.5)(0.5,1)(0,0.5)(0.5,0)
        \psline[ArrowInside=->](0.5,0)(0,0.5)\rput(.05,0.25){$a$}
        \psline[ArrowInside=->](0.5,-0.5)(0.5,0)\rput(0.7,-0.25){$e$}\rput(0.7,0){$\mu$}
    \end{pspicture},
\end{equation}
where $e$ can be any anyon such that $N_{a\bar{a}}^e \neq 0$. The symbol $\otimes$ represents a non-contractible cycle associated with a handle of the surface, for either the inside or outside basis. The topological charge line $a$ circling the non-contractible cycle is written in bra/ket notation as $(a)$ in order to distinguish it from the charges labeling boundaries or quasiparticles. The dual space $V^e_{(a)}$ is spanned by the covectors
\begin{equation}
    \bra{(a);e,\mu}=d_e^{1/4}
    \begin{pspicture}[shift=-0.8](-0.2,-0.2)(1.2,1.5)
        \scriptsize
        \rput(0.5,0.5){$\otimes$}
        \psline[border=1.5pt](0.5,0)(1,0.5)(0.5,1)(0,0.5)(0.5,0)
        \psline[ArrowInside=->](0.5,0)(0,0.5)\rput(0.05,0.25){$a$}
        \psline[ArrowInside=->](0.5,1)(0.5,1.5)\rput(0.7,1.25){$e$}\rput(0.7,1){$\mu$}
    \end{pspicture},
\end{equation}

Larger spaces are constructed by taking tensor products. For example, consider the anyonic Hilbert space $V^{(a) (b) c}_0$ of a genus $g=2$ surface with topological charge lines $a$ and $b$ wrapping around its two handles and an anyon $c$ on its surface.

\begin{center}
\includegraphics[width=0.8\linewidth]{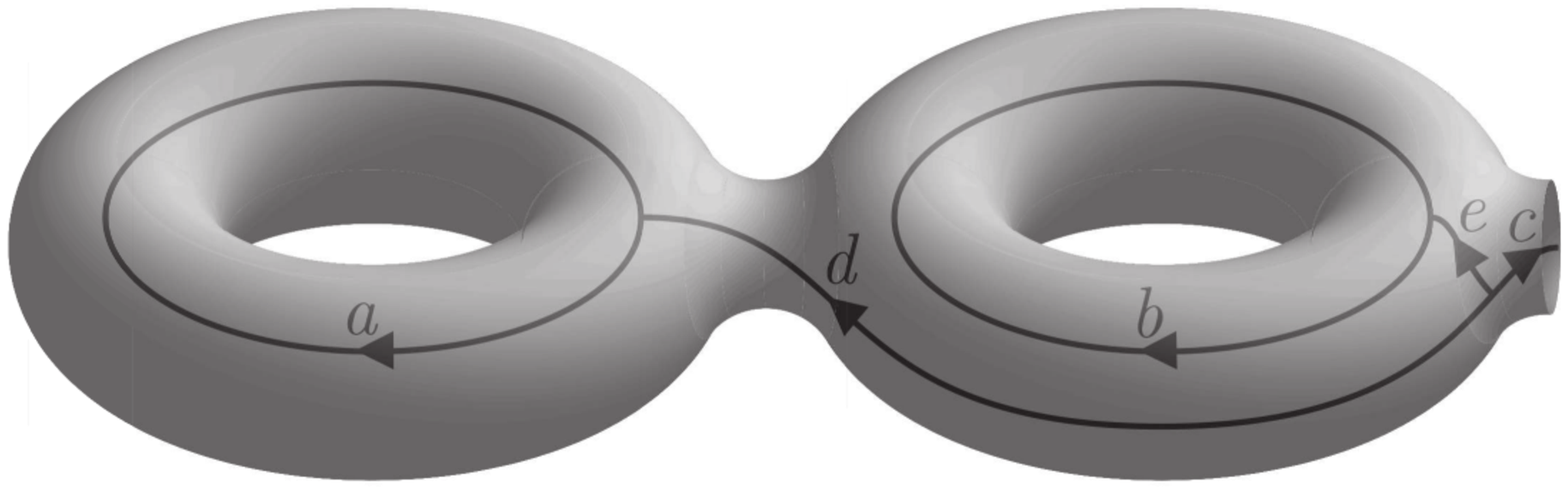}
\end{center}

This Hilbert space can be constructed as
\begin{equation}\label{eq:Vabc0}
    V^{(a) (b) c}_0\cong \bigoplus_{d,e} V^{(a)}_d \otimes V^{(b)}_e \otimes V^{de}_{\bar{c}} \otimes V^{\bar{c}c}_0,
\end{equation}
which is spanned by the vectors
\begin{align}\label{eq:Vbasis}
    \ket{(a);d,\mu}&\ket{(b);e,\nu}\ket{d,e;\bar{c},\alpha}\ket{\bar{c},c;0}
    =\frac{1}{d_c^{1/4}}
    \begin{pspicture}[shift=-1.1](-.2,-1.4)(3.6,1.2)
        \scriptsize
        \rput(0.5,0.5){$\otimes$}
        \psline[border=1.5pt](0.5,0)(1,0.5)(0.5,1)(0,0.5)(0.5,0)
        \psline[ArrowInside=->](0.5,0)(0,0.5)\rput(0.05,0.25){$a$}\rput(0.7,0){$\mu$}
        \rput(1.2,0){
            \rput(0.5,0.5){$\otimes$}
            \psline[border=1.5pt](0.5,0)(1,0.5)(0.5,1)(0,0.5)(0.5,0)
            \psline[ArrowInside=->](0.5,0)(0,0.5)\rput(0.05,0.25){$b$}\rput(0.7,0){$\nu$}}
        \psline[border=1.5pt](1.1,-0.6)(1.7,0)\psline[border=1.5pt](3.2,0.5)(1.6,-1.1)(1.1,-0.6)
        \psline[ArrowInside=->](1.1,-0.6)(0.5,0)\rput(0.6,-0.3){$d$}
        \psline[ArrowInside=->](1.1,-0.6)(1.7,0)\rput(1.6,-0.3){$e$}\rput(0.9,-0.6){$\alpha$}
        \psline[ArrowInside=->](1.6,-1.1)(1.1,-0.6)\rput(1.2,-0.9){$\bar{c}$}
        \psline[ArrowInside=->](1.6,-1.1)(3.2,0.5)\rput(2.6,-0.3){$c$}
    \end{pspicture},
\end{align}
where $\mu=1, \dots, N_{a \bar{a}}^{d}$, $\nu =1,\dots, N_{b \bar{b}}^{e}$, and $d$ and $e$ are any anyons such that $N_{a \bar{a}}^{d} \ge 1$, $N_{b \bar{b}}^{e} \ge 1$, and $N_{d e }^{\bar{c}} \ge 1$.

In general, the space $V^{(z_1)\dots (z_g) a_1 \dots a_n}_e$ for a subsystem containing anyons $a_1, \dots, a_n$ and genus $g$ is spanned by
\begin{align}
    &\ket{\vec{z};\vec{x},\vec{\omega}}\ket{\vec{x},\vec{y},\vec{\chi};d}\ket{\vec{a},\vec{b},\vec{\alpha};c}\ket{d,c;e,\mu} \notag \\
    &=\left(\frac{d_e}{d_{\vec{a}}}\right)^{1/4}
    \begin{pspicture}[shift=-2](-0.2,-2.75)(5.5,1.2)
        \scriptsize
        \rput(0.5,0.5){$\otimes$}
        \psline[border=1.5pt](0.5,0)(1,0.5)(0.5,1)(0,0.5)(0.5,0)
        \psline[ArrowInside=->](0.5,0)(0,0.5)\rput(.05,0.25){$z_1$}\rput(0.8,0){$\omega_1$}
        \rput(1.25,0.5){\dots}
        \rput(0.85,-0.25){$\ddots$}
        \rput(1.5,0){
            \rput(0.5,0.5){$\otimes$}
            \psline[border=1.5pt](0.5,0)(1,0.5)(0.5,1)(0,0.5)(0.5,0)
            \psline[ArrowInside=->](0.5,0)(0,0.5)\rput(.05,0.25){$z_g$}\rput(0.8,0){$\omega_g$}}
        \psline[border=1.5pt](2.75,-2.25)(1.25,-0.75)\psline[border=1.5pt](1.25,-0.75)(2,0)
        \psline[ArrowInside=->](0.7,-0.2)(0.5,0)\rput(0.5,-0.2){$x_1$}
        \psline[ArrowInside=->](1.25,-0.75)(1,-0.5)\rput(0.9,-0.8){$y_{g-1}$}
        \psline[ArrowInside=->](1.25,-0.75)(2,0)\rput(1.35,-0.4){$x_g$}
        \psline[ArrowInside=->](2.75,-2.25)(1.25,-0.75)\rput(1.9,-1.6){$d$}
        \rput(1.55,-0.75){$\chi_g$}
        \rput(3.5,-1){
            \psline[ArrowInside=->](0.5,1)(0,1.5)\rput(0.05,1.25){$a_1$}
            \psline[ArrowInside=->](0.5,1)(1,1.5)\rput(1,1.25){$a_2$}
            \psline[ArrowInside=->](1,0.5)(2,1.5)\rput(1.8,1){$a_n$}
            \psline[ArrowInside=->](0.7,0.8)(0.5,1)\rput(0.5,0.75){$b_2$}
            \psline[ArrowInside=->](1,0.5)(0.8,0.7)\rput(0.6,0.5){$b_{n-1}$}
            \rput(0.25,1){$\alpha_2$}
            \rput(1.3,0.5){$\alpha_n$}
            \rput(0.75,0.75){.}
            \rput(1.5,1.5){\dots}
            \psline[ArrowInside=->](-0.75,-1.25)(1,0.5)\rput(0.2,-0.5){$c$}}
        \psline[ArrowInside=->](2.75,-2.75)(2.75,-2.25)\rput(2.9,-2.5){$e$}
        \rput(3,-2.25){$\mu$}
    \end{pspicture},
\end{align}

We only use the bra/ket notation when the system is in the canonical basis written above.  When applying $F$-moves that take the state out of the canonical basis, the diagrammatic representation of the topological Hilbert space is much easier to use, see e.g., the entropy calculations of Section~\ref{sec:ATEE}.

Finally, we note that, when considering states on compact surfaces, the overall topological charge of each connected component of the surface (including their boundaries) is always the trivial charge $0$.  We return to this point in Section~\ref{sec:higher-g-trace} when discussing subtleties of performing the partial quantum trace.

\subsubsection{Dimension}

The dimension of $V^{(z_1)\dots (z_g) a_1 \dots a_n}_0$ is given by
\begin{equation}
    \dim(V^{(z_1)\dots (z_g) a_1 \dots a_n}_0)=N_{z_1 \bar{z}_1\dots z_g \bar{z}_g a_1 \dots a_n}^0,
\end{equation}
The dimension of the space of anyons $a_1, \dots, a_n$ on a surface with genus $g$ is
\begin{align}
   \mathcal{N}_{g;a_1\dots a_n} \equiv  \sum_{\vec{z}} \dim(V^{(z_1)\dots (z_g) a_1 \dots a_n}_0)&=\sum_{\vec{z}} N_{z_1 \bar{z}_1\dots z_g \bar{z}_g a_1 \dots a_n}^0,
\end{align}
which can also be expressed in terms of the $\mathcal{S}$-matrix (see Section~\ref{sec:S-matrix}) as
\begin{equation}
\mathcal{N}_{g;a_1\dots a_n}=\sum_x \left(\frac{d_x}{\mathcal{D}}\right)^{2-n-2g} \mathcal{S}_{a_1 x} \dots \mathcal{S}_{a_n x}
.
\end{equation}
In particular, if there are no anyons present, then
\begin{equation}
\mathcal{N}_{g;0} = \sum_x \left(\frac{d_x}{\mathcal{D}}\right)^{2-2g}  \sim |\mathcal{C}_{\text{Abelian}}|\mathcal{D}^{2g-2}
\end{equation}
for large $g$, where $|\mathcal{C}_{\text{Abelian}}|$ is the number of distinct Abelian topological charges in $\mathcal{C}$.

\subsubsection{Inner Product}

Inner products of states on surfaces with non-contractible cycles can be evaluated in the diagrammatic representation by cutting open the anyon lines encircling the non-contractible cycle, introducing a factor of $1/\sqrt{d_a}$ for each anyon line $a$ that is cut, and then stacking the diagrams. For example, consider a ground state on the torus
\begin{equation}
\ket{(a)}= \begin{pspicture}[shift=-.45](-0.7,-.5)(.5,.5)
        \scriptsize
        \psline[ArrowInside=->](0,-.5)(-.5,0)
        \psline(.5,0)(0,.5)(-.5,0)(0,-.5)(.5,0)
        \rput(0,0){$\otimes$}
        \rput(-.45,-.3){$a$}
    \end{pspicture}
\end{equation}
In order to compute the inner product of such states in the diagrammatic formalism, we first cut open the diagram, as though we are cutting open the corresponding handle of the surface (the torus), and multiply by a normalization factor for each of the new leaves of the diagram, giving
\begin{equation}
\ket{(a)_{\text{cut}}}= \ket{a,\bar{a};0 }= \frac{1}{\sqrt{d_a }}
\begin{pspicture}[shift=-.35](-0.7,-.6)(.7,.2)
        \scriptsize
        \psline[ArrowInside=->](0,-.5)(.5,0)\rput(.5,.1){$\bar{a}$}
        \psline[ArrowInside=->](0,-.5)(-.5,0)\rput(-.5,.1){$a$}
\end{pspicture}
.
\end{equation}
Then, the inner product $\bra{(b)}(a)\rangle$ can be expressed as
\begin{equation}
\bra{(b)}(a)\rangle = \bra{(b)_{\text{cut}} }(a)_{\text{cut}} \rangle  =  \frac{1}{\sqrt{d_a d_b}}
\begin{pspicture}[shift=-.45](-0.8,-.5)(.8,.5)
        \scriptsize
        \psline[ArrowInside=->](.5,0)(0,.5)
        \psline[ArrowInside=->](-.5,0)(0,.5)
        \psline[ArrowInside=->](0,-.5)(.5,0)
        \psline[ArrowInside=->](0,-.5)(-.5,0)
        \rput(.5,.3){$\bar{b}$}
        \rput(.5,-.3){$\bar{a}$}
        \rput(-.5,.3){$b$}
        \rput(-.5,-.3){$a$}
        \psline[linestyle=dashed](-.6,0)(.6,0)
\end{pspicture}
= \delta_{a,b}\frac{1}{d_a}
\begin{pspicture}[shift=-.45](-0.7,-.5)(.5,.5)
        \scriptsize
        \psline[ArrowInside=->](0,-.5)(-.5,0)
        \psline(.5,0)(0,.5)(-.5,0)(0,-.5)(.5,0)
        \rput(-.45,-.3){$a$}
    \end{pspicture}
     = \delta_{a,b}.
\end{equation}
In the above, we have included a dashed line to indicate where the topological charge lines were cut and glued together.

Similarly, for the states of a punctured torus,
\begin{equation}
    \bra{(a);c,\mu}=d_c^{1/4}
    \begin{pspicture}[shift=-0.8](-0.2,-0.2)(1.2,1.5)
        \scriptsize
        \rput(0.5,0.5){$\otimes$}
        \psline[border=1.5pt](0.5,0)(1,0.5)(0.5,1)(0,0.5)(0.5,0)
        \psline[ArrowInside=->](0.5,0)(0,0.5)\rput(0.05,0.25){$a$}
        \psline[ArrowInside=->](0.5,1)(0.5,1.5)\rput(0.7,1.25){$c$}\rput(0.7,1){$\mu$}
    \end{pspicture},
\end{equation}
the corresponding states when the handle is cut open are given by
\begin{equation}
\ket{(a)_{\text{cut}}; c,\mu} = \ket{a,\bar{a}; c, \mu}= \left(\frac{d_{e}}{d_a^2}\right)^{1/4}
\begin{pspicture}[shift=-.45](-0.7,-1.1)(.7,.2)
        \scriptsize
        \psline[ArrowInside=->](0,-.5)(.5,0)\rput(.5,.1){$\bar{a}$}
        \psline[ArrowInside=->](0,-.5)(-.5,0)\rput(-.5,.1){$a$}
        \psline[ArrowInside=->](0,-1)(0,-.5)\rput(.15,-.9){$c$} \rput(.15,-.6){$\mu$}
\end{pspicture}
.
\end{equation}
The inner product of two basis states of the punctured torus is
\begin{eqnarray}
    \langle (b);e,\nu|(a);c,\mu\rangle &=& \langle (b)_{\text{cut}};e,\nu|(a)_{\text{cut}};c,\mu\rangle
=\left(\frac{d_cd_e}{d_a^2 d_{b}^2}\right)^{1/4}
    \psscalebox{1}{
    \begin{pspicture}[shift=-0.9](-0.1,0)(1.2,2)
        \scriptsize
        \psline[ArrowInside=->](0.5,0.5)(0,1)\rput(0.05,0.75){$a$}
        \psline[ArrowInside=->](0.5,0.5)(1,1)\rput(0.95,0.75){$\bar{a}$}
        \psline[ArrowInside=->](0.5,0)(0.5,0.5)\rput(0.65,0.2){$c$}
        \rput(0.65,0.45){$\mu$}
        \psline[linestyle=dashed](-.1,1)(1.1,1)
        \rput(0,1){
        \psline[ArrowInside=->](0,0)(0.5,0.5)\rput(0.05,0.3){$b$}
        \psline[ArrowInside=->](1,0)(0.5,0.5)\rput(1,0.25){$\bar{b}$}
        \psline[ArrowInside=->](0.5,0.5)(0.5,1)\rput(0.7,0.8){$e$}
        \rput(0.7,0.55){$\nu$}}
    \end{pspicture}}
\notag \\
&=&
    \delta_{a,b}\delta_{c,e}\delta_{\mu,\nu}
    \begin{pspicture}[shift=-0.25](-0.2,0)(0.3,.9)
        \scriptsize
        \psline[ArrowInside=->](0,0)(0,.7)\rput(0.2,.5){$c$}
    \end{pspicture}    =\delta_{a,b}\delta_{c,e}\delta_{\mu,\nu}\ket{c}\bra{c}.
\end{eqnarray}
More complicated diagrams can be similarly evaluated.
In the general case, each additional endpoint in the diagram (boundary of the surface) of charge $a$ that results from cutting open a handle requires a normalization factor of $d_{a}^{-1/4}$ in the diagrammatic representation of the ``cut'' state.

\subsubsection{Operators}

The space $V_{(Z_1')\dots (Z_g') A_1'\dots A_n'}^{(Z_1)\dots (Z_g) A_1\dots A_n}$ of operators acting on $n$ anyons on a surface of genus $g$ can be constructed as
\begin{align}
    &V_{(Z_1')\dots (Z_g') A_1'\dots A_n'}^{(Z_1)\dots (Z_g) A_1\dots A_n}
   =\sum_{\vec{z},\vec{z}',\vec{a},\vec{a}'}\bigoplus_c V^c_{(z_1')\dots (z_g') a_1'\dots a_n'}\otimes V^{(z_1)\dots (z_g) a_1\dots a_n}_c.
\end{align}

For example, the identity operator acting on the state space of a punctured torus is
\begin{equation}
  \mathbb{1} = \sum_a  \mathbb{1}_{(a)}=\sum_{a,c,\mu} \ket{(a);c,\mu}\bra{(a);c,\mu}
    =\sum_{a,c,\mu} \sqrt{d_c}
    \begin{pspicture}[shift=-1.4](-0.2,-1.7)(1.2,1.2)
        \scriptsize
        \rput(0.5,-1){$\otimes$}
        \rput(0.5,0.5){$\otimes$}
        \psline[border=1.5pt](0.5,0)(1,0.5)(0.5,1)(0,0.5)(0.5,0)
        \psline[ArrowInside=->](0.5,0)(0,0.5)\rput(0.05,0.25){$a$}
        \rput(0,-1.5){
            \psline[border=1.5pt](0.5,0)(1,0.5)(0.5,1)(0,0.5)(0.5,0)
            \psline[ArrowInside=->](0.5,0)(0,0.5)\rput(0.05,0.25){$a$}}
        \psline[ArrowInside=->](0.5,-0.5)(0.5,0)\rput(0.65,-0.25){$c$}\rput(0.7,0){$\mu$}\rput(0.7,-0.5){$\mu$}
    \end{pspicture}.
\end{equation}

\subsubsection{Trace}\label{sec:higher-g-trace}

The trace of an operator involving non-contractible cycles is defined, as usual, to be the sum of its diagonal elements, e.g.
\begin{equation}\label{eq:egtrace1}
    \text{Tr}(\ket{(a);c,\mu}\bra{(a');c,\mu'})=\delta_{a,a'}\delta_{\mu,\mu'}.
\end{equation}
To evaluate the quantum trace $\aTr$ for a system with charge lines circling non-contractible cycles, cut open the anyon lines circling the non-contractible cycle, introduce a factor $1/\sqrt{d_a}$ for every cut charge line $a$, and join the outgoing charge lines of the operator's diagram back onto the incoming charge lines.  In doing so, we remove the non-contractible cycles, which can be understood as mapping the system to the sphere with certain charge lines identified~\cite{Pfeifer10}. As an example,
\begin{align}
   \aTr\left( \ket{(a);c,\mu} \bra{(a');c,\mu'} \right)
   &=  \aTr\Bigg(\sqrt{d_c}
   \psscalebox{.9}{
    \begin{pspicture}[shift=-1.4](.05,-1.7)(1.2,1.2)
        \scriptsize
        \rput(0.5,-1){$\otimes$}
        \rput(0.5,0.5){$\otimes$}
        \psline[border=1.5pt](0.5,0)(1,0.5)(0.5,1)(0,0.5)(0.5,0)
        \psline[ArrowInside=->](0.5,0)(0,0.5)\rput(0.05,0.25){$a$}
        \rput(0,-1.5){
            \psline[border=1.5pt](0.5,0)(1,0.5)(0.5,1)(0,0.5)(0.5,0)
            \psline[ArrowInside=->](0.5,0)(0,0.5)\rput(0.05,0.2){$a'$}}
        \psline[ArrowInside=->](0.5,-0.5)(0.5,0)\rput(0.65,-0.25){$c$}\rput(0.7,0){$\mu$}\rput(0.75,-0.45){$\mu'$}
    \end{pspicture}
    }
    \Bigg)
    =\sqrt{\frac{d_c}{d_a d_{a'}}}\delta_{a,a'} \delta_{\mu,\mu'}
    \psscalebox{.9}{
    \begin{pspicture}[shift=-1.2](-.2,-0.6)(1.8,1.7)
        \scriptsize
        \psline[ArrowInside=->](0,0)(0.5,0.5)\rput(0,0.25){$a$}
        \psline[ArrowInside=->](1,0)(0.5,0.5)\rput(1,0.25){$\bar{a}$}
        \rput(0.7,0.5){$\mu$}
        \psline[ArrowInside=->](0.5,0.5)(0.5,1)\rput(0.65,0.75){$c$}
        \rput(0.7,1){$\mu$}
        \psline[ArrowInside=->](0.5,1)(0,1.5)
        \psline(0,1.5)(0.75,2.25)(1.75,1.25)(1.75,0.25)(0.75,-0.75)
        \psline(0.75,-0.75)(0,0)\rput(0,1.25){$a$}
        \psline(0.5,1)(1,1.5)(1.5,1)(1.5,0.5)(1,0)(1.5,0.5)
        \psline[ArrowInside=->](0.5,1)(1,1.5)\rput(1,1.25){$\bar{a}$}
    \end{pspicture}
    } \notag
    \\ &=d_c \delta_{a,a'}\delta_{\mu,\mu'}.
\end{align}
The above agrees with Eq.~(\ref{eq:egtrace1}) up to a factor of $d_c$. This corresponds to the general relation between the anyonic trace of an operator ${X\in V^{(z_1)\dots (z_g) a_1\dots a_n}_{(z_1')\dots (z_g') a_1'\dots a_n'}}$ and the ordinary trace, given by
\begin{align}
    \aTr(X) =\sum_c d_c\text{Tr}([X]_c),\\
    \text{Tr}(X) =\sum_c \frac{1}{d_c}\aTr([X]_c),
\end{align}
where
\begin{equation}
    [X]_c = \Pi_c X \Pi_c\in  V_c^{(z_1)\dots (z_g) a_1 \dots a_n} \otimes V_{(z_1')\dots (z_g') a_1' \dots a_n'}^c
\end{equation}
is the projection of $X$ onto definite total charge $c$, with $X=\sum_c [X]_c$.

One can also compute the partial quantum trace of a surface of genus $g$ by joining the charge lines and cycles of only the subset of anyons being traced out.  First, one must specify which regions of the surface are being traced out, thereby identifying which anyons and cycles are being traced over.  In doing so, one is implicitly specifying the path through which one performs the trace over anyonic charge lines~\footnote{When considering anyons in a planar surface, one sometimes traces out anyons by ``taking the anyons to infinity."  This amounts to moving the anyons to the edge of the diagram by braiding them past other anyons, a process that is not necessarily unique when the partition is not specified.  One must be more careful to specify the partition and to keep track of the boundary charges in a connected surface of higher genus, as will be further discussed in the next section.}. In general, the partial quantum trace of $X\in V^{(z_1)\dots (z_g)(v_1)\dots(v_h)a_1\dots a_n b_1\dots b_m}_{(z_1')\dots (z_g')(v_1')\dots (v_h') a_1'\dots a_n' b_1'\dots b_m'}$ over the anyons $b_1,\dots, b_m$ and handles $v_1, \dots , v_h$ is related to the ordinary partial trace by
\begin{align}
\aTr_{(v_1)\dots (v_h)b_1\dots b_m}(X)&=\sum_{c,a} \frac{d_c}{d_a} \left[\text{Tr}_{(v_1)\dots(v_h)b_1\dots b_m}\left([X_c]\right) \right]_a,
\\ \text{Tr}_{(v_1)\dots (v_h)b_1\dots b_m}(X)&= \sum_{c,a} \frac{d_a}{d_c} \left[\aTr_{(v_1)\dots (v_h)b_1\dots b_m}\left([X]_c\right) \right]_a.
\end{align}

\subsection{Anyonic Density Matrices} \label{sec:DSg}

An anyonic density matrix is a Hermitian, positive semi-definite anyonic operator normalized by the quantum trace, $\aTr \arho =1$, that describes the topological state of the system.  For any connected component of a compact surface, the overall topological charge, including boundary charges and quasiparticles, is 0.  Thus, if one includes the boundaries (and their corresponding topological charges) that arise when tracing out portions of the system, the corresponding anyonic density matrix calculated from the quantum trace is equivalent to the ordinary density matrix calculated from the regular trace.

The anyonic density matrix determines the expectation value of anyonic operators acting on the system, ${\langle X \rangle=\aTr \left(\arho X\right)}$. On a higher genus surface, $\arho$ can involve anyons living in the bulk or on the boundary of the surface, as well as anyonic charge lines circling non-contractible cycles of the surface.

The reduced anyonic density matrix $\arho_A$ for a subsystem $A$ is calculated by taking the partial quantum trace over the degrees of freedom belonging to the complement $\bar{A}$. For any operator $X_A$ acting solely on degrees of freedom in $A$,
\begin{equation}
\langle X_A \rangle= \aTr \left( \arho X_A \right)=\aTr_A\left(\arho_A X_A\right).
\end{equation}
That is, the expectation value of $X_A$ can be equivalently computed with the density matrix for the full system or with the reduced density matrix for $A$.

One must be careful to include boundary charges when computing reduced density matrices for surfaces with genus and multiple boundaries.  In Section~\ref{sec:anyonicEntropy}, we only considered states on genus zero surfaces with one partition boundary. To compute the reduced density matrix for a region $A$, we specified which topological charge lines belonged to $A$ and which belonged to $\bar{A}$, then moved the charge lines in $\bar{A}$ to the outside of the diagram and joined the incoming and outgoing lines.  In doing so, we did not keep track of the charge associated with the boundary of $A$, which meant that we sometimes found a density matrix with nontrivial overall charge.  This can be reconciled with conservation of topological charge by recognizing that, in the sphere or planar case, one is implicitly specifying a disk-like region $A$ and tracing out the complementary region $\bar{A}$. Since there is a single boundary component for the disk, quasiparticles inside region $A$ cannot braid with the boundary charge and, as long as the quasiparticles are kept far away from the boundary, they cannot fuse with it either. Therefore, one can safely trace out the boundary charge (or the charge at infinity), since the quasiparticles do not interact with the boundary charge topologically. If one wishes to treat the states of more general systems involving genus and boundaries, one must be careful to only trace out the parts of the states corresponding to regions of the surface that will be considered ``inaccessible.''

The following method allows computation of the anyonic reduced density matrix for a region $A$ on a general compact surface, assuming that the full system is in a pure state $\ket{\psi}$:
\begin{enumerate}
\item Write the density matrix $\ket{\psi}\bra{\psi}$ for the full system $A\cup \bar{A}$ in a basis such that the charge lines for each connected component of region $A$ are grouped together and there is a single charge line threading each boundary component connecting $A$ with $\bar{A}$.
\item Cut the system along the boundary $\partial A \cap \partial \bar{A}$ between $A$ and $\bar{A}$ to form disjoint compact surfaces $A$ and $\bar{A}$. For each charge line $a_j$ that is cut, introduce a factor of $d_{a_j}^{-1/2}$ to normalize the state in the basis $\ket{\psi_{\text{cut}}}\bra{\psi_{\text{cut}}}$.  Each charge line that is cut corresponds to a new pair of boundaries (carrying the corresponding charge) produced by cutting the surface, one of which belongs to $A$ and the other to $\bar{A}$.
\item Perform a partial quantum trace over the portion of the anyonic state corresponding $\bar{A}$. The resulting state $\arho_A=\aTr_{\bar{A}} \ket{\psi_{\text{cut}}}\bra{\psi_{\text{cut}}}$ is the reduced anyonic density matrix for $A$.
\end{enumerate}
In step 1, the requirement that only one charge line threads each boundary component of $\partial A$ comes from the TQFT statement that the charge associated with a puncture is equivalent to the charge line threading it.  As it is not well-defined to think of multiple charges associated with the same puncture, before we introduce new punctures by cutting the surface, we must apply $F$-moves so that there is a single charge line threading each boundary component.
In step 2, we again emphasize that each connected component of the surface, both before and after cutting, has total charge $0$, when including the boundary charges.  As a result, the partial quantum trace in step 3 will be equivalent to the regular partial trace.
Our construction of the reduced density matrix differs from that of Ref.~\cite{Pfeifer14} in that we do not trace over the (new) boundary charges of $A$ (see the discussion at the beginning of Section~\ref{sec:anyonmodelshigher}).

As a demonstrative example, we compute the anyonic reduced density matrices obtained from the state (suppressing vertex labels)
\begin{equation}
\ket{\psi}=\sum_{a,b,c,d,e} \frac{\psi_{a,b,c,d,e}}{d_c^{1/4}}
\begin{pspicture}[shift=-1.1](-.2,-1.4)(3.6,1.2)
        \scriptsize
        \rput(0.5,0.5){$\otimes$}
        \psline(0.5,0)(1,0.5)(0.5,1)(0,0.5)(0.5,0)
        \psline[ArrowInside=->](0.5,0)(0,0.5)\rput(0.05,0.25){$a$}
        \rput(1.2,0){
            \rput(0.5,0.5){$\otimes$}
            \psline(0.5,0)(1,0.5)(0.5,1)(0,0.5)(0.5,0)
            \psline[ArrowInside=->](0.5,0)(0,0.5)\rput(0.05,0.25){$b$}}
        \psline(1.1,-0.6)(1.7,0)\psline(3.2,0.5)(1.6,-1.1)(1.1,-0.6)
        \psline[ArrowInside=->](1.1,-0.6)(0.5,0)\rput(0.6,-0.3){$d$}
        \psline[ArrowInside=->](1.1,-0.6)(1.7,0)\rput(1.6,-0.3){$e$}
        \psline[ArrowInside=->](1.6,-1.1)(1.1,-0.6)\rput(1.2,-0.9){$\bar{c}$}
        \psline[ArrowInside=->](1.6,-1.1)(3.2,0.5)\rput(2.6,-0.3){$c$}
    \end{pspicture}
\end{equation}
of a surface with genus $g=2$ and $n=1$ puncture, when it is partitioned into the regions $A$ and $\bar{A}$ indicated by the dashed lines drawn on the surface:
\begin{center}
\includegraphics[width=0.8\linewidth]{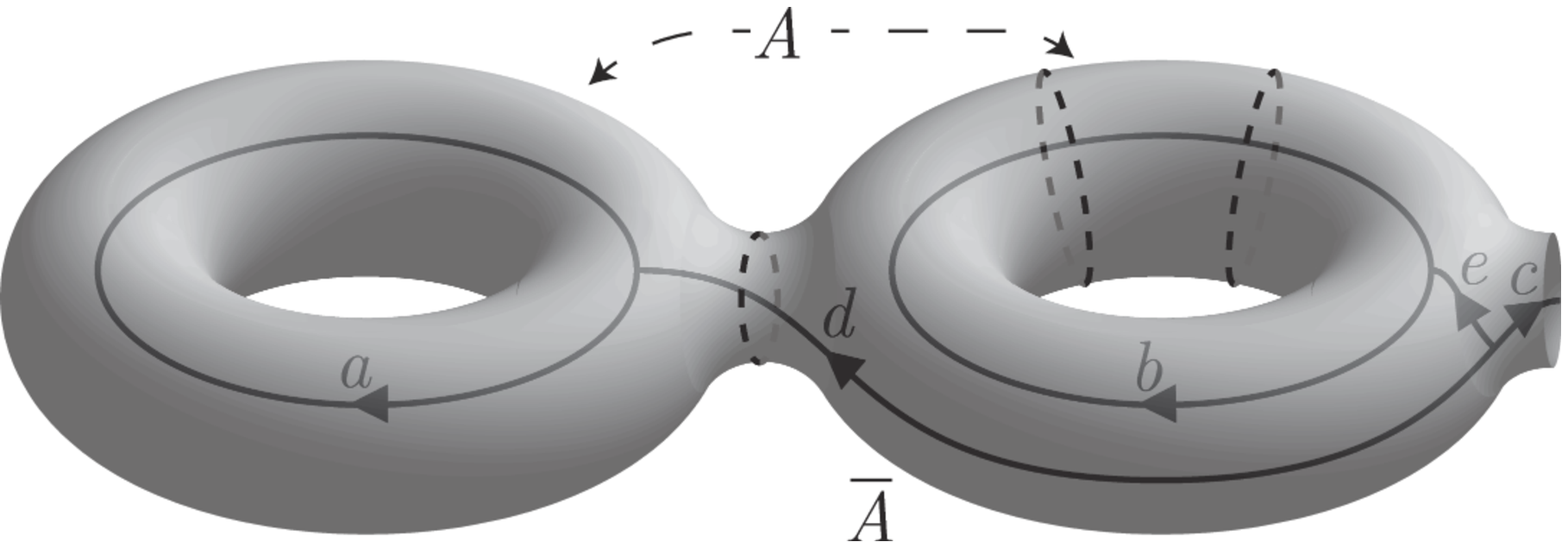}
.
\end{center}

Following the steps outlined above:
\begin{enumerate}

\item We write the full density matrix
\begin{equation}
\arho = \ket{\psi}\bra{\psi} = \sum_{\substack{a,b,c,d,e\\ a',b',c',d',e'}} \frac{\psi_{a,b,c,d,e}\psi^*_{a',b',c',d',e'}}{\left(d_cd_{c'}\right)^{1/4}}
\begin{pspicture}[shift=-2.4](-.2,-3.9)(3.6,1.2)
        \scriptsize
        \rput(0.5,0.5){$\otimes$}
        \psline(0.5,0)(1,0.5)(0.5,1)(0,0.5)(0.5,0)
        \psline[ArrowInside=->](0.5,0)(0,0.5)\rput(0.05,0.25){$a$}
        \rput(1.2,0){
            \rput(0.5,0.5){$\otimes$}
            \psline(0.5,0)(1,0.5)(0.5,1)(0,0.5)(0.5,0)
            \psline[ArrowInside=->](0.5,0)(0,0.5)\rput(0.05,0.25){$b$}}
        \psline(1.1,-0.6)(1.7,0)\psline(3.2,0.5)(1.6,-1.1)(1.1,-0.6)
        \psline[ArrowInside=->](1.1,-0.6)(0.5,0)\rput(0.6,-0.3){$d$}
        \psline[ArrowInside=->](1.1,-0.6)(1.7,0)\rput(1.6,-0.3){$e$}
        \psline[ArrowInside=->](1.6,-1.1)(1.1,-0.6)\rput(1.2,-0.9){$\bar{c}$}
        \psline[ArrowInside=->](1.6,-1.1)(3.2,0.5)\rput(2.6,-0.3){$c$}
        \psline[ArrowInside=->](1.1,-2.1)(1.6,-1.6) \rput(1.1,-1.9){$\bar{c}'$}
        \psline[ArrowInside=->](3.2,-3.2)(1.6,-1.6) \rput(2.65,-2.4){$c'$}
        \psline[ArrowInside=->](.5,-2.7)(1.1,-2.1)\rput(0.5,-2.4){$d'$}
        \psline[ArrowInside=->](1.7,-2.7)(1.1,-2.1)\rput(1.65,-2.4){$e'$}
        \psline(0,-3.2)(.5,-2.7)(1,-3.2)(.5,-3.7)(0,-3.2)\psline[ArrowInside=->](.5,-3.7)(0,-3.2)\rput(.5,-3.2){$\otimes$}\rput(.15,-3.6){$a'$}
        \psline(1.7,-2.7)(1.2,-3.2)(1.7,-3.7)(2.2,-3.2)(1.7,-2.7)\psline[ArrowInside=->](1.7,-3.7)(1.2,-3.2)\rput(1.7,-3.2){$\otimes$}\rput(1.35,-3.6){$b'$}
    \end{pspicture}
.
\end{equation}

\item We cut the surface:
\begin{equation}\label{eq:ex-psi-cut}
\begin{split}
\ket{\psi_{\text{cut}}}\bra{\psi_{\text{cut}}} = \sum_{\substack{a,b,c,d,e \\ a',b',c',d',e'}}& \frac{\psi_{a,b,c,d,e}\psi^*_{a',b',c',d',e'}}{\left(d_c d_{c'}\right)^{1/4} d_b d_{b'} \sqrt{d_d d_{d'}}}
\\ &\psscalebox{.8}{
\begin{pspicture}[shift=-2.4](-.2,-3.9)(7.2,1.2)
        \scriptsize
        \rput(0.5,0.5){$\otimes$}
        \psline(0.5,0)(1,0.5)(0.5,1)(0,0.5)(0.5,0)
        \psline[ArrowInside=->](0.5,0)(0,0.5)\rput(0.05,0.25){$a$}
        \psline[ArrowInside=->](1,-.5)(0.5,0)\rput(.55,-.3){$d$}
        \psline[ArrowInside=->](1,-.5)(2,.5)\rput(1.7,0){$\bar{d}$}
        \psline[ArrowInside=->](2.75,0)(2.25,.5)\rput(2.4,.2){$b$}
        \psline[ArrowInside=->](2.75,0)(3.25,.5)\rput(3.15,.2){$\bar{b}$}
        \rput(4,0){
        \psline[ArrowInside=->](1.1,-0.6)(0,.5)\rput(0.6,-0.3){$d$}
        \psline[ArrowInside=->](1.1,-0.6)(1.7,0)\rput(1.6,-0.3){$e$}
        \psline[ArrowInside=->](1.7,0)(1.2,.5) \rput(1.35,.2){$b$}
        \psline[ArrowInside=->](1.7,0)(2.2,.5)\rput(2.1,.2){$\bar{b}$}
        \psline[ArrowInside=->](1.6,-1.1)(1.1,-0.6)\rput(1.2,-0.9){$\bar{c}$}
        \psline[ArrowInside=->](1.6,-1.1)(3.2,.5)\rput(2.2,-0.7){$c$}}
\rput(2,-1.4){$(A)$}
        \psline(0,-3.2)(.5,-2.7)(1,-3.2)(.5,-3.7)(0,-3.2) \psline[ArrowInside=->](.5,-3.7)(0,-3.2)\rput(.5,-3.2){$\otimes$}\rput(.15,-3.6){$a'$}
        \psline[ArrowInside=->](.5,-2.7)(1,-2.2)\rput(.55,-2.4){$d'$}
        \psline[ArrowInside=->](2,-3.2)(1,-2.2)\rput(1.7,-2.6){$\bar{d}'$}
        \psline[ArrowInside=->](2.25,-3.2)(2.75,-2.7)\rput(2.4,-2.8){$b'$}
        \psline[ArrowInside=->](3.25,-3.2)(2.75,-2.7)\rput(3.15,-2.8){$\bar{b}'$}
        \rput(4,0){
        \rput(1.6,-1.4){$(\bar{A})$}
        \psline[ArrowInside=->](1.1,-2.1)(1.6,-1.6) \rput(1.1,-1.9){$\bar{c}'$}
        \psline[ArrowInside=->](3.2,-3.2)(1.6,-1.6) \rput(2.65,-2.4){$c'$}
        \psline[ArrowInside=->](0,-3.2)(1.1,-2.1)\rput(0.5,-2.4){$d'$}
        \psline[ArrowInside=->](1.7,-2.7)(1.1,-2.1)\rput(1.65,-2.4){$e'$}
        \psline[ArrowInside=->](1.2,-3.2)(1.7,-2.7)\rput(2.1,-2.8){$\bar{b}'$}
        \psline[ArrowInside=->](2.2,-3.2)(1.7,-2.7)\rput(1.35,-2.8){$b'$}}
\end{pspicture}}
.
\end{split}
\end{equation}

\item We trace over region $\bar{A}$:
\begin{equation}
\begin{split}
\aTr_{\bar{A}} \left(\psscalebox{.85}{\begin{pspicture}[shift=-2.4](-.2,-3.9)(3.2,1.2)
        \scriptsize
        \psline[ArrowInside=->](1.1,-0.6)(0,.5)\rput(0.6,-0.3){$d$}
        \psline[ArrowInside=->](1.1,-0.6)(1.7,0)\rput(1.6,-0.3){$e$}
        \psline[ArrowInside=->](1.7,0)(1.2,.5) \rput(1.35,.2){$b$}
        \psline[ArrowInside=->](1.7,0)(2.2,.5)\rput(2.1,.2){$\bar{b}$}
        \psline[ArrowInside=->](1.6,-1.1)(1.1,-0.6)\rput(1.2,-0.9){$\bar{c}$}
        \psline[ArrowInside=->](1.6,-1.1)(3.2,.5)\rput(2.2,-0.7){$c$}
        \psline[ArrowInside=->](1.1,-2.1)(1.6,-1.6) \rput(1.1,-1.9){$\bar{c}'$}
        \psline[ArrowInside=->](3.2,-3.2)(1.6,-1.6) \rput(2.65,-2.4){$c'$}
        \psline[ArrowInside=->](0,-3.2)(1.1,-2.1)\rput(0.5,-2.4){$d'$}
        \psline[ArrowInside=->](1.7,-2.7)(1.1,-2.1)\rput(1.65,-2.4){$e'$}
        \psline[ArrowInside=->](1.2,-3.2)(1.7,-2.7)\rput(2.1,-2.8){$\bar{b}'$}
        \psline[ArrowInside=->](2.2,-3.2)(1.7,-2.7)\rput(1.35,-2.8){$b'$}
\end{pspicture}} \right)
&= \delta_{b,b'}\delta_{c,c'}\delta_{d,d'} \psscalebox{.85}{\begin{pspicture}[shift=-2.4](.7,-3.9)(3.2,1.2)
        \scriptsize
        \psline[ArrowInside=->](1.1,-0.6)(.5,0)\rput(0.6,-0.3){$d$}
        \psline(1.1,-.6)(.5,0)(.5,.25)(1.2,.95)(2.2,.95)(3.15,0) (3.15,-2.7)(2.2,-3.7)(1.2,-3.7)(.5,-3)(.5,-2.7)(1.1,-2.1)
        \psline[ArrowInside=->](1.1,-0.6)(1.7,0)\rput(1.6,-0.3){$e$}
        \psline[ArrowInside=->](1.7,0)(1.45,.25)\rput(1.45,0.05){$b$}
        \psline(1.7,0)(1.45,.25)(1.95,.75)(2.95,-.25)(2.95,-2.45) (1.95,-3.45)(1.45,-2.95)(1.7,-2.7)
        \psline[ArrowInside=->](1.7,0)(1.95,.25)\rput(1.95,0.05){$\bar{b}$}
        \psline(1.7,0)(1.95,.25)(2.7,-.5)(2.7,-2.2)(1.95,-2.95)(1.7,-2.7)
        \psline[ArrowInside=->](1.6,-1.1)(1.1,-0.6)\rput(1.2,-0.9){$\bar{c}$}
        \psline[ArrowInside=->](1.6,-1.1)(2.2,-.5)\rput(2.,-0.9){$c$}
        \psline(1.5,-1)(1.6,-1.1)(1.7,-1)
        \psline(1.5,-1.7)(1.6,-1.6)(1.7,-1.7)
        \psline(1.6,-1.1)(2.2,-.5)(2.45,-.75)(2.45,-1.95)(2.2,-2.2)(1.6,-1.6)
        \psline[ArrowInside=->](1.1,-2.1)(1.6,-1.6) \rput(1.1,-1.9){$\bar{c}$}
        \rput(2.1,-1.85){$c$}
        \rput(1.45,-2.7){$b$}
        \rput(1.95,-2.7){$\bar{b}$}
        \rput(.6,-2.4){$d$}
        \psline[ArrowInside=->](2.2,-2.2)(1.6,-1.6)
        \psline[ArrowInside=->](.5,-2.7)(1.1,-2.1)
        \psline[ArrowInside=->](1.7,-2.7)(1.1,-2.1)\rput(1.65,-2.4){$e'$}
        \psline[ArrowInside=->](1.45,-2.95)(1.7,-2.7)
        \psline[ArrowInside=->](1.95,-2.95)(1.7,-2.7)
\end{pspicture}}
\\ &= d_b \sqrt{d_c d_d} \delta_{b,b'}\delta_{c,c'}\delta_{d,d'}\delta_{e,e'}
\end{split}
\end{equation}
to find the reduced density matrix for $A$:
\begin{equation}
\begin{split}
\arho_A &= \aTr_{\bar{A}} \ket{\psi_{\text{cut}}}\bra{\psi_{\text{cut}}} = \sum_{\substack{a,b,c,\\d,e,a'}} \frac{\psi_{a,b,c,d,e}\psi^*_{a',b,c,d,e}}{ d_b \sqrt{d_d}}
\psscalebox{1}{
\begin{pspicture}[shift=-2.2](.2,-3.3)(3.5,1.2)
        \scriptsize
        \rput(0.5,0.5){$\otimes$}
        \psline(0.5,0)(1,0.5)(0.5,1)(0,0.5)(0.5,0)
        \psline[ArrowInside=->](0.5,0)(0,0.5)\rput(0.05,0.25){$a$}
        \psline[ArrowInside=->](1,-.5)(0.5,0)\rput(.55,-.3){$d$}
        \psline[ArrowInside=->](1,-.5)(2,.5)\rput(1.7,0){$\bar{d}$}
        \psline[ArrowInside=->](2.75,0)(2.25,.5)\rput(2.4,.2){$b$}
        \psline[ArrowInside=->](2.75,0)(3.25,.5)\rput(3.15,.2){$\bar{b}$}
\rput(0,1){
        \psline(0,-3.2)(.5,-2.7)(1,-3.2)(.5,-3.7)(0,-3.2) \psline[ArrowInside=->](.5,-3.7)(0,-3.2)\rput(.5,-3.2){$\otimes$}\rput(.15,-3.6){$a'$}
        \psline[ArrowInside=->](.5,-2.7)(1,-2.2)\rput(.5,-2.4){$d$}
        \psline[ArrowInside=->](2,-3.2)(1,-2.2)\rput(1.7,-2.6){$\bar{d}$}
        \psline[ArrowInside=->](2.25,-3.2)(2.75,-2.7)\rput(2.4,-2.8){$b$}
        \psline[ArrowInside=->](3.25,-3.2)(2.75,-2.7)\rput(3.15,-2.8){$\bar{b}$}}
\end{pspicture}}
.
\end{split}
\end{equation}

Alternatively, we can trace over the region $A$:
\begin{equation}
\begin{split}
\aTr_A \left(
\psscalebox{.85}{
\begin{pspicture}[shift=-2.2](0,-3.3)(3.4,1.2)
        \scriptsize
        \rput(0.5,0.5){$\otimes$}
        \psline(0.5,0)(1,0.5)(0.5,1)(0,0.5)(0.5,0)
        \psline[ArrowInside=->](0.5,0)(0,0.5)\rput(0.05,0.25){$a$}
        \psline[ArrowInside=->](1,-.5)(0.5,0)\rput(.55,-.3){$d$}
        \psline[ArrowInside=->](1,-.5)(2,.5)\rput(1.7,0){$\bar{d}$}
        \psline[ArrowInside=->](2.75,0)(2.25,.5)\rput(2.4,.2){$b$}
        \psline[ArrowInside=->](2.75,0)(3.25,.5)\rput(3.15,.2){$\bar{b}$}
\rput(0,1){
        \psline(0,-3.2)(.5,-2.7)(1,-3.2)(.5,-3.7)(0,-3.2) \psline[ArrowInside=->](.5,-3.7)(0,-3.2)\rput(.5,-3.2){$\otimes$}\rput(.15,-3.6){$a'$}
        \psline[ArrowInside=->](.5,-2.7)(1,-2.2)\rput(.5,-2.4){$d'$}
        \psline[ArrowInside=->](2,-3.2)(1,-2.2)\rput(1.7,-2.6){$\bar{d}'$}
        \psline[ArrowInside=->](2.25,-3.2)(2.75,-2.7)\rput(2.4,-2.8){$b'$}
        \psline[ArrowInside=->](3.25,-3.2)(2.75,-2.7)\rput(3.15,-2.8){$\bar{b}'$}}
\end{pspicture}}
 \right)
&=\frac{\delta_{a,a'}\delta_{d,d'}\delta_{b,b'}}{d_a}
\psscalebox{.85}{
\begin{pspicture}[shift=-2.2](-.1,-3.3)(3.75,1.2)
        \scriptsize
        \psline[ArrowInside=->](0.5,0)(1,.5)\rput(.95,.25){$\bar{a}$}
        \psline[ArrowInside=->](0.5,0)(0,0.5)\rput(0.05,0.25){$a$}
        \psline[ArrowInside=->](1,-.5)(0.5,0)\rput(.55,-.3){$d$}
        \psline[ArrowInside=->](1,-.5)(1.5,0)\rput(1.4,-.35){$\bar{d}$}
        \psline(1,-.5)(1.5,0)(1.75,-.25)(1.75,-1.5)(1.5,-1.75)(1,-1.25) (.5,-1.75)(1,-2.25)(1.5,-2.25)(2,-1.75)(2,0)(1.5,.5)(1,.5)(.5,0) (0,0.5)(.5,1)(1.5,1)(2.25,.25)(2.25,-2)(1.5,-2.75)(.5,-2.75) (0,-2.25)(.5,-1.75)
        \psline[ArrowInside=->](0,-2.25)(.5,-1.75)\rput(.0,-2){$a$}
        \psline[ArrowInside=->](1,-2.25)(.5,-1.75)\rput(1,-2){$\bar{a}$}
        \psline[ArrowInside=->](1.5,-1.75)(1,-1.25)\rput(1.4,-1.35){$\bar{d}$}
        \psline[ArrowInside=->](.5,-1.75)(1,-1.25)\rput(.5,-1.4){$d$}
\rput(.5,-.5){ \psline[ArrowInside=->](2.75,0)(2.5,.25)\rput(2.45,.1){$b$}
       \psline[ArrowInside=->](2.75,0)(3.,.25)\rput(3.,.1){$\bar{b}$}
       \psline(2.75,0)(3,.25)(3.25,0)(3.25,-.5)(3,-.75)(2.75,-.5) (2.5,-.75)(3,-1.25)(3.5,-.75)(3.5,.25)(3,.75)(2.5,.25)(2.75,0)(2.85,.1)
       \psline[ArrowInside=->](2.5,-.75)(2.75,-.5)\rput(2.45,-.5){$b$}
       \psline[ArrowInside=->](3,-.75)(2.75,-.5)\rput(3,-.5){$\bar{b}$}
       }
\end{pspicture}}
\\ &= d_b \sqrt{d_d}\delta_{a,a'}\delta_{d,d'}\delta_{b,b'}
\end{split}
\end{equation}
to find the reduced density matrix for $\bar{A}$:
\begin{equation}
\begin{split}
\arho_{\bar{A}} &= \aTr_{A} \ket{\psi_{\text{cut}}}\bra{\psi_{\text{cut}}} = \sum_{\substack{a,b,c,d,e,\\c',e'}} \frac{\psi_{a,b,c,d,e}\psi^*_{a,b,c',d,e'}}{d_b \sqrt{d_c d_d} }
\psscalebox{1}{\begin{pspicture}[shift=-2.4](.2,-3.9)(3.2,1.2)
        \scriptsize
        \psline[ArrowInside=->](1.1,-0.6)(0,.5)\rput(0.6,-0.3){$d$}
        \psline[ArrowInside=->](1.1,-0.6)(1.7,0)\rput(1.6,-0.3){$e$}
        \psline[ArrowInside=->](1.7,0)(1.2,.5) \rput(1.35,.2){$b$}
        \psline[ArrowInside=->](1.7,0)(2.2,.5)\rput(2.1,.2){$\bar{b}$}
        \psline[ArrowInside=->](1.6,-1.1)(1.1,-0.6)\rput(1.2,-0.9){$\bar{c}$}
        \psline[ArrowInside=->](1.6,-1.1)(3.2,.5)\rput(2.2,-0.7){$c$}
        \psline[ArrowInside=->](1.1,-2.1)(1.6,-1.6) \rput(1.1,-1.9){$\bar{c}'$}
        \psline[ArrowInside=->](3.2,-3.2)(1.6,-1.6) \rput(2.65,-2.4){$c'$}
        \psline[ArrowInside=->](0,-3.2)(1.1,-2.1)\rput(0.5,-2.4){$d$}
        \psline[ArrowInside=->](1.7,-2.7)(1.1,-2.1)\rput(1.65,-2.4){$e'$}
        \psline[ArrowInside=->](1.2,-3.2)(1.7,-2.7)\rput(2.1,-2.8){$\bar{b}$}
        \psline[ArrowInside=->](2.2,-3.2)(1.7,-2.7)\rput(1.35,-2.8){$b$}
\end{pspicture}}
.
\end{split}
\end{equation}
\end{enumerate}

\subsection{Framing}
\label{sec:framing}

Finally, when working with anyon models on a higher genus surface it is necessary to specify a framing of the charge lines.  That is, charge lines should be thickened into ribbons, so that the diagram accurately keeps tracks of twists in a ribbon.  These twists correspond to the phase a particle with fractional statistics picks up when undergoing a $2\pi$ rotation.  There is no canonical choice of framing for a general three manifold.  There is, however, a definite law for how partition functions transform under a change of framing, i.e. under the modular $\mathcal{T}$ transformations, known as Dehn twists. Thus, we must simply pick some framing and be consistent~\cite{Witten89}. The framing can be defined as the continuous map from the topological charge line inside the surface to a projection of the charge line on the surface, which defined a ribbon. One can think of the projection of the line onto the surface as being specified by the path along which quasiparticles were transported and fused in order to generate the corresponding state. Note that a Dehn twist of the surface will put a corresponding twist in the ribbon.

While the framing is technically necessary, we note that it will have no effect on the entanglement entropies we calculate in the following section.  Similar to the conventional entanglement entropy of Section~\ref{sec:EE}, the AEE is only a well-defined entanglement measure if the full system is in a pure state ${\arho_{A\bar{A}}=\ket{\psi}\bra{\psi}}$.  Writing the Schmidt decomposition of the state as ${\ket{\psi}=\sum_\alpha \lambda_{\alpha} \ket{\psi^A_\alpha}\ket{\psi^{\bar{A}}_\alpha}}$, we see the anyonic reduced density matrix for $A$ will take the form ${\arho_A = \sum_\alpha |\lambda_\alpha|^2 \ket{\psi^A_\alpha}\bra{\psi^A_\alpha}}$.  The framing keeps track of twists in the diagram, which contribute a phase to the untwisted diagram.  This phase of $\ket{\psi^A_\alpha}$ will always be paired with its complex conjugate when considering the density matrix $\arho_A$, and thus will cancel out of the AEE calculations.  We simplify our expressions in the next section by neglecting the framing, which should be interpreted as some implicit choice having been made.

\section{Topological Entanglement Entropy in Anyon Models II}
\label{sec:ATEE}

We are now in a position to compute the AEE for a bipartition of a topological state on a compact orientable surface with arbitrary genus and number of boundaries.  Central to our method is the derivation of the reduced density matrix from the partitioning of the surface such that we account for correlations across the boundary. Our approach may be viewed as a generalization of the Kitaev-Preskill derivation of the TEE.

We first review the Kitaev-Preskill method for calculating the TEE, which used a geometric cancellation argument to isolate the TEE from the entanglement entropies of seven geometrically different partitions of the plane into a disk and its complement (we refer the reader to Ref.~\cite{Kitaev06b} for more details):
\begin{enumerate}
\item Pair the plane with its time-reversal conjugate surface.
\item Join the two surfaces by adiabatically inserting four wormholes that connect the surfaces and gluing the two planes together along a circle at infinity. ``Adiabatic insertion" means that the system remains in its ground state during the entire process of inserting the wormholes.  Thus, an anyon circling a wormhole should detect no difference from an anyon circling a region in the plane containing no topological excitations, i.e., each wormhole is threaded by a trivial topological charge line. The location of the wormholes corresponds to the ``corners'' of the different disk partitions of the plane.
\item For each choice of geometric partition, cut the surface along the partition boundary, which now runs along the regions between wormholes, i.e. around the tubes connecting the different partition regions. A partition cut divides the surface into disjoint compact, orientable surfaces with either three or four punctures, depending on the choice of partition.
\item Compute the state (reduced density matrix) and entanglement entropy of the resulting surfaces using standard TQFT methods. More specifically, this involves rewriting the state of the uncut doubled system in a basis that is more suitable to the ensuing cut by (a) applying modular $\mathcal{S}$-transformations to rewrite the trivial charge line through each wormhole as an $\omega_0$-loop circling the throat of the wormhole, and (b) applying $F$-moves to all the topological charge lines threading the tubes that will be cut, so that there is a single topological charge line threading each boundary component generated by the partition cut (i.e. to obtain the basis states in which each resulting puncture has a definite value of topological charge).
\item Add and subtract the entanglement entropies of the seven geometric partitions such that their linear dependence cancels and the topological contribution survives.
\end{enumerate}

We generalize the Kitaev-Preskill method to enable the computation of all topological contributions to the entanglement entropy, including the TEE and anyonic entanglement, for any compact region $A$ of a 2D topological phase living on a compact, orientable surface $M$ with any genus and number of punctures and/or quasiparticles using the following steps, which will be illustrated in detail for several examples:
\begin{enumerate}
\item Pair the surface $M$ with its time-reversal conjugate $M^{\ast}$. (When embedded in 3D, we assume the original surface is enclosed by the conjugate surface.)
\item Adiabatically insert $n$ wormholes along the original partition boundary $\partial A$. Each wormhole is threaded by a trivial topological charge line. The system will now look like two parallel surfaces connected by a series of tubes.~\footnote{Not a big truck.~\cite{Stevens06}} We denote this new surface by $\mathbb{M}$ and the doubled regions corresponding to $A$ and $\bar{A}$ of the un-doubled system are denoted by $\mathbb{A}$ and $\bar{\mathbb{{A}}}$, respectively. The partition boundary $\partial \mathbb{A}$ has $n$ connected components, each running along the regions between two wormholes, i.e. around the tubes connecting $\mathbb{A}$ and $\bar{\mathbb{{A}}}$.
\item Cut $\mathbb{M}$ along the partition boundary $\partial \mathbb{A}$. The partition cut divides the surface into disjoint compact, orientable surfaces $\mathbb{A}$ and $\bar{\mathbb{A}}$, each of which obtains $n$ new punctures from the cut, corresponding to the boundary components where regions $\mathbb{A}$ and $\bar{\mathbb{{A}}}$ were formerly connected.
\item Compute the state (reduced density matrix) and AEE entropy of the resulting surface $\mathbb{A}$. More specifically, this involves rewriting the state of the uncut doubled system in a basis that is more suitable to the ensuing cut by (a) applying modular $\mathcal{S}$-transformations to rewrite the trivial charge line through each wormhole as an $\omega_0$-loop circling the throat of the wormhole, and (b) applying $F$-moves to all the topological charge lines threading the tubes that will be cut, so that there is a single topological charge line threading each boundary component generated by the partition cut.
\item Taking $n$ large,~\footnote{Taking $n$ large corresponds to inserting as many wormholes along the boundary as possible. In other words, one inserts roughly one wormhole per regularization length, so $n \sim L / \ell$, as before.} the AEE of region $\mathbb{A}$ will exhibit a term that is linear in $n$, which is identified as the contribution that is linear in the boundary length, and a constant term, which is identified as the topological contribution. The contributions from the boundary (i.e. the linear term and the TEE) are divided by two for the contribution to the entanglement entropy of $A$, the original (un-doubled) system.
\end{enumerate}

Given the topological reduced density matrix for region $\mathbb{A}$, the AEE can be evaluated using the anyonic formalism discussed in Section~\ref{sec:anyonmodelshigher}. When there are punctures and/or quasiparticles in the system, one can choose whether or not to also double this content of the system, as long as one is careful to correctly attribute the corresponding contributions when accounting for the doubling. Similarly, if there is genus, one can choose different states (topological charge lines winding around the non-contractible cycles). We will utilize these options in our analysis when it simplifies the computations.

When writing the topological state of the doubled system with wormholes, one must be careful to identify the correct total number of non-contractible cycles of the surface $\mathbb{M}$. On the doubled infinite plane, there is a one-to-one correspondence between wormholes and non-contractible cycles.  However, on the doubled sphere, the first wormhole inserted does not create a non-contractible cycle, but simply yields the ``connected sum'' of the two spheres, which is a single sphere. Each subsequent wormhole inserted will then increase the genus of the resulting surface by one. A consequence of this is the the normalization on the doubled sphere will differ from the normalization on the doubled plane by a factor of $\mathcal{D}$, when written with $\omega_0$-loops encircling every wormhole. More generally, when we double a connected, compact surface of genus $g$ and insert $n$ wormholes attaching the doubled surfaces, the resulting surface will have genus $2g+n-1$. This is, again, because the first wormhole inserted simply creates a connected sum of the two surfaces, and each subsequent wormhole increases the genus by one. We will restrict our attention to compact surfaces in order to make the analysis more rigorous, but similar methods can be used for non-compact surfaces.

One might be worried that inserting a large number of closely spaced wormholes would introduce non-contractible cycles whose lengths are too small to provide topological protection of the corresponding state degeneracies associated with them. In particular, if a cycle in $\mathbb{M}$ is not long compared with the correlation length $\xi$, non-universal microscopic effects will generically lead to an energy splitting that favors different values of topological charge lines threading that cycle. This is, however, not a problem for our construction for the following reasons. The potentially small cycles introduced by inserting the wormholes are $L_{\text{throat}}$, the circumference of a given wormhole's throat, and $L_{\text{tube}}$, the circumference of the tubes connecting regions $\AD$ and $\bar{\AD}$. It is perfectly acceptable for $L_{\text{throat}}$ to be small, because we are already requiring a specific value of topological charge line threading the throat of the wormhole, namely the trivial charge $0$. As long as the Hamiltonian of the system is such that trivial charge line threading the wormhole is energetically favored by the adiabatic insertion of the wormhole, its throat circumference can be arbitrarily small (meaning down to the regularization length). In fact, this condition may be viewed as part of the definition of the process of adiabatically inserting a wormhole. On the other hand, it is important that $L_{\text{tube}}$ be much larger than $\xi$, because the ground state of $\mathbb{M}$ will require superpositions of the values of topological charge line threading these cycles. At first glance, one might think that this should dissuade us from inserting wormholes separated by a distance $d\approx \ell$. However, the circumference of the tube is roughly $L_{\text{tube}} \sim d+h$, where $h$ is the ``height" of the wormholes, i.e. the spacing between conjugate surfaces. Since we are free to choose $h$, we can let it be arbitrarily large, which allows us to also have arbitrarily small $d$ without sacrificing the necessary topological degeneracy.

There are several benefits of the method we present: (1) It applies beyond the ground state, to states containing anyonic excitations, i.e., boundaries and/or quasiparticles carrying topological charge. (2) It makes the origin of the TEE more explicit.  (3) It captures the topological contribution to the boundary law term in the entanglement entropy.  (4) It may be used to extract the TEE from the R\'enyi entropy for arbitrary topological phases.

In this section, we use our method (described above) to calculate the topological contribution to the entanglement entropy. We first illustrate the approach in the simplest example of the ground state on a sphere partitioned into two disks, Section~\ref{sec:unpunctured-disk}. We analyze this example in greater detail than subsequent examples, as it exhibits most of the crucial methodology that will be repeated. In Sections~\ref{sec:disk}-\ref{sec:three-punctured-sphere}, we apply the same method to an excited state of a disk cut from the sphere, an annulus cut from the sphere, an annular segment cut from the torus, and a 3-punctured sphere cut from the sphere.  In Section~\ref{sec:general}, we discuss the general form of the entanglement entropy for a subregion of a compact, orientable surface of arbitrary genus and number of punctures/quasiparticles.

\subsection{Sphere Partitioned into Two Disks}
\label{sec:unpunctured-disk}

Before diving into the derivation of the reduced density matrix, we first comment on how to visualize the surfaces discussed in this section.  Consider a sphere partitioned into two disks, $A$ and $\bar{A}$.  For ease of illustration, we zoom in so that locally the surface looks planar.

\begin{center}
\noindent\includegraphics[width=0.6\linewidth]{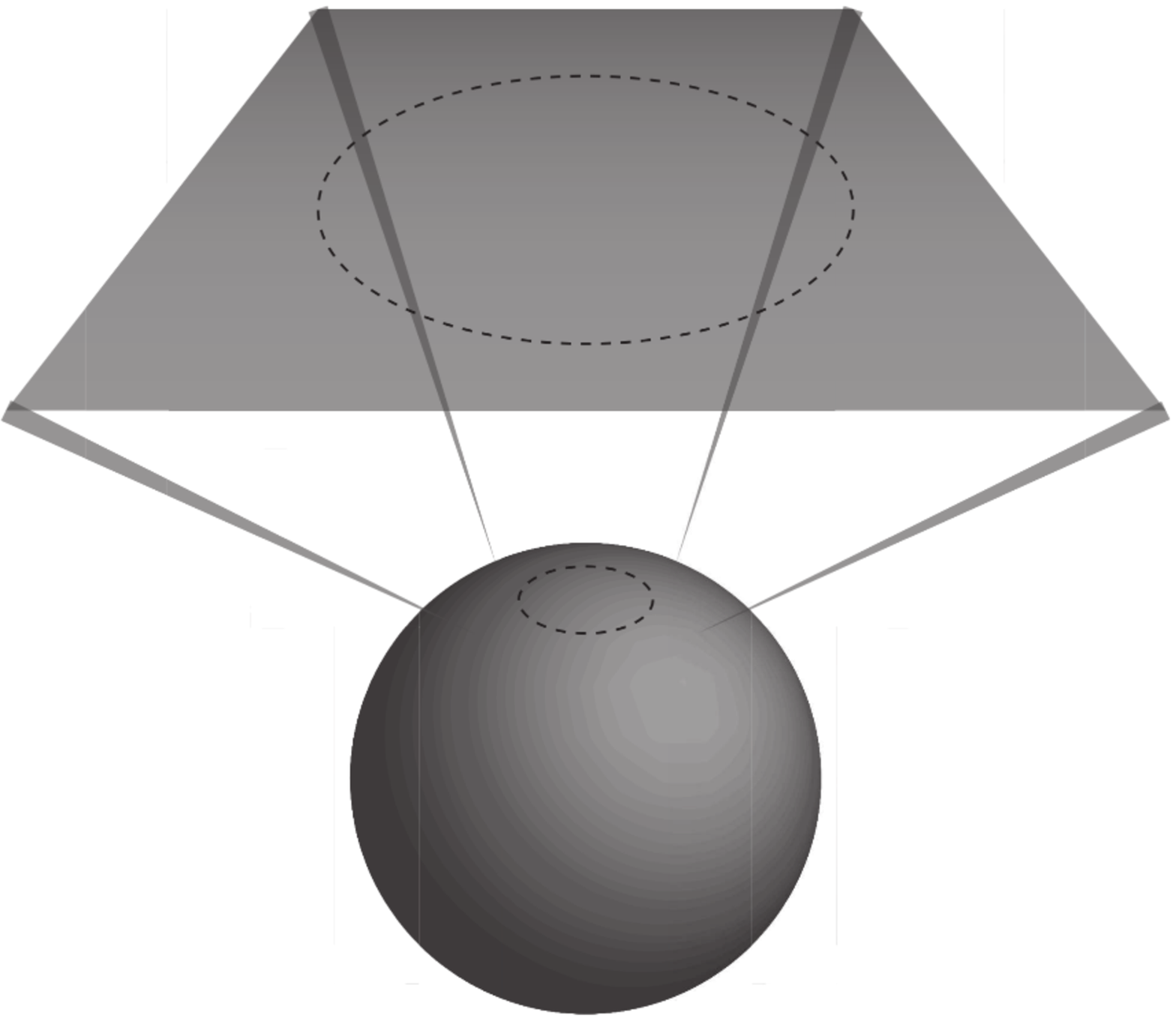}
\end{center}
We pair the original surface $M$ with its time-reversal conjugate $M^{\ast}$, and join the two surfaces by adiabatically inserting $n$ wormholes along the partition boundary separating $A$ from $\bar{A}$.  The resulting surface $\mathbb{M}$ has genus $g=n-1$.

\begin{center}
\noindent\includegraphics[width=.9\linewidth]{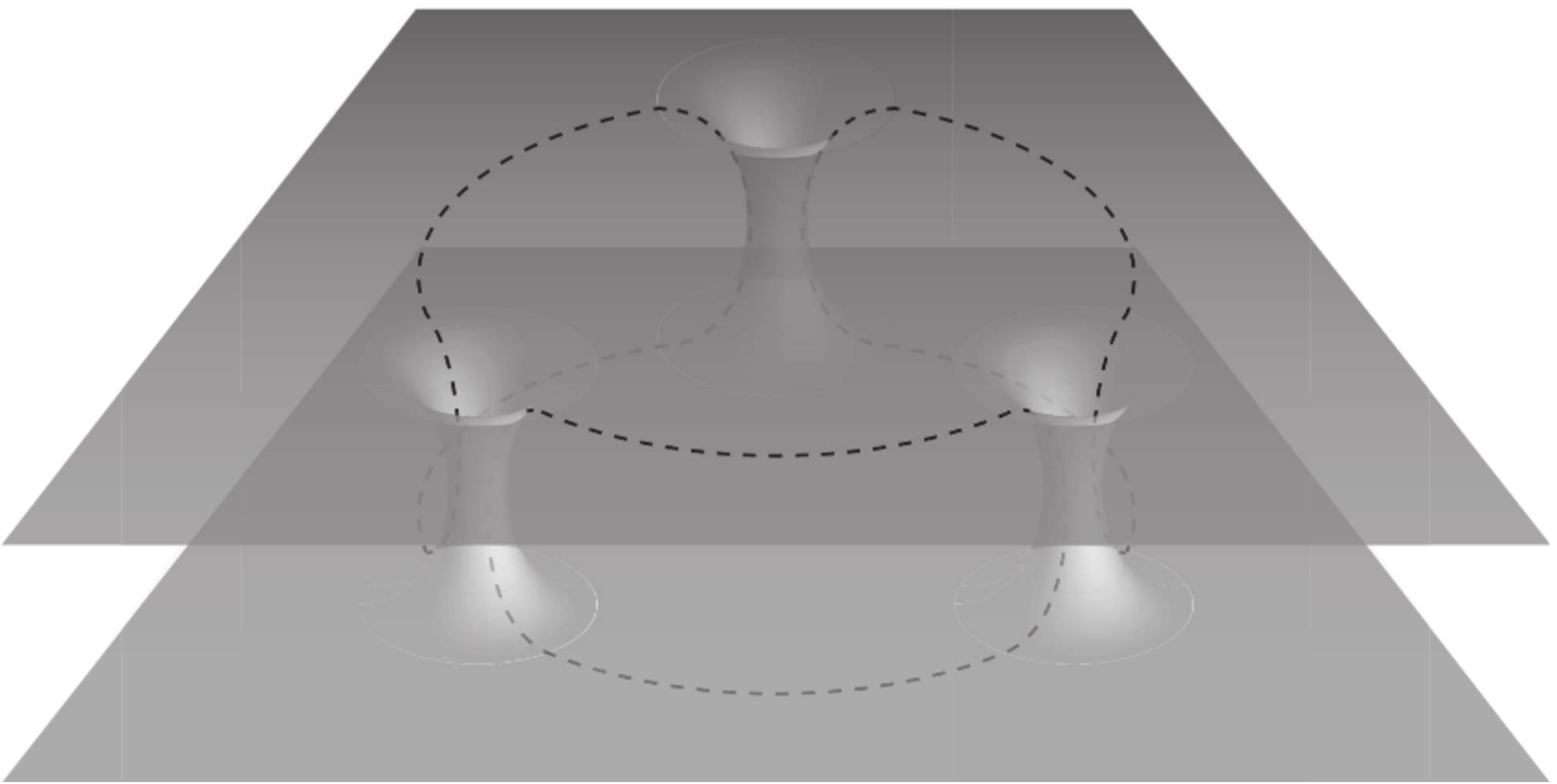}
\end{center}

\noindent Here, we show the case with $n=3$ wormholes. The partition boundary is now broken into segments, each of which runs between two wormholes and pass through the wormholes between the upper layer region of $\mathbb{M}$ and the lower layer region, as indicated in the above by dashed lines. In order to find the reduced density matrix for the doubled region $\mathbb{A}$, we cut the surface along the new partition boundary, resulting in the following surfaces for $\mathbb{A}$ and $\bar{\mathbb{A}}$:

\begin{center}
\noindent\includegraphics[width=.4\linewidth]{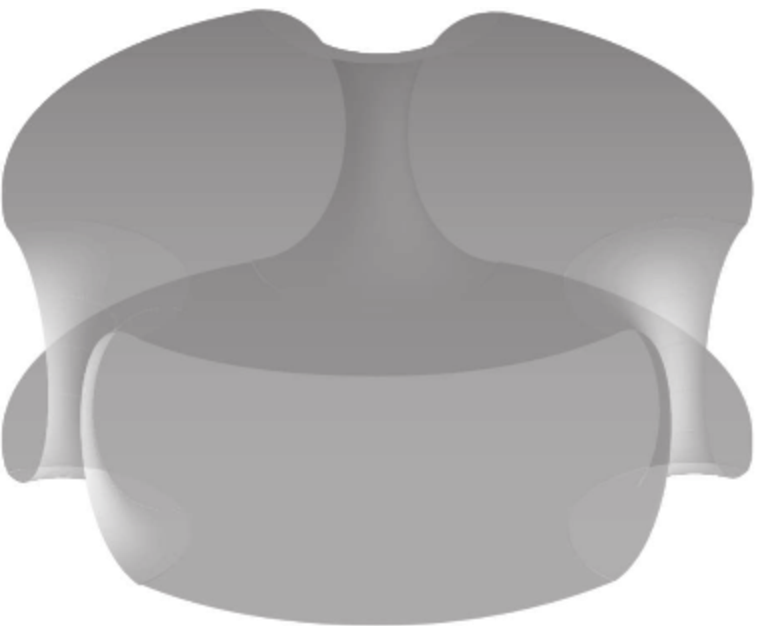}
\end{center}

\begin{center}
\noindent\includegraphics[width=.9\linewidth]{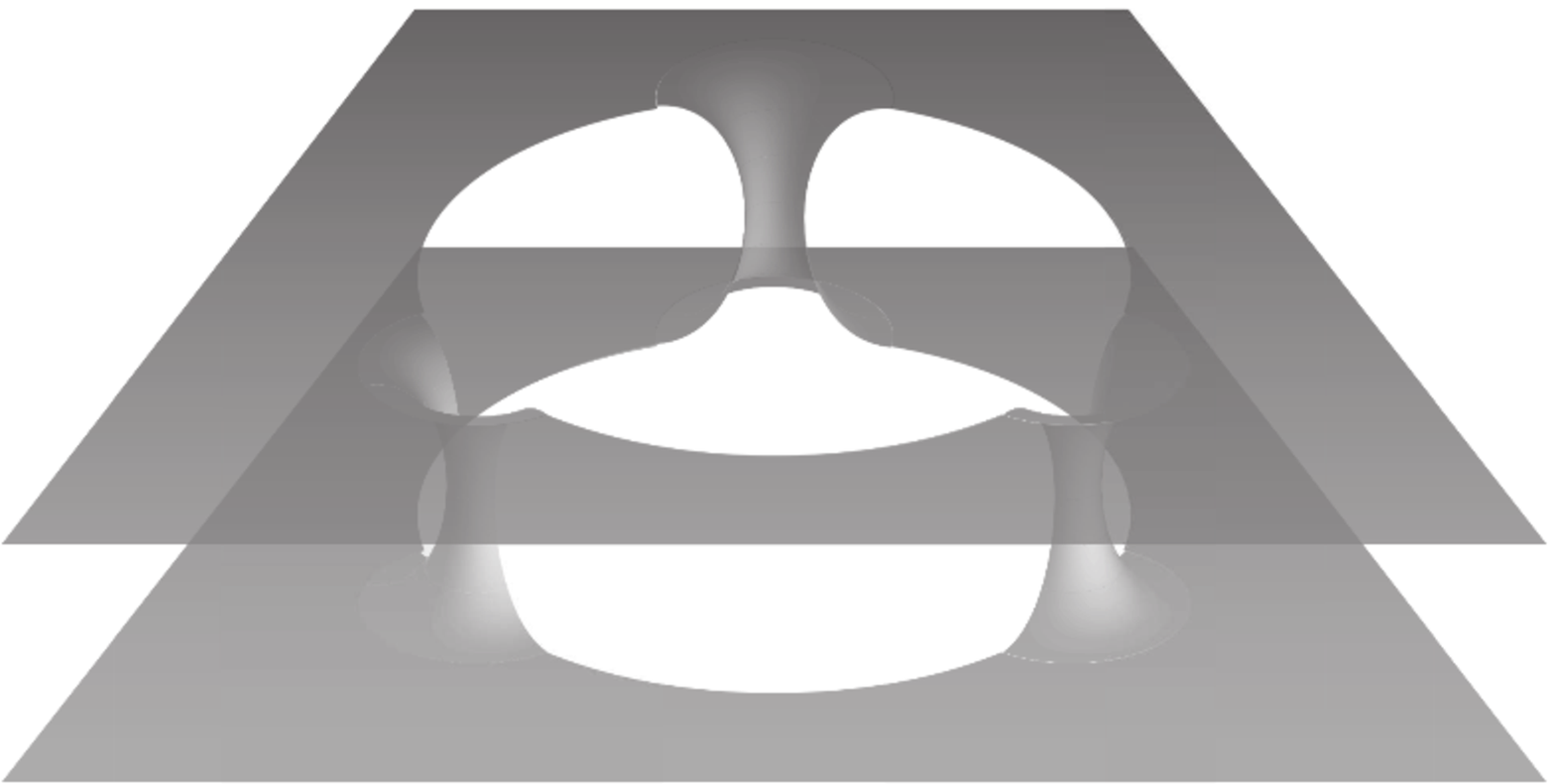}
\end{center}
Each of these regions are topologically equivalent to a sphere with $n$ punctures:

\begin{center}
\noindent\includegraphics[width=.4\linewidth]{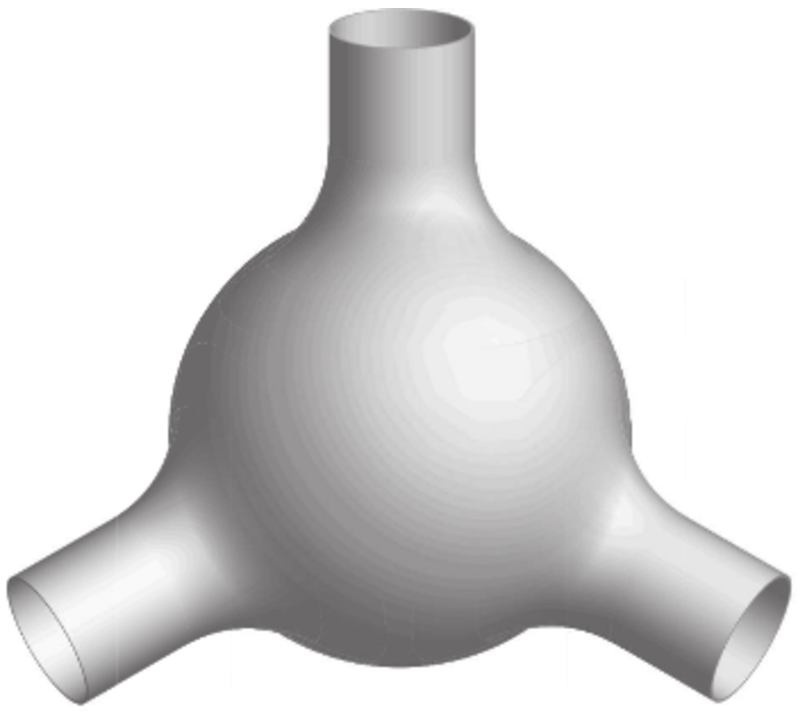}
\end{center}

In the remainder of this section, we will omit the dashed lines indicating the partition boundary in the pictures of the surfaces, but we will include them in the corresponding anyon diagram representation of the state.

Having oriented ourselves to what the three-dimensional embedding of our surfaces look like, we are now ready to derive the corresponding anyonic reduced density matrix for $\mathbb{A}$. First, recall that adiabatic insertion implies that each wormhole is threaded by a trivial topological charge line.

\begin{center}
\noindent\includegraphics[width=.9\linewidth]{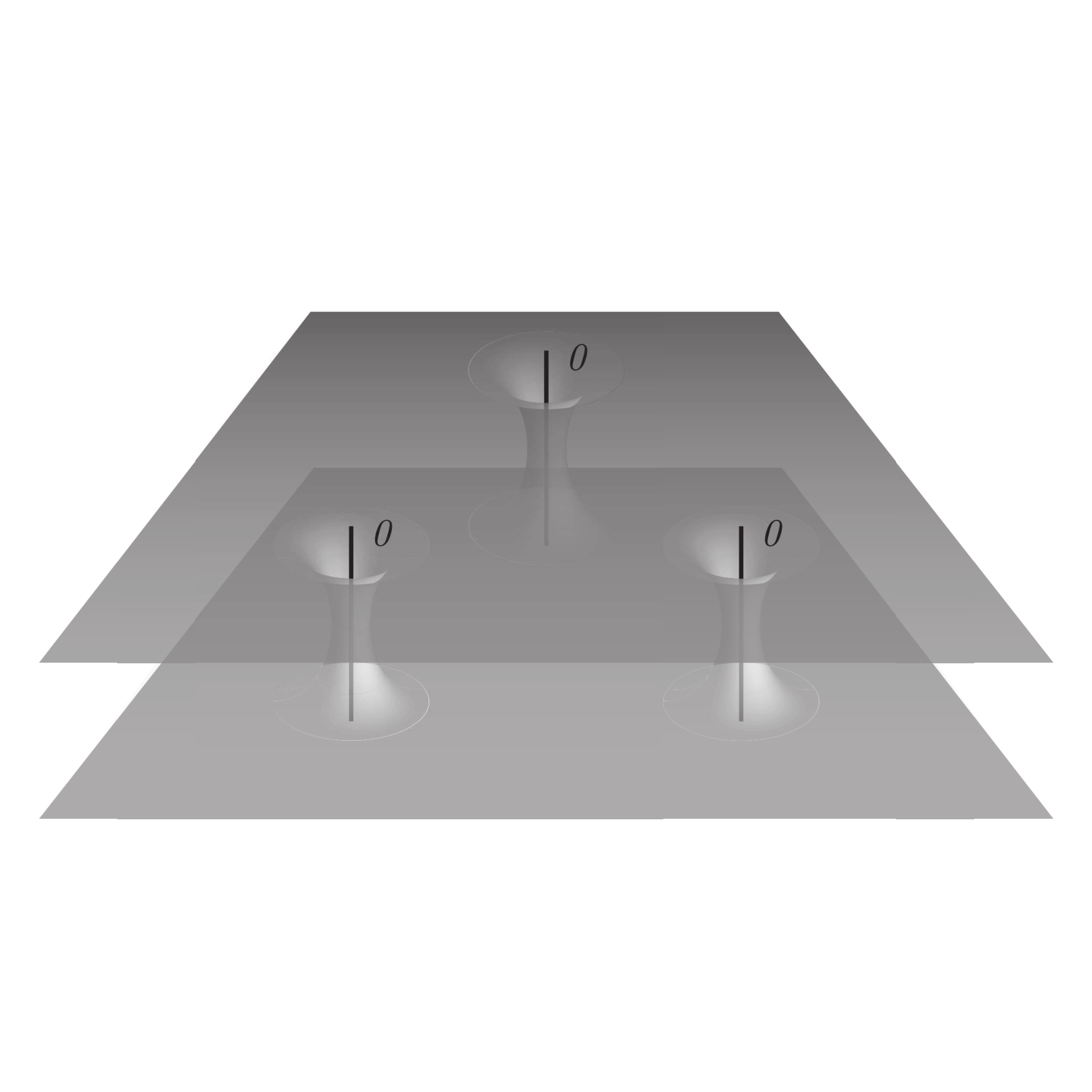}
\end{center}
We can use the modular $\mathcal{S}$-transformation to rewrite the topological charge line threading a given wormhole in terms of an $\omega_0$-loop circling the throat of that wormhole, up to an overall normalization factor of the state, essentially converting between the inside and outside bases (see Section~\ref{sec:inside-outside-basis}).
\begin{center}
\noindent\includegraphics[width=.9\linewidth]{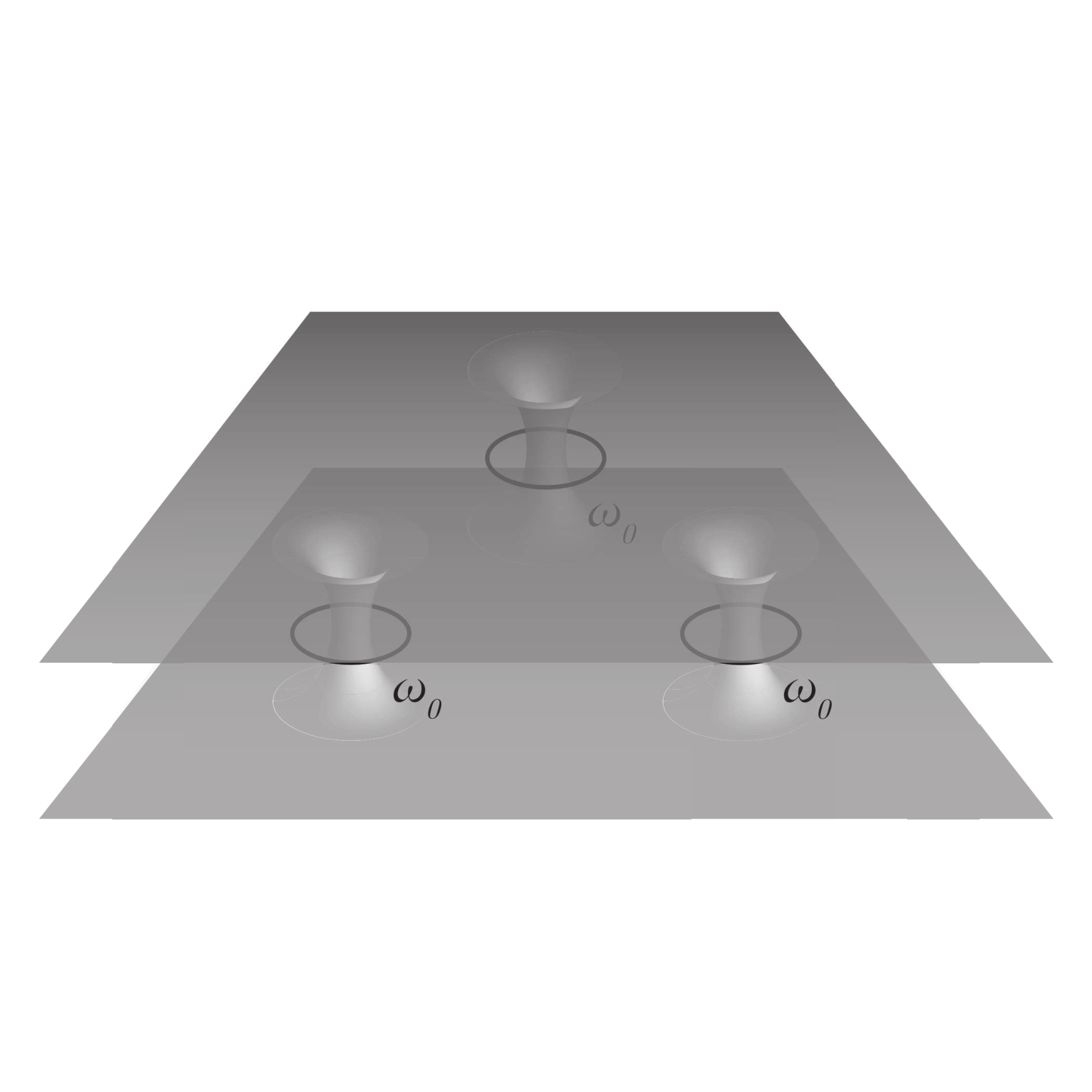}
\end{center}
This claim is justified by first isolating a given wormhole of the surface (when there is more than one wormhole), which locally takes the form of a punctured torus with charge line $0$ through the handle. Next, one can apply the modular $\mathcal{S}$-transformation for a punctured torus, described in Section~\ref{sec:inside-outside-basis}. When the topological charge threading the handle is $b=0$ for a punctured torus, the charge on the puncture it is necessarily $c=0$, i.e. the punctured torus state in the outside basis is $\ket{ (0);0 }_{\text{outside}}$. Applying the modular $\mathcal{S}$-transformation to the punctured torus state in the outside basis gives the state in terms of the inside basis
\begin{equation}
\ket{ (0);0 }_{\text{outside}} = \sum_{a} \mathcal{S}_{0a} \ket{ (a);0 }_{\text{inside}} = \sum_{a} \frac{d_a }{\mathcal{D}} \ket{ (a);0 }_{\text{inside}}
,
\end{equation}
which is the same as representing the state by having an $\omega_0$-loop circling the throat of that wormhole, up to an overall normalization factor.
Thus, the state of the system can be re-expressed in the basis represented by topological charge lines that thread the region inside the surface $\mathbb{M}$.

Using the diagrammatic formalism, we can write the state as
\begin{equation}\label{eq:omega-loop-psi}
\ket{\psi} = \mathcal{D}^{n-1}
 \begin{pspicture}[shift=-2](0,-1.8)(4,1.4)
 \scriptsize
 \psline[linestyle=dashed](1.35,1)(1.75,1)
 \rput(2,1){$\dots$}
 \psline[linestyle=dashed](2.65,1)(2.25,1)
\psline[linestyle=dashed](1.15,1)(.75,.1)
\psline[linestyle=dashed](1.15,-1)(.75,-.1)
\psline[linestyle=dashed](1.35,-1)(2.65,-1)
\psline[linestyle=dashed](2.85,-1)(3.25,-.1)
\psline[linestyle=dashed](2.85,1)(3.25,.1)
 \psellipse[border=1.5pt](1.25,1)(.25,.5)\psline[ArrowInside=->](1.015,0.9)(1.015,1.05)\rput(.8,.75){$\omega_0$}
 \psellipse[border=1.5pt](2.75,1)(.25,.5)\psline[ArrowInside=->](2.515,.9)(2.515,1.05)\rput(2.3,.75){$\omega_0$}
 \psellipse[border=1.5pt](.75,0)(.25,.5)\psline[ArrowInside=->](.515,-.1)(.515,.05)\rput(.3,-.25){$\omega_0$}
 \psellipse[border=1.5pt](1.25,-1)(.25,.5)\psline[ArrowInside=->](1.015,-1.1)(1.015,-.95)\rput(.8,-1.2){$\omega_0$}
 \psellipse[border=1.5pt](2.75,-1)(.25,.5)\psline[ArrowInside=->](2.515,-1.1)(2.515,-.95)\rput(2.3,-1.2){$\omega_0$}
 \psellipse[border=1.5pt](3.25,0)(.25,.5)\psline[ArrowInside=->](3.015,-.1)(3.015,.05)\rput(2.8,-.25){$\omega_0$}
 \rput(1.25,1){$\odot$}
 \rput(2.75,1){$\odot$}
 \rput(.75,0){$\odot$}
 \rput(1.25,-1){$\odot$}
 \rput(2.75,-1){$\odot$}
 \rput(3.25,0){$\odot$}
 \end{pspicture}
,
\end{equation}
where the dashed line indicates the partition boundary, and we have introduced the notation $\odot$ to represent the throats of the wormholes around which the $\omega_0$-loops wind. This notation will be more convenient than expressing the state in terms of the non-contractible cycles associated with the genus, because the ensuing boundary partition cut is more naturally represented with respect to the wormholes. It is, however, straightforward to represent this state using the non-contractible cycles associated with the genus, and doing so makes clear the extra factor of $\mathcal{D}^{-1}$ necessary for proper normalization.
In particular, because the genus of the surface is $g=n-1$ (see the discussion at the beginning of Section~\ref{sec:ATEE}), one of the $\omega_0$-loops encircling a wormhole is redundant.  This can be seen using the handle-slide property of the $\omega_0$-loop, which states that a topological charge line may be passed through a nontrivial cycle (or other charge lines) if the cycle is encircled by an $\omega_0$-loop:
\begin{equation}\label{eq:handle-slide}
    \begin{pspicture}[shift=-1](-1,-1)(1,1)
        \scriptsize
        \psellipse(0,0)(.5,.4)\psline[ArrowInside=->](0,-.39)(-.15,-.368)
        \rput(0.45,-0.4){$\omega_0$}\rput(0,0){$\otimes$}
		\psline(0,-1)(0,-0.8)(-0.8,0)(0,0.8)(0,1)
		\psline[ArrowInside=->](0,-0.8)(-0.8,0)\rput(-0.6,-0.6){$a$}
    \end{pspicture}
    =
    \begin{pspicture}[shift=-1](-1,-1)(1,1)
        \scriptsize
        \psellipse(0,0)(.5,.4)\psline[ArrowInside=->](0,-.39)(-.15,-.368)
        \rput(-0.5,-0.4){$\omega_0$}\rput(0,0){$\otimes$}
		\psline(0,-1)(0,-0.8)(0.8,0)(0,0.8)(0,1)
		\psline[ArrowInside=->](0,-0.8)(0.8,0)\rput(0.6,-0.6){$a$}
    \end{pspicture}.
\end{equation}
One of the $\omega_0$-loops circling a wormhole can be deformed around the surface using handle-slide moves until it encircles nothing and can then simply be removed. If we treat that same wormhole as the one that is not contributing to the genus (i.e. the one responsible for first connecting the conjugate surfaces $M$ and $M^{\ast}$), then the state may be re-expressed in the notation of Section~\ref{sec:anyonmodelshigher} for the state of a genus $g=n-1$ surface as
\begin{eqnarray}
\ket{\psi}
&=& \mathcal{D}^{n-1} \begin{pspicture}[shift=-0.6](-0.8,-0.2)(4.0,1.4)
        \scriptsize
\rput(-1,0){        \rput(1.3,0.5){$\odot$}
        \psellipse[border=1.5pt](1.3,.5)(.25,.5)
\rput(.8,.65){$\omega_0$}
\psline[ArrowInside=->](1.065,.4)(1.065,.55)
}
        \rput(1.3,0.5){$\odot$}
        \psellipse[border=1.5pt](1.3,.5)(.25,.5)
\rput(.8,.65){$\omega_0$}
\psline[ArrowInside=->](1.065,.4)(1.065,.55)
        \rput(2.2,0.5){$\dots$}
        \rput(3,0){
        \rput(0.5,0.5){$\odot$}
        \psellipse[border=1.5pt](.5,.5)(.25,.5)\psline[ArrowInside=->](.265,.4)(.265,.55)\rput(0,.65){$\omega_0$}}
    \end{pspicture}
\notag \\
&=& \mathcal{D}^{n-1} \begin{pspicture}[shift=-0.6](-0.8,-0.2)(4.0,1.4)
        \scriptsize
        \rput(1.3,0.5){$\otimes$}
        \psellipse[border=1.5pt](1.3,.5)(.25,.5)
\rput(.8,.65){$\omega_0$}
\psline[ArrowInside=->](1.065,.4)(1.065,.55)
        \rput(2.2,0.5){$\dots$}
        \rput(3,0){
        \rput(0.5,0.5){$\otimes$}
        \psellipse[border=1.5pt](.5,.5)(.25,.5)\psline[ArrowInside=->](.265,.4)(.265,.55)\rput(0,.65){$\omega_0$}}
    \end{pspicture}
\notag \\
&=& \sum_{x_1 ,\ldots , x_{n-1}} \frac{d_{x_1} \cdots d_{x_{n-1}} }{\mathcal{D}^{n-1}}
\begin{pspicture}[shift=-0.6](0.2,-0.2)(4.0,1.4)
        \scriptsize
        \psline(1.3,1)(1,.7)(1,.3)(1.3,0)(1.6,.3)(1.6,.7)(1.3,1)
        \psline[ArrowInside=->](1,.3)(1,.7)\rput(.8,.65){$x_1$}
        \rput(2.2,0.5){$\dots$}
        \rput(3,0){
       \psline[border=1.5pt](0.5,0)(0.8,0.3)(0.8,0.7)(0.5,1) (0.2,0.7)(0.2,0.3)(0.5,0)
       \psline[ArrowInside=->](0.2,0.3)(0.2,0.7)\rput(-0.2,.65){$x_{n-1}$}}
        \rput(1.3,0.5){$\otimes$}
        \rput(3.5,0.5){$\otimes$}
    \end{pspicture}
.
\label{eq:psi_wormhole_to_genus}
\end{eqnarray}

Deforming the wormhole representation of the state of Eq.~(\ref{eq:omega-loop-psi}) and using Eq.~(\ref{eq:identity-op}), we can fuse together the charge lines threading the same boundary region, so that it is expressed as (suppressing the fusion vertex labels)
\begin{equation}
\label{eq:unpunct-disk-fuse}
\begin{split}
\ket{\psi} &= \mathcal{D}^{n-1} \begin{pspicture}[shift=-2](-0.6,-1.8)(4.8,1.4)
        \scriptsize
        \psline[linestyle=dashed](0.25,.4)(0.25,-.2)(1.4,-1.4)(3,-1.4)(4.35,-.2)(4.35,.5)(3.6,.5)
        \psline[linestyle=dashed](.35,.5)(1.2,.5)
        \psline[linestyle=dashed](1.4,.5)(1.8,.5)
        \psline[linestyle=dashed](2.6,.5)(3.4,.5)
        \psellipse[border=1.5pt](1.3,.5)(.25,.5)\psline[ArrowInside=->](1.065,.4)(1.065,.55)\rput(.8,.65){$\omega_0$}
        \psarc(.25,.7){.25}{0}{180}
        \psarc(4.35,.7){.25}{0}{180}
        \psline(0,0.7)(0,-0.3)(1.4,-1.7)(3,-1.7)(4.6,-.3)(4.6,.7)
        \psline[ArrowInside=->](0,.3)(0,.7)\rput(-.2,.65){$\omega_0$}
        \psline[border=1.5pt](0.5,0.7)(0.5,-0.1)(1.4,-1.1)(2.9,-1.1)(4.1,-.1)(4.1,.7)
        \rput(2.2,0.5){$\dots$}
        \rput(3,0){
        \psellipse[border=1.5pt](.5,.5)(.25,.5)\psline[ArrowInside=->](.265,.4)(.265,.55)\rput(0,.65){$\omega_0$}}
        \rput(0.25,0.5){$\odot$}
        \rput(3.5,0.5){$\odot$}
        \rput(1.3,0.5){$\odot$}
    \end{pspicture}
\\ &= \frac{1}{\mathcal{D}} \sum_{\vec{a}} \frac{d_{\vec{a}}}{\mathcal{D}^n} \begin{pspicture}[shift=-2](-0.6,-1.8)(4.8,1.4)
        \scriptsize
        \psline[linestyle=dashed](0.25,.4)(0.25,-.2)(1.4,-1.4)(3,-1.4)(4.35,-.2)(4.35,.5)(3.6,.5)
        \psline[linestyle=dashed](.35,.5)(1.2,.5)
        \psline[linestyle=dashed](1.4,.5)(1.8,.5)
        \psline[linestyle=dashed](2.6,.5)(3.4,.5)
        \psline(1.3,1)(1,.7)(1,.3)(1.3,0)(1.6,.3)(1.6,.7)(1.3,1)
        \psline[ArrowInside=->](1,.3)(1,.7)\rput(.8,.65){$a_2$}
        \psline[border=1.5pt](.5,0.7)(.25,1)(0,0.7)(0,-0.3)(1.4,-1.7)(3,-1.7)(4.6,-.3)
        (4.6,.7)(4.35,1)
        \psline[ArrowInside=->](0,.3)(0,.7)\rput(-.2,.65){$a_1$}
        \psline[border=1.5pt](0.5,0.7)(0.5,-0.1)(1.4,-1.1)(2.9,-1.1)(4.1,-.1)(4.1,.7)(4.35,1)
        \psline(4.1,.7)(4.35,1)
        \rput(2.2,0.5){$\dots$}
        \rput(3,0){
       \psline[border=1.5pt](0.5,0)(0.8,0.3)(0.8,0.7)(0.5,1) (0.2,0.7)(0.2,0.3)(0.5,0)
       \psline[ArrowInside=->](0.2,0.3)(0.2,0.7)\rput(0,.65){$a_n$}}
        \rput(0.25,0.5){$\odot$}
        \rput(1.3,0.5){$\odot$}
        \rput(3.5,0.5){$\odot$}
    \end{pspicture}
\\ &=\frac{1}{\mathcal{D}}\sum_{\vec{a}, \vec{b}} \frac{\sqrt{d_{\vec{b}}}}{\mathcal{D}^n}
    \begin{pspicture}[shift=-2.5](-0.2,-2.6)(5,1.4)
        \psline[linestyle=dashed](0.25,.4)(0.25,-.2)(1.4,-1.4)(3,-1.4)(4.35,-.2)(4.35,.5)(4.1,.5)
        \psline[linestyle=dashed](.4,.5)(.5,.5)
        \psline[linestyle=dashed](1,.5)(.9,.5)
        \psline[linestyle=dashed](1.2,.5)(1.3,.5)
        \psline[linestyle=dashed](1.75,.5)(1.9,.5)
        \scriptsize
        \psline[border=1.5pt](0.8,0.7)(0.5,1)(0,0.5)(0,-0.3)(1.4,-1.7)(3,-1.7)(4.6,-.3)
        (4.6,.5)(4.1,1)
        \psline[border=1.5pt](0.8,0.3)(0.5,0)(0.5,-0.1)(1.4,-1.1)(2.9,-1.1)(4.1,-.1)
        \psline[border=1.5pt](1.1,0)(1.4,0.3)(1.4,0.7)(1.1,1)(0.8,0.7)(0.8,0.3)(1.1,0)
        \psline[ArrowInside=->](0.5,0)(0.8,0.3)\rput(0.5,0.25){$\bar{a}_1$}
        \psline[ArrowInside=->](0.8,0.3)(0.8,0.7)\rput(0.6,0.5){$b_1$}
        \psline[ArrowInside=->](0.8,0.7)(0.5,1)\rput(0.5,0.75){$\bar{a}_1$}
        \psline[ArrowInside=->](1.1,0)(.8,.3)\rput(1.1,0.25){$a_2$}
        \psline[ArrowInside=->](1.4,0.3)(1.4,0.7)\rput(1.6,0.5){$b_2$}
        \psline[ArrowInside=->](.8,0.7)(1.1,1)\rput(1.1,0.7){$a_2$}
        \psline[ArrowInside=->](1.7,0)(1.4,0.3)\rput(1.7,0.2){$a_3$}
        \psline[ArrowInside=->](1.4,0.7)(1.7,1)\rput(1.7,0.7){$a_3$}
        \rput(2.2,0.5){$\dots$}
        \rput(3,0){
        \psline[linestyle=dashed](.6,.5)(.7,.5)
        \psline[linestyle=dashed](.4,.5)(.3,.5)
       \psline[border=1.5pt](0.5,0)(0.8,0.3)(0.8,0.7)(0.5,1)(0.2,0.7)(0.2,0.3)(0.5,0)
        \psline[ArrowInside=->](-0.1,0)(0.2,0.3)\rput(-0.3,0.25){$\bar{a}_{n-1}$}
        \psline[ArrowInside=->](0.2,0.3)(0.2,0.7)\rput(-0.15,0.5){$b_{n-1}$}
        \psline[ArrowInside=->](0.2,0.7)(-0.1,1)\rput(-0.3,0.85){$\bar{a}_{n-1}$}
        \psline[ArrowInside=->](0.5,0)(0.2,0.3)\rput(0.5,0.25){$a_n$}
        \psline[ArrowInside=->](0.8,0.3)(0.8,0.7)\rput(1,0.5){$b_n$}
        \psline[ArrowInside=->](0.2,0.7)(0.5,1)\rput(0.5,0.7){$a_n$}
        \psline[ArrowInside=->](1.1,0)(0.8,0.3)\rput(1.1,0.25){$a_1$}
        \psline[ArrowInside=->](0.8,0.7)(1.1,1)\rput(1.1,0.7){$a_1$}}
        \psline(4.1,0)(4.1,-.1)
        \rput(0.25,0.5){$\odot$}
        \rput(1.1,0.5){$\odot$}
        \rput(3.5,0.5){$\odot$}
    \end{pspicture}
.
\end{split}
\end{equation}
The corresponding anyon diagram embedded in three dimensions looks like
\begin{center}
\noindent\includegraphics[width=.9\linewidth]{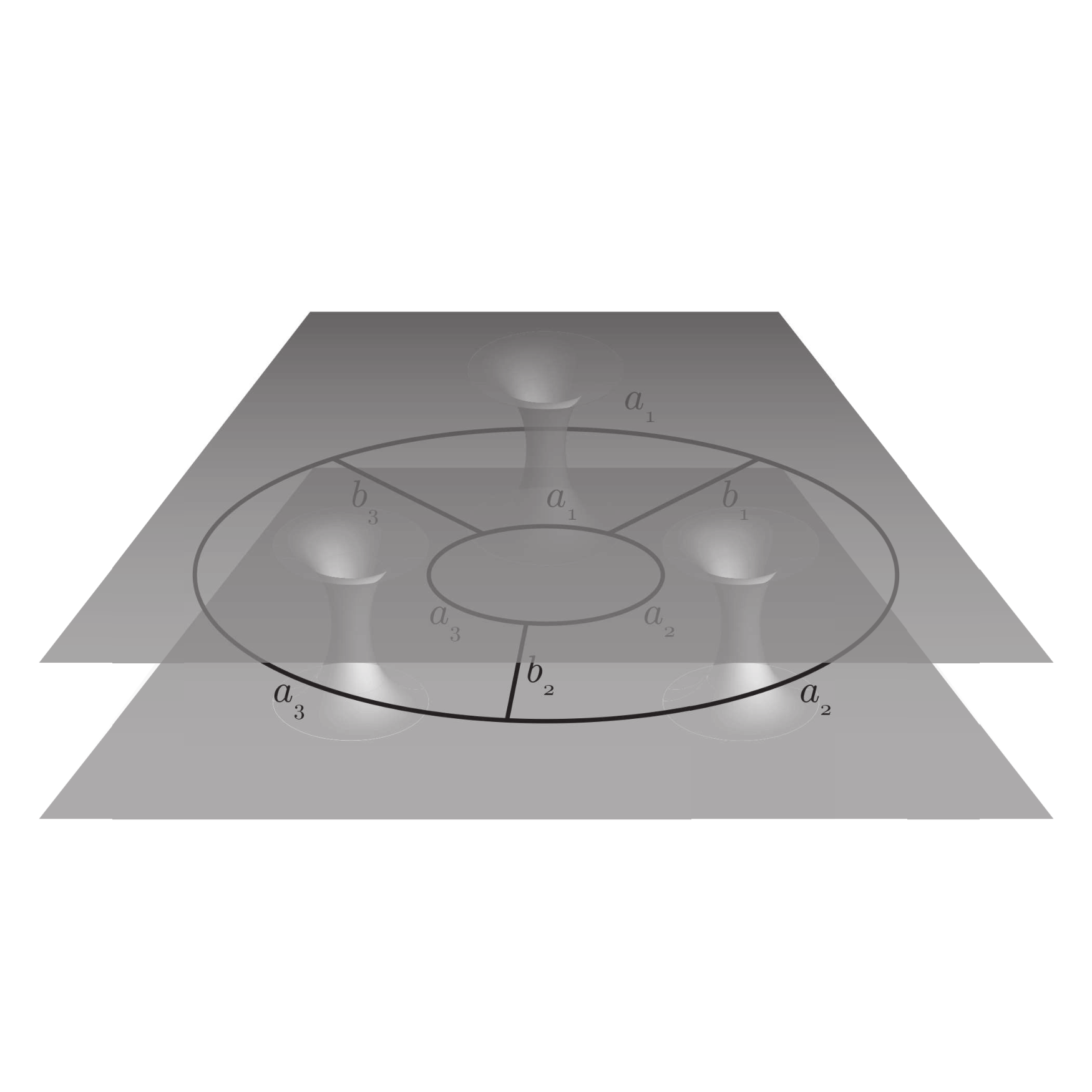}
\end{center}

We can rewrite the state in a tree-like form using a series of $F$-moves~\footnote{This series of transformations also involves ``bending'' moves, i.e., vertex rotations~\cite{Bonderson07b}. The bending transformations also cancel out, so we leave them implicit to avoid excessive clutter.}:
\begin{equation}\label{eq:F-move-steps}
\begin{split}
&\ket{\psi} =  \sum_{\vec{a},\vec{b},e_2,e_2'} \frac{\sqrt{d_{\vec{b}}}}{\mathcal{D}^{n+1}} \left[F^{a_1 b_1 b_2}_{a_3}\right]^\dagger_{e_2' a_2}\left[F^{a_1 b_1 b_2}_{a_3} \right]_{a_2 e_2}
\psscalebox{.9}{
\begin{pspicture}[shift=-2.3](-0.2,-2.1)(5,1.7)
        \scriptsize
        \psline[linestyle=dashed](0.25,.4)(0.25,-.3)(1.4,-1.4)(3.1,-1.4)(4.25,-.3)(4.25,.5)(4.1,.5)
        \rput(0.25,0.5){$\odot$}
        \rput(1.1,0.5){$\odot$}
        \psline[linestyle=dashed](1,.5)(.9,.5)
        \psline[linestyle=dashed](1.2,.5)(1.3,.5)
        \psline[linestyle=dashed](.35,.5)(.45,.5)
        \psline(.5,1.1)(0,0.6)(0,-0.3)(1.4,-1.6)(3.3,-1.6)(4.6,-.3)
        (4.6,.5)(4.1,1)\rput(0.75,0){$a_1$} \rput(1.15,-.2){$e_2$}
        \psline(.5,0)(.5,-.3)(1.4,-1.2)(2.9,-1.2)(4.1,-.1)
        \psline(1.1,0)(1.4,0.3)(1.4,0.7)(1.1,1)(0.8,0.7)(0.8,0.3)(1.1,0)
        \psline[ArrowInside=->](.5,1.1)(.8,1.4)\rput(0.5,1.35){$a_1$}
        \psline[ArrowInside=->](0.8,0.3)(0.8,0.7)\rput(0.6,0.5){$b_1$}
        \psline[ArrowInside=->](1.1,0)(1.4,0.3)
        \psline[ArrowInside=->](1.4,0.3)(1.4,0.7)\rput(1.6,0.5){$b_2$}
        \psline[ArrowInside=->](1.4,0.7)(1.1,1)
        \psline[ArrowInside=->](1.1,1)(.8,1.4)\rput(1.15,1.3){$e_2'$}
        \psline[ArrowInside=->](.8,1.4)(1.1,1.7)
        \psline[ArrowInside=->](1.7,1.1)(1.1,1.7)\rput(1.7,1.4){$\bar{a}_3$}
        \psline[ArrowInside=->](.8,-.3)(.5,0)
        \psline[ArrowInside=->](.8,-.3)(1.1,0)
        \psline[ArrowInside=->](1.1,-.6)(.8,-.3)
        \psline[ArrowInside=->](1.1,-.6)(1.7,0)\rput(1.7,-.3){$\bar{a}_3$}
        \rput(2.2,0.5){$\dots$}
        \psline[linestyle=dashed](1.8,.5)(1.9,.5)
        \rput(3,0){
        \rput(0.5,0.5){$\odot$}
        \psline[linestyle=dashed](.6,.5)(.7,.5)
        \psline[linestyle=dashed](.4,.5)(.3,.5)
       \psline(0.5,0)(0.8,0.3)(0.8,0.7)(0.5,1)(0.2,0.7)(0.2,0.3)(0.5,0)
        \psline[ArrowInside=->](-0.1,0)(0.2,0.3)\rput(-0.3,0.2){$\bar{a}_{n-1}$}
        \psline[ArrowInside=->](0.2,0.3)(0.2,0.7)\rput(-0.15,0.5){$b_{n-1}$}
        \psline[ArrowInside=->](0.2,0.7)(-0.1,1)\rput(-0.3,0.8){$\bar{a}_{n-1}$}
        \psline[ArrowInside=->](0.5,0)(0.2,0.3)\rput(0.5,0.25){$a_n$}
        \psline[ArrowInside=->](0.8,0.3)(0.8,0.7)\rput(1,0.5){$b_n$}
        \psline[ArrowInside=->](0.2,0.7)(0.5,1)\rput(0.5,0.7){$a_n$}
        \psline[ArrowInside=->](1.1,0)(0.8,0.3)\rput(1.1,0.25){$a_1$}
        \psline[ArrowInside=->](0.8,0.7)(1.1,1)\rput(1.1,0.7){$a_1$}}
        \psline(4.1,0)(4.1,-.1)
    \end{pspicture}}
\\ &= \sum_{\substack{a_1,a_3,\dots, a_n\\ \vec{b},e_2,e_{2}',e_3,e_3'}} \frac{\sqrt{d_{\vec{b}}}}{\mathcal{D}^{n+1}} \delta_{e_2,e_2'} \Big[F^{a_1 e_2' b_3}_{a_4}\Big]^\dagger_{e_3' a_3} \Big[F^{a_1 e_2 b_3}_{a_4}\Big]_{a_3 e_3 }
\psscalebox{.9}{
\begin{pspicture}[shift=-2.65](-0.2,-2.1)(5,1.7)
        \scriptsize
        \psline[linestyle=dashed](0.3,.4)(0.3,-.3)(1.4,-1.4)(3.1,-1.4)(4.25,-.3)(4.25,.5)(4.1,.5)
        \psline[linestyle=dashed](.4,.5)(.5,.5)
        \psline[linestyle=dashed](1,.5)(.9,.5)
        \psline[linestyle=dashed](1.2,.5)(1.3,.5)
        \rput(0.3,0.5){$\odot$}
        \rput(1.1,0.5){$\odot$}
        \psline(1.1,1.6)(0.1,0.6)(0.1,-0.3)(1.4,-1.6)(3.3,-1.6)(4.6,-.3)
        (4.6,.5)(4.1,1)\rput(0.7,-.2){$a_1$} \rput(1.1,-.2){$e_2$}
        \rput(1.45,-.5){$e_3$}
        \psline(.8,-.3)(.8,-.6)(1.4,-1.2)(2.9,-1.2)(4.1,-.1)
        \psline(1.1,0)(1.4,0.3)(1.4,0.7)(1.1,1)(0.8,0.7)(0.8,0.3)(1.1,0)
        \psline[ArrowInside=->](.5,1.)(.8,1.3)\rput(0.5,1.35){$a_1$}
        \psline[ArrowInside=->](0.8,0.3)(0.8,0.7)\rput(0.65,0.55){$b_1$}
        \psline[ArrowInside=->](1.1,0)(1.4,0.3)
        \psline[ArrowInside=->](1.4,0.3)(1.4,0.7)\rput(1.6,0.5){$b_2$}
        \psline[ArrowInside=->](1.4,0.7)(1.1,1)
        \psline[ArrowInside=->](1.1,1)(1.4,1.3)\rput(1.1,1.3){$e_2'$}
        \psline[ArrowInside=->](1.7,1)(1.4,1.3)\rput(1.7,.85){$b_3$}
        \psline[ArrowInside=->](1.4,1.3)(1.1,1.6)\rput(1.4,1.6){$e_3'$}
        \psline[ArrowInside=->](1.1,1.6)(1.4,1.9)
        \psline[ArrowInside=->](1.4,1.9)(2,1.3)\rput(2,1.15){$\bar{a}_4$}
        \psline[ArrowInside=->](1.4,-.3)(1.1,0)
        \psline[ArrowInside=->](1.1,-.6)(.8,-.3)
        \psline[ArrowInside=->](1.4,-.9)(1.1,-.6)
        \psline[ArrowInside=->](1.4,-.9)(2,-.3)\rput(2,-.2){$\bar{a}_4$}
        \psline[ArrowInside=->](1.1,-.6)(1.4,-.3)
        \psline[ArrowInside=->](1.4,-.3)(1.7,0)\rput(1.7,.15){$b_3$}
        \rput(2.2,0.5){$\dots$}
        \psline[linestyle=dashed](1.8,.5)(1.9,.5)
        \rput(3,0){
        \psline[linestyle=dashed](.6,.5)(.7,.5)
        \psline[linestyle=dashed](.4,.5)(.3,.5)
        \rput(0.5,0.5){$\odot$}
       \psline(0.5,0)(0.8,0.3)(0.8,0.7)(0.5,1)(0.2,0.7)(0.2,0.3)(0.5,0)
        \psline[ArrowInside=->](-0.1,0)(0.2,0.3)\rput(-0.3,0.2){$\bar{a}_{n-1}$}
        \psline[ArrowInside=->](0.2,0.3)(0.2,0.7)\rput(-0.15,0.5){$b_{n-1}$}
        \psline[ArrowInside=->](0.2,0.7)(-0.1,1)\rput(-0.3,0.8){$\bar{a}_{n-1}$}
        \psline[ArrowInside=->](0.5,0)(0.2,0.3)\rput(0.5,0.25){$a_n$}
        \psline[ArrowInside=->](0.8,0.3)(0.8,0.7)\rput(1,0.5){$b_n$}
        \psline[ArrowInside=->](0.2,0.7)(0.5,1)\rput(0.5,0.7){$a_n$}
        \psline[ArrowInside=->](1.1,0)(0.8,0.3)\rput(1.1,0.25){$a_1$}
        \psline[ArrowInside=->](0.8,0.7)(1.1,1)\rput(1.1,0.7){$a_1$}}
        \psline(4.1,0)(4.1,-.1)
    \end{pspicture}}
\\ &= \sum_{a_1, \vec{b}, \vec{e}} \frac{\sqrt{d_{\vec{b}}}}{\mathcal{D}^{n+1}}
\psscalebox{.9}{
 \begin{pspicture}[shift=-2.8](0,-2.5)(4.1,3.2)
        \scriptsize
        \rput(0.5,0.5){$\odot$}
        \rput(1.1,0.5){$\odot$}
        \rput(1.7,0.5){$\odot$}
        \psline(0.8,0.7)(1.7,1.6)
        \psline[border=1.5pt](1.7,-0.6)(0.8,0.3)
        \psline(1.4,1.3)(2,.7)(2,.3)(1.4,-0.3)
        \psline[ArrowInside=->](2,.3)(2,.7)\rput(2.2,.75){$b_3$}
        \psline[ArrowInside=->](1.1,1)(1.4,1.3)\rput(1.15,1.3){$e_2$}
        \psline[ArrowInside=->](1.4,1.3)(1.7,1.6)\rput(1.45,1.6){$e_3$}
        \psline[linestyle=dotted](1.7,1.6)(2,1.9)
        \psline[ArrowInside=->](0.8,0.3)(0.8,0.7)\rput(0.65,0.75){$b_1$}
        \psline[ArrowInside=->](2.2,2.1)(2.5,2.4)\rput(2.25,1.9){$e_{n-1}$}
        \psline(2.5,2.4)(2.8,2.1)(2.8,-1)(2.5,-1.3)
        \psline[ArrowInside=->](2.8,-1.6)(2.5,-1.3)\rput(2.8,-1.3){$e_{n}$}
        \psline(2.8,-1.6)(3.1,-1.3)(3.4,-1.6)(3.1,-1.9)(2.8,-1.6)\rput(3.35,-1.3){$a_1$}
        \psline[ArrowInside=->](2.5,2.4)(2.8,2.7)\rput(2.5,2.7){$e_{n}$}
        \psline(.2,.3)(.2,.7)(2.5,3)(4,1.5)(4,-1.6)(3.4,-2.2)(2.7,-2.2)(.2,.3)\rput(3.2,2.6){$a_1$}
        \psline[linestyle=dashed](.5,.4)(.5,.3)(2.85,-2.05)(3.3,-2.05)(3.75,-1.6)(3.75,.5)(2.9,.5)
        \psline[linestyle=dashed](2.7,.5)(2.6,.5)
        \psline[linestyle=dashed](2.1,.5)(2.17,.5)
        \psline[linestyle=dashed](1.9,.5)(1.8,.5)
        \psline[linestyle=dashed](1.6,.5)(1.5,.5)
        \psline[linestyle=dashed](1.3,.5)(1.2,.5)
        \psline[linestyle=dashed](1,.5)(.9,.5)
        \psline[linestyle=dashed](.7,.5)(.6,.5)
        \psline(1.1,0)(1.4,0.3)(1.4,0.7)(1.1,1)
        \psline[ArrowInside=->](1.4,0.3)(1.4,0.7)\rput(1.6,0.75){$b_2$}
        \psline[ArrowInside=->](2.8,0.3)(2.8,0.7)
        \rput(3,0.75){$b_{n}$}
        \psline[ArrowInside=->](1.4,-0.3)(1.1,0)\rput(1.45,0){$e_2$}
        \psline[ArrowInside=->](1.7,-0.6)(1.4,-0.3)\rput(1.75,-0.3){$e_3$}
        \psline[linestyle=dotted](1.7,-0.6)(2,-.9)
        \psline[ArrowInside=->](2.5,-1.3)(2.2,-1)\rput(2.4,-.9){$e_{n-1}$}
        \rput(2.4,0.5){$\dots$}
        \rput(2.5,0.5){$\odot$}
    \end{pspicture}}
= \sum_{\substack{\vec{b},\\ e_2,\dots,e_{n-2}}} \frac{\sqrt{d_{\vec{b}}}}{\mathcal{D}^{n-1}}
\psscalebox{.9}{
  \begin{pspicture}[shift=-2](.2,-1.6)(3.2,2.4)
        \scriptsize
        \rput(0.5,0.5){$\odot$}
        \rput(1.1,0.5){$\odot$}
        \rput(1.7,0.5){$\odot$}
        \psline(0.8,0.7)(1.7,1.6)
        \psline[border=1.5pt](1.7,-0.6)(0.8,0.3)
        \psline(1.4,1.3)(2,.7)(2,.3)(1.4,-0.3)
        \psline[ArrowInside=->](2,.3)(2,.7)\rput(2.2,.75){$b_3$}
        \psline[ArrowInside=->](1.1,1)(1.4,1.3)\rput(1.15,1.3){$e_2$}
        \psline[ArrowInside=->](1.4,1.3)(1.7,1.6)\rput(1.45,1.6){$e_3$}
        \psline[linestyle=dotted](1.7,1.6)(2,1.9)
        \psline[ArrowInside=->](0.8,0.3)(0.8,0.7)\rput(0.65,0.75){$b_1$}
        \psline[ArrowInside=->](2.2,2.1)(2.5,2.4)\rput(2.25,1.9){$\bar{b}_{n}$}
        \psline(2.5,2.4)(2.8,2.1)(2.8,-1)(2.5,-1.3)
        \psline(1.1,0)(1.4,0.3)(1.4,0.7)(1.1,1)
        \psline[ArrowInside=->](1.4,0.3)(1.4,0.7)\rput(1.6,0.75){$b_2$}
        \psline[ArrowInside=->](2.8,0.3)(2.8,0.7)
        \rput(3,0.75){$b_{n}$}
        \psline[ArrowInside=->](1.4,-0.3)(1.1,0)\rput(1.45,-0){$e_2$}
        \psline[ArrowInside=->](1.7,-0.6)(1.4,-0.3)\rput(1.75,-0.3){$e_3$}
        \psline[linestyle=dotted](1.7,-0.6)(2,-.9)
        \psline[ArrowInside=->](2.5,-1.3)(2.2,-1)\rput(2.4,-.9){$\bar{b}_{n}$}
        \rput(2.31,0.5){$\dots$}
        \psline[linestyle=dashed](.5,.4)(.5,.3)(2.3,-1.5)(2.7,-1.5)(3,-1.2)(3,.5)(2.9,.5)
        \psline[linestyle=dashed](2.75,.5)(2.71,.5)
        \psline[linestyle=dashed](1.9,.5)(1.8,.5)
        \psline[linestyle=dashed](1.6,.5)(1.5,.5)
        \psline[linestyle=dashed](1.3,.5)(1.2,.5)
        \psline[linestyle=dashed](1,.5)(.9,.5)
        \psline[linestyle=dashed](.7,.5)(.6,.5)
      \rput(2.61,0.5){$\odot$}
    \end{pspicture}}.
\end{split}
\end{equation}
The unitarity of the $F$-symbols together with the summation over $a_j$ results in $\delta_{e_j , e_j'}$ factors (and similarly for the suppressed vertex labels).  In the last line, we collapse a tadpole diagram in both $\mathbb{A}$ and $\bar{\mathbb{A}}$,~\footnote{Note that the outer loop in the second to last expression of Eq.~(\ref{eq:F-move-steps}) can be deformed around the surface until it no longer encloses anything, i.e. it is truly a tadpole diagram.} which sets $e_n=0$ and $e_{n-1}=\bar{b}_n$ and results in a factor of $\mathcal{D}^2$ when $a_1$ is summed over. In the following, we write $\vec{e}$ to mean $e_2,\dots, e_{n-2}$. We note that the final expression could have alternatively been obtained from the state written as $n$ wormholes with $\omega_0$-loops around only $n-1$ of the wormholes.

When embedded in three dimensional space, the anyon diagram corresponding to the final representation of the state in Eq.~(\ref{eq:F-move-steps}) looks like

\begin{center}
\noindent\includegraphics[width=.9\linewidth]{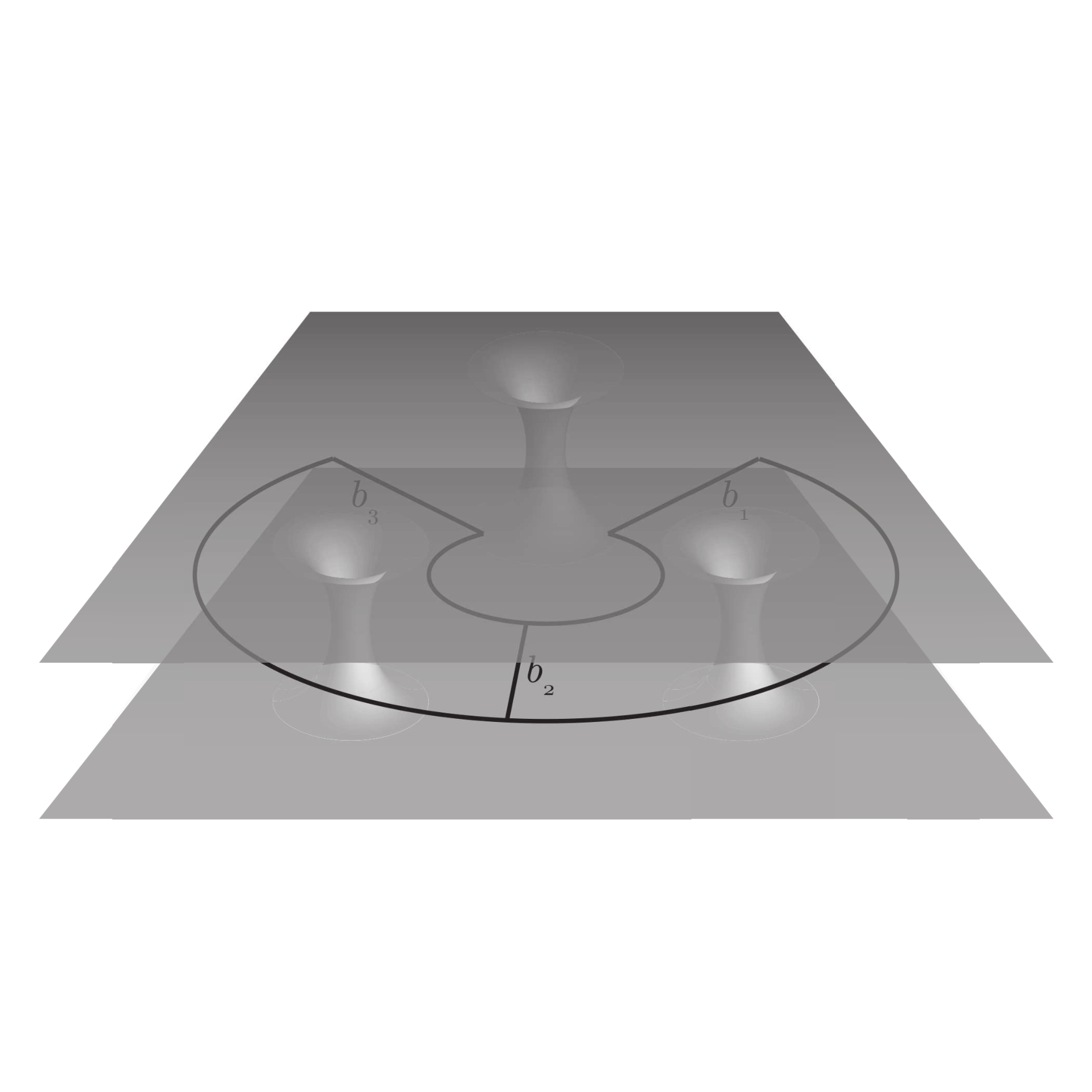}
\end{center}

Now that each partition boundary component, i.e. each tube connecting $\AD$ to $\bar{\AD}$, is threaded by a single topological charge line, when we cut the surface along the partition boundary between $\AD$ and $\bar{\AD}$, indicated by the dashed lines in Eq.~(\ref{eq:F-move-steps}), each resulting boundary components of $\AD$ will correspondingly be ascribed the topological charge $b_j$ of the charge line threading it, and similarly for the boundaries of $\bar{\AD}$. The resulting state after cutting is
\begin{equation}
\begin{split}
   & |\psi_{\text{cut}}\rangle=\sum_{ \vec{b}, \vec{e}} \frac{1}{\mathcal{D}^{n-1}}
\begin{pspicture}[shift=-1](-.3,-2)(7,.3)
        \scriptsize
        \rput(1.7,-2){$(\AD)$}
        \psline[ArrowInside=->](.3,-.3)(0,0)\rput(0,.2){$b_1$}
        \psline[ArrowInside=->](.6,-.6)(.3,-.3)\rput(.6,-.35){$e_2$}
        \psline[ArrowInside=->](.3,-.3)(.6,0)\rput(.6,.2){$b_2$}
        \psline[ArrowInside=->](.9,-.9)(.6,-.6)\rput(.9,-.65){$e_3$}
        \psline(.6,-.6)(1.2,0)\psline[ArrowInside=->](.9,-.3)(1.2,0)\rput(1.2,.2){$b_3$}
        \psline[linestyle=dotted](.9,-.9)(1.3,-1.3)
        \psline[ArrowInside=->](1.6,-1.6)(1.3,-1.3)\rput(1.55,-1.2){$\bar{b}_n$}
        \psline(1.6,-1.6)(3.2,0)\psline[ArrowInside=->](2.9,-.3)(3.2,0)\rput(3.2,.2){$b_n$}
        \rput(-.7,0){\rput(6.3,-2){$(\bar{\AD})$}
        \psline[ArrowInside=->](7.7,-.3)(8,0)\rput(8,.2){$b_1$}
        \psline[ArrowInside=->](7.4,-.6)(7.7,-.3)\rput(7.4,-.35){$e_2$}
        \psline[ArrowInside=->](7.7,-.3)(7.4,0)\rput(7.4,.2){$b_2$}
        \psline[ArrowInside=->](7.1,-.9)(7.4,-.6)\rput(7.1,-.65){$e_3$}
        \psline(7.4,-.6)(6.8,0)\psline[ArrowInside=->](7.1,-.3)(6.8,0)\rput(6.8,.2){$b_3$}
        \psline[linestyle=dotted](7.1,-.9)(6.7,-1.3)
        \psline[ArrowInside=->](6.4,-1.6)(6.7,-1.3)\rput(6.4,-1.2){$\bar{b}_{n}$}
        \psline(6.4,-1.6)(4.8,0)\psline[ArrowInside=->](5.1,-.3)(4.8,0)\rput(4.8,.2){$b_{n}$}}
    \end{pspicture},
\end{split}
\end{equation}
where the diagram for $\AD$ embedded in three dimensional space looks like

\begin{center}
\includegraphics[width=.4\linewidth]{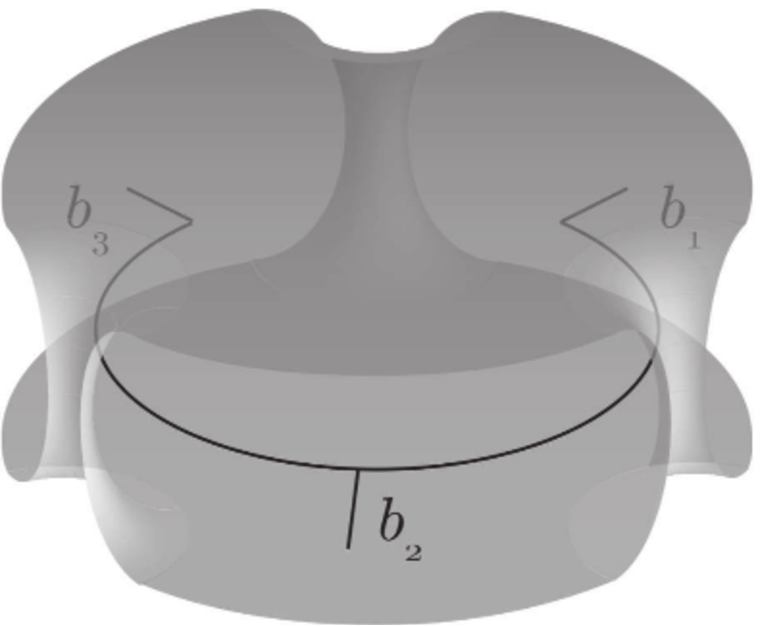}
\end{center}
or, equivalently,

\begin{center}
\includegraphics[width=.5\linewidth]{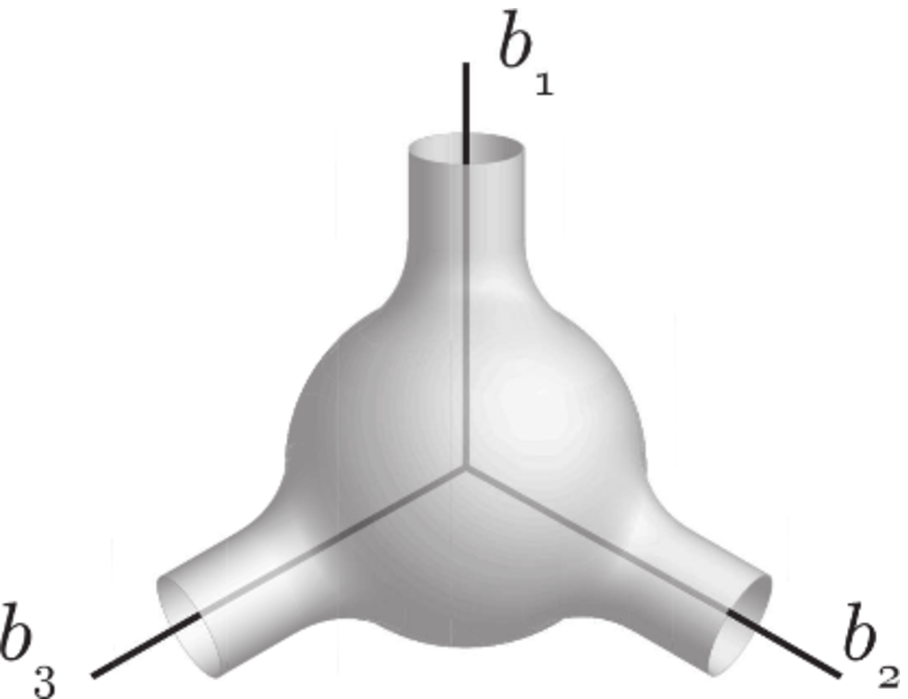}
\end{center}

We can now find the reduced density matrix for $\AD$ by tracing over $\bar{\AD}$,
\begin{equation}\label{eq:disk-trace}
\begin{split}
&\rhoD= \aTr_{\bar{\AD}}\Big[ \ket{\psi_{\text{cut}}}\bra{\psi_{\text{cut}}}\Big]
\\ &= \sum_{\vec{b},\vec{e}, \vec{b}',\vec{e}'} \frac{1}{\mathcal{D}^{2n-2}}
\psscalebox{.9}{
\begin{pspicture}[shift=-2.4](-.3,-4.8)(8.5,1)
        \scriptsize
        \rput(1.6,-1.95){$(\AD)$}
        \psline[ArrowInside=->](.3,-.3)(0,0)\rput(0,.2){$b_1$}
        \psline[ArrowInside=->](.6,-.6)(.3,-.3)\rput(.6,-.35){$e_2$}
        \psline[ArrowInside=->](.3,-.3)(.6,0)\rput(.6,.2){$b_2$}
        \psline[ArrowInside=->](.9,-.9)(.6,-.6)\rput(.9,-.65){$e_3$}
        \psline(.6,-.6)(1.2,0)\psline[ArrowInside=->](.9,-.3)(1.2,0)\rput(1.2,.2){$b_3$}
        \psline[linestyle=dotted](.9,-.9)(1.3,-1.3)
        \psline[ArrowInside=->](1.6,-1.6)(1.3,-1.3)\rput(1.55,-1.2){$\bar{b}_n$}
        \psline(1.6,-1.6)(3.2,0)\psline[ArrowInside=->](2.9,-.3)(3.2,0)\rput(3.2,.2){$b_n$}
        \rput(0,.6){
        \psline[ArrowInside=->](0,-4.4)(.3,-4.1)\rput(0,-4.6){$b_1'$}
        \psline[ArrowInside=->](.3,-4.1)(.6,-3.8)\rput(.6,-4.1){$e_2'$}
        \psline[ArrowInside=->](.6,-4.4)(.3,-4.1)\rput(.6,-4.6){$b_2'$}
        \psline[ArrowInside=->](.6,-3.8)(.9,-3.5)\rput(.9,-3.8){$e_3'$}
        \psline(.6,-3.8)(1.2,-4.4)\psline[ArrowInside=->](1.2,-4.4)(.9,-4.1)\rput(1.2,-4.6){$b_3'$}
        \psline[linestyle=dotted](.9,-3.5)(1.3,-3.1)
        \psline[ArrowInside=->](1.3,-3.1)(1.6,-2.8)\rput(1.55,-3.3){$\bar{b}_{n}'$}
        \psline(3.2,-4.4)(1.6,-2.8)\psline[ArrowInside=->](3.2,-4.4)(2.9,-4.1)\rput(3.2,-4.6){$b_n'$}}
        \rput(-.7,0){
        \rput(6.4,-1.95){$(\bar{\AD})$}
        \psline[ArrowInside=->](7.7,-.3)(8,0)
        \psline(8,0)(8.2,-.2)(8.2,-3.6)(8,-3.8)\rput(8,-0.3){$b_1$}\rput(8,-3.45){$b_1'$}
        \psline[ArrowInside=->](8,-3.8)(7.7,-3.5)
        \psline[ArrowInside=->](7.4,-3.8)(7.7,-3.5)
        \psline[ArrowInside=->](6.8,-3.8)(7.1,-3.5)\psline(7.1,-3.5)(7.4,-3.1)
        \psline[ArrowInside=->](7.7,-3.5)(7.4,-3.2)
        \psline[ArrowInside=->](7.4,-3.2)(7.1,-2.9)
        \psline[linestyle=dotted](7.1,-2.9)(6.7,-2.5)
        \psline[ArrowInside=->](6.7,-2.5)(6.4,-2.2)
        \psline(4.8,-3.8)(6.4,-2.2)\psline[ArrowInside=->](4.8,-3.8)(5.1,-3.5)
        \psline[ArrowInside=->](7.4,-.6)(7.7,-.3)\rput(7.4,-.35){$e_2$}
        \psline[ArrowInside=->](7.7,-.3)(7.4,0)\rput(7.4,.3){$b_2$} \rput(7.4,-4){$b_2'$}
        \psline(7.4,0)(7.7,.3)(8,.3)(8.5,-.2)(8.5,-3.6)(8,-4.1)(7.7,-4.1)(7.4,-3.8)
        \psline[ArrowInside=->](7.1,-.9)(7.4,-.6)\rput(7.1,-.65){$e_3$}
        \psline(7.4,-.6)(6.8,0)\psline[ArrowInside=->](7.1,-.3)(6.8,0)\rput(6.8,.3){$b_3$} \rput(6.8,-4){$b_3'$}
        \psline(6.8,0)(7.4,.6)(8,.6)(8.8,-.2)(8.8,-3.6)(8,-4.4)(7.4,-4.4)(6.8,-3.8)
        \psline[linestyle=dotted](7.1,-.9)(6.7,-1.3)
        \psline[ArrowInside=->](6.4,-1.6)(6.7,-1.3)\rput(6.4,-1.2){$\bar{b}_{n}$}
        \psline(4.8,0)(5.7,.9)(8,.9)(9.1,-.2)(9.1,-3.6)(8,-4.7)(5.7,-4.7)(4.8,-3.8)
        \psline(6.4,-1.6)(4.8,0)\psline[ArrowInside=->](5.1,-.3)(4.8,0)\rput(4.8,.3){$b_{n}$} \rput(4.8,-4){$b_{n}'$}
        \rput(7.4,-3.45){$e_2'$}
        \rput(7.1,-3.15){$e_3'$}
        \rput(6.4,-2.5){$\bar{b}_n'$}}
    \end{pspicture}}.
\end{split}
\end{equation}
The quantum trace over $\bar{\AD}$ sets $b_j=b_j'$ and $e_j=e'_j$, and evaluating the inner product yields a factor of $\sqrt{d_{\vec{b}}}$. Therefore, the anyonic reduced density matrix for $A$ (restoring the vertex labels) is

\begin{equation}
\rhoD=\sum_{\vec{b}, \vec{e},\vec{\mu}} \frac{\sqrt{d_{\vec{b}}}}{\mathcal{D}^{2n-2}}
\begin{pspicture}[shift=-2](0,-4.1)(3.5,.3)
        \scriptsize
        \psline[ArrowInside=->](.3,-.3)(0,0)\rput(0,.2){$b_1$}
        \psline[ArrowInside=->](.6,-.6)(.3,-.3)\rput(.6,-.35){$e_2$}
        \rput(.1,-.45){$\mu_2$}
        \psline[ArrowInside=->](.3,-.3)(.6,0)\rput(.6,.2){$b_2$}
        \psline[ArrowInside=->](.9,-.9)(.6,-.6)\rput(.9,-.65){$e_3$}
        \rput(.4,-.75){$\mu_3$}
        \psline(.6,-.6)(1.2,0)\psline[ArrowInside=->](.9,-.3)(1.2,0)\rput(1.2,.2){$b_3$}
        \psline[linestyle=dotted](.9,-.9)(1.3,-1.3)
        \psline[ArrowInside=->](1.6,-1.6)(1.3,-1.3)\rput(1.55,-1.2){$\bar{b}_n$}
        \psline(1.6,-1.6)(3.2,0)\psline[ArrowInside=->](2.9,-.3)(3.2,0)\rput(3.2,.2){$b_n$}
        \rput(0,.6){
        \psline[ArrowInside=->](0,-4.4)(.3,-4.1)\rput(0,-4.6){$b_1$}
        \psline[ArrowInside=->](.3,-4.1)(.6,-3.8)\rput(.6,-4.1){$e_2$}
        \rput(.1,-4){$\mu_2$}
        \psline[ArrowInside=->](.6,-4.4)(.3,-4.1)\rput(.6,-4.6){$b_2$}
        \psline[ArrowInside=->](.6,-3.8)(.9,-3.5)\rput(.9,-3.8){$e_3$}
        \rput(.4,-3.7){$\mu_3$}
        \psline(.6,-3.8)(1.2,-4.4)\psline[ArrowInside=->](1.2,-4.4)(.9,-4.1)\rput(1.2,-4.6){$b_3$}
        \psline[linestyle=dotted](.9,-3.5)(1.3,-3.1)
        \psline[ArrowInside=->](1.3,-3.1)(1.6,-2.8)\rput(1.55,-3.3){$\bar{b}_{n}$}
        \psline(3.2,-4.4)(1.6,-2.8)\psline[ArrowInside=->](3.2,-4.4)(2.9,-4.1)\rput(3.2,-4.6){$b_n$}}
    \end{pspicture}.
\end{equation}
We note that this is precisely equal to the reduced density matrix $\rhoH$ from Eq.~(\ref{eq:rhoheuristic}).

From the reduced density matrix, we can calculate the anyonic R\'enyi entropy.  First, consider $\left(\rhoD\right)^2$:
\begin{equation}\label{eq:disk-ARE-start}
\begin{split}
\left( \rhoD\right)^2 &=
 \sum_{\substack{
\vec{b},\vec{e},\vec{\mu},\\ \vec{e}',\vec{\mu}'}} \frac{d_{\vec{b}}}{ \mathcal{D}^{4n-4}}
\psscalebox{.7}{
\begin{pspicture}[shift=-5.5](-.5,-9.5)(3.5,.3)
        \scriptsize
        \psline[ArrowInside=->](.3,-.3)(0,0)\rput(0,.2){$b_1$}
        \psline[ArrowInside=->](.6,-.6)(.3,-.3)\rput(.6,-.35){$e_2$}
        \rput(.1,-.45){$\mu_2$}
        \psline[ArrowInside=->](.3,-.3)(.6,0)\rput(.6,.2){$b_2$}
        \psline[ArrowInside=->](.9,-.9)(.6,-.6)\rput(.9,-.65){$e_3$}
        \rput(.4,-.75){$\mu_3$}
        \psline(.6,-.6)(1.2,0)\psline[ArrowInside=->](.9,-.3)(1.2,0)\rput(1.2,.2){$b_3$}
        \psline[linestyle=dotted](.9,-.9)(1.3,-1.3)
        \psline[ArrowInside=->](1.6,-1.6)(1.3,-1.3)\rput(1.55,-1.2){$\bar{b}_n$}        \psline(1.6,-1.6)(3.2,0)\psline[ArrowInside=->](2.9,-.3)(3.2,0)\rput(3.2,.2){$b_n$}
        \rput(0,.6){
        \psline[ArrowInside=->](0,-4.4)(.3,-4.1)
        \psline[ArrowInside=->](.3,-4.1)(.6,-3.8)\rput(.6,-4.1){$e_2$}
        \rput(.1,-4){$\mu_2$}
        \psline[ArrowInside=->](.6,-4.4)(.3,-4.1)
        \psline[ArrowInside=->](.6,-3.8)(.9,-3.5)\rput(.9,-3.8){$e_3$}
        \rput(.4,-3.7){$\mu_3$}
        \psline(.6,-3.8)(1.2,-4.4)\psline[ArrowInside=->](1.2,-4.4)(.9,-4.1)
        \psline[linestyle=dotted](.9,-3.5)(1.3,-3.1)
        \psline[ArrowInside=->](1.3,-3.1)(1.6,-2.8)\rput(1.55,-3.3){$\bar{b}_{n}$}
        \psline(3.2,-4.4)(1.6,-2.8)\psline[ArrowInside=->](3.2,-4.4)(2.9,-4.1)}
\rput(0,-3.8){
         \psline[ArrowInside=->](.3,-.3)(0,0)\rput(-.2,0){$b_1$}
        \psline[ArrowInside=->](.6,-.6)(.3,-.3)\rput(.6,-.35){$e_2'$}
        \rput(.1,-.45){$\mu_2'$}
        \psline[ArrowInside=->](.3,-.3)(.6,0)\rput(.8,0){$b_2$}
        \psline[ArrowInside=->](.9,-.9)(.6,-.6)\rput(.9,-.65){$e_3'$}
        \rput(.4,-.75){$\mu_3'$}
        \psline(.6,-.6)(1.2,0)\psline[ArrowInside=->](.9,-.3)(1.2,0)\rput(1.4,0){$b_3$}
        \psline[linestyle=dotted](.9,-.9)(1.3,-1.3)
        \psline[ArrowInside=->](1.6,-1.6)(1.3,-1.3)\rput(1.55,-1.2){$\bar{b}_{n}$}
        \psline(1.6,-1.6)(3.2,0)\psline[ArrowInside=->](2.9,-.3)(3.2,0)\rput(3.4,0){$b_n$}
        \rput(0,.6){
        \psline[ArrowInside=->](0,-4.4)(.3,-4.1)\rput(0,-4.6){$b_1$}
        \psline[ArrowInside=->](.3,-4.1)(.6,-3.8)\rput(.6,-4.1){$e_2'$}
        \rput(.1,-4){$\mu_2'$}
        \psline[ArrowInside=->](.6,-4.4)(.3,-4.1)\rput(.6,-4.6){$b_2$}
        \psline[ArrowInside=->](.6,-3.8)(.9,-3.5)\rput(.9,-3.8){$e_3'$}
        \rput(.4,-3.7){$\mu_3'$}
        \psline(.6,-3.8)(1.2,-4.4)\psline[ArrowInside=->](1.2,-4.4)(.9,-4.1)\rput(1.2,-4.6){$b_3$}
        \psline[linestyle=dotted](.9,-3.5)(1.3,-3.1)
        \psline[ArrowInside=->](1.3,-3.1)(1.6,-2.8)\rput(1.55,-3.3){$\bar{b}_n$}
        \psline(3.2,-4.4)(1.6,-2.8)\psline[ArrowInside=->](3.2,-4.4)(2.9,-4.1)\rput(3.2,-4.6){$b_n$}}
        }
    \end{pspicture}}
= \sum_{\vec{b},\vec{e},\vec{\mu}} \frac{\sqrt{d_{\vec{b}}}}{\mathcal{D}^{2n-2}}\left( \frac{d_{\vec{b}}}{ \mathcal{D}^{2n-2}}\right)
\psscalebox{.7}{
\begin{pspicture}[shift=-2.6](.6,-4.7)(3,.5)
        \scriptsize
        \psline[ArrowInside=->](.3,-.3)(0,0)\rput(0,.2){$b_1$}
        \psline[ArrowInside=->](.6,-.6)(.3,-.3)\rput(.6,-.35){$e_2$}
        \rput(.1,-.45){$\mu_2$}
        \psline[ArrowInside=->](.3,-.3)(.6,0)\rput(.6,.2){$b_2$}
        \psline[ArrowInside=->](.9,-.9)(.6,-.6)\rput(.9,-.65){$e_3$}
        \rput(.4,-.75){$\mu_3$}
        \psline(.6,-.6)(1.2,0)\psline[ArrowInside=->](.9,-.3)(1.2,0)\rput(1.2,.2){$b_3$}
        \psline[linestyle=dotted](.9,-.9)(1.3,-1.3)
        \psline[ArrowInside=->](1.6,-1.6)(1.3,-1.3)\rput(1.55,-1.2){$\bar{b}_{n}$}
        \psline(1.6,-1.6)(3.2,0)\psline[ArrowInside=->](2.9,-.3)(3.2,0)\rput(3.2,.2){$b_n$}
        \rput(0,.6){
        \psline[ArrowInside=->](0,-4.4)(.3,-4.1)\rput(0,-4.6){$b_1$}
        \psline[ArrowInside=->](.3,-4.1)(.6,-3.8)\rput(.6,-4.1){$e_2$}
        \rput(.1,-4){$\mu_2$}
        \psline[ArrowInside=->](.6,-4.4)(.3,-4.1)\rput(.6,-4.6){$b_2$}
        \psline[ArrowInside=->](.6,-3.8)(.9,-3.5)\rput(.9,-3.8){$e_3$}
        \rput(.4,-3.7){$\mu_3$}
        \psline(.6,-3.8)(1.2,-4.4)\psline[ArrowInside=->](1.2,-4.4)(.9,-4.1)\rput(1.2,-4.6){$b_3$}
        \psline[linestyle=dotted](.9,-3.5)(1.3,-3.1)
        \psline[ArrowInside=->](1.3,-3.1)(1.6,-2.8)\rput(1.55,-3.3){$\bar{b}_{n}$}
        \psline(3.2,-4.4)(1.6,-2.8)\psline[ArrowInside=->](3.2,-4.4)(2.9,-4.1)\rput(3.2,-4.6){$b_n$}}
    \end{pspicture}}.
\end{split}
\end{equation}

It is then easy to generalize to $\rhoD$ raised to an arbitrary power:
\begin{equation}\label{eq:rho-A-alpha}
\begin{split}
\left( \rhoD\right)^{\alpha}= \sum_{\vec{b},\vec{e},\vec{\mu}} &\frac{\sqrt{d_{\vec{b}}}}{\mathcal{D}^{2n-2}}\left( \frac{d_{\vec{b}}}{ \mathcal{D}^{2n-2}}\right)^{\alpha-1}
\begin{pspicture}[shift=-2.6](0,-4.7)(3.3,.5)
        \scriptsize
        \psline[ArrowInside=->](.3,-.3)(0,0)\rput(0,.2){$b_1$}
        \psline[ArrowInside=->](.6,-.6)(.3,-.3)\rput(.6,-.35){$e_2$}
        \rput(.1,-.45){$\mu_2$}
        \psline[ArrowInside=->](.3,-.3)(.6,0)\rput(.6,.2){$b_2$}
        \psline[ArrowInside=->](.9,-.9)(.6,-.6)\rput(.9,-.65){$e_3$}
        \rput(.4,-.75){$\mu_3$}
        \psline(.6,-.6)(1.2,0)\psline[ArrowInside=->](.9,-.3)(1.2,0)\rput(1.2,.2){$b_3$}
        \psline[linestyle=dotted](.9,-.9)(1.3,-1.3)
        \psline[ArrowInside=->](1.6,-1.6)(1.3,-1.3)\rput(1.55,-1.2){$\bar{b}_{n}$}        \psline(1.6,-1.6)(3.2,0)\psline[ArrowInside=->](2.9,-.3)(3.2,0)\rput(3.2,.2){$b_n$}
        \rput(0,.6){
        \psline[ArrowInside=->](0,-4.4)(.3,-4.1)\rput(0,-4.6){$b_1$}
        \psline[ArrowInside=->](.3,-4.1)(.6,-3.8)\rput(.6,-4.1){$e_2$}
        \rput(.1,-4){$\mu_2$}
        \psline[ArrowInside=->](.6,-4.4)(.3,-4.1)\rput(.6,-4.6){$b_2$}
        \psline[ArrowInside=->](.6,-3.8)(.9,-3.5)\rput(.9,-3.8){$e_3$}
        \rput(.4,-3.7){$\mu_3$}
        \psline(.6,-3.8)(1.2,-4.4)\psline[ArrowInside=->](1.2,-4.4)(.9,-4.1)\rput(1.2,-4.6){$b_3$}
        \psline[linestyle=dotted](.9,-3.5)(1.3,-3.1)
        \psline[ArrowInside=->](1.3,-3.1)(1.6,-2.8)\rput(1.55,-3.3){$\bar{b}_{n}$}
        \psline(3.2,-4.4)(1.6,-2.8)\psline[ArrowInside=->](3.2,-4.4)(2.9,-4.1)\rput(3.2,-4.6){$b_n$}
        }
    \end{pspicture}.
\end{split}
\end{equation}

Performing the quantum trace over Eq.~(\ref{eq:rho-A-alpha}) yields
\begin{equation} \label{eq:disk-ARE-end}
\begin{split}
\aTr \left( \rhoD\right)^\alpha &= \sum_{\vec{b},\vec{e},\vec{\mu}} \frac{\sqrt{d_{\vec{b}}}}{\mathcal{D}^{2n-2}} \left( \frac{d_{\vec{b}}}{ \mathcal{D}^{2n-2}}\right)^{\alpha-1}
\begin{pspicture}[shift=-2.6](0,-4.7)(3.3,.5)
        \scriptsize
        \psline[ArrowInside=->](.3,-.3)(0,0)\rput(-.2,0){$b_1$}
        \psline(0,0)(.3,.3)(1.9,.3)(3.1,-.9)(3.1,-2.9)(1.9,-4.1)(.3,-4.1)(0,-3.8)
        \psline[ArrowInside=->](.6,-.6)(.3,-.3)\rput(.6,-.35){$e_2$}
        \rput(.1,-.45){$\mu_2$}
        \psline[ArrowInside=->](.3,-.3)(.6,0)\rput(.35,0){$b_2$}
        \psline(.6,0)(1.9,0)(2.8,-.9)(2.8,-2.9)(1.9,-3.8)(.6,-3.8)
        \psline[ArrowInside=->](.9,-.9)(.6,-.6)\rput(.9,-.65){$e_3$}
        \rput(.4,-.75){$\mu_3$}
        \psline[ArrowInside=->](.6,-.6)(.9,-.3)\rput(.8,-.1){$b_3$}
        \psline(.9,-.3)(1.9,-.3)(2.5,-.9)(2.5,-2.9)(1.9,-3.5)(.9,-3.5)
        \psline[linestyle=dotted](.9,-.9)(1.3,-1.3)
        \psline[ArrowInside=->](1.6,-1.6)(1.3,-1.3)\rput(1.2,-1.3){$\bar{b}_{n}$}
        \psline[ArrowInside=->](1.6,-1.6)(1.9,-1.3)\rput(1.6,-1.3){$b_n$}
        \psline(1.9,-1.3)(2.2,-1.6)(2.2,-2.2)(1.9,-2.5)
        \rput(0,.6){
        \psline[ArrowInside=->](0,-4.4)(.3,-4.1)
        \psline[ArrowInside=->](.3,-4.1)(.6,-3.8)\rput(.6,-4.1){$e_2$}
        \rput(.1,-4){$\mu_2$}
        \psline[ArrowInside=->](.6,-4.4)(.3,-4.1)
        \psline[ArrowInside=->](.6,-3.8)(.9,-3.5)\rput(.9,-3.8){$e_3$}
        \rput(.4,-3.7){$\mu_3$}
        \psline[ArrowInside=->](.9,-4.1)(.6,-3.8)
        \psline[linestyle=dotted](.9,-3.5)(1.3,-3.1)
        \psline[ArrowInside=->](1.3,-3.1)(1.6,-2.8)\rput(1.55,-3.3){$\bar{b}_{n}$}
        \psline[ArrowInside=->](1.9,-3.1)(1.6,-2.8)
        }
    \end{pspicture}
\\ &=\sum_{\vec{b},\vec{e},\vec{\mu}} \left( \frac{d_{\vec{b}}}{\mathcal{D}^{2n-2}}\right)^\alpha
= \sum_{\vec{b}} N_{b_1\dots b_n}^{0} \left( \frac{d_{\vec{b}}}{\mathcal{D}^{2n-2}}\right)^{\alpha},
\end{split}
\end{equation}
from which we see the anyonic R\'enyi entropy is
\begin{equation}\label{eq:ARE-disk-unpunctured}
\aS^{(\alpha)}\left(\rhoD\right) = \frac{1}{1-\alpha} \log\Big[ \sum_{\vec{b}} N^{0}_{b_1\dots b_n} \left( \frac{d_{\vec{b}}}{\mathcal{D}^{2n-2}}\right)^\alpha \Big].
\end{equation}

Taking the limit $\alpha \to 1$ yields the (von Neumann) AEE:
\begin{equation}
\label{eq:S_A_doubled_disk}
\begin{split}
&\aS \left(\rhoD\right) = \lim_{\alpha \to 1} \aS^{(\alpha)}\left(\rhoD\right) = -\sum_{\vec{b}} N^{0}_{b_1\dots b_n} \left( \frac{d_{\vec{b}}}{\mathcal{D}^{2n-2}}\right) \log \left( \frac{d_{\vec{b}}}{\mathcal{D}^{2n-2}}\right)
\\ &= -\sum_{\vec{b},\vec{e}} N_{b_1 b_2}^{e_2}N_{e_2 b_3}^{e_3}\dots N^0_{\bar{b}_n,b_n} \frac{d_{\vec{b}}}{\mathcal{D}^{2n-2}} \Big[ \log\left( \frac{d_{b_1}}{\mathcal{D}^2} \right) + \dots + \log \left( \frac{d_{b_n}}{\mathcal{D}^2}\right) +2\log \mathcal{D} \Big]
\\ &= -n \sum_b \frac{d_b^2}{\mathcal{D}^2} \log \left( \frac{d_b}{\mathcal{D}^2}\right) - 2\log \mathcal{D}
\\ &= n\aS \left(\arho_{\partial \mathbb{A}_j}\right) +2 S_{\text{topo}}.
\end{split}
\end{equation}
In the second to last equality, we used Eq.~(\ref{eq:Nabc}) to sum over the multiplicities.  In the last equality, we used $S_{\text{topo}} \equiv -\log \mathcal{D}$ and the definition of the anyonic entropy of a ``boundary anyon'' given in Eq.~(\ref{eq:random-anyon-entropy}), which now applies to the anyonic state of the topological charge on $\partial \mathbb{A}_j$, the $j$th connected component of $\partial \mathbb{A}$, i.e. ${\aS\left(\arho_{\partial \mathbb{A}_j}\right)=\aS\left( \arho_{\partial \mathcal{A}_j}\right)}$.

At this point, the reason for the doubling of the topological contribution to the entanglement entropy coming from the partition boundary should be clear: we doubled the original region $A$ and the original partition boundary in this method of computation. Thus, the topological contribution to the entanglement entropy for the original region $A$ is given by
\begin{equation}\label{eq:AEE-disk-unpunctured}
\aS_A = \frac{n}{2} \aS \left(\arho_{\partial \mathbb{A}_j}\right) +S_{\text{topo}}.
\end{equation}

As shown in Ref.~\cite{Flammia09} for string-net and quantum double models, using the R\'enyi entropy produces the same value of the TEE for any index $\alpha$. This can be seen for more general topological phases using our approach by rewriting Eq.~(\ref{eq:ARE-disk-unpunctured}) in powers of the boundary length.
Consider the matrix
\begin{equation}\label{eq:start-ARE}
    [K_\alpha]_{e e'} \equiv \sum_{b} N_{e b}^{e'} d_{b}^{\alpha}.
\end{equation}
Since $d_b = d_{\bar{b}}$ and $N_{e b}^{e'}=N_{e' \bar{b}}^{e}$, it follows that $K_\alpha$ is normal and can, thus, be unitarily diagonalized, allowing us to write it as
\begin{equation}
[K_\alpha]_{e e'} = \sum_{\mu} \kappa_{\alpha,\mu} [v_{\alpha,\mu}]_{e} [v_{\alpha,\mu}]_{e'}^*.
\end{equation}
where $\kappa_{\alpha,\mu}$ is the $\mu$th eigenvalue with corresponding normalized eigenvector $v_{\alpha,\mu}$.  We note that $[K_\alpha]_{e e'}>0$ for all $e$ and $e'$, since there must be some value of $b$ such that $N_{e b}^{e'}\neq 0$. Thus, $K_\alpha$ obeys the Perron-Frobenius theorem, which implies that there is a unique eigenvector which has all positive real components (up to an overall scalar), and the corresponding eigenvalue of this eigenvector is positive and larger in magnitude than all other eigenvalues. We label this eigenvector by $\mu=0$. It is straightforward to check that $[v_{\alpha}]_e = d_e/\mathcal{D}$ is a normalized eigenvector, so it must be the $\mu=0$ eigenvector. Its corresponding eigenvalue is
\begin{equation}
\kappa_{\alpha,0}=\displaystyle{\sum_{e}} d_{e}^{1+\alpha}
.
\end{equation}
Thus, we find that
\begin{align} \label{eq:AREint}
    \log \left( \sum_{\vec{b}} N_{b_1\dots b_n}^{0} d_{b_1}^\alpha \dots d_{b_n}^\alpha \right)
 &= \log \left( [ (K_\alpha)^n]_{00} \right)
 = \log \left( \sum_\mu \kappa_{\alpha,\mu}^n [v_{\alpha,\mu}]_0 [v_{\alpha,\mu}]_{0}^* \right) \notag \\
&= \log \left( \frac{\kappa_{\alpha,0}^n}{\mathcal{D}^2} + \sum_{\mu \neq 0} \kappa_{\alpha,\mu}^n [v_{\alpha,\mu}]_0 [v_{\alpha,\mu}]_{0}^* \right)
\notag \\
 &= n \log \kappa_{\alpha,0} - \log \mathcal{D}^{2}  + F\left( n, 0, K_\alpha \right).
\end{align}
Here, we have defined
\begin{equation}\label{eq:end-ARE}
    F(n, c, K_\alpha)\equiv\log \left( 1 + \frac{\mathcal{D}^2}{d_c} \sum_{\mu \neq 0}\left(\frac{\kappa_{\alpha,\mu}}{\kappa_{\alpha,0} } \right)^n [v_{\alpha, \mu}]_0 [v_{\alpha,\mu}]_{c}^* \right),
\end{equation}
which is exponentially suppressed in $n$ for large $n$, since $\kappa_{\alpha,\mu} < \kappa_{\alpha,0}$ for all $\mu \neq 0$. More specifically, $|F(n, c, K_\alpha)| = \mathcal{O}(e^{- \lambda n})$, where $\lambda = - \log ( \max\limits_{\mu \neq 0} | \kappa_{\alpha,\mu} / \kappa_{\alpha,0}  |)$ is a constant that only depends on the TQFT.

Plugging Eq.~(\ref{eq:AREint}) back into Eq.~(\ref{eq:ARE-disk-unpunctured}), we have
\begin{align}\label{eq:ARE-disk-TEE}
    \aS^{(\alpha)}\left(\rhoD\right) &= n \aS^{(\alpha)}(\arho_{\partial \mathbb{A}_j})+2S_{\text{topo}} +\frac{F(n, 0, K_\alpha)}{1-\alpha},
\end{align}
where we have denoted the anyonic R\'enyi entropy of a boundary anyon as
\begin{align}
    \aS^{(\alpha)}(\arho_{\partial \mathbb{A}_j}) &= \aS^{(\alpha)}(\arho_{\partial \mathcal{A}_j}) = \frac{1}{1-\alpha}\log\left(\frac{\kappa_{\alpha, 0}}{\mathcal{D}^{2\alpha}}\right).
\end{align}
Eq.~(\ref{eq:ARE-disk-TEE}) has the same form as Eq.~(\ref{eq:ententropy}): a term that is linear in the length of the boundary ($n \sim L/\ell$), a universal constant topological contribution, and sub-constant corrections. Again, for the topological contribution to the entanglement R\'enyi entropy of the original (un-doubled) system, this should be divided by two
\begin{align}\label{eq:ARE-disk-TEE}
    \aS^{(\alpha)}_{A} &= \frac{n}{2} \aS^{(\alpha)}(\arho_{\partial \mathbb{A}_j})+S_{\text{topo}} +\frac{F(n, 0, K_\alpha)}{2(1-\alpha)}.
\end{align}

Finally, we clarify why the original Kitaev-Preskill method of computing the TEE must me modified when using the R\'enyi entropy. Let $\aS_n$ and $\aS^{(\alpha)}_n$ denote the anyonic von Neumann and R\'enyi entanglement entropies, respectively, of the doubled region $\AD$ when $n$ wormholes were inserted along the partition boundary in the doubling process, i.e. $\AD$ is an $n$-punctured sphere. The method of Ref.~\cite{Kitaev06b} utilized different geometric partitions of the systems into disks that resulted in 3-punctured and 4-punctured spheres after doubling and cutting, and showed that
\begin{equation}
S_{\text{topo}}= 2 \aS_3 -\frac{3}{2} \aS_4.
\end{equation}
We see that this result holds given the form of Eq.~(\ref{eq:AEE-disk-unpunctured}).  However, this result does not extend to the anyonic R\'enyi entropies, as can be seen from the form of Eq.~(\ref{eq:ARE-disk-unpunctured}):
\begin{equation}
2\aS^{(\alpha)}_{3}-\frac{3}{2}\aS^{(\alpha)}_4 = S_{\text{topo}} + \frac{2 F(3,0,K_\alpha) -\frac{3}{2} F(4, 0, K_\alpha)}{1-\alpha},
\end{equation}
as the second term is some constant that depends on the TQFT, with no dependence on the boundary length. Our method recovers the boundary-law (linear length dependence) of the entanglement entropy and the TEE when utilizing R\'enyi entropies.

\subsection{2-Punctured Sphere Partitioned into Two 1-Punctured Disks}\label{sec:disk}

We now extend the results of the previous section to the case when the disk $A$ hosts an anyon $c$.

\begin{center}
\includegraphics[width=.6\linewidth]{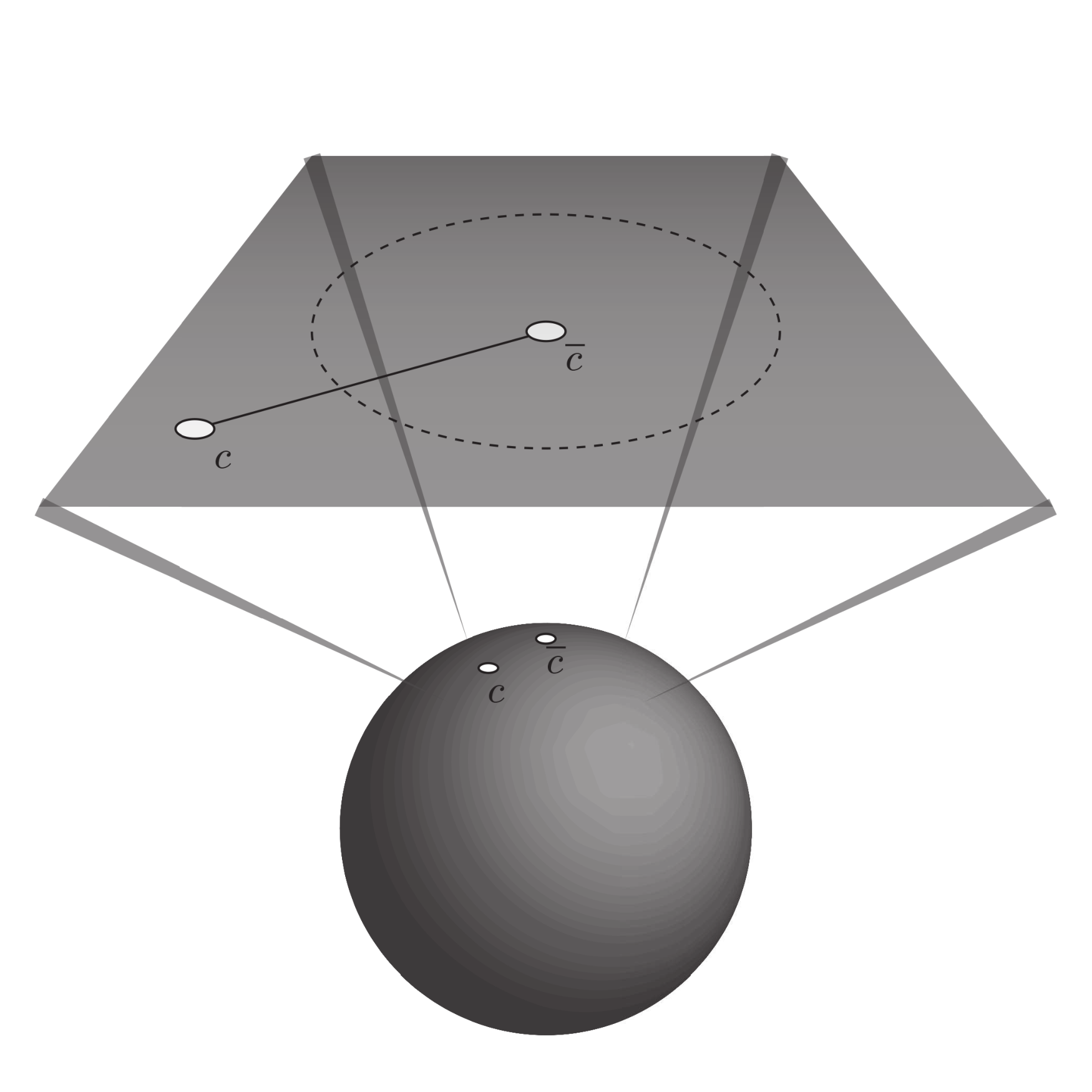}
\end{center}

\noindent The line connecting $c$ and $\bar{c}$ along the surface can be thought of as the path through which the topological charges were created and moved to the shown positions.

As before, we pair the system with its time-reversal conjugate, joining them by adiabatically inserting $n$ wormholes along the boundary partition ($n=3$ in the following picture).

\begin{center}
\includegraphics[width=.9\linewidth]{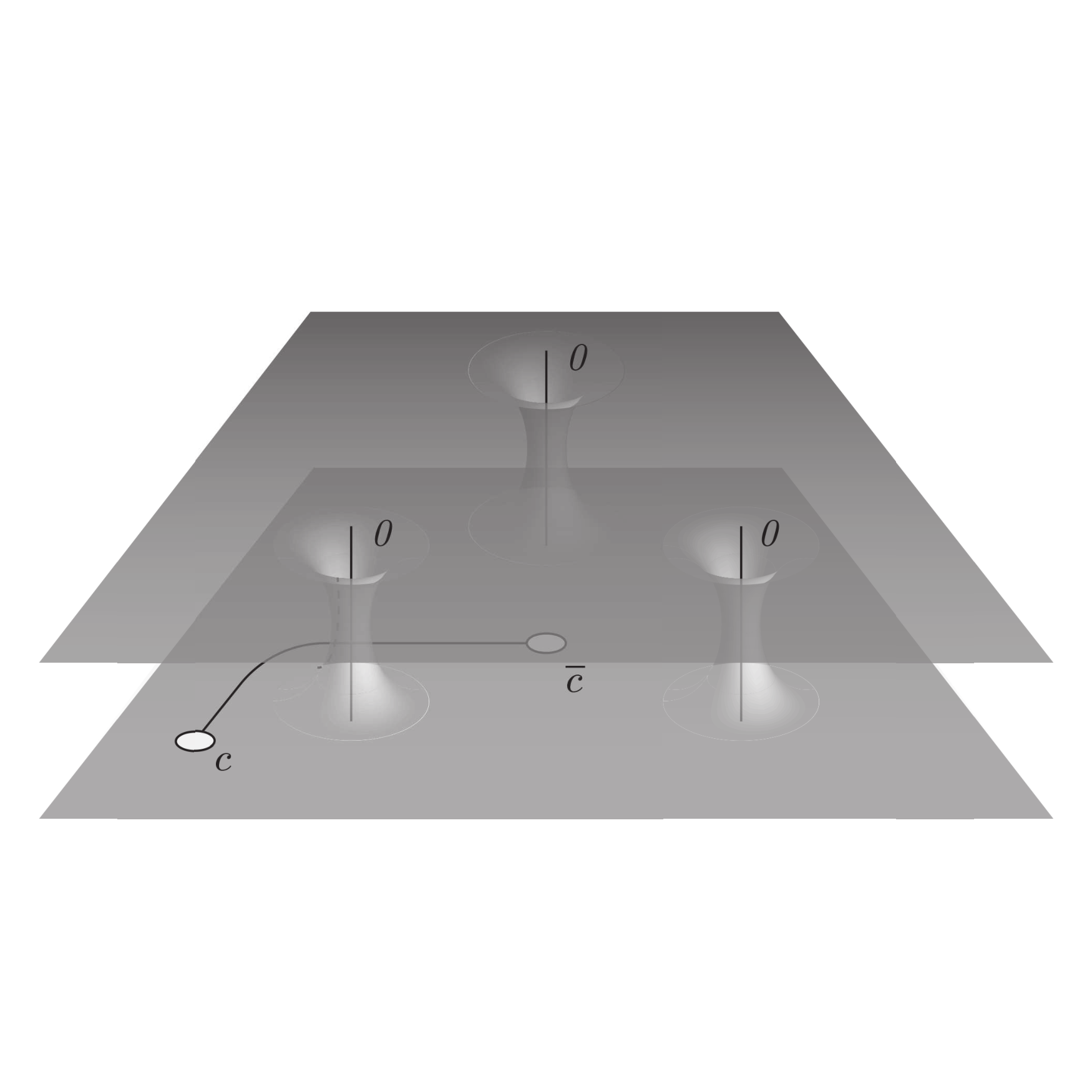}
\end{center}

The derivation of the anyonic reduced density matrix for $\AD$ proceeds in much the same way as for the unpunctured disk in Section~\ref{sec:unpunctured-disk}, with the only difference being that the topological charges $c$ and $\bar{c}$ are present.  For instance, Eq.~(\ref{eq:unpunct-disk-fuse}) is modified to
\begin{equation}
    |\psi\rangle=\frac{1}{\mathcal{D}}\sum_{\vec{a}, \vec{b}} \frac{\sqrt{d_{\vec{b}}}}{\sqrt{d_c}\mathcal{D}^n}
    \begin{pspicture}[shift=-2.5](-0.2,-2.6)(5,1.4)
        \scriptsize
        \psline[linestyle=dashed](0.25,.4)(0.25,-.2)(1.65,-1.6)(2.95,-1.6) (4.35,-.2)(4.35,.5)(4.1,.5)
        \psline[linestyle=dashed](.35,.5)(.45,.5)
        \psline[linestyle=dashed](1,.5)(.9,.5)
        \psline[linestyle=dashed](1.2,.5)(1.3,.5)
        \rput(0.25,0.5){$\odot$}
        \rput(1.1,0.5){$\odot$}
        \psline[ArrowInside=->](4.05,-1.15)(3.3,-.4)\rput(3.35,-.25){$c$}
        \psline[ArrowInside=->](4.05,-1.15)(4.8,-.4)\rput(4.85,-.25){$\bar{c}$}
        \psline[border=1.5pt](0.8,0.7)(0.5,1)(0,0.5)(0,-0.3)(1.6,-1.9)(3,-1.9)(4.6,-.3)
        (4.6,.5)(4.1,1)
        \psline[border=1.5pt](0.8,0.3)(0.5,0)(0.5,-0.1)(1.7,-1.3)(2.9,-1.3)(4.1,-.1)
        \psline[border=1.5pt](1.1,0)(1.4,0.3)(1.4,0.7)(1.1,1)(0.8,0.7)(0.8,0.3)(1.1,0)
        \psline[ArrowInside=->](0.5,0)(0.8,0.3)\rput(0.5,0.2){$\bar{a}_1$}
        \psline[ArrowInside=->](0.8,0.3)(0.8,0.7)\rput(0.6,0.5){$b_1$}
        \psline[ArrowInside=->](0.8,0.7)(0.5,1)\rput(0.5,0.8){$\bar{a}_1$}
        \psline[ArrowInside=->](1.1,0)(.8,0.3)\rput(1.1,0.3){$a_2$}
        \psline[ArrowInside=->](1.4,0.3)(1.4,0.7)\rput(1.6,0.5){$b_2$}
        \psline[ArrowInside=->](.8,0.7)(1.1,1)\rput(1.1,0.7){$a_2$}
        \psline[ArrowInside=->](1.7,0)(1.4,0.3)\rput(1.7,0.2){$a_3$}
        \psline[ArrowInside=->](1.4,0.7)(1.7,1)\rput(1.7,0.7){$a_3$}
        \rput(2.2,0.5){$\dots$}
        \rput(3,0){
        \rput(0.5,0.5){$\odot$}
        \psline[linestyle=dashed](.4,.5)(.3,.5)
        \psline[linestyle=dashed](.6,.5)(.7,.5)
       \psline[border=1.5pt](0.5,0)(0.8,0.3)(0.8,0.7)(0.5,1)(0.2,0.7)(0.2,0.3)(0.5,0)
        \psline[ArrowInside=->](-0.1,0)(0.2,0.3)\rput(-0.3,0.2){$\bar{a}_{n-1}$}
        \psline[ArrowInside=->](0.2,0.3)(0.2,0.7)\rput(-0.15,0.5){$b_{n-1}$}
        \psline[ArrowInside=->](0.2,0.7)(-0.1,1)\rput(-0.3,0.8){$\bar{a}_{n-1}$}
        \psline[ArrowInside=->](0.5,0)(0.2,0.3)\rput(0.5,0.3){$a_n$}
        \psline[ArrowInside=->](0.8,0.3)(0.8,0.7)\rput(1,0.5){$b_n$}
        \psline[ArrowInside=->](0.2,0.7)(0.5,1)\rput(0.5,0.7){$a_n$}
        \psline[ArrowInside=->](1.1,0)(0.8,0.3)\rput(1.1,0.25){$a_1$}
        \psline[ArrowInside=->](0.8,0.7)(1.1,1)\rput(1.1,0.7){$a_1$}}
        \psline(4.1,0)(4.1,-.1)
    \end{pspicture},
\end{equation}
where the charge lines embedded in the doubled surface  look like

\begin{center}
\includegraphics[width=.9\linewidth]{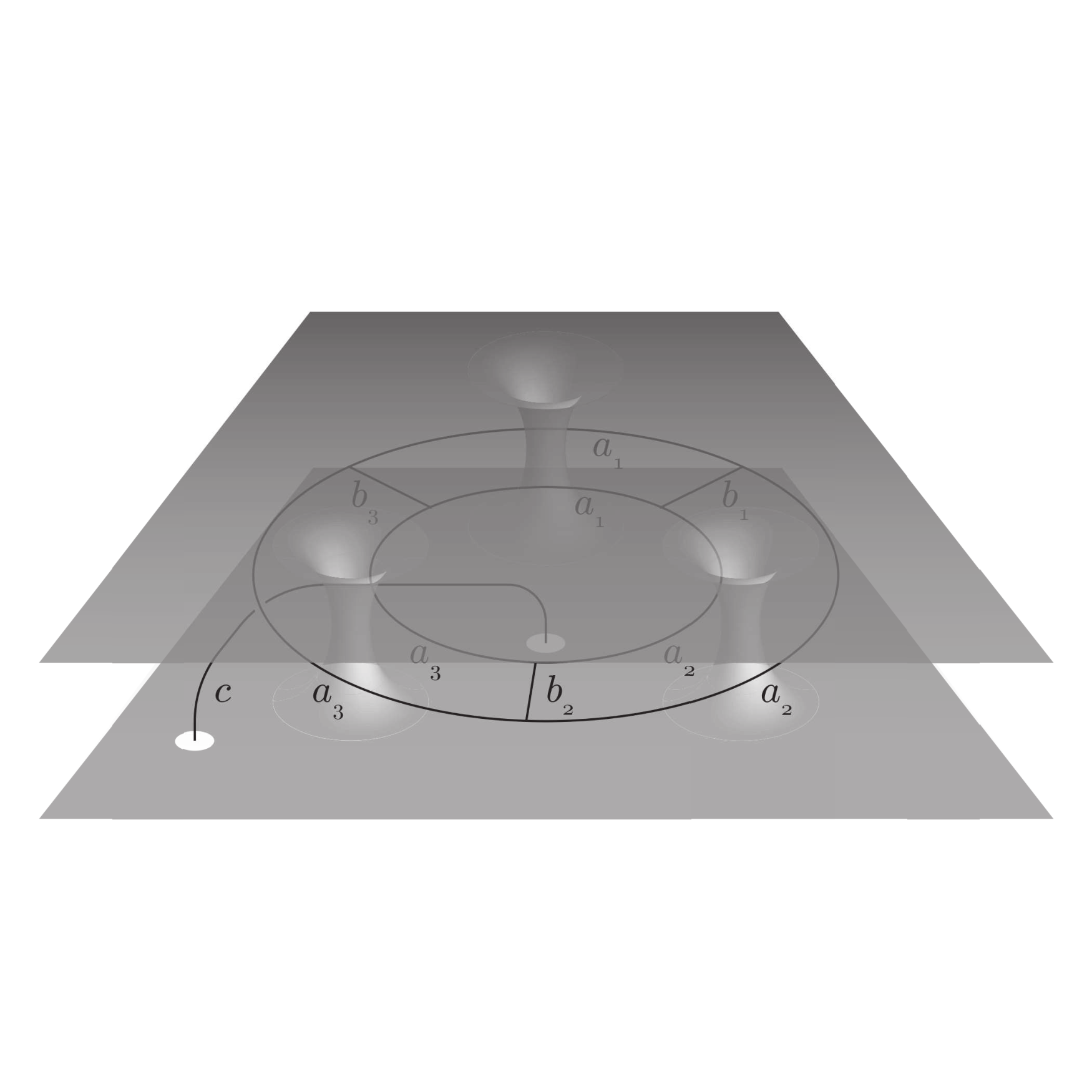}
\end{center}
As the charge line connecting $c$ and $\bar{c}$ lies below all other charge lines in the above picture, the steps illustrated in Eq.~(\ref{eq:F-move-steps}) (i.e., $F$-moves to rewrite the state in tree-like form and collapsing the tadpoles in $\AD$ and $\bar{\AD}$) also apply to the excited state considered here.  After applying these steps, we are left with the state in the form
\begin{equation}
|\psi\rangle=
\sum_{\vec{b}, \vec{e}} \frac{\sqrt{d_{\vec{b}}}}{\sqrt{d_c}\mathcal{D}^{n-1}}
\psscalebox{.8}{
\begin{pspicture}[shift=-2.4](.1,-2.2)(4.7,2.2)
        \scriptsize
        \psline[linestyle=dashed](3.6,.5)(4.25,.5)(4.25,-1.05)(3.25,-2.05) (2.65,-2.05)(.3,.3)(.3,.5)(.4,.5)
        \psline[linestyle=dashed](.6,.5)(.7,.5)
        \psline[linestyle=dashed](.9,.5)(1,.5)
        \psline[linestyle=dashed](1.2,.5)(1.3,.5)
        \psline[linestyle=dashed](1.5,.5)(1.6,.5)
        \psline[linestyle=dashed](1.8,.5)(1.9,.5)
        \psline[linestyle=dashed](2.9,.5)(3,.5)
        \psline[linestyle=dashed](3.2,.5)(3.3,.5)
        \rput(0.5,0.5){$\odot$}
        \rput(1.1,0.5){$\odot$}
        \rput(1.7,0.5){$\odot$}
        \rput(3.1,0.5){$\odot$}
        \psline(4.25,-.9)(3.75,-.4)\psline[ArrowInside=->](3.95,-.6)(3.75,-.4) \rput(3.75,-.25){$c$}
        \psline(4.25,-.9)(4.75,-.4)\psline[ArrowInside=->](4.55,-.6)(4.75,-.4) \rput(4.75,-.25){$\bar{c}$}
        \psline(0.8,0.7)(1.7,1.6)
        \psline[border=1.5pt](1.7,-0.6)(0.8,0.3)
        \psline(1.4,1.3)(2,.7)(2,.3)(1.4,-0.3)
        \psline[ArrowInside=->](2,.3)(2,.7)\rput(2.2,.75){$b_3$}
        \psline[ArrowInside=->](1.1,1)(1.4,1.3)\rput(1.15,1.3){$e_2$}
        \psline[ArrowInside=->](1.4,1.3)(1.7,1.6)\rput(1.45,1.6){$e_3$}
        \psline[linestyle=dotted](1.7,1.6)(2,1.9)
        \psline[ArrowInside=->](0.8,0.3)(0.8,0.7)\rput(0.65,0.75){$b_1$}
        \psline[ArrowInside=->](2.2,2.1)(2.5,2.4)\rput(2.25,2){$e_{n-2}$}
        \psline[ArrowInside=->](2.5,2.4)(2.8,2.7)
        \psline(2.5,2.4)(2.8,2.1)(2.8,-1)(2.5,-1.3)
        \psline(2.8,2.7)(3.4,2.1)(3.4,-1.)(2.8,-1.6)
        \psline(1.1,0)(1.4,0.3)(1.4,0.7)(1.1,1)
        \psline[ArrowInside=->](1.4,0.3)(1.4,0.7)\rput(1.6,0.75){$b_2$}
        \psline[ArrowInside=->](2.8,0.3)(2.8,0.7)\rput(2.45,.2){$b_{n-1}$}
        \psline[ArrowInside=->](3.4,0.3)(3.4,0.7)\rput(3.2,0.75){$b_{n}$}
        \psline[ArrowInside=->](1.4,-0.3)(1.1,0)\rput(1.15,-0.3){$e_2$}
        \psline[ArrowInside=->](1.7,-0.6)(1.4,-0.3)\rput(1.45,-0.6){$e_3$}
        \psline[linestyle=dotted](1.7,-0.6)(2,-.9)
        \psline[ArrowInside=->](2.5,-1.3)(2.2,-1)\rput(2.4,-.9){$e_{n-2}$}
        \psline[ArrowInside=->](2.8,-1.6)(2.5,-1.3)
        \rput(2.4,0.5){$\dots$}
    \end{pspicture}}
,
\end{equation}
with corresponding three dimensional embedding

\begin{center}
\noindent\includegraphics[width=.9\linewidth]{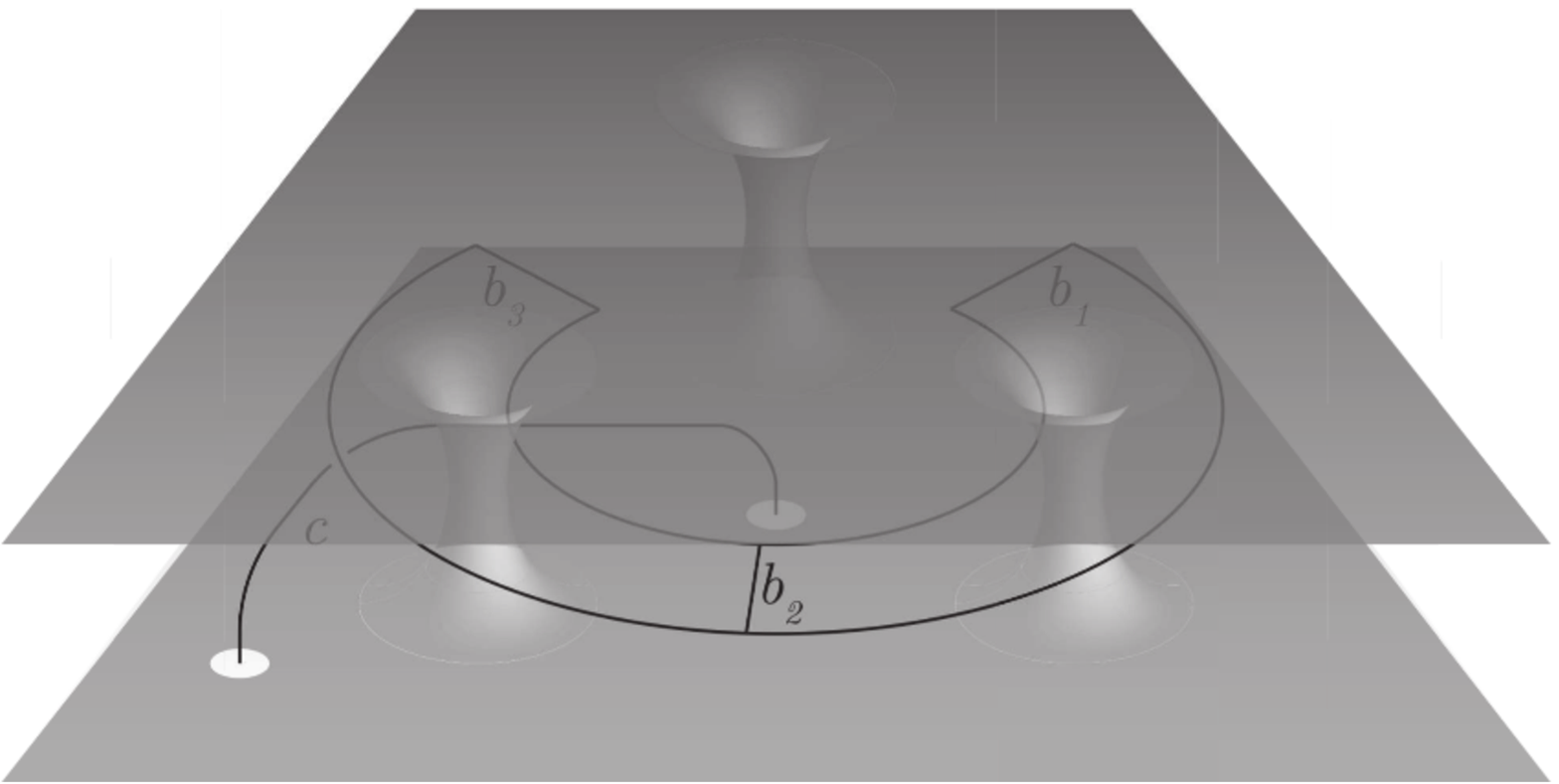}
\end{center}

Now, before cutting the surface into $\AD$ and $\bar{\AD}$, we must first fuse $c$ with the topological charge line running through the same boundary region (taken in the picture to be $b_3$ and in the diagram to be $b_n$):
\begin{equation}\label{eq:psiIntDis}
|\psi\rangle=
\sum_{ \vec{b}, \vec{e}} \frac{\sqrt{d_{\vec{b}}}}{d_c\mathcal{D}^{n-1}}
\psscalebox{.8}{
\begin{pspicture}[shift=-2.8](.1,-2.5)(4.5,3.5)
        \scriptsize
        \psline[linestyle=dashed](3.6,.5)(4.3,.5)(4.3,-1.3)(3.25,-2.35) (2.95,-2.35)(.3,.3)(.3,.5)(.4,.5)
        \psline[linestyle=dashed](.6,.5)(.7,.5)
        \psline[linestyle=dashed](.9,.5)(1,.5)
        \psline[linestyle=dashed](1.2,.5)(1.3,.5)
        \psline[linestyle=dashed](1.5,.5)(1.6,.5)
        \psline[linestyle=dashed](1.8,.5)(1.9,.5)
        \psline[linestyle=dashed](2.9,.5)(3,.5)
        \psline[linestyle=dashed](3.2,.5)(3.3,.5)
        \rput(0.5,0.5){$\odot$}
        \rput(1.1,0.5){$\odot$}
        \rput(1.7,0.5){$\odot$}
        \rput(3.1,0.5){$\odot$}
        \psline(2.8,2.7)(3.3,3.2)\psline[ArrowInside=->](2.8,2.7)(3.1,3)\rput(3.4,3.3){$\bar{c}$}
        \psline(0.8,0.7)(1.7,1.6)
        \psline[border=1.5pt](1.7,-0.6)(0.8,0.3)
        \psline(1.4,1.3)(2,.7)(2,.3)(1.4,-0.3)
        \psline[ArrowInside=->](2,.3)(2,.7)\rput(2.2,.75){$b_3$}
        \psline[ArrowInside=->](1.1,1)(1.4,1.3)\rput(1.15,1.3){$e_2$}
        \psline[ArrowInside=->](1.4,1.3)(1.7,1.6)\rput(1.45,1.6){$e_3$}
        \psline[linestyle=dotted](1.7,1.6)(2,1.9)
        \psline[ArrowInside=->](0.8,0.3)(0.8,0.7)\rput(0.65,0.75){$b_1$}
        \psline[ArrowInside=->](2.2,2.1)(2.5,2.4)\rput(2.25,2){$e_{n-2}$}
        \psline[ArrowInside=->](2.5,2.4)(2.8,2.7)\rput(2.35,2.7){$e_{n-1}$}
        \psline(2.5,2.4)(2.8,2.1)(2.8,-1)(2.5,-1.3)
        \psline(2.8,2.7)(3.4,2.1)(3.4,-1.)(2.8,-1.6)
        \psline[ArrowInside=->](3.1,-1.9)(2.8,-1.6)
        \psline(3.1,-1.9)(4,-1)\psline[ArrowInside=->](3.7,-1.3)(4,-1)\rput(4,-.85){$c$}
        \psline(1.1,0)(1.4,0.3)(1.4,0.7)(1.1,1)
        \psline[ArrowInside=->](1.4,0.3)(1.4,0.7)\rput(1.6,0.75){$b_2$}
        \psline[ArrowInside=->](2.8,0.3)(2.8,0.7)\rput(2.45,.2){$b_{n-1}$}
        \psline[ArrowInside=->](3.4,0.3)(3.4,0.7)\rput(3.2,0.75){$b_{n}$}
        \psline[ArrowInside=->](1.4,-0.3)(1.1,0)\rput(1.15,-0.3){$e_2$}
        \psline[ArrowInside=->](1.7,-0.6)(1.4,-0.3)\rput(1.45,-0.6){$e_3$}
        \psline[linestyle=dotted](1.7,-0.6)(2,-.9)
        \psline[ArrowInside=->](2.5,-1.3)(2.2,-1)\rput(2.4,-.9){$e_{n-2}$}
        \psline[ArrowInside=->](2.8,-1.6)(2.5,-1.3) \rput(2.55,-1.65){$e_{n-1}$}
        \rput(2.4,0.5){$\dots$}
    \end{pspicture}}.
\end{equation}
In rewriting the state into this final form, we have used a braiding transformation that only contributes an overall phase to the state, which we therefore can drop. Additionally, we applied a partition of identity and have relabeled $b_n$ as $\bar{e}_{n-1}$ and instead used $b_n$ to denote the fusion channel of $\bar{e}_{n-1}$ and $\bar{c}$ in the partition of identity. The shorthand notation $\vec{e}$ now means $e_2\dots e_{n-1}$. This diagrammatic state embedded in the doubled surface looks like

\begin{center}
\includegraphics[width=.9\linewidth]{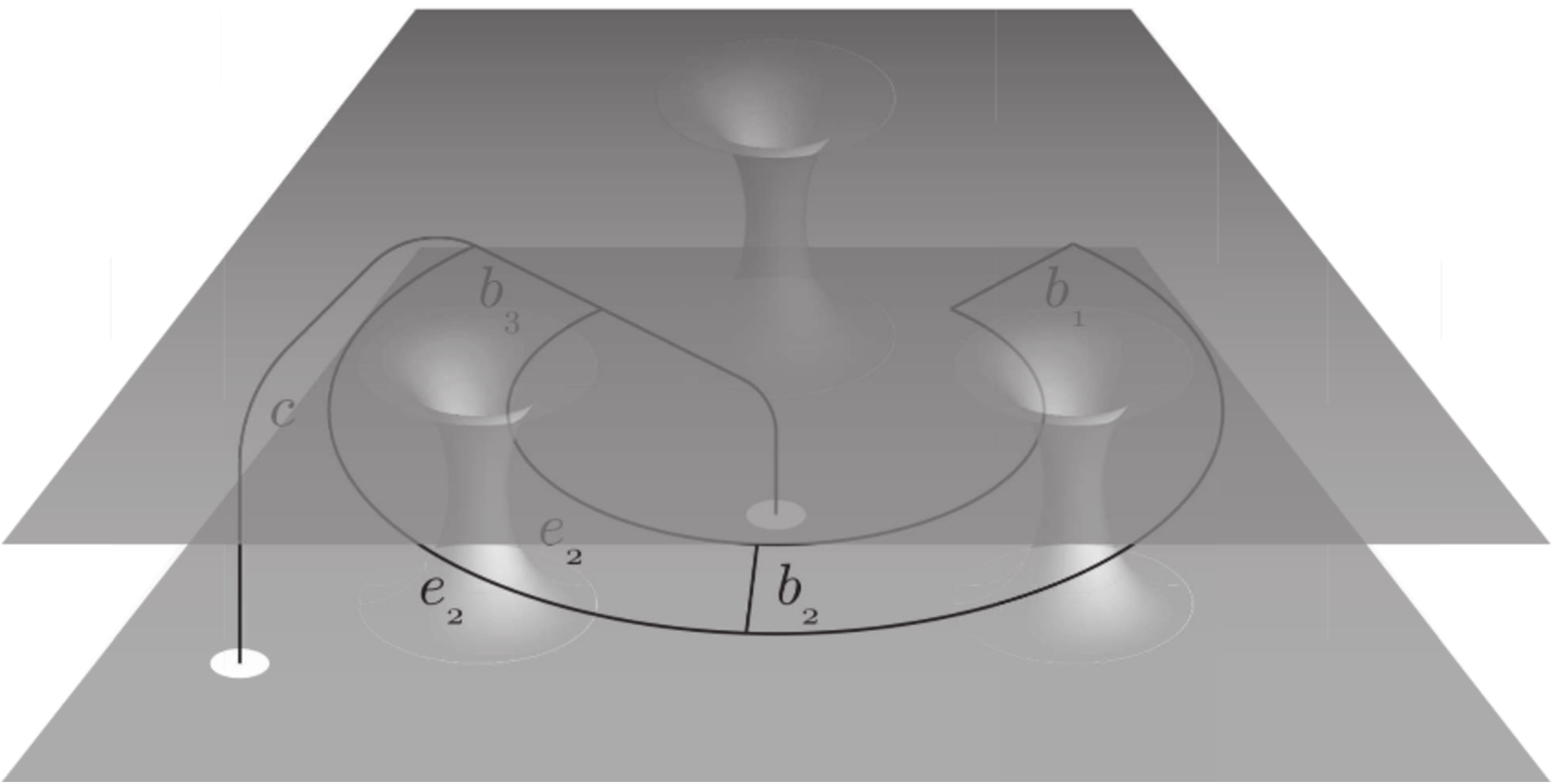}
\end{center}

We can now cut the doubled surface, resulting in the state
\begin{equation}
\begin{split}
   & |\psi_{\text{cut}}\rangle=\sum_{ \vec{b}, \vec{e}} \frac{1}{d_c\mathcal{D}^{n-1}}
\begin{pspicture}[shift=-1](-.3,-2)(8,.3)
        \scriptsize
        \rput(1.9,-2.2){$(\AD)$}
        \psline[ArrowInside=->](.3,-.3)(0,0)\rput(0,.2){$b_1$}
        \psline[ArrowInside=->](.6,-.6)(.3,-.3)\rput(.6,-.35){$e_2$}
        \psline[ArrowInside=->](.3,-.3)(.6,0)\rput(.6,.2){$b_2$}
        \psline[ArrowInside=->](.9,-.9)(.6,-.6)\rput(.9,-.65){$e_3$}
        \psline(.6,-.6)(1.2,0)\psline[ArrowInside=->](.9,-.3)(1.2,0)\rput(1.2,.2){$b_3$}
        \psline[linestyle=dotted](.9,-.9)(1.3,-1.3)
        \psline[ArrowInside=->](1.6,-1.6)(1.3,-1.3)\rput(1.55,-1.2){$e_{n-2}$}
        \psline(1.6,-1.6)(3.2,0)\psline[ArrowInside=->](2.9,-.3)(3.2,0)\rput(3.2,.2){$b_n$}
        \psline[ArrowInside=->](1.9,-1.9)(1.6,-1.6)\rput(1.9,-1.65){$\bar{c}$}
        \psline(1.9,-1.9)(3.8,0)\psline[ArrowInside=->](3.5,-.3)(3.8,0)\rput(3.8,.2){$c$}
        \rput(.3,0){\rput(6.1,-2.2){$(\bar{\AD})$}
        \psline[ArrowInside=->](7.7,-.3)(8,0)\rput(8,.2){$b_1$}
        \psline[ArrowInside=->](7.4,-.6)(7.7,-.3)\rput(7.4,-.35){$e_2$}
        \psline[ArrowInside=->](7.7,-.3)(7.4,0)\rput(7.4,.2){$b_2$}
        \psline[ArrowInside=->](7.1,-.9)(7.4,-.6)\rput(7.1,-.65){$e_3$}
        \psline(7.4,-.6)(6.8,0)\psline[ArrowInside=->](7.1,-.3)(6.8,0)\rput(6.8,.2){$b_3$}
        \psline[linestyle=dotted](7.1,-.9)(6.7,-1.3)
        \psline[ArrowInside=->](6.4,-1.6)(6.7,-1.3)\rput(6.4,-1.2){$e_{n-2}$}
        \psline(6.4,-1.6)(4.8,0)\psline[ArrowInside=->](5.1,-.3)(4.8,0)\rput(4.8,.2){$b_{n}$}
        \psline[ArrowInside=->](6.1,-1.9)(6.4,-1.6)\rput(6.1,-1.65){$c$}
        \psline(6.1,-1.9)(4.2,0)\psline[ArrowInside=->](4.5,-.3)(4.2,0)\rput(4.2,.2){$\bar{c}$}}
    \end{pspicture}
\end{split}
\end{equation}

Finally, we write the density matrix in the cut Hilbert space, $\arho=\ket{\psi_{\text{cut}}}\bra{\psi_{\text{cut}}}$, and then trace over $\bar{\AD}$ to find the reduced anyonic density matrix (restoring the vertex labels):
\begin{equation}\label{eq:Rhodisk}
\rhoD=\sum_{\vec{b}, \vec{e}} \frac{\sqrt{d_{\vec{b}}}}{d_c^{3/2}\mathcal{D}^{2n-2}}
\begin{pspicture}[shift=-2.25](0,-4.7)(4,.3)
        \scriptsize
        \psline[ArrowInside=->](.3,-.3)(0,0)\rput(0,.2){$b_1$}
        \psline[ArrowInside=->](.6,-.6)(.3,-.3)\rput(.6,-.35){$e_2$}
        \rput(.1,-.45){$\mu_2$}
        \psline[ArrowInside=->](.3,-.3)(.6,0)\rput(.6,.2){$b_2$}
        \psline[ArrowInside=->](.9,-.9)(.6,-.6)\rput(.9,-.65){$e_3$}
        \rput(.4,-.75){$\mu_3$}
        \psline(.6,-.6)(1.2,0)\psline[ArrowInside=->](.9,-.3)(1.2,0)\rput(1.2,.2){$b_3$}
        \psline[linestyle=dotted](.9,-.9)(1.3,-1.3)
        \psline[ArrowInside=->](1.6,-1.6)(1.3,-1.3)\rput(1.55,-1.2){$e_{n-2}$}
        \rput(1.2,-1.75){$\mu_{n-1}$}
        \psline(1.6,-1.6)(3.2,0)\psline[ArrowInside=->](2.9,-.3)(3.2,0)\rput(3.2,.2){$b_n$}
        \psline[ArrowInside=->](1.9,-1.9)(1.6,-1.6)\rput(1.9,-1.65){$\bar{c}$}
        \psline(1.9,-1.9)(3.8,0)\psline[ArrowInside=->](3.5,-.3)(3.8,0)\rput(3.8,.2){$c$}
        \psline[ArrowInside=->](0,-4.4)(.3,-4.1)\rput(0,-4.6){$b_1$}
        \psline[ArrowInside=->](.3,-4.1)(.6,-3.8)\rput(.6,-4.1){$e_2$}
        \rput(.1,-4){$\mu_2$}
        \psline[ArrowInside=->](.6,-4.4)(.3,-4.1)\rput(.6,-4.6){$b_2$}
        \psline[ArrowInside=->](.6,-3.8)(.9,-3.5)\rput(.9,-3.8){$e_3$}
        \rput(.4,-3.7){$\mu_3$}
        \psline(.6,-3.8)(1.2,-4.4)\psline[ArrowInside=->](1.2,-4.4)(.9,-4.1)\rput(1.2,-4.6){$b_3$}
        \psline[linestyle=dotted](.9,-3.5)(1.3,-3.1)
        \psline[ArrowInside=->](1.3,-3.1)(1.6,-2.8)\rput(1.55,-3.3){$e_{n-2}$}
        \rput(1.2,-2.65){$\mu_{n-1}$}
        \psline(3.2,-4.4)(1.6,-2.8)\psline[ArrowInside=->](3.2,-4.4)(2.9,-4.1)\rput(3.2,-4.6){$b_n$}
        \psline[ArrowInside=->](1.6,-2.8)(1.9,-2.5)\rput(1.9,-2.8){$\bar{c}$}
        \psline(1.9,-2.5)(3.8,-4.4)\psline[ArrowInside=->](3.8,-4.4)(3.5,-4.1)\rput(3.8,-4.6){$c$}
    \end{pspicture}.
\end{equation}
Comparing Eq.~(\ref{eq:rho_c_heuristic}) with Eq.~(\ref{eq:Rhodisk}), we see that the heuristic argument of Section~\ref{sec:ATEEI} produced the same reduced anyonic density matrix $\rhoH$ for a disk containing a puncture or quasiparticle of topological charge $c$ as did our method (generalizing the Kitaev-Preskill method) using a doubled surface connected by wormholes.

Using the same steps outlined in Eqs.~(\ref{eq:start-ARE})-(\ref{eq:end-ARE}) for the unpunctured disk, we can calculate the anyonic R\'enyi entropy
\begin{equation}\label{eq:Renyidisk}
\aS^{(\alpha)}\left(\rhoD\right) = \frac{1}{1-\alpha} \log \left( \sum_{\vec{b}} N_{b_1...b_n}^{\bar{c}} \left( \frac{d_{\vec{b}}}{d_c \mathcal{D}^{2n-2} } \right)^\alpha \right).
\end{equation}
Taking the limit $\alpha\to 1$ yields the (von Neumann) AEE
\begin{equation}\label{eq:AEEdisk}
\aS(\rhoD)=\lim_{\alpha\to 1}\aS^{(\alpha)}\left( \rhoD \right) =n\aS \left( \arho_{\partial \mathbb{A}_j} \right)+ 2S_{\text{topo}}+\aS_c,
\end{equation}
which agrees with Eq.~(\ref{eq:Srhonc}). Again, since we doubled the original surface in this method, both the area law term and the TEE appear with an extra factor of two.  The $\aS_c$ term is \emph{not} doubled, because we did not double the punctures carrying charge $c$ and $\bar{c}$ of the original surface.  Therefore, the topological contribution to the entanglement entropy of the original system in region $A$ is
\begin{equation}
\aS_{A} =\frac{n}{2}\aS \left( \arho_{\partial \mathbb{A}_j} \right)+ S_{\text{topo}}+\aS_c,
\end{equation}
agreeing with Eq.~(\ref{eq:Sactual_general}).

As before, we can extract the topological contributions to the anyonic R\'enyi entropy by rewriting Eq.~(\ref{eq:Renyidisk}) in powers of the boundary length.  We find
\begin{eqnarray}
\aS^{(\alpha)}\left(\rhoD\right) &=& n \aS^{(\alpha)} \left(\arho_{\partial \mathbb{A}_j}\right) +2 S_{\text{topo}}+\aS_c+\frac{F(n,\bar{c},K_\alpha)}{1-\alpha} , \\
\aS^{(\alpha)}_{A} &=& \frac{n}{2} \aS^{(\alpha)} \left(\arho_{\partial \mathbb{A}_j}\right) + S_{\text{topo}}+\aS_c+\frac{F(n,\bar{c},K_\alpha)}{2(1-\alpha)},
\end{eqnarray}
where $F(n,\bar{c},K_\alpha)$ is exponentially suppressed in $n$ for large $n$, which is essentially the regime in which the boundary length is large ($n \sim L/\ell$).

We note that the geometric cancelation method used in Ref.~\cite{Kitaev06b} to isolate $S_{\text{topo}}$ also cancels the $\aS_c$ contribution due to a topological charge $c$ in the region, so it does not isolate this term as well.

\subsection{Punctured Sphere Partitioned into an Annulus and Two 1-Punctured Disks}
\label{sec:annulus}

We now consider a sphere with a pair of punctures (or quasiparticles) carrying topological charge $c$ and $\bar{c}$. We apply our method for a partition of the system into an annular region $A$, chosen such that $c$ and $\bar{c}$ lie outside and on opposite sides of the annulus, i.e. each of the disks that form $\bar{A}$ contains one of the punctures.

\begin{center}
\includegraphics[width=.6\linewidth]{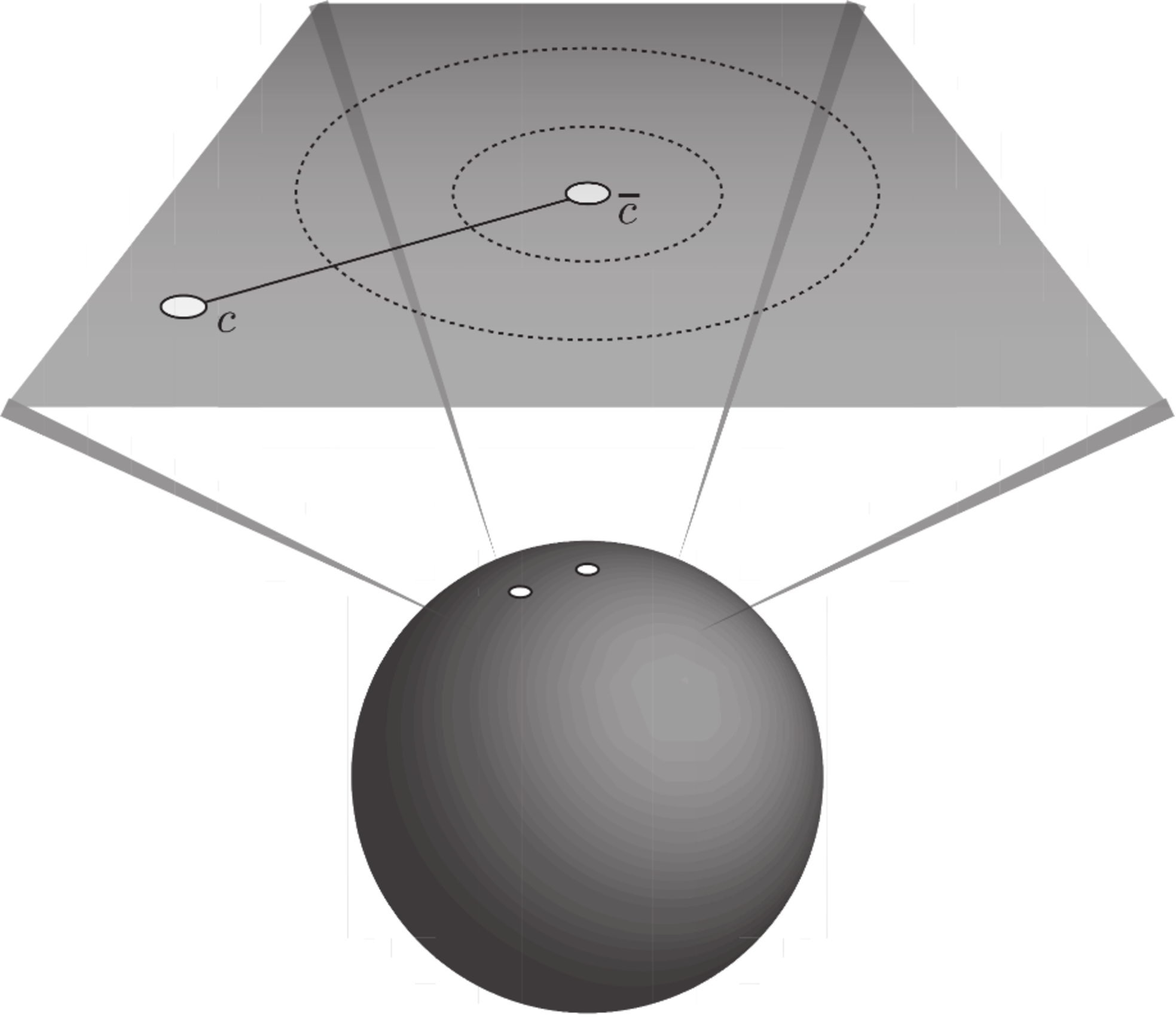}
\end{center}

We follow the same approach as for the previous example of the disk.  We create the manifold $\mathbb{M}$ by pairing the system with its time-reversal conjugate and connecting the surfaces through an array of wormholes adiabatically inserted along the partition boundary, which in this case is delineated by two concentric circles. We insert $n$ wormholes along one boundary component and $m$ wormholes along the other. Each wormhole is threaded by a trivial topological charge line. Then, analogous to Eq.~(\ref{eq:omega-loop-psi}) for the un-punctured disk, we apply a modular $\mathcal{S}$-transformation to express the state in the basis represented by topological charge lines in between the two surfaces, i.e. the inside basis.  We then use $F$-moves to fuse the charge lines threading each new partition boundary component of the doubled surface with wormholes, similar to Eq.~(\ref{eq:unpunct-disk-fuse}). The charge lines embedded in $\mathbb{M}$ look like:

\begin{center}
\includegraphics[width=\linewidth]{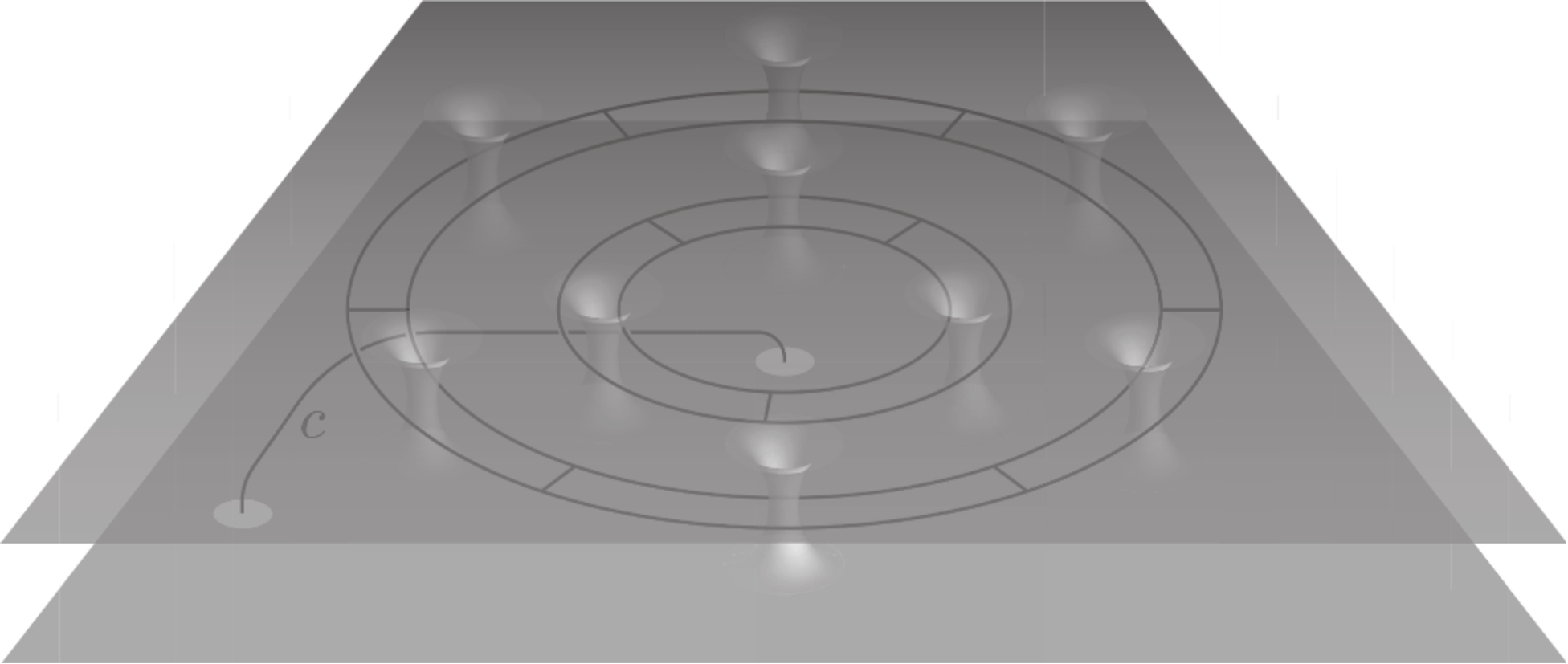}
\end{center}
We now apply the same series of $F$-moves outlined in the first three equalities of Eq.~(\ref{eq:F-move-steps}).  The state can be written as (suppressing vertex labels)
\begin{equation}\label{eq:annulus-after-F-moves}
\ket{\psi}= \sum_{\substack{\vec{a},\vec{b},\\ \vec{e},\vec{f},\\ g_1,h_1}} \frac{\sqrt{d_{\vec{a}} d_{\vec{b}}}}{\mathcal{D}^{n+m+1}} \frac{1}{\sqrt{d_c}}
\psscalebox{.9}{
 \begin{pspicture}[shift=-3.1](-2.3,-2.9)(5.4,5.5)
        \scriptsize
        \psline(3.15,-.1)(4.15,-1.1)(5.15,-.1)
        \psline[ArrowInside=->](3.25,-.2)(3.05,0)\rput(3.05,.15){$c$}
        \psline[ArrowInside=->](5.05,-.2)(5.25,0)\rput(5.25,.15){$\bar{c}$}
 \rput(-2.6,2.7){
        \rput(0.5,0.7){$\odot$}
        \rput(1.1,0.7){$\odot$}
        \rput(1.7,0.7){$\odot$}
        \psline(0.8,0.7)(1.7,1.6)
        \psline[border=1.5pt](1.7,-0.6)(0.8,0.3)
        \psline(1.4,1.3)(2,.7)(2,.3)(1.4,-0.3)
        \psline[ArrowInside=->](2,.3)(2,.7)\rput(2.2,.9){$a_3$}
        \psline[ArrowInside=->](1.1,1)(1.4,1.3)\rput(1.15,1.3){$e_2$}
        \psline[ArrowInside=->](1.4,1.3)(1.7,1.6)\rput(1.45,1.6){$e_3$}
        \psline[linestyle=dotted](1.7,1.6)(2,1.9)
        \psline[ArrowInside=->](0.8,0.3)(0.8,0.7)\rput(0.65,0.9){$a_1$}
        \psline[ArrowInside=->](2.2,2.1)(2.5,2.4)\rput(2.25,1.9){$e_{n-1}$}
        \psline(2.5,2.4)(2.8,2.1)(2.8,-1)(2.5,-1.3)
        \rput(2.3,-1.8){$g_1$}\psline[border=1.5pt](2.8,-1.6)(4.9,.5)(5.3,.5)(6.9,-1.1)(6.9,-4.4)(6.2,-5.1)(5.2,-5.1)(2.5,-2.4)(2.5,-1.9)(2.8,-1.6)        \rput(3.2,2.7){$g_1$}\psline[border=1.5pt](.2,.7)(2.4,2.9)(3.7,2.9)(7.5,-.9)(7.5,-4.6)(6.6,-5.5)(4.9,-5.5)(.2,-.8)(.2,.7)
        \psline[ArrowInside=->](2.8,-1.6)(2.5,-1.3)\rput(2.8,-1.3){$e_{n}$}
        \psline[ArrowInside=->](2.5,2.4)(2.2,2.7)\rput(2.2,2.45){$e_{n}$}
        \psline[linestyle=dashed,border=1.5pt](.5,.6)(.5,-.8)(5,-5.3)(6.4,-5.3)(7.2,-4.5)(7.2,-1)(5.5,.7)(2.9,.7)
        \psline[linestyle=dashed](2.05,.7)(2.15,.7)
        \psline[linestyle=dashed](2.65,.7)(2.75,.7)
        \psline[linestyle=dashed](1.9,.7)(1.8,.7)
        \psline[linestyle=dashed](1.6,.7)(1.5,.7)
        \psline[linestyle=dashed](1.3,.7)(1.2,.7)
        \psline[linestyle=dashed](1,.7)(.9,.7)
        \psline[linestyle=dashed](.7,.7)(.6,.7)
        \psline(1.1,0)(1.4,0.3)(1.4,0.7)(1.1,1)
        \psline[ArrowInside=->](1.4,0.3)(1.4,0.7)\rput(1.45,0.9){$a_2$}
        \psline[ArrowInside=->](2.8,0.3)(2.8,0.7)
        \rput(3,0.9){$a_{n}$}
        \psline[ArrowInside=->](1.4,-0.3)(1.1,0)\rput(1.45,0){$e_2$}
        \psline[ArrowInside=->](1.7,-0.6)(1.4,-0.3)\rput(1.75,-0.3){$e_3$}
        \psline[linestyle=dotted](1.7,-0.6)(2,-.9)
        \psline[ArrowInside=->](2.5,-1.3)(2.2,-1)\rput(2.4,-.9){$e_{n-1}$}
        \rput(2.4,0.7){$\dots$}
        } 
        \rput(0.5,0.5){$\odot$}
        \rput(1.1,0.5){$\odot$}
        \rput(1.7,0.5){$\odot$}
        \psline(0.8,0.7)(1.7,1.6)
        \psline[border=1.5pt](1.7,-0.6)(0.8,0.3)
        \psline(1.4,1.3)(2,.7)(2,.3)(1.4,-0.3)
        \psline[ArrowInside=->](2,.3)(2,.7)\rput(2.2,.75){$b_3$}
        \psline[ArrowInside=->](1.1,1)(1.4,1.3)\rput(1.15,1.3){$f_2$}
        \psline[ArrowInside=->](1.4,1.3)(1.7,1.6)\rput(1.45,1.6){$f_3$}
        \psline[linestyle=dotted](1.7,1.6)(2,1.9)
        \psline[ArrowInside=->](0.8,0.3)(0.8,0.7)\rput(0.65,0.75){$b_1$}
        \psline[ArrowInside=->](2.2,2.1)(2.5,2.4)\rput(2.25,1.9){$f_{m-1}$}
        \psline(2.5,2.4)(2.8,2.1)(2.8,-1)(2.5,-1.3)
        \psline[border=1.5pt](2.8,-1.6)(3.1,-1.3)(3.4,-1.6)(3.1,-1.9)(2.8,-1.6)\rput(3.35,-1.3){$h_1$}
        \psline[border=1.5pt](.2,.3)(.2,.7)(2.5,3)(4,1.5)(4,-1.6)(3.4,-2.2)(2.7,-2.2)(.2,.3)\rput(3.2,2){$h_1$}
        \psline[ArrowInside=->](2.5,2.4)(2.8,2.7)\rput(2.5,2.7){$f_{m}$}
        \psline[ArrowInside=->](2.8,-1.6)(2.5,-1.3)\rput(2.8,-1.3){$f_{m}$}
        \psline[linestyle=dashed,border=1.5pt](.5,.4)(.5,.3)(2.85,-2.05)(3.3,-2.05)(3.75,-1.6)(3.75,.5)(2.85,.5)
        \psline[linestyle=dashed](1.9,.5)(1.8,.5)
        \psline[linestyle=dashed](1.6,.5)(1.5,.5)
        \psline[linestyle=dashed](1.3,.5)(1.2,.5)
        \psline[linestyle=dashed](1,.5)(.9,.5)
        \psline[linestyle=dashed](.7,.5)(.6,.5)
        \psline[linestyle=dashed](2.05,.5)(2.15,.5)
        \psline[linestyle=dashed](2.65,.5)(2.75,.5)
        \psline(1.1,0)(1.4,0.3)(1.4,0.7)(1.1,1)
        \psline[ArrowInside=->](1.4,0.3)(1.4,0.7)\rput(1.6,0.75){$b_2$}
        \psline[ArrowInside=->](2.8,0.3)(2.8,0.7)
        \rput(3,0.75){$b_{m}$}
        \psline[ArrowInside=->](1.4,-0.3)(1.1,0)\rput(1.45,0){$f_2$}
        \psline[ArrowInside=->](1.7,-0.6)(1.4,-0.3)\rput(1.75,-0.3){$f_3$}
        \psline[linestyle=dotted](1.7,-0.6)(2,-.9)
        \psline[ArrowInside=->](2.5,-1.3)(2.2,-1)\rput(2.4,-.9){$f_{m-1}$}
        \rput(2.4,0.5){$\dots$}
    \end{pspicture}},
    \end{equation}
where the dashed lines indicate the partition boundary between $\AD$ (corresponding to the annulus $A$ of the un-doubled system) and $\bar{\AD}$ (corresponding to the two disks comprising $\bar{A}$ of the un-doubled system).  We can collapse the two tadpole diagrams in region $\bar{\AD}$ (the outermost $g_1$ loop and the innermost $h_1$ loop).  In doing so, $e_n$ and $f_m$ are both required to equal the trivial charge $0$.  The remaining $g_1$ and $h_1$ loops in $\AD$ (which both encircle a non-contractible cycle) can be fused together, resulting in the state:
\begin{equation}\label{eq:annulus-tadpole-collapsed}
\ket{\psi}= \sum_{\substack{\vec{a},\vec{b}, k, \\ e_2,\dots, e_{n-2},\\ f_2, \dots, f_{m-2}}} \frac{\sqrt{d_{\vec{a}} d_{\vec{b}}}}{\mathcal{D}^{n+m-2}} \frac{1}{\sqrt{d_c}} \frac{d_k}{\mathcal{D}}
\psscalebox{.9}{
 \begin{pspicture}[shift=-3.1](-2,-2.6)(5,5)
        \scriptsize
        \psline(3.15,-.1)(4.15,-1.1)(5.15,-.1)
        \psline[ArrowInside=->](3.25,-.2)(3.05,0)\rput(3.05,.15){$c$}
        \psline[ArrowInside=->](5.05,-.2)(5.25,0)\rput(5.25,.15){$\bar{c}$}
 \rput(-2.6,2.7){
        \rput(0.5,0.7){$\odot$}
        \rput(1.1,0.7){$\odot$}
        \rput(1.7,0.7){$\odot$}
        \psline(0.8,0.7)(1.7,1.6)
        \psline[border=1.5pt](1.7,-0.6)(0.8,0.3)
        \psline(1.4,1.3)(2,.7)(2,.3)(1.4,-0.3)
        \psline[ArrowInside=->](2,.3)(2,.7)\rput(2.2,.9){$a_3$}
        \psline[ArrowInside=->](1.1,1)(1.4,1.3)\rput(1.15,1.3){$e_2$}
        \psline[ArrowInside=->](1.4,1.3)(1.7,1.6)\rput(1.45,1.6){$e_3$}
        \psline[linestyle=dotted](1.7,1.6)(2,1.9)
        \psline[ArrowInside=->](0.8,0.3)(0.8,0.7)\rput(0.65,0.9){$a_1$}
        \psline[ArrowInside=->](2.2,2.1)(2.5,2.4)\rput(2.25,1.9){$\bar{a}_{n}$}
        \psline(2.5,2.4)(2.8,2.1)(2.8,-1)(2.5,-1.3)
        \rput(2.3,-1.8){$k$}\psline[border=1.5pt](2.8,-1.6)(4.9,.5)(5.3,.5)(6.9,-1.1)(6.9,-4.4)(6.2,-5.1)(5.2,-5.1)(2.5,-2.4)(2.5,-1.9)(2.8,-1.6)        
        \psline[linestyle=dashed,border=1.5pt](.5,.6)(.5,-.8)(5,-5.3)(6.4,-5.3)(7.2,-4.5)(7.2,-1)(5.5,.7)(2.9,.7)
        \psline[linestyle=dashed](2.05,.7)(2.15,.7)
        \psline[linestyle=dashed](2.65,.7)(2.75,.7)
        \psline[linestyle=dashed](1.9,.7)(1.8,.7)
        \psline[linestyle=dashed](1.6,.7)(1.5,.7)
        \psline[linestyle=dashed](1.3,.7)(1.2,.7)
        \psline[linestyle=dashed](1,.7)(.9,.7)
        \psline[linestyle=dashed](.7,.7)(.6,.7)
        \psline(1.1,0)(1.4,0.3)(1.4,0.7)(1.1,1)
        \psline[ArrowInside=->](1.4,0.3)(1.4,0.7)\rput(1.45,0.9){$a_2$}
        \psline[ArrowInside=->](2.8,0.3)(2.8,0.7)
        \rput(3,0.9){$a_{n}$}
        \psline[ArrowInside=->](1.4,-0.3)(1.1,0)\rput(1.45,0){$e_2$}
        \psline[ArrowInside=->](1.7,-0.6)(1.4,-0.3)\rput(1.75,-0.3){$e_3$}
        \psline[linestyle=dotted](1.7,-0.6)(2,-.9)
        \psline[ArrowInside=->](2.5,-1.3)(2.2,-1)\rput(2.4,-.9){$\bar{a}_{n}$}
        \rput(2.4,0.7){$\dots$}
        } 
        \rput(0.5,0.5){$\odot$}
        \rput(1.1,0.5){$\odot$}
        \rput(1.7,0.5){$\odot$}
        \psline(0.8,0.7)(1.7,1.6)
        \psline[border=1.5pt](1.7,-0.6)(0.8,0.3)
        \psline(1.4,1.3)(2,.7)(2,.3)(1.4,-0.3)
        \psline[ArrowInside=->](2,.3)(2,.7)\rput(2.2,.75){$b_3$}
        \psline[ArrowInside=->](1.1,1)(1.4,1.3)\rput(1.15,1.3){$f_2$}
        \psline[ArrowInside=->](1.4,1.3)(1.7,1.6)\rput(1.45,1.6){$f_3$}
        \psline[linestyle=dotted](1.7,1.6)(2,1.9)
        \psline[ArrowInside=->](0.8,0.3)(0.8,0.7)\rput(0.65,0.75){$b_1$}
        \psline[ArrowInside=->](2.2,2.1)(2.5,2.4)\rput(2.25,1.9){$\bar{b}_{m}$}
        \psline(2.5,2.4)(2.8,2.1)(2.8,-1)(2.5,-1.3)
        \psline[linestyle=dashed,border=1.5pt](.5,.4)(.5,.3)(2.85,-2.05)(3.3,-2.05)(3.75,-1.6)(3.75,.5)(2.85,.5)
        \psline[linestyle=dashed](1.9,.5)(1.8,.5)
        \psline[linestyle=dashed](1.6,.5)(1.5,.5)
        \psline[linestyle=dashed](1.3,.5)(1.2,.5)
        \psline[linestyle=dashed](1,.5)(.9,.5)
        \psline[linestyle=dashed](.7,.5)(.6,.5)
        \psline[linestyle=dashed](2.05,.5)(2.15,.5)
        \psline[linestyle=dashed](2.65,.5)(2.75,.5)
        \psline(1.1,0)(1.4,0.3)(1.4,0.7)(1.1,1)
        \psline[ArrowInside=->](1.4,0.3)(1.4,0.7)\rput(1.6,0.75){$b_2$}
        \psline[ArrowInside=->](2.8,0.3)(2.8,0.7)
        \rput(3,0.75){$b_{m}$}
        \psline[ArrowInside=->](1.4,-0.3)(1.1,0)\rput(1.45,0){$f_2$}
        \psline[ArrowInside=->](1.7,-0.6)(1.4,-0.3)\rput(1.75,-0.3){$f_3$}
        \psline[linestyle=dotted](1.7,-0.6)(2,-.9)
        \psline[ArrowInside=->](2.5,-1.3)(2.2,-1)\rput(2.4,-.9){$\bar{b}_{m}$}
        \rput(2.4,0.5){$\dots$}
    \end{pspicture}}
\end{equation}
Here, we have used the property
\begin{eqnarray}
\sum_{g,h} d_g d_h
\begin{pspicture}[shift=-.6](-.3,-.2)(1.3,1.2)
        \scriptsize
        \rput(0.5,0.5){$\otimes$}
        \psline(0.5,0)(1,0.5)(0.5,1)(0,0.5)(0.5,0)
        \psline[ArrowInside=->](0.5,0)(0,0.5)\rput(0.35,0.38){$h$}
        \psline(0.5,-.2)(-.2,.5)(0.5,1.2)(1.2,.5)(0.5,-.2)
        \psline[ArrowInside=->](0.5,-.2)(-.2,.5)\rput(0.07,-0.05){$g$}
\end{pspicture}
&=& \sum_{g,h,k} d_g d_h N_{gh}^{k}     \begin{pspicture}[shift=-.6](-.3,-.2)(1.3,1.2)
        \scriptsize
        \rput(0.5,0.5){$\otimes$}
        \psline(0.5,-.2)(-.2,.5)(0.5,1.2)(1.2,.5)(0.5,-.2)
        \psline[ArrowInside=->](0.5,-.2)(-.2,.5)\rput(0.05,-0.05){$k$}
\end{pspicture}
= \sum_{h,k} d_h^2 d_k     \begin{pspicture}[shift=-.6](-.3,-.2)(1.3,1.2)
        \scriptsize
        \rput(0.5,0.5){$\otimes$}
        \psline(0.5,-.2)(-.2,.5)(0.5,1.2)(1.2,.5)(0.5,-.2)
        \psline[ArrowInside=->](0.5,-.2)(-.2,.5)\rput(0.05,-0.05){$k$}
\end{pspicture} \notag\\
 &=& \mathcal{D}^{2} \sum_{k} d_k     \begin{pspicture}[shift=-.6](-.3,-.2)(1.3,1.4)
        \scriptsize
        \rput(0.5,0.5){$\otimes$}
        \psline(0.5,-.2)(-.2,.5)(0.5,1.2)(1.2,.5)(0.5,-.2)
        \psline[ArrowInside=->](0.5,-.2)(-.2,.5)\rput(0.05,-0.05){$k$}
    \end{pspicture}.
\end{eqnarray}
Note that the loop labeled by $k$ in Eq.~(\ref{eq:annulus-tadpole-collapsed}) is actually an $\omega_0$-loop circling one of the connected components of $\bar{\mathbb{A}}$ (wrapping around a non-contractible cycle), because it is weighted by $d_k$ in the sum over $k$. Similar to Eq.~(\ref{eq:psiIntDis}), we fuse the topological charge $c$ line to the charge lines threading the same boundary regions, taken here to be $b_m$ and $a_n$.  The state embedded in the doubled surface looks like:

\begin{center}
\includegraphics[width=\linewidth]{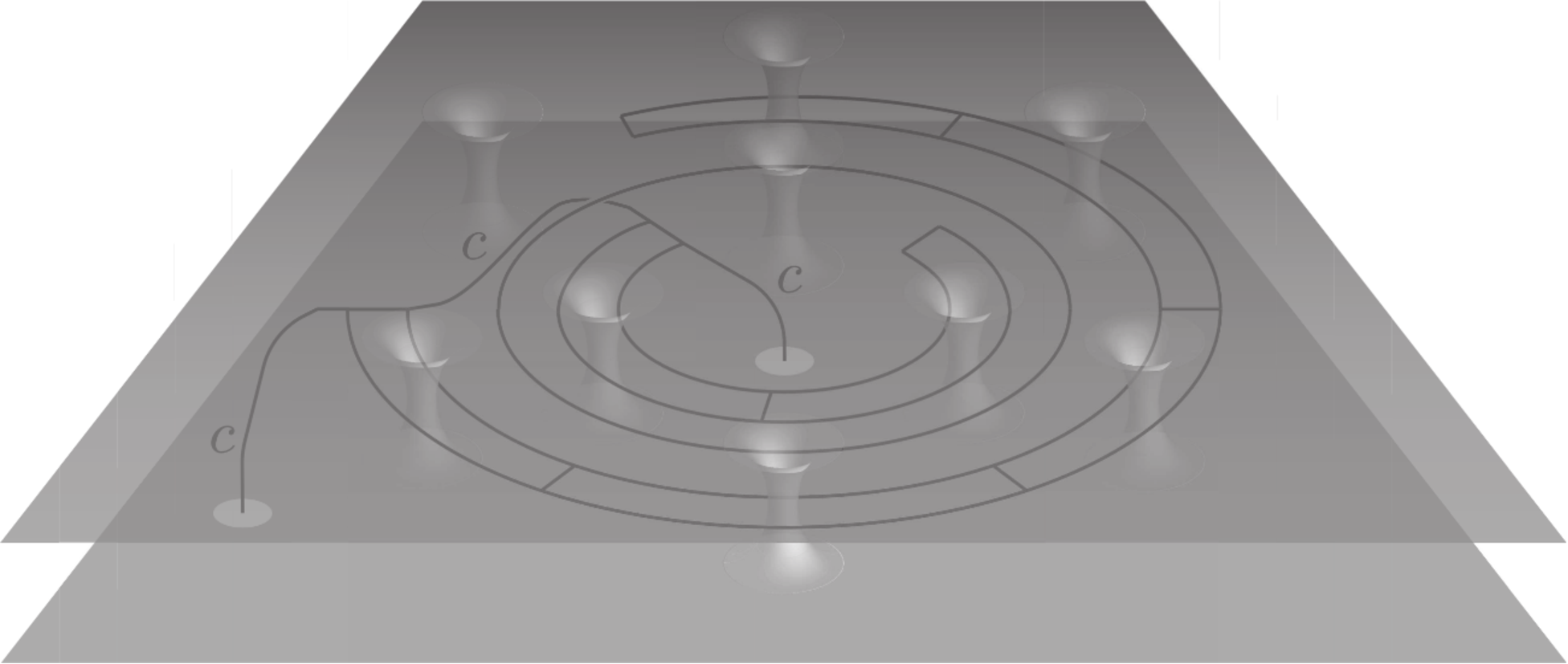}
\end{center}
Finally, we cut along the partition boundary. After cutting, the region $\mathbb{A}$ of the doubled system, which is an $(n+m)$-punctured torus (genus $g=1$), looks like

\begin{center}
\includegraphics[width=.6\linewidth]{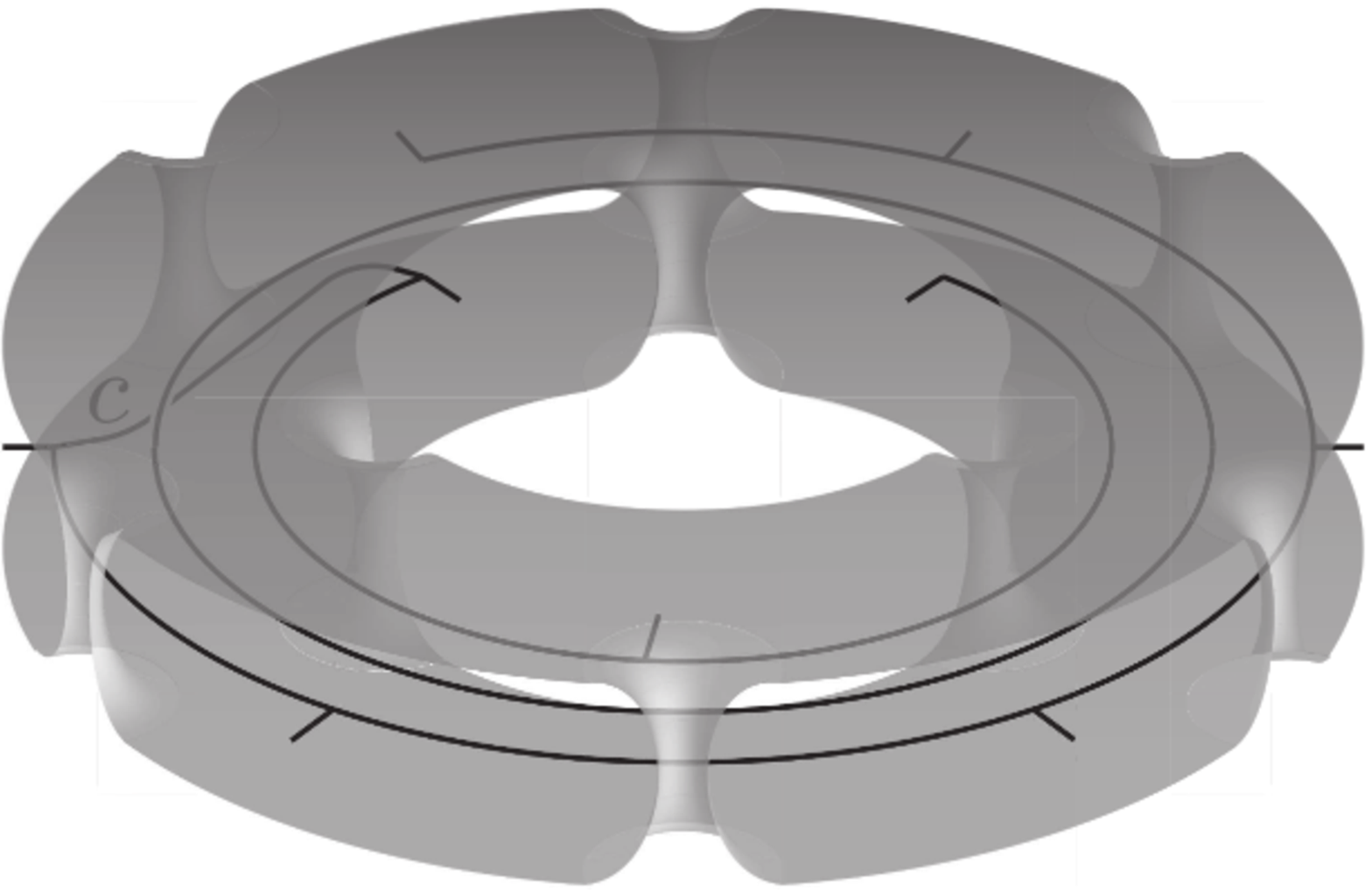}
\end{center}
and the state $\ket{\psi_{\text{cut}}}$ of the cut system (including $\mathbb{A}$ and $\bar{\mathbb{A}}$) can be represented diagrammatically as

\begin{equation}\label{eq:psi-cut-annulus}
\ket{\psi_{\text{cut}}} = \sum_{\substack{\vec{a},\vec{b},\\ \vec{e},\vec{f}}} \frac{1}{\mathcal{D}^{n+m-3}} \frac{1}{d_c^{3/2}}
\psscalebox{.85}{
 \begin{pspicture}[shift=.5](-2.8,-.5)(6.5,.3)
        \scriptsize
 \rput(-4.5,0){ \psline[ArrowInside=->](3.5,-.3)(3.8,0)\rput(3.8,.2){$a_1$}
        \psline[ArrowInside=->](3.5,-.3)(3.2,0)\rput(3.2,.2){$a_2$}
        \psline(3.5,-.3)(3.4,-.4)
        \rput(3.15,-.35){$e_2$}
        \psline[linestyle=dotted](3.4,-.4)(3.1,-.7)
        \psline(3,-.8)(2.2,0)\psline[ArrowInside=->](2.5,-.3)(2.2,0)\rput(2.2,.2){$a_n$}
        \psline(3,-.8)(3.1,-.7)
        \psline(3,-.8)(2.7,-1.1)(1.6,0)\psline[ArrowInside=->](1.9,-.3)(1.6,0)\rput(1.6,.2){$\bar{c}$}
        \rput(2.7,-1.6){$(\bar{\mathbb{A}})$}}
 \rput(4.5,0){
        \psline[ArrowInside=->](.3,-.3)(0,0)\rput(0,.2){$b_1$}
        \psline[ArrowInside=->](.3,-.3)(.6,0)\rput(.6,.2){$b_2$}
        \psline(.4,-.4)(.3,-.3)
        \psline[linestyle=dotted](.7,-.7)(.4,-.4)
        \rput(.6,-.4){$f_2$}
        \psline(.8,-.8)(.7,-.7)
        \psline(.8,-.8)(1.6,0)\psline[ArrowInside=->](1.3,-.3)(1.6,0)\rput(1.6,.2){$b_m$}
        \psline(.8,-.8)(1.1,-1.1)(2.2,0)\psline[ArrowInside=->](1.9,-.3)(2.2,0)\rput(2.2,.2){$c$}
        \rput(1.1,-1.6){$(\bar{\mathbb{A}})$}}
        \psline[ArrowInside=->](.3,-.3)(0,0)\rput(0,.2){$a_1$}
        \psline[ArrowInside=->](.3,-.3)(.6,0)\rput(.6,.2){$a_2$}
        \psline(.4,-.4)(.3,-.3)
        \psline[linestyle=dotted](.7,-.7)(.4,-.4)
        \rput(.6,-.4){$e_2$}
        \psline(.8,-.8)(.7,-.7)
        \psline(.8,-.8)(1.6,0)\psline[ArrowInside=->](1.3,-.3)(1.6,0)\rput(1.6,.2){$a_n$}
        \psline(.8,-.8)(1.9,-1.9)
        \psline[ArrowInside=->](3.5,-.3)(3.8,0)\rput(3.8,.2){$b_1$}
        \psline[ArrowInside=->](3.5,-.3)(3.2,0)\rput(3.2,.2){$b_2$}
        \psline(3.5,-.3)(3.4,-.4)
        \rput(3.15,-.35){$f_2$}
        \psline[linestyle=dotted](3.4,-.4)(3.1,-.7)
        \psline(3,-.8)(2.2,0)\psline[ArrowInside=->](2.5,-.3)(2.2,0)\rput(2.2,.2){$b_m$}
        \psellipse[border=1.5pt](1.4,-1.25)(.4,.2)\psline[ArrowInside=->](1.4,-1.45)(1.25,-1.42)\psline[border=1.5pt](1,-1)(1.3,-1.3)
        \rput(1.5,-1.25){$\otimes$}\rput(1.2,-1.55){$\omega_0$}
        \psline(3.1,-.7)(1.9,-1.9)
        \psline[ArrowInside=->](1.8,-1.8)(1.5,-1.5)\rput(1.7,-1.5){$c$}
        \psline[ArrowInside=->](2,-1.8)(2.3,-1.5)\rput(2.1,-1.5){$\bar{c}$}
        \rput(1.9,-2.2){$(\mathbb{A})$}
    \end{pspicture}}
\end{equation}
or, alternatively, as
\begin{equation}\label{eq:psi-cut-alt}
\ket{\psi_{\text{cut}}} = \sum_{\substack{\vec{a},\vec{b},\\ \vec{e},\vec{f}}} \frac{1}{\mathcal{D}^{n+m-3}} \frac{1}{d_c^{3/2}}
\psscalebox{.85}{
 \begin{pspicture}[shift=.5](-2.8,-.5)(6.5,.3)
        \scriptsize
 \rput(-4.5,0){ \psline[ArrowInside=->](3.5,-.3)(3.8,0)\rput(3.8,.2){$a_1$}
        \psline[ArrowInside=->](3.5,-.3)(3.2,0)\rput(3.2,.2){$a_2$}
        \psline(3.5,-.3)(3.4,-.4)
        \rput(3.15,-.35){$e_2$}
        \psline[linestyle=dotted](3.4,-.4)(3.1,-.7)
        \psline(3,-.8)(2.2,0)\psline[ArrowInside=->](2.5,-.3)(2.2,0)\rput(2.2,.2){$a_n$}
        \psline(3,-.8)(3.1,-.7)
        \psline(3,-.8)(2.7,-1.1)(1.6,0)\psline[ArrowInside=->](1.9,-.3)(1.6,0)\rput(1.6,.2){$\bar{c}$}
        \rput(2.7,-1.6){$(\bar{\mathbb{A}})$}}
 \rput(4.5,0){
        \psline[ArrowInside=->](.3,-.3)(0,0)\rput(0,.2){$b_1$}
        \psline[ArrowInside=->](.3,-.3)(.6,0)\rput(.6,.2){$b_2$}
        \psline(.4,-.4)(.3,-.3)
        \psline[linestyle=dotted](.7,-.7)(.4,-.4)
        \rput(.6,-.4){$f_2$}
        \psline(.8,-.8)(.7,-.7)
        \psline(.8,-.8)(1.6,0)\psline[ArrowInside=->](1.3,-.3)(1.6,0)\rput(1.6,.2){$b_m$}
        \psline(.8,-.8)(1.1,-1.1)(2.2,0)\psline[ArrowInside=->](1.9,-.3)(2.2,0)\rput(2.2,.2){$c$}
        \rput(1.1,-1.6){$(\bar{\mathbb{A}})$}}
        \psline[ArrowInside=->](.3,-.3)(0,0)\rput(0,.2){$a_1$}
        \psline[ArrowInside=->](.3,-.3)(.6,0)\rput(.6,.2){$a_2$}
        \psline(.4,-.4)(.3,-.3)
        \psline[linestyle=dotted](.7,-.7)(.4,-.4)
        \rput(.6,-.4){$e_2$}
        \psline(.8,-.8)(.7,-.7)
        \psline(.8,-.8)(1.6,0)\psline[ArrowInside=->](1.3,-.3)(1.6,0)\rput(1.6,.2){$a_n$}
        \psline(.8,-.8)(1.9,-1.9)
        \psline[ArrowInside=->](3.5,-.3)(3.8,0)\rput(3.8,.2){$b_1$}
        \psline[ArrowInside=->](3.5,-.3)(3.2,0)\rput(3.2,.2){$b_2$}
        \psline(3.5,-.3)(3.4,-.4)
        \rput(3.15,-.35){$f_2$}
        \psline[linestyle=dotted](3.4,-.4)(3.1,-.7)
        \psline(3,-.8)(2.2,0)\psline[ArrowInside=->](2.5,-.3)(2.2,0)\rput(2.2,.2){$b_m$}
        \psellipse[border=1.5pt](1.9,-1)(.4,.2)\psline[ArrowInside=->](1.95,-1.2)(1.75,-1.17)\psline[border=1.5pt](1,-1)(1.3,-1.3)
        \rput(1.9,-1){$\otimes$}\rput(1.7,-1.3){$\omega_0$}
        \psline(3.1,-.7)(1.9,-1.9)
        \psline[ArrowInside=->](1.8,-1.8)(1.5,-1.5)\rput(1.7,-1.5){$c$}
        \psline[ArrowInside=->](2,-1.8)(2.3,-1.5)\rput(2.1,-1.5){$\bar{c}$}
        \rput(1.9,-2.2){$(\mathbb{A})$}
    \end{pspicture}}.
\end{equation}
The choice to represent $\ket{\psi_{\text{cut}}}$ as Eq.~(\ref{eq:psi-cut-annulus}) or Eq.~(\ref{eq:psi-cut-alt}) amounts to a highly non-trivial change of basis, or a mental exercise in topology (essentially turning the embedding of region $\mathbb{A}$ inside-out).  It is instructive to work with the more complicated looking representation in Eq.~(\ref{eq:psi-cut-annulus}) to convince oneself that the remaining steps of the computation for the AEE are equally simple in either representation, provided one does not attempt to transform to the canonical basis.

Given the density matrix of the cut state $\arho_{\text{cut}}=\ket{\psi_{\text{cut}}}\bra{\psi_{\text{cut}}}$, we can take the trace over each of the disks of region $\bar{\mathbb{A}}$ in the same way as shown in Eq.~(\ref{eq:disk-trace}). The reduced density matrix for $\mathbb{A}$ is (restoring the vertex labels):
\begin{equation}\label{eq:rhoannulus}
\rhoD=\sum_{\vec{a},\vec{b}, \vec{e}, \vec{f},\vec{\mu}, \vec{\nu}} \frac{1}{\mathcal{D}^{2(n+m-3)}} \frac{\sqrt{d_{\vec{a}}d_{\vec{b}}}}{d_c^{2}}
 \begin{pspicture}[shift=-2.25](.3,-4.7)(4,.3)
        \scriptsize
        \psline[ArrowInside=->](.3,-.3)(0,0)\rput(0,.2){$a_1$}
        \psline[ArrowInside=->](.3,-.3)(.6,0)\rput(.6,.2){$a_2$}
        \psline(.4,-.4)(.3,-.3)
        \psline[linestyle=dotted](.7,-.7)(.4,-.4)
        \rput(.1,-.45){$\mu_2$}\rput(.6,-.4){$e_2$}
        \psline(.8,-.8)(.7,-.7)
        \psline(.8,-.8)(1.6,0)\psline[ArrowInside=->](1.3,-.3)(1.6,0)\rput(1.6,.2){$a_n$}
        \rput(.6,-.85){$\mu_n$}
        \psline(.8,-.8)(1.9,-1.9)
        \psline[ArrowInside=->](3.5,-.3)(3.8,0)\rput(3.8,.2){$b_1$}
        \psline[ArrowInside=->](3.5,-.3)(3.2,0)\rput(3.2,.2){$b_2$}
        \psline(3.5,-.3)(3.4,-.4)
        \rput(3.7,-.45){$\nu_2$}\rput(3.15,-.35){$f_2$}
        \psline[linestyle=dotted](3.4,-.4)(3.1,-.7)
        \psline(3,-.8)(2.2,0)\psline[ArrowInside=->](2.5,-.3)(2.2,0)\rput(2.2,.2){$b_m$}
        \rput(3.3,-.85){$\nu_m$}
        \psline(3.1,-.7)(1.9,-1.9)
        \psline[ArrowInside=->](1.8,-1.8)(1.5,-1.5)\rput(1.7,-1.5){$c$}
        \psline[ArrowInside=->](2,-1.8)(2.3,-1.5)\rput(2.1,-1.5){$\bar{c}$}
        \psline[ArrowInside=->](0,-4.4)(.3,-4.1)\rput(0,-4.6){$a_1$}
        \psline[ArrowInside=->](.6,-4.4)(.3,-4.1)\rput(.6,-4.6){$a_2$}
        \psline(.4,-4)(.3,-4.1)
        \psline[linestyle=dotted](.4,-4)(.7,-3.7)
        \rput(.1,-4){$\mu_2$}\rput(.6,-4){$e_2$}
        \psline(.8,-3.6)(.7,-3.7)
        \psline(.8,-3.6)(1.6,-4.4)\psline[ArrowInside=->](1.6,-4.4)(1.3,-4.1)\rput(1.6,-4.6){$a_n$}
        \rput(.6,-3.55){$\mu_n$}
        \psline(.8,-3.6)(1.9,-2.5)
        \psline[ArrowInside=->](3.8,-4.4)(3.5,-4.1)\rput(3.8,-4.6){$b_1$}
        \psline[ArrowInside=->](3.2,-4.4)(3.5,-4.1)\rput(3.2,-4.6){$b_2$}
        \psline(3.5,-4.1)(3.4,-4)
        \rput(3.7,-4){$\nu_2$}\rput(3.2,-4){$f_2$}
        \psline[linestyle=dotted](3.4,-4)(3.1,-3.7)
        \psline(3,-3.6)(2.2,-4.4)\psline[ArrowInside=->](2.2,-4.4)(2.5,-4.1)\rput(2.2,-4.6){$b_m$}
        \rput(3.3,-3.6){$\nu_m$}
        \psline(3.1,-3.7)(1.9,-2.5)
        \psline[ArrowInside=->](1.5,-2.9)(1.8,-2.6)\rput(1.7,-2.9){$c$}
        \psline[ArrowInside=->](2.3,-2.9)(2,-2.6)\rput(2.1,-2.9){$\bar{c}$}
        \psellipse[border=1.5pt](1.4,-1.25)(.4,.2)\psline[ArrowInside=->](1.4,-1.45)(1.25,-1.42)\psline[border=1.5pt](1,-1)(1.3,-1.3)
        \rput(1.5,-1.25){$\otimes$}\rput(1.2,-1.55){$\omega_0$}
        \psellipse[border=1.5pt](1.4,-3.15)(.4,.2)\psline[ArrowInside=->](1.4,-3.35)(1.25,-3.32)\psline[border=1.5pt](1.2,-3.2)(1.5,-2.9)
        \rput(1.5,-3.15){$\otimes$}\rput(1.65,-3.5){$\omega_0$}
    \end{pspicture}.
\end{equation}

As in the previous sections, in order to calculate the anyonic R\'enyi entropy and the AEE we consider powers of the reduced density matrix.  We square $\rhoD$ by stacking the diagrams.  Note that
 \begin{equation}
\aTr \left(
\begin{pspicture}[shift=-1.25](1,-3.5)(2.8,-.5)
\scriptsize
        \psellipse[border=1.5pt](1.4,-3.1)(.4,.2)\psline[ArrowInside=->](1.4,-3.3)(1.25,-3.27) \psline[border=1.5pt](1.2,-3.2)(1.5,-2.9)\psline(1.1,-3.3)(1,-3.4)
        \psline(2.8,-3.4)(1.9,-2.5)
        \psline(1.2,-3.2)(1.9,-2.5)(2.1,-2.7)
        \psline(1.7,-1.7)(1.9,-1.9)(2.1,-1.7)
        \psline[border=1.5pt](1.5,-2.9)(1.8,-2.6)
        \psline[ArrowInside=->](1.5,-2.9)(1.8,-2.6)\rput(1.55,-2.6){$c$}
        \psline[ArrowInside=->](2.3,-2.9)(2,-2.6)\rput(2.2,-2.65){$\bar{c}$}
        \psellipse[border=1.5pt](1.4,-1.2)(.4,.2)\psline[ArrowInside=->](1.4,-1.4)(1.25,-1.37)\psline[border=1.5pt](1,-1)(1.3,-1.3)
        \rput(1.5,-1.2){$\otimes$}\rput(1.2,-1.55){$\omega_0$}
        \rput(1.5,-3.1){$\otimes$}\rput(1.65,-3.5){$\omega_0$}
        \psline(2.8,-1)(1.9,-1.9)
        \psline[ArrowInside=->](1.8,-1.8)(1.5,-1.5)\rput(1.55,-1.85){$c$}
        \psline[ArrowInside=->](2,-1.8)(2.3,-1.5)\rput(2.2,-1.85){$\bar{c}$}
\end{pspicture} \right)
= \sum_k \frac{d_k}{\mathcal{D}^4} \begin{pspicture}[shift=-1.25](1,-3.5)(3.1,-.5)
\scriptsize
        \psline(1,-1)(1.9,-1.9)(2.2,-1.6)(2.3,-1.7)(2.3,-2.7)(2.2,-2.8)(1.9,-2.5)(1,-3.4)(1.45,-3.85)(2.35,-3.85)(3.05,-3.15) (3.05,-1.25)(2.35,-.55)(1.45,-.55)(1,-1)
        \psline[ArrowInside=->](1.6,-2.8)(1.8,-2.6)\rput(1.55,-2.6){$c$}
        \psline[ArrowInside=->](2.15,-2.75)(2,-2.6)\rput(2.1,-2.55){$\bar{c}$}
        \psline[ArrowInside=->](1.8,-1.8)(1.6,-1.6)\rput(1.55,-1.85){$c$}
        \psline[ArrowInside=->](2,-1.8)(2.2,-1.6)\rput(2.1,-1.9){$\bar{c}$}
        \psline[border=1.5pt](1.15,-1.25)(1.1,-1.3)(1.3,-1.5)(1.7,-1.1)(2.1,-1.1)(2.5,-1.5)(2.5,-2.85) (2.1,-3.25)(1.7,-3.25)(1.3,-2.85)(1.1,-3.05)(1.15,-3.1)
        \psline(1.25,-3.2)(1.55,-3.5)(2.25,-3.5)(2.75,-3)(2.75,-1.35)(2.25,-.85)(1.55,-.85)(1.25,-1.15)
        \psline[ArrowInside=->](1.25,-1.45)(1.1,-1.3)\rput(1.1,-1.5){$k$} \psline[ArrowInside=->](1.15,-3.0)(1.3,-2.85)\rput(1.1,-2.8){$k$}
\end{pspicture}
= \aTr \left(
\begin{pspicture}[shift=-1.25](1,-3.5)(2.8,-.5)
\scriptsize
        \psline(2.8,-3.4)(1.9,-2.5)
        \psline(1,-1)(1.9,-1.9)(2.8,-1)
        \psline(1,-3.4)(1.9,-2.5)(2.8,-3.4)
        \psline[ArrowInside=->](1.5,-2.9)(1.8,-2.6)\rput(1.55,-2.6){$c$}
        \psline[ArrowInside=->](2.3,-2.9)(2,-2.6)\rput(2.2,-2.65){$\bar{c}$}
        \psellipse[border=1.5pt](1.9,-1.2)(.4,.2)\psline[ArrowInside=->](1.9,-1.4)(1.75,-1.37)\psline[border=1.5pt](1,-1)(1.3,-1.3)
        \rput(1.9,-1.2){$\otimes$}\rput(1.72,-1.5){$\omega_0$}
        \psellipse[border=1.5pt](1.9,-3.1)(.4,.2)\psline[ArrowInside=->](1.9,-3.3)(1.75,-3.27) \psline[border=1.5pt](1.2,-3.2)(1.5,-2.9)\psline(1.1,-3.3)(1,-3.4)
        \rput(1.9,-3.1){$\otimes$}\rput(1.7,-3.45){$\omega_0$}
        \psline(2.8,-1)(1.9,-1.9)
        \psline[ArrowInside=->](1.8,-1.8)(1.5,-1.5)\rput(1.55,-1.85){$c$}
        \psline[ArrowInside=->](2,-1.8)(2.3,-1.5)\rput(2.2,-1.85){$\bar{c}$}
\end{pspicture}
 \right)
= \frac{d_c}{\mathcal{D}^{2}}
.
 \end{equation}
Therefore,
 \begin{equation}\label{eq:rhoannulus-squared}
 \begin{split}
\left(\rhoD\right)^2=\sum_{\substack{\vec{a},\vec{b}, \\ \vec{e}, \vec{f},\\ \vec{\mu}, \vec{\nu},\\ \vec{e}', \vec{f}', \\ \vec{\mu}', \vec{\nu}'\\ k}}
&\frac{1}{\mathcal{D}^{4(n+m-3)}}\frac{d_{\vec{a}}d_{\vec{b}}}{d_c^{4}}
 \frac{d_k^2}{\mathcal{D}^4} \frac{1}{d_k}
\psscalebox{.9}{\begin{pspicture}[shift=-5](-1.8,-9.5)(4.2,.3)
        \scriptsize
        \psline(-.8,-4.1)(-.5,-4.4)(-.8,-4.7)(-1.1,-4.4)(-.8,-4.1)
        \rput(-1,-4.1){$k$}
        \psline[ArrowInside=->](.3,-.3)(0,0)\rput(0,.2){$a_1$}
        \psline[ArrowInside=->](.3,-.3)(.6,0)\rput(.6,.2){$a_2$}
        \psline(.4,-.4)(.3,-.3)
        \psline[linestyle=dotted](.7,-.7)(.4,-.4)
        \rput(.1,-.45){$\mu_2$}\rput(.6,-.4){$e_2$}
        \psline(.8,-.8)(.7,-.7)
        \psline(.8,-.8)(1.6,0)\psline[ArrowInside=->](1.3,-.3)(1.6,0)\rput(1.6,.2){$a_n$}
        \rput(.6,-.85){$\mu_n$}
        \psline(.8,-.8)(1.9,-1.9)
        \psline[ArrowInside=->](3.5,-.3)(3.8,0)\rput(3.8,.2){$b_1$}
        \psline[ArrowInside=->](3.5,-.3)(3.2,0)\rput(3.2,.2){$b_2$}
        \psline(3.5,-.3)(3.4,-.4)
        \rput(3.7,-.45){$\nu_2$}\rput(3.15,-.35){$f_2$}
        \psline[linestyle=dotted](3.4,-.4)(3.1,-.7)
        \psline(3,-.8)(2.2,0)\psline[ArrowInside=->](2.5,-.3)(2.2,0)\rput(2.2,.2){$b_m$}
        \rput(3.3,-.85){$\nu_m$}
        \psline(3.1,-.7)(1.9,-1.9)
        \psline[ArrowInside=->](1.8,-1.8)(1.5,-1.5)\rput(1.7,-1.5){$c$}
        \psline[ArrowInside=->](2,-1.8)(2.3,-1.5)\rput(2.1,-1.5){$\bar{c}$}
        \psline[ArrowInside=->](0,-4.4)(.3,-4.1)\rput(-.2,-4.4){$a_1$}
        \psline[ArrowInside=->](.6,-4.4)(.3,-4.1)\rput(.8,-4.4){$a_2$}
        \psline(.4,-4)(.3,-4.1)
        \psline[linestyle=dotted](.4,-4)(.7,-3.7)
        \rput(.1,-4){$\mu_2$}\rput(.6,-4){$e_2$}
        \psline(.8,-3.6)(.7,-3.7)
        \psline(.8,-3.6)(1.6,-4.4)\psline[ArrowInside=->](1.6,-4.4)(1.3,-4.1)\rput(1.3,-4.4){$a_n$}
        \rput(.6,-3.55){$\mu_n$}
        \psline(.8,-3.6)(1.9,-2.5)
        \psline[ArrowInside=->](3.8,-4.4)(3.5,-4.1)\rput(4.,-4.4){$b_1$}
        \psline[ArrowInside=->](3.2,-4.4)(3.5,-4.1)\rput(3,-4.4){$b_2$}
        \psline(3.5,-4.1)(3.4,-4)
        \rput(3.7,-4){$\nu_2$}\rput(3.2,-4){$f_2$}
        \psline[linestyle=dotted](3.4,-4)(3.1,-3.7)
        \psline(3,-3.6)(2.2,-4.4)\psline[ArrowInside=->](2.2,-4.4)(2.5,-4.1)\rput(2.5,-4.4){$b_m$}
        \rput(3.3,-3.6){$\nu_m$}
        \psline(3.1,-3.7)(1.9,-2.5)
        \psline[ArrowInside=->](1.5,-2.9)(1.8,-2.6)\rput(1.7,-2.9){$c$}
        \psline[ArrowInside=->](2.3,-2.9)(2,-2.6)\rput(2.1,-2.9){$\bar{c}$}
        \psellipse[border=1.5pt](1.4,-1.25)(.4,.2)\psline[ArrowInside=->](1.4,-1.45)(1.25,-1.42)\psline[border=1.5pt](1,-1)(1.3,-1.3)
        \rput(1.5,-1.25){$\otimes$}\rput(1.2,-1.55){$\omega_0$}
\rput(0,-4.4){
        \psline[ArrowInside=->](-.8,-.3)(-.5,0)
        \psline[ArrowInside=->](.3,-.3)(0,0)
        \psline[ArrowInside=->](.3,-.3)(.6,0)
        \psline(.4,-.4)(.3,-.3)
        \psline[linestyle=dotted](.7,-.7)(.4,-.4)
        \rput(.1,-.45){$\mu_2'$}\rput(.6,-.4){$e_2'$}
        \psline(.8,-.8)(.7,-.7)
        \psline(.8,-.8)(1.6,0)\psline[ArrowInside=->](1.3,-.3)(1.6,0)
        \rput(.6,-.85){$\mu_n'$}
        \psline(.8,-.8)(1.9,-1.9)
        \psline[ArrowInside=->](3.5,-.3)(3.8,0)
        \psline[ArrowInside=->](3.5,-.3)(3.2,0)
        \psline(3.5,-.3)(3.4,-.4)
        \rput(3.7,-.45){$\nu_2'$}\rput(3.15,-.35){$f_2'$}
        \psline[linestyle=dotted](3.4,-.4)(3.1,-.7)
        \psline(3,-.8)(2.2,0)\psline[ArrowInside=->](2.5,-.3)(2.2,0)
        \rput(3.3,-.85){$\nu_m'$}
        \psline(3.1,-.7)(1.9,-1.9)
        \psline[ArrowInside=->](1.8,-1.8)(1.5,-1.5)\rput(1.7,-1.5){$c$}
        \psline[ArrowInside=->](2,-1.8)(2.3,-1.5)\rput(2.1,-1.5){$\bar{c}$}
        \psline[ArrowInside=->](0,-4.4)(.3,-4.1)\rput(0,-4.6){$a_1$}
        \psline[ArrowInside=->](.6,-4.4)(.3,-4.1)\rput(.6,-4.6){$a_2$}
        \psline(.4,-4)(.3,-4.1)
        \psline[linestyle=dotted](.4,-4)(.7,-3.7)
        \rput(.1,-4){$\mu_2'$}\rput(.6,-4){$e_2'$}
        \psline(.8,-3.6)(.7,-3.7)
        \psline(.8,-3.6)(1.6,-4.4)\psline[ArrowInside=->](1.6,-4.4)(1.3,-4.1)\rput(1.6,-4.6){$a_n$}
        \rput(.6,-3.55){$\mu_n'$}
        \psline(.8,-3.6)(1.9,-2.5)
        \psline[ArrowInside=->](3.8,-4.4)(3.5,-4.1)\rput(3.8,-4.6){$b_1$}
        \psline[ArrowInside=->](3.2,-4.4)(3.5,-4.1)\rput(3.2,-4.6){$b_2$}
        \psline(3.5,-4.1)(3.4,-4)
        \rput(3.7,-4){$\nu_2'$}\rput(3.2,-4){$f_2'$}
        \psline[linestyle=dotted](3.4,-4)(3.1,-3.7)
        \psline(3,-3.6)(2.2,-4.4)\psline[ArrowInside=->](2.2,-4.4)(2.5,-4.1)\rput(2.2,-4.6){$b_m$}
        \rput(3.3,-3.6){$\nu_m'$}
        \psline(3.1,-3.7)(1.9,-2.5)
        \psline[ArrowInside=->](1.5,-2.9)(1.8,-2.6)\rput(1.7,-2.9){$c$}
        \psline[ArrowInside=->](2.3,-2.9)(2,-2.6)\rput(2.1,-2.9){$\bar{c}$}
        \psellipse[border=1.5pt](1.4,-3.15)(.4,.2)\psline[ArrowInside=->](1.4,-3.35)(1.25,-3.32)\psline[border=1.5pt](1.2,-3.2)(1.5,-2.9)
        \rput(1.5,-3.15){$\otimes$}\rput(1.65,-3.5){$\omega_0$}
        }
    \end{pspicture}},
\end{split}
\end{equation}
where the $k$ loop with prefactor $d_k/\mathcal{D}^4$ comes from taking the inner product of two $\omega_0$-loops.  Evaluating the middle diagram, we find
\begin{equation}
\label{eq:rhoannulus-squared-simplified}
\begin{split}
 \left( \rhoD\right)^2=\sum_{\substack{\vec{a},\vec{e}, \vec{\mu} \\ \vec{b}, \vec{f}, \vec{\nu}}}
&\frac{1}{\mathcal{D}^{2(n+m-3)}}\frac{\sqrt{d_{\vec{a}}d_{\vec{b}}}}{d_c^{2}}\left( \frac{d_{\vec{a}} d_{\vec{b}}}{\mathcal{D}^{2(n+m-2)} d_c^2}\right)
\psscalebox{1}{\begin{pspicture}[shift=-2.25](.3,-4.7)(4,.3)
        \scriptsize
        \psline[ArrowInside=->](.3,-.3)(0,0)\rput(0,.2){$a_1$}
        \psline[ArrowInside=->](.3,-.3)(.6,0)\rput(.6,.2){$a_2$}
        \psline(.4,-.4)(.3,-.3)
        \psline[linestyle=dotted](.7,-.7)(.4,-.4)
        \rput(.1,-.45){$\mu_2$}\rput(.6,-.4){$e_2$}
        \psline(.8,-.8)(.7,-.7)
        \psline(.8,-.8)(1.6,0)\psline[ArrowInside=->](1.3,-.3)(1.6,0)\rput(1.6,.2){$a_n$}
        \rput(.6,-.85){$\mu_n$}
        \psline(.8,-.8)(1.9,-1.9)
        \psline[ArrowInside=->](3.5,-.3)(3.8,0)\rput(3.8,.2){$b_1$}
        \psline[ArrowInside=->](3.5,-.3)(3.2,0)\rput(3.2,.2){$b_2$}
        \psline(3.5,-.3)(3.4,-.4)
        \rput(3.7,-.45){$\nu_2$}\rput(3.15,-.35){$f_2$}
        \psline[linestyle=dotted](3.4,-.4)(3.1,-.7)
        \psline(3,-.8)(2.2,0)\psline[ArrowInside=->](2.5,-.3)(2.2,0)\rput(2.2,.2){$b_m$}
        \rput(3.3,-.85){$\nu_m$}
        \psline(3.1,-.7)(1.9,-1.9)
        \psline[ArrowInside=->](1.8,-1.8)(1.5,-1.5)\rput(1.7,-1.5){$c$}
        \psline[ArrowInside=->](2,-1.8)(2.3,-1.5)\rput(2.1,-1.5){$\bar{c}$}
        \psline[ArrowInside=->](0,-4.4)(.3,-4.1)\rput(0,-4.6){$a_1$}
        \psline[ArrowInside=->](.6,-4.4)(.3,-4.1)\rput(.6,-4.6){$a_2$}
        \psline(.4,-4)(.3,-4.1)
        \psline[linestyle=dotted](.4,-4)(.7,-3.7)
        \rput(.1,-4){$\mu_2$}\rput(.6,-4){$e_2$}
        \psline(.8,-3.6)(.7,-3.7)
        \psline(.8,-3.6)(1.6,-4.4)\psline[ArrowInside=->](1.6,-4.4)(1.3,-4.1)\rput(1.6,-4.6){$a_n$}
        \rput(.6,-3.55){$\mu_n$}
        \psline(.8,-3.6)(1.9,-2.5)
        \psline[ArrowInside=->](3.8,-4.4)(3.5,-4.1)\rput(3.8,-4.6){$b_1$}
        \psline[ArrowInside=->](3.2,-4.4)(3.5,-4.1)\rput(3.2,-4.6){$b_2$}
        \psline(3.5,-4.1)(3.4,-4)
        \rput(3.7,-4){$\nu_2$}\rput(3.2,-4){$f_2$}
        \psline[linestyle=dotted](3.4,-4)(3.1,-3.7)
        \psline(3,-3.6)(2.2,-4.4)\psline[ArrowInside=->](2.2,-4.4)(2.5,-4.1)\rput(2.2,-4.6){$b_m$}
        \rput(3.3,-3.6){$\nu_m$}
        \psline(3.1,-3.7)(1.9,-2.5)
        \psline[ArrowInside=->](1.5,-2.9)(1.8,-2.6)\rput(1.7,-2.9){$c$}
        \psline[ArrowInside=->](2.3,-2.9)(2,-2.6)\rput(2.1,-2.9){$\bar{c}$}
        \psellipse[border=1.5pt](1.4,-1.25)(.4,.2)\psline[ArrowInside=->](1.4,-1.45)(1.25,-1.42)\psline[border=1.5pt](1,-1)(1.3,-1.3)
        \rput(1.5,-1.25){$\otimes$}\rput(1.2,-1.55){$\omega_0$}
        \psellipse[border=1.5pt](1.4,-3.15)(.4,.2)\psline[ArrowInside=->](1.4,-3.35)(1.25,-3.32)\psline[border=1.5pt](1.2,-3.2)(1.5,-2.9)
        \rput(1.5,-3.15){$\otimes$}\rput(1.65,-3.5){$\omega_0$}
    \end{pspicture}}.
 \end{split}
 \end{equation}

From the previous equation, it is straightforward to see that
\begin{equation}\label{eq:rhoannulus-arbpower}
\begin{split}
\left(\rhoD\right)^{\alpha}= \sum_{\substack{\vec{a},\vec{e}, \vec{\mu} \\ \vec{b}, \vec{f}, \vec{\nu}}}
&\frac{1}{\mathcal{D}^{2(n+m-3)}}\frac{\sqrt{d_{\vec{a}}d_{\vec{b}}}}{d_c^{2}}\left( \frac{d_{\vec{a}} d_{\vec{b}}}{\mathcal{D}^{2(n+m-2)} d_c^2}\right)^{\alpha-1}
\psscalebox{1}{\begin{pspicture}[shift=-2.25](.5,-4.7)(3.6,.3)
        \scriptsize
        \psline[ArrowInside=->](.3,-.3)(0,0)\rput(0,.2){$a_1$}
        \psline[ArrowInside=->](.3,-.3)(.6,0)\rput(.6,.2){$a_2$}
        \psline(.4,-.4)(.3,-.3)
        \psline[linestyle=dotted](.7,-.7)(.4,-.4)
        \rput(.1,-.45){$\mu_2$}\rput(.6,-.4){$e_2$}
        \psline(.8,-.8)(.7,-.7)
        \psline(.8,-.8)(1.6,0)\psline[ArrowInside=->](1.3,-.3)(1.6,0)\rput(1.6,.2){$a_n$}
        \rput(.6,-.85){$\mu_n$}
        \psline(.8,-.8)(1.9,-1.9)
        \psline[ArrowInside=->](3.5,-.3)(3.8,0)\rput(3.8,.2){$b_1$}
        \psline[ArrowInside=->](3.5,-.3)(3.2,0)\rput(3.2,.2){$b_2$}
        \psline(3.5,-.3)(3.4,-.4)
        \rput(3.7,-.45){$\nu_2$}\rput(3.15,-.35){$f_2$}
        \psline[linestyle=dotted](3.4,-.4)(3.1,-.7)
        \psline(3,-.8)(2.2,0)\psline[ArrowInside=->](2.5,-.3)(2.2,0)\rput(2.2,.2){$b_m$}
        \rput(3.3,-.85){$\nu_m$}
        \psline(3.1,-.7)(1.9,-1.9)
        \psline[ArrowInside=->](1.8,-1.8)(1.5,-1.5)\rput(1.7,-1.5){$c$}
        \psline[ArrowInside=->](2,-1.8)(2.3,-1.5)\rput(2.1,-1.5){$\bar{c}$}
        \psline[ArrowInside=->](0,-4.4)(.3,-4.1)\rput(0,-4.6){$a_1$}
        \psline[ArrowInside=->](.6,-4.4)(.3,-4.1)\rput(.6,-4.6){$a_2$}
        \psline(.4,-4)(.3,-4.1)
        \psline[linestyle=dotted](.4,-4)(.7,-3.7)
        \rput(.1,-4){$\mu_2$}\rput(.6,-4){$e_2$}
        \psline(.8,-3.6)(.7,-3.7)
        \psline(.8,-3.6)(1.6,-4.4)\psline[ArrowInside=->](1.6,-4.4)(1.3,-4.1)\rput(1.6,-4.6){$a_n$}
        \rput(.6,-3.55){$\mu_n$}
        \psline(.8,-3.6)(1.9,-2.5)
        \psline[ArrowInside=->](3.8,-4.4)(3.5,-4.1)\rput(3.8,-4.6){$b_1$}
        \psline[ArrowInside=->](3.2,-4.4)(3.5,-4.1)\rput(3.2,-4.6){$b_2$}
        \psline(3.5,-4.1)(3.4,-4)
        \rput(3.7,-4){$\nu_2$}\rput(3.2,-4){$f_2$}
        \psline[linestyle=dotted](3.4,-4)(3.1,-3.7)
        \psline(3,-3.6)(2.2,-4.4)\psline[ArrowInside=->](2.2,-4.4)(2.5,-4.1)\rput(2.2,-4.6){$b_m$}
        \rput(3.3,-3.6){$\nu_m$}
        \psline(3.1,-3.7)(1.9,-2.5)
        \psline[ArrowInside=->](1.5,-2.9)(1.8,-2.6)\rput(1.7,-2.9){$c$}
        \psline[ArrowInside=->](2.3,-2.9)(2,-2.6)\rput(2.1,-2.9){$\bar{c}$}
        \psellipse[border=1.5pt](1.4,-1.25)(.4,.2)\psline[ArrowInside=->](1.4,-1.45)(1.25,-1.42)\psline[border=1.5pt](1,-1)(1.3,-1.3)
        \rput(1.5,-1.25){$\otimes$}\rput(1.2,-1.55){$\omega_0$}
        \psellipse[border=1.5pt](1.4,-3.15)(.4,.2)\psline[ArrowInside=->](1.4,-3.35)(1.25,-3.32)\psline[border=1.5pt](1.2,-3.2)(1.5,-2.9)
        \rput(1.5,-3.15){$\otimes$}\rput(1.65,-3.5){$\omega_0$}
    \end{pspicture}}.
    \end{split}
\end{equation}

Performing the quantum trace and summing over the vertex labels we find
 \begin{equation}\label{eq:annulus-ARE-end}
 \begin{split}
 \aTr \left(\rhoD\right)^\alpha = \sum_{\substack{\vec{a},\vec{b}}} \left( \frac{d_{\vec{a}}d_{\vec{b}}}{\mathcal{D}^{2(n+m-2)} d_c^2 }\right)^\alpha N^c_{a_1\dots a_n} N^{\bar{c}}_{b_1\dots b_m}
 \end{split}
 \end{equation}
The anyonic R\'enyi entropy is therefore
\begin{equation}\label{eq:Renyiannulus}
\aS^{(\alpha)} \left( \rhoD \right)=
\frac{1}{1-\alpha} \log \left( \sum_{\vec{a},\vec{b}} N_{a_1 ... a_n}^c N_{b_1 ... b_m}^{\bar{c}}  \left(\frac{d_{\vec{a}}d_{\vec{b}}}{ \mathcal{D}^{2(n+m-2)}d_c^2} \right)^{\alpha}
 \right).
\end{equation}
Taking the limit $\alpha\to 1$ yields
\begin{equation}
\aS \left( \rhoD \right) = -(n+m) \sum_{a} \frac{d_{a}^2}{\mathcal{D}^2} \log \left( \frac{d_{a}}{\mathcal{D}^2}\right) -4 \log \mathcal{D}+2 \log d_c .
\end{equation}
Taking into account the doubling of the surface, the topological contribution to the entanglement entropy of the original (un-doubled) system is
\begin{eqnarray}
\aS_A &=& -\frac{n+m}{2} \sum_{a} \frac{d_{a}^2}{\mathcal{D}^2} \log \left( \frac{d_{a}}{\mathcal{D}^2}\right) -2 \log \mathcal{D} +2 \log d_c
\notag \\
&=& -\frac{n+m}{2} \aS \left( \arho_{\partial \mathbb{A}_j} \right) + 2 S_{\text{topo}} +2 \aS_c
.
\label{eq:AEEannulus}
\end{eqnarray}

\subsection{Torus Partitioned into Two Cylinders (Two Annuli)}

We now consider a torus in the ground state $\ket{(c);0}_{\text{inside}}$, corresponding to a topological charge line $c$ running in the longitudinal direction, i.e. in  the inside basis, and apply our method for a partition the system into two cylindrical regions $A$ and $\bar{A}$.

\begin{center}
\noindent\includegraphics[width=.65\linewidth]{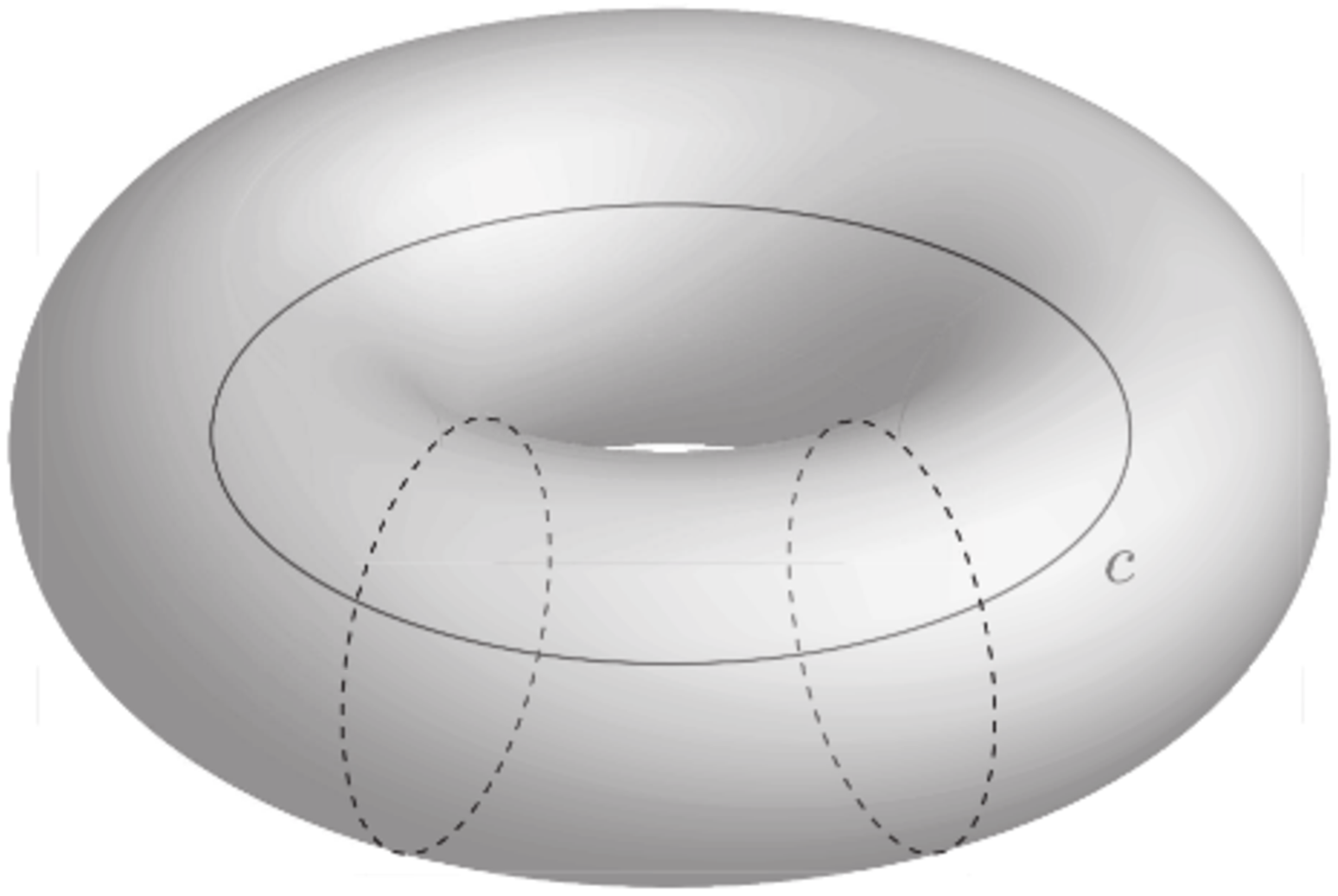}
\end{center}
As with the previous examples, we pair the system with its time-reversal conjugate. Specifically, we introduce the conjugate inside the original torus, and choose it to be in the its ground state $\ket{(0);0}_{\text{inside}}$, as signified by the $\omega_0$-loops in the figure. (We draw two $\omega_0$-loops instead of just one, the utility of which will become clear later.)

\begin{center}
\noindent\includegraphics[width=.85\linewidth]{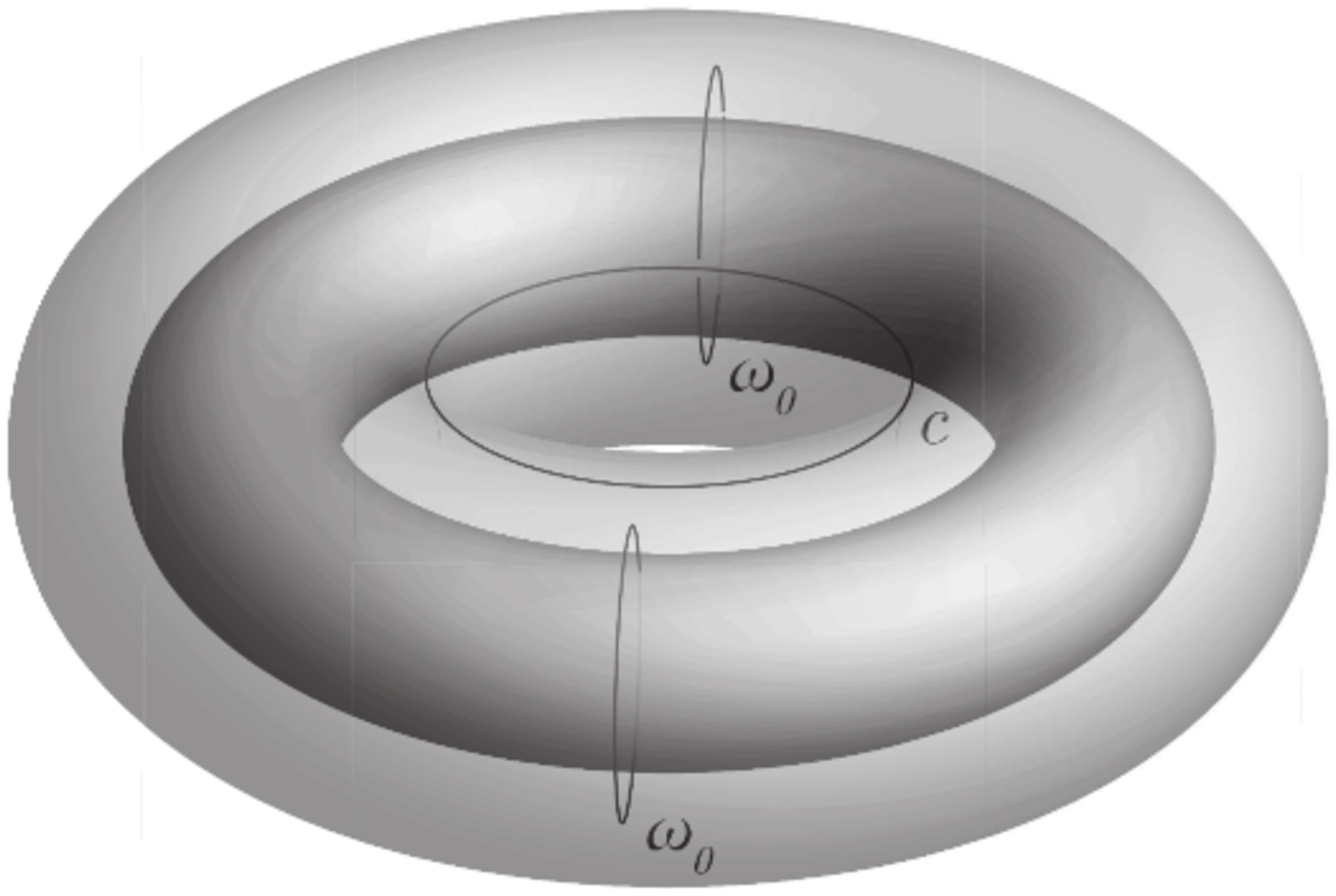}
\end{center}
As before, to construct $\mathbb{M}$ we adiabatically insert wormholes (threaded by trivial topological charge lines) along the partition boundary, with $n$ wormholes along one of the boundary components and $m$ wormholes along the other. Then we use the modular $\mathcal{S}$-transformation to re-express the state in the inside basis (where all the charge lines are between the two surfaces).

\begin{center}
\noindent\includegraphics[width=.85\linewidth]{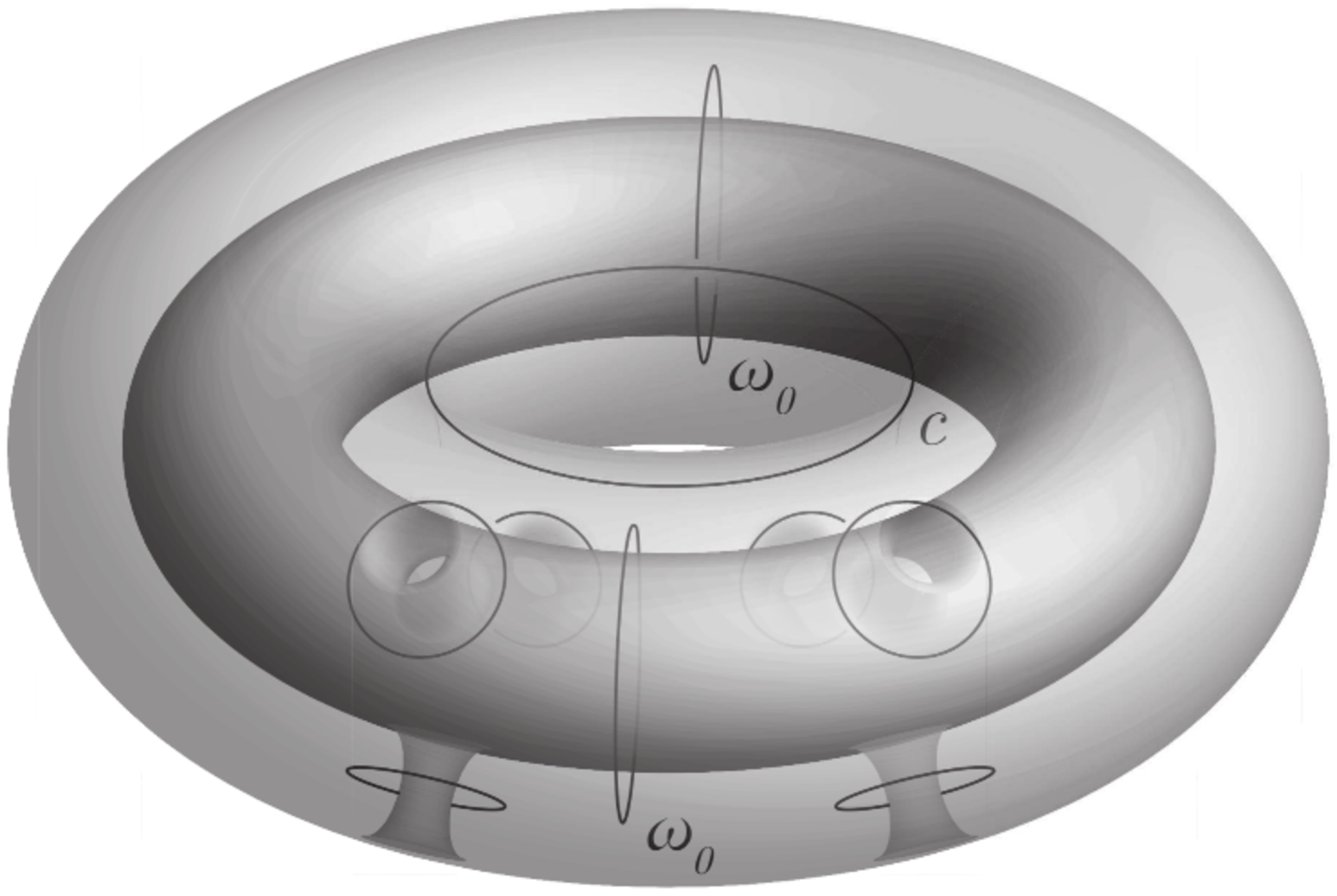}
\end{center}
Analogous to Eq.~(\ref{eq:unpunct-disk-fuse}) for the disk cut from the sphere, we apply a series of $F$-moves to fuse topological charge lines that thread the new boundary components between regions $\AD$ and $\bar{\AD}$.

\begin{center}
\noindent\includegraphics[width=.85\linewidth]{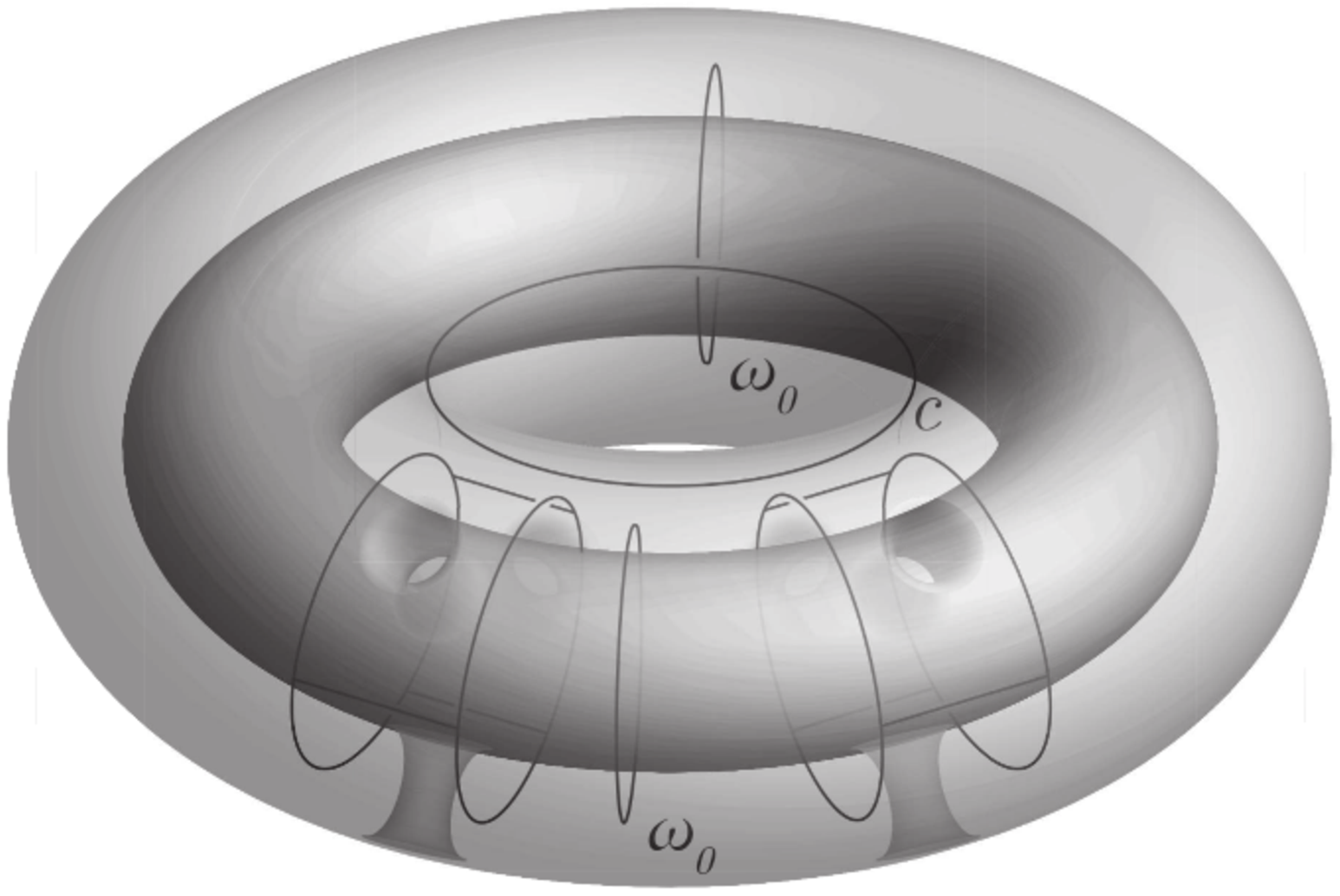}
\end{center}
Similarly to the first three equalities in Eq.~(\ref{eq:F-move-steps}), we rewrite the state with further use of $F$-moves.

\begin{center}
\noindent\includegraphics[width=.85\linewidth]{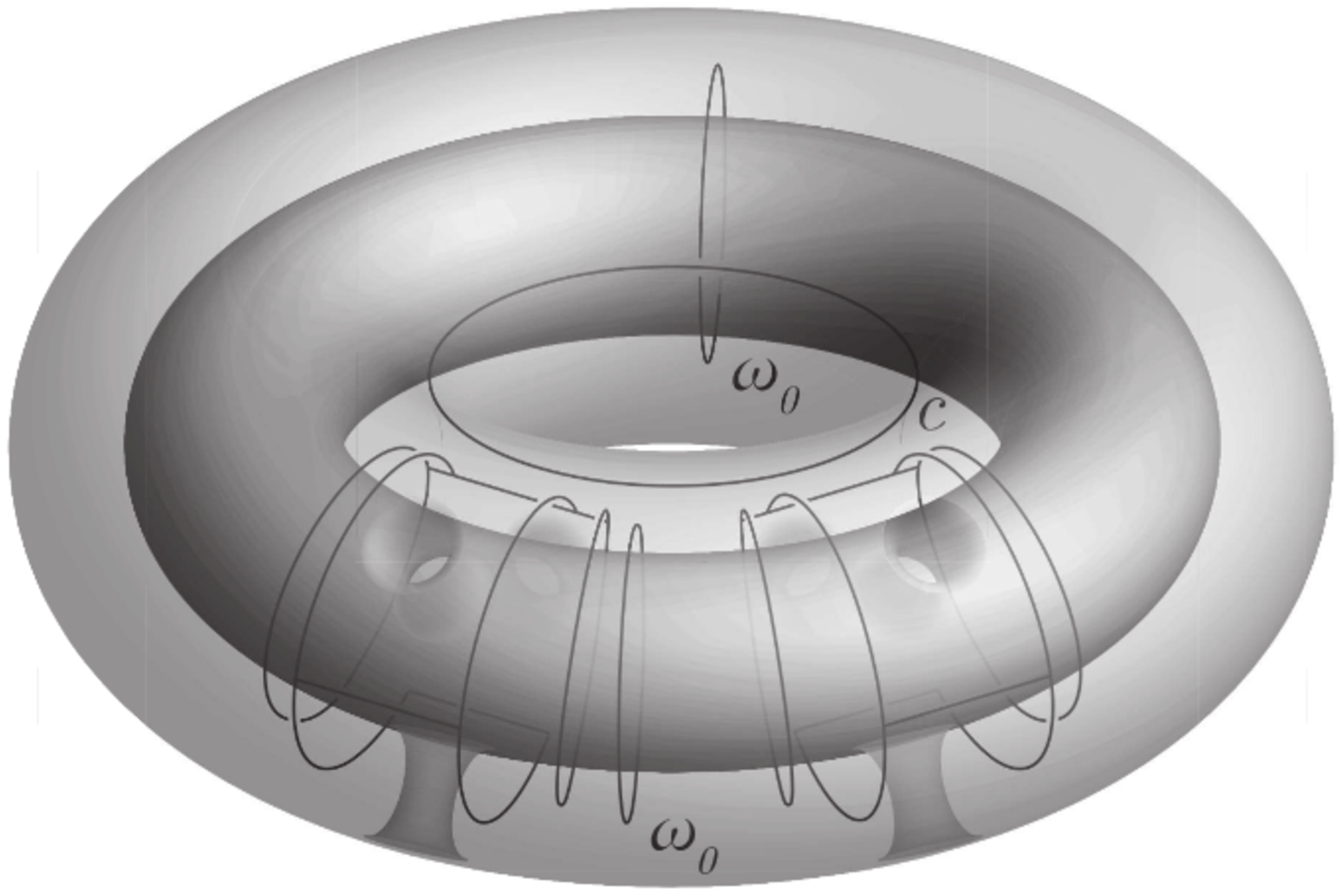}
\end{center}
In the last equality of Eq.~(\ref{eq:F-move-steps}), we collapsed a tadpole diagram in both $\mathbb{A}$ and $\bar{\mathbb{A}}$.  In the present situation, the analogous ``tadpole-like" diagrams now enclose a non-contractible cycle, i.e. the inner torus.  Nonetheless, we can contract these loops using the handle-slide property of the $\omega_0$-loop, see Eq.~(\ref{eq:handle-slide}).
Thus, even though the tadpoles encircle a nontrivial cycle,  they can be passed through it due to the presence of the $\omega_0$-loop. In this way, they become true tadpoles, and can be subsequently collapsed. (This step reveals the reason for beginning with two $\omega_0$-loops: there needs to be one on either side of the wormholes to help collapse the tadpoles.) The result is:

\begin{center}
\noindent\includegraphics[width=.85\linewidth]{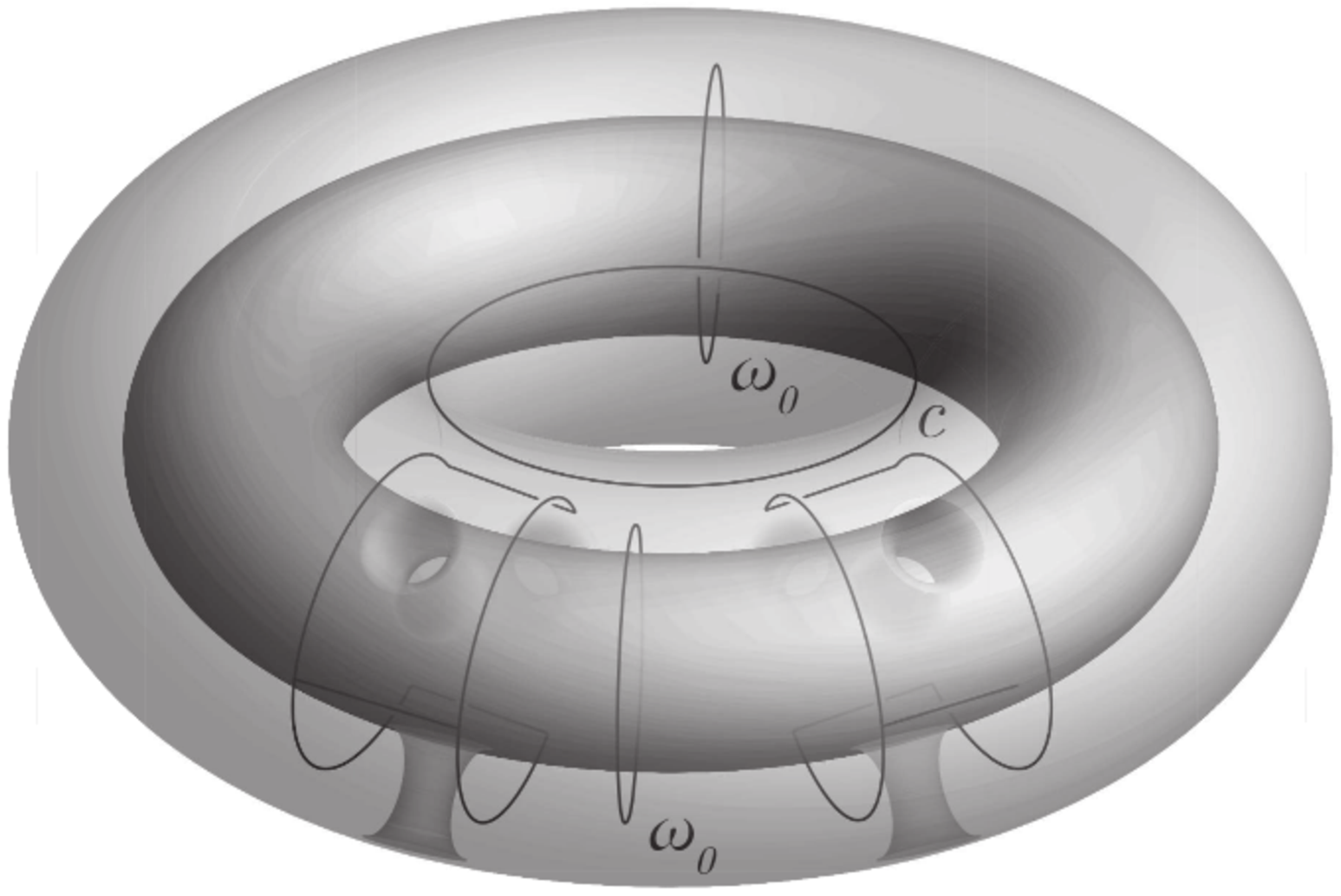}
\end{center}
Fusing the topological charge $c$ line into the other charge lines crossing the partition boundary, similar to Eq.~(\ref{eq:psiIntDis}), we have the state

\begin{center}
\noindent\includegraphics[width=.85\linewidth]{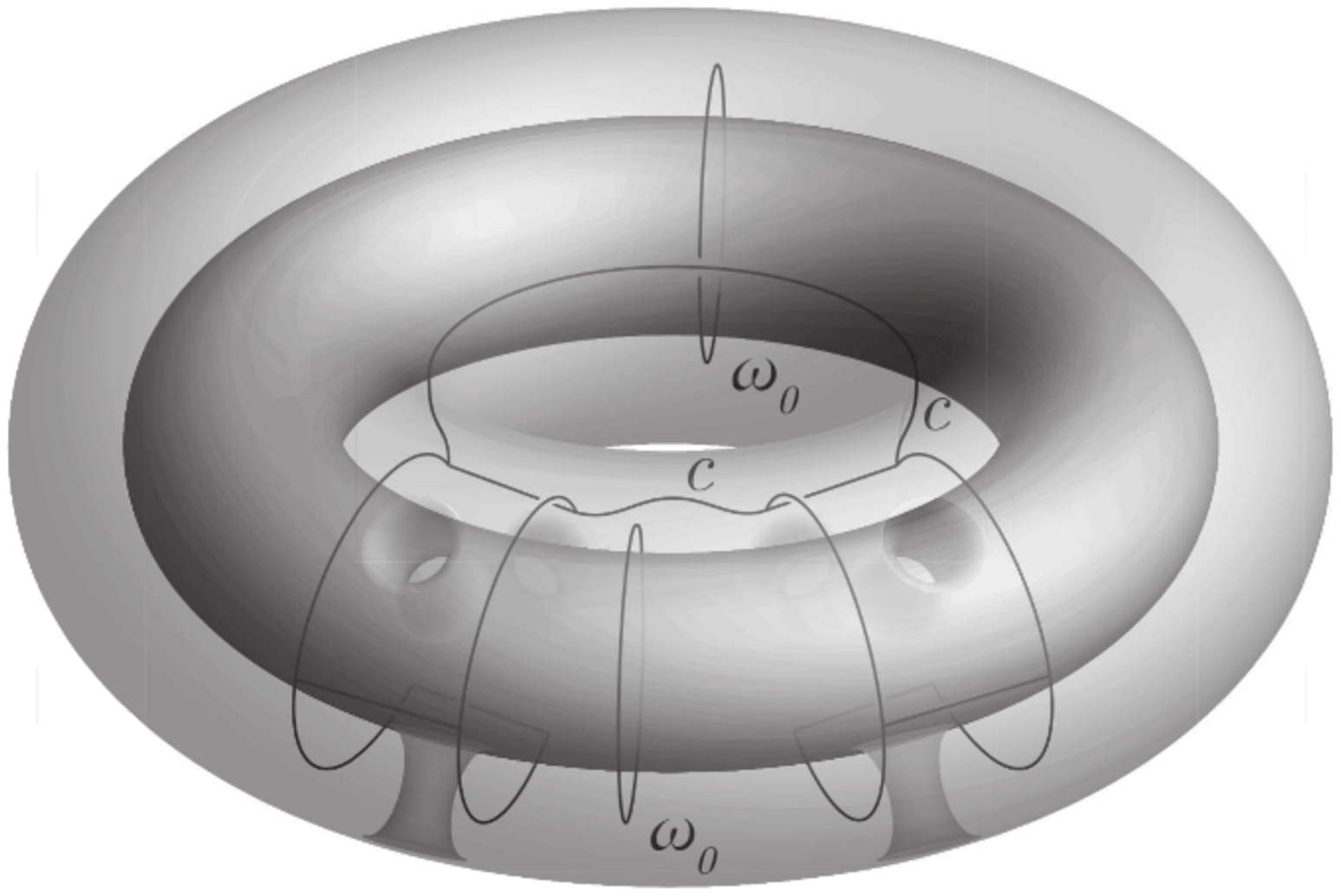}
\end{center}
Finally, we cut along the partition boundary to produce the cut state. Each of the resulting regions $\mathbb{A}$ and $\bar{\mathbb{A}}$ after cutting is a surface with genus $g=1$ and $n+m$ punctures, and looks like:

\begin{center}
\noindent\includegraphics[width=.35\linewidth]{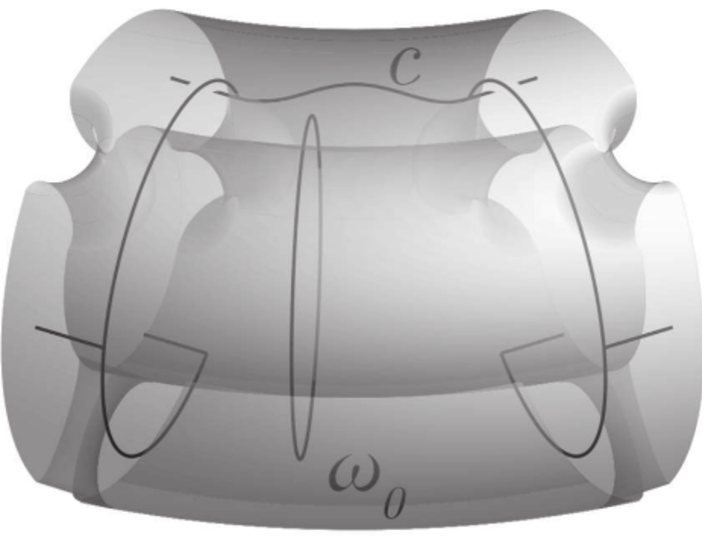}
\end{center}
Calculating the density matrix $\arho=\ket{\psi_{\text{cut}}}\bra{\psi_{\text{cut}}}$ and tracing out region $\bar{\mathbb{A}}$ yields the reduced density matrix for region $\mathbb{A}$.  Once again, there is a choice of basis for how to diagrammatically represent the region $\mathbb{A}$, which essentially amounts to either projecting the above picture to the plane as drawn, or turning the picture inside out so that the center tube becomes external to the region, resulting in the $\omega_0$-loop circling the $c$ charge line.  The former results in the reduced density matrix
\begin{equation}\label{eq:rhotorus}
\rhoD=\sum_{\substack{\vec{a}, \vec{e},\vec{\mu} \\ \vec{b}, \vec{f}, \vec{\nu}}} \frac{1}{\mathcal{D}^{2(n+m-3)}}\frac{\sqrt{d_{\vec{a}}d_{\vec{b}}}}{d_c^{2}}
 \begin{pspicture}[shift=-2.25](.3,-4.7)(4,.3)
        \scriptsize
        \psline[ArrowInside=->](.3,-.3)(0,0)\rput(0,.2){$a_1$}
        \psline[ArrowInside=->](.3,-.3)(.6,0)\rput(.6,.2){$a_2$}
        \psline(.4,-.4)(.3,-.3)
        \psline[linestyle=dotted](.7,-.7)(.4,-.4)
        \rput(.1,-.45){$\mu_2$}\rput(.6,-.4){$e_2$}
        \psline(.8,-.8)(.7,-.7)
        \psline(.8,-.8)(1.6,0)\psline[ArrowInside=->](1.3,-.3)(1.6,0)\rput(1.6,.2){$a_n$}
        \rput(.6,-.85){$\mu_n$}
        \psline(.8,-.8)(1.9,-1.9)
        \psline[ArrowInside=->](3.5,-.3)(3.8,0)\rput(3.8,.2){$b_1$}
        \psline[ArrowInside=->](3.5,-.3)(3.2,0)\rput(3.2,.2){$b_2$}
        \psline(3.5,-.3)(3.4,-.4)
        \rput(3.7,-.45){$\nu_2$}\rput(3.15,-.35){$f_2$}
        \psline[linestyle=dotted](3.4,-.4)(3.1,-.7)
        \psline(3,-.8)(2.2,0)\psline[ArrowInside=->](2.5,-.3)(2.2,0)\rput(2.2,.2){$b_m$}
        \rput(3.3,-.85){$\nu_m$}
        \psline(3.1,-.7)(1.9,-1.9)
        \psline[ArrowInside=->](1.8,-1.8)(1.5,-1.5)\rput(1.7,-1.5){$c$}
        \psline[ArrowInside=->](2,-1.8)(2.3,-1.5)\rput(2.1,-1.5){$\bar{c}$}
        \psline[ArrowInside=->](0,-4.4)(.3,-4.1)\rput(0,-4.6){$a_1$}
        \psline[ArrowInside=->](.6,-4.4)(.3,-4.1)\rput(.6,-4.6){$a_2$}
        \psline(.4,-4)(.3,-4.1)
        \psline[linestyle=dotted](.4,-4)(.7,-3.7)
        \rput(.1,-4){$\mu_2$}\rput(.6,-4){$e_2$}
        \psline(.8,-3.6)(.7,-3.7)
        \psline(.8,-3.6)(1.6,-4.4)\psline[ArrowInside=->](1.6,-4.4)(1.3,-4.1)\rput(1.6,-4.6){$a_n$}
        \rput(.6,-3.55){$\mu_n$}
        \psline(.8,-3.6)(1.9,-2.5)
        \psline[ArrowInside=->](3.8,-4.4)(3.5,-4.1)\rput(3.8,-4.6){$b_1$}
        \psline[ArrowInside=->](3.2,-4.4)(3.5,-4.1)\rput(3.2,-4.6){$b_2$}
        \psline(3.5,-4.1)(3.4,-4)
        \rput(3.7,-4){$\nu_2$}\rput(3.2,-4){$f_2$}
        \psline[linestyle=dotted](3.4,-4)(3.1,-3.7)
        \psline(3,-3.6)(2.2,-4.4)\psline[ArrowInside=->](2.2,-4.4)(2.5,-4.1)\rput(2.2,-4.6){$b_m$}
        \rput(3.3,-3.6){$\nu_m$}
        \psline(3.1,-3.7)(1.9,-2.5)
        \psline[ArrowInside=->](1.5,-2.9)(1.8,-2.6)\rput(1.7,-2.9){$c$}
        \psline[ArrowInside=->](2.3,-2.9)(2,-2.6)\rput(2.1,-2.9){$\bar{c}$}
        \psellipse[border=1.5pt](1.9,-1)(.4,.2)\psline[ArrowInside=->](1.9,-1.2)(1.75,-1.17)
        \rput(1.9,-1){$\otimes$}\rput(1.7,-1.3){$\omega_0$}
        \psellipse[border=1.5pt](1.9,-3.3)(.4,.2)\psline[ArrowInside=->](1.9,-3.5)(1.75,-3.48)
        \rput(1.9,-3.3){$\otimes$}\rput(2.15,-3.6){$\omega_0$}
    \end{pspicture},
\end{equation}
while the latter results in the reduced density matrix given in Eq.~(\ref{eq:rhoannulus}). Therefore, the reduced density matrix for the region $\mathbb{A}$ of the doubled torus with wormholes is equivalent to that of the doubled region $\mathbb{A}$ corresponding to when $A$ was an annulus cut from a sphere, as we would expect from topological considerations.  It follows that the anyonic R\'enyi entropy of region $\mathbb{A}$ is given by Eq.~(\ref{eq:Renyiannulus}) and the topological contribution to the entanglement entropy of $A$, the original (un-doubled) system,  is given by Eq.~(\ref{eq:AEEannulus}).

\subsection{3-Punctured Sphere Partitioned into a $3$-Punctured Sphere and Three $1$-Punctured Disks}
\label{sec:three-punctured-sphere}

As a final example, we consider a sphere containing three punctures (or quasiparticles) carrying topological charges $x$, $y$, and $z$.  We partition the region so that each puncture is contained in a separate disk, and apply our method for the three punctured sphere $A$ that remains when the three disks are removed.

\begin{center}
\noindent\includegraphics[width=.6\linewidth]{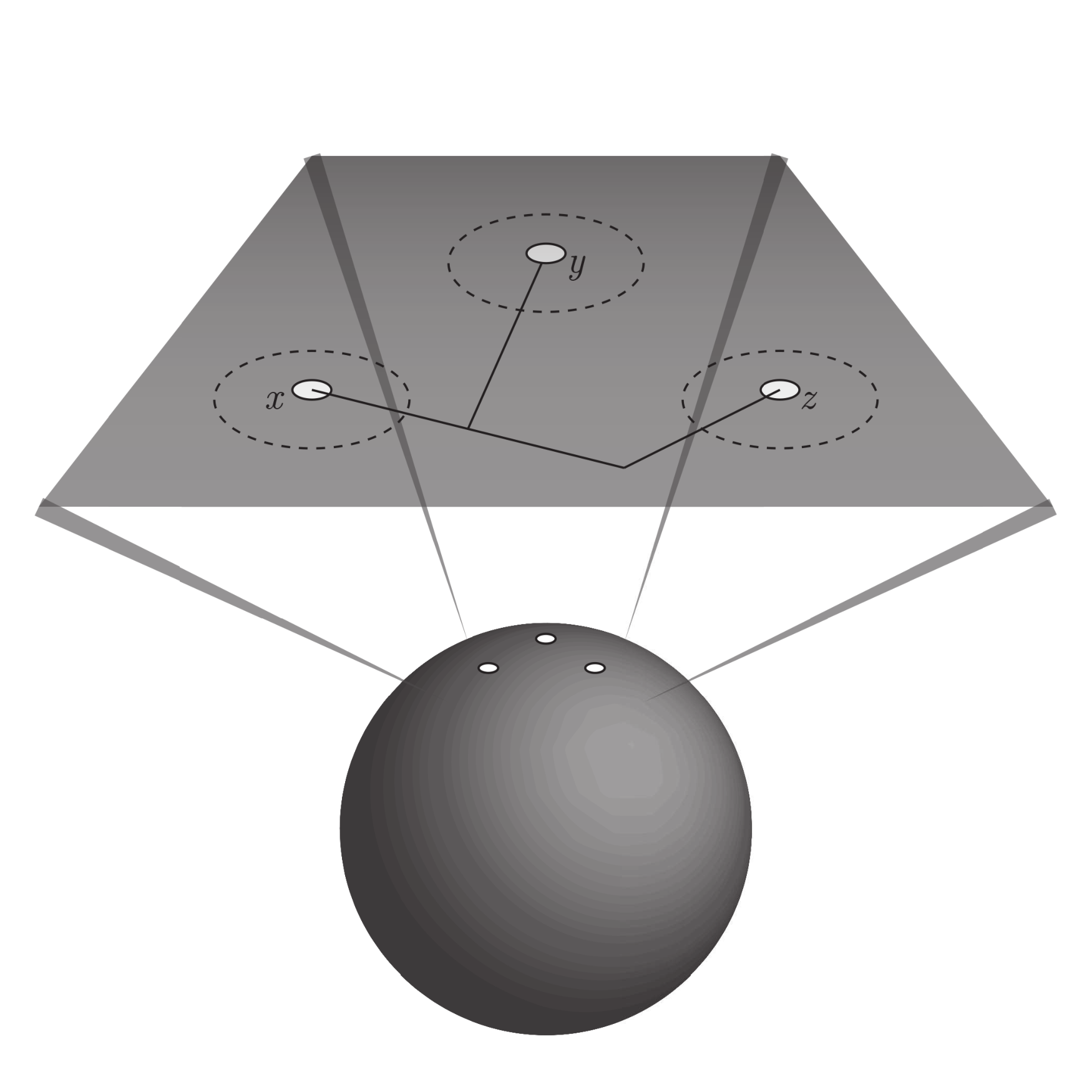}
\end{center}

We follow the same steps as in the previous examples: (1) pair the system with its time-reversal conjugate, (2) insert wormholes threaded by trivial charge lines along the partition boundary, with $l$, $m$, and $n$ wormholes along the three different boundary components, respectively, (3) apply modular $\mathcal{S}$-transformations to express the state in the inside basis (all topological charge lines are between the two surfaces), (4) use $F$-moves to fuse topological charge lines that thread each new partition boundary component, (5) use further $F$-moves to write the state in a tree-like form, and (6) fuse the $x$, $y$, and $z$ charge lines to the topological charge line threading the same boundary component.  Analogous to Eq.~(\ref{eq:annulus-after-F-moves}) for the annulus, after step (5) each disk in $\bar{\mathbb{A}}$ will contain a tadpole that can be collapsed.  Similar to Eq.~(\ref{eq:annulus-tadpole-collapsed}), collapsing this tadpole results in an $\omega_0$-loop in $\mathbb{A}$ encircling the corresponding region of $\bar{\mathbb{A}}$. Region $\mathbb{A}$ is a surface with genus $g=2$ and $l+m+n$ punctures. After performing steps (1)-(6), the state embedded in $\mathbb{M}$ is

\begin{center}
\includegraphics[width=.9\linewidth]{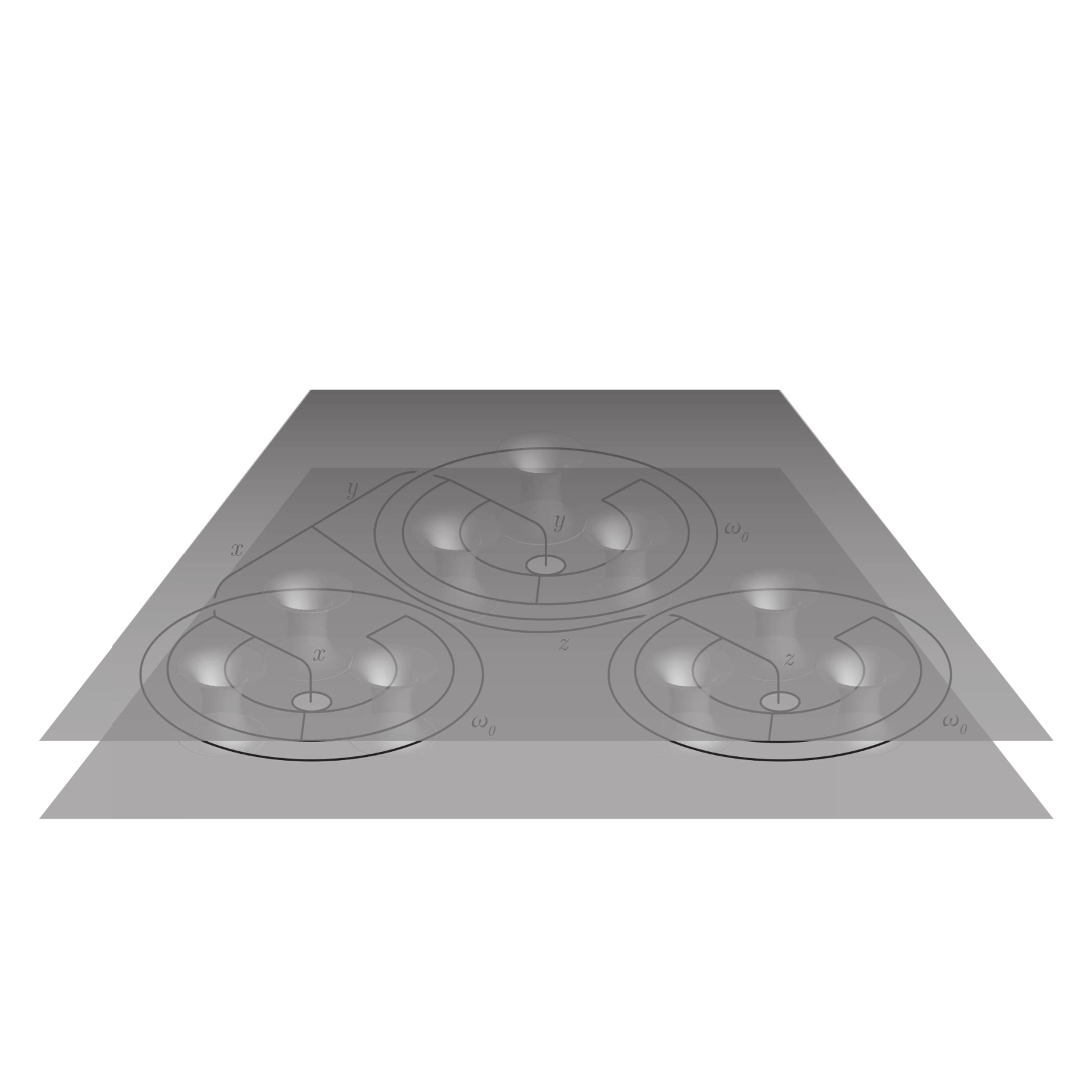}
\end{center}
with the corresponding diagrammatic representation
\begin{equation}
\ket{\psi}=\sum_{\substack{\vec{a},\vec{b}, \vec{c},\\ \vec{e}, \vec{f},\vec{g}\\
}} \frac{\sqrt{d_{\vec{a}}d_{\vec{b}}d_{\vec{c}}}}{\mathcal{D}^{l+m+n-5}\left(d_x d_y d_z\right)^{3/4}}
\psscalebox{.9}{
 \begin{pspicture}[shift=-1.75](.9,-4)(8,2)
        \scriptsize
        \psline[ArrowInside=->](.3,-.3)(0,0)\psline(0,0)(0,.3)
        \psline[ArrowInside=->](0,.3)(.3,.6)
        \rput(-.1,-.1){$a_1$}
        \psline[ArrowInside=->](.3,-.3)(.6,0)\psline(.6,0)(.6,.3)
        \psline[ArrowInside=->](.6,.3)(.3,.6)
        \rput(.75,-.1){$a_2$}
        \psline(.4,-.4)(.3,-.3)
        \psline[linestyle=dotted](.7,-.7)(.4,-.4)
        \rput(.6,-.4){$e_2$}\rput(.6,.7){$e_2$}
        \psline(.8,-.8)(.7,-.7)
        \psline(.8,-.8)(1.6,0)
        \psline[ArrowInside=->](1.3,-.3)(1.6,0)\psline(1.6,0)(1.6,.3)
        \psline[ArrowInside=->](1.6,.3)(1.3,.6)\psline(1.3,.6)(.8,1.1)
        \psline(.8,1.1)(.7,1)\psline[linestyle=dotted](.7,1)(.4,.7)\psline(.4,.7)(.3,.6)
        \rput(1.75,-.1){$a_l$}\rput(1.2,1.2){$x$}
        \psline[ArrowInside=->](.8,1.1)(1.1,1.4)
        \rput(-.3,.15){$\odot$}\rput(.3,.15){$\odot$}\rput(.95,.15){$\dots$}\rput(1.3,.15){$\odot$}
        \psline[linestyle=dashed](-.3,.25)(-.3,.35)(.95,1.6)(1.2,1.6)(1.75,1.05)(1.75,.15)(1.6,.15)
        \psline[linestyle=dashed](-.2,.15)(-.1,.15)
        \psline[linestyle=dashed](.1,.15)(.2,.15)\psline[linestyle=dashed](.4,.15)(.5,.15)
        \psline[linestyle=dashed](.65,.15)(.75,.15)
        \psline[linestyle=dashed](1.1,.15)(1.2,.15)
        \psline[linestyle=dashed](1.4,.15)(1.5,.15)
        \psline(.8,-.8)(2.25,-2.25)\psline[ArrowInside=->](2.1,-2.1)(1.9,-1.9)\psline(2.25,-2.25)(3.7,-3.7)\rput(1.8,-2){$x$}
        \psellipse[border=1.5pt](.8,.3)(1.3,1.5)\psline[ArrowInside=->](-.5,.1)(-.5,.25)\rput(-.6,-.15){$\omega_0$}
 \rput(2.9,0){        \psline[ArrowInside=->](.3,-.3)(0,0)\psline(0,0)(0,.3)
        \psline[ArrowInside=->](0,.3)(.3,.6)
        \rput(-.1,-.1){$b_1$}
        \psline[ArrowInside=->](.3,-.3)(.6,0)\psline(.6,0)(.6,.3)
        \psline[ArrowInside=->](.6,.3)(.3,.6)
        \rput(.75,-.1){$b_2$}
        \psline(.4,-.4)(.3,-.3)
        \psline[linestyle=dotted](.7,-.7)(.4,-.4)
        \rput(.6,-.4){$f_2$}\rput(.6,.7){$f_2$}
        \psline(.8,-.8)(.7,-.7)
        \psline(.8,-.8)(1.6,0)
        \psline[ArrowInside=->](1.3,-.3)(1.6,0)\psline(1.6,0)(1.6,.3)
        \psline[ArrowInside=->](1.6,.3)(1.3,.6)\psline(1.3,.6)(.8,1.1)
        \psline(.8,1.1)(.7,1)\psline[linestyle=dotted](.7,1)(.4,.7)\psline(.4,.7)(.3,.6)
        \rput(1.75,-.15){$b_m$}\rput(1.2,1.2){$y$}
        \psline[ArrowInside=->](.8,1.1)(1.1,1.4)
        \rput(-.3,.15){$\odot$}\rput(.3,.15){$\odot$}\rput(.9,.15){$\dots$}\rput(1.3,.15){$\odot$}
        \psline[linestyle=dashed](-.3,.25)(-.3,.35)(.95,1.6)(1.2,1.6)(1.75,1.05)(1.75,.15)(1.6,.15)
        \psline[linestyle=dashed](-.2,.15)(-.1,.15)
        \psline[linestyle=dashed](.1,.15)(.2,.15)\psline[linestyle=dashed](.4,.15)(.5,.15)
        \psline[linestyle=dashed](.65,.15)(.75,.15)
        \psline[linestyle=dashed](1.1,.15)(1.2,.15)
        \psline[linestyle=dashed](1.4,.15)(1.5,.15)
        \psline(.8,-.8)(-.65,-2.25)\psline[ArrowInside=->](-.5,-2.1)(-.2,-1.8)\rput(-.2,-2){$y$}
        \psellipse[border=1.5pt](.8,.3)(1.3,1.5)\psline[ArrowInside=->](-.5,.1)(-.5,.25)\rput(-.6,-.15){$\omega_0$}}
\rput(5.8,0){        \psline[ArrowInside=->](.3,-.3)(0,0)\psline(0,0)(0,.3)
        \psline[ArrowInside=->](0,.3)(.3,.6)
        \rput(-.1,-.1){$c_1$}
        \psline[ArrowInside=->](.3,-.3)(.6,0)\psline(.6,0)(.6,.3)
        \psline[ArrowInside=->](.6,.3)(.3,.6)
        \rput(.75,-.1){$c_2$}
        \psline(.4,-.4)(.3,-.3)
        \psline[linestyle=dotted](.7,-.7)(.4,-.4)
        \rput(.6,-.4){$g_2$}\rput(.6,.7){$g_2$}
        \psline(.8,-.8)(.7,-.7)
        \psline(.8,-.8)(1.6,0)
        \psline[ArrowInside=->](1.3,-.3)(1.6,0)\psline(1.6,0)(1.6,.3)
        \psline[ArrowInside=->](1.6,.3)(1.3,.6)\psline(1.3,.6)(.8,1.1)
        \psline(.8,1.1)(.7,1)\psline[linestyle=dotted](.7,1)(.4,.7)\psline(.4,.7)(.3,.6)
        \rput(1.75,-.1){$c_n$}\rput(1.2,1.2){$z$}
        \psline[ArrowInside=->](.8,1.1)(1.1,1.4)
        \rput(-.3,.15){$\odot$}\rput(.3,.15){$\odot$}\rput(.9,.15){$\dots$}\rput(1.3,.15){$\odot$}
        \psline[linestyle=dashed](-.3,.25)(-.3,.35)(.95,1.6)(1.2,1.6)(1.75,1.05)(1.75,.15)(1.6,.15)
        \psline[linestyle=dashed](-.2,.15)(-.1,.15)
        \psline[linestyle=dashed](.1,.15)(.2,.15)\psline[linestyle=dashed](.4,.15)(.5,.15)
        \psline[linestyle=dashed](.65,.15)(.75,.15)
        \psline[linestyle=dashed](1.1,.15)(1.2,.15)
        \psline[linestyle=dashed](1.4,.15)(1.5,.15)
        \psline(.8,-.8)(-2.1,-3.7)\psline[ArrowInside=->](-.5,-2.1)(-.2,-1.8)\rput(-.2,-2){$z$}
        \psellipse[border=1.5pt](.8,.3)(1.3,1.5)\psline[ArrowInside=->](-.5,.1)(-.5,.25)\rput(-.6,-.15){$\omega_0$}
        }
    \end{pspicture}}.
\end{equation}
The third $\omega_0$-loop can be brought around the other side of the sphere, so that it encloses the other two $\omega_0$-loops.  Then, using the handle-slide property of Eq.~(\ref{eq:handle-slide}), it can be slid over the other two $\omega_0$-loops, so that it does not enclose any non-contractible cycles.  Finally, we can collapse this $\omega_0$-loop, using
\begin{equation}
\begin{pspicture}[shift=-.7](-.6,-.7)(.6,.7)
\scriptsize
\psellipse(0,0)(.5,.25)
\psline[ArrowInside=->](.15,-.24)(-.1,-.23) \rput(0,-.05){$\omega_0$}
\end{pspicture}
= \sum_a \frac{d_a}{\mathcal{D}^2}
\begin{pspicture}[shift=-.7](-.6,-.7)(.6,.7)
\scriptsize
\psline(-.5,0)(0,.5)(.5,0)(0,-.5)(-.5,0)
\psline[ArrowInside=->](0,-.5)(-.5,0) \rput(-.1,-.15){$a$}
\end{pspicture}
=1
.
\end{equation}
Thus, the state can be written as
\begin{equation}\label{eq:3-punc-sphere-omega}
\ket{\psi}=\sum_{\substack{\vec{a},\vec{b}, \vec{c},\\ \vec{e}, \vec{f},\vec{g}
 }} \frac{\sqrt{d_{\vec{a}}d_{\vec{b}}d_{\vec{c}}}}{\mathcal{D}^{l+m+n-5}\left(d_x d_y d_z\right)^{3/4}}
 \psscalebox{.9}{
\begin{pspicture}[shift=-2.5](.3,-4)(8,2)
        \scriptsize
        \psline[ArrowInside=->](.3,-.3)(0,0)\psline(0,0)(0,.3)
        \psline[ArrowInside=->](0,.3)(.3,.6)
        \rput(-.1,-.1){$a_1$}
        \psline[ArrowInside=->](.3,-.3)(.6,0)\psline(.6,0)(.6,.3)
        \psline[ArrowInside=->](.6,.3)(.3,.6)
        \rput(.75,-.1){$a_2$}
        \psline(.4,-.4)(.3,-.3)
        \psline[linestyle=dotted](.7,-.7)(.4,-.4)
        \rput(.6,-.4){$e_2$}\rput(.6,.7){$e_2$}
        \psline(.8,-.8)(.7,-.7)
        \psline(.8,-.8)(1.6,0)
        \psline[ArrowInside=->](1.3,-.3)(1.6,0)\psline(1.6,0)(1.6,.3)
        \psline[ArrowInside=->](1.6,.3)(1.3,.6)\psline(1.3,.6)(.8,1.1)
        \psline(.8,1.1)(.7,1)\psline[linestyle=dotted](.7,1)(.4,.7)\psline(.4,.7)(.3,.6)
        \rput(1.75,-.1){$a_l$}\rput(1.2,1.2){$x$}
        \psline[ArrowInside=->](.8,1.1)(1.1,1.4)
        \rput(-.3,.15){$\odot$}\rput(.3,.15){$\odot$}\rput(.9,.15){$\dots$}\rput(1.3,.15){$\odot$}
        \psline[linestyle=dashed](-.3,.25)(-.3,.35)(.95,1.6)(1.2,1.6)(1.75,1.05)(1.75,.15)(1.6,.15)
        \psline[linestyle=dashed](-.2,.15)(-.1,.15)
        \psline[linestyle=dashed](.1,.15)(.2,.15)\psline[linestyle=dashed](.4,.15)(.5,.15)
        \psline[linestyle=dashed](.65,.15)(.75,.15)
        \psline[linestyle=dashed](1.1,.15)(1.2,.15)
        \psline[linestyle=dashed](1.4,.15)(1.5,.15)
        \psline(.8,-.8)(2.25,-2.25)\psline[ArrowInside=->](2.1,-2.1)(1.9,-1.9)\psline(2.25,-2.25)(3.7,-3.7)\rput(1.8,-2){$x$}
        \psellipse[border=1.5pt](.8,.3)(1.3,1.5)\psline[ArrowInside=->](-.5,.1)(-.5,.25)\rput(-.6,-.15){$\omega_0$}
 \rput(2.9,0){        \psline[ArrowInside=->](.3,-.3)(0,0)\psline(0,0)(0,.3)
        \psline[ArrowInside=->](0,.3)(.3,.6)
        \rput(-.1,-.1){$b_1$}
        \psline[ArrowInside=->](.3,-.3)(.6,0)\psline(.6,0)(.6,.3)
        \psline[ArrowInside=->](.6,.3)(.3,.6)
        \rput(.75,-.1){$b_2$}
        \psline(.4,-.4)(.3,-.3)
        \psline[linestyle=dotted](.7,-.7)(.4,-.4)
        \rput(.6,-.4){$f_2$}\rput(.6,.7){$f_2$}
        \psline(.8,-.8)(.7,-.7)
        \psline(.8,-.8)(1.6,0)
        \psline[ArrowInside=->](1.3,-.3)(1.6,0)\psline(1.6,0)(1.6,.3)
        \psline[ArrowInside=->](1.6,.3)(1.3,.6)\psline(1.3,.6)(.8,1.1)
        \psline(.8,1.1)(.7,1)\psline[linestyle=dotted](.7,1)(.4,.7)\psline(.4,.7)(.3,.6)
        \rput(1.8,-.1){$b_m$}\rput(1.2,1.2){$y$}
        \psline[ArrowInside=->](.8,1.1)(1.1,1.4)
        \rput(-.3,.15){$\odot$}\rput(.3,.15){$\odot$}\rput(.9,.15){$\dots$}\rput(1.3,.15){$\odot$}
        \psline[linestyle=dashed](-.3,.25)(-.3,.35)(.95,1.6)(1.2,1.6)(1.75,1.05)(1.75,.15)(1.6,.15)
        \psline[linestyle=dashed](-.2,.15)(-.1,.15)
        \psline[linestyle=dashed](.1,.15)(.2,.15)\psline[linestyle=dashed](.4,.15)(.5,.15)
        \psline[linestyle=dashed](.65,.15)(.75,.15)
        \psline[linestyle=dashed](1.1,.15)(1.2,.15)
        \psline[linestyle=dashed](1.4,.15)(1.5,.15)
        \psline(.8,-.8)(-.65,-2.25)\psline[ArrowInside=->](-.5,-2.1)(-.2,-1.8)\rput(-.2,-2){$y$}
        \psellipse[border=1.5pt](.8,.3)(1.3,1.5)\psline[ArrowInside=->](-.5,.1)(-.5,.25)\rput(-.6,-.15){$\omega_0$}}
\rput(5.8,0){        \psline[ArrowInside=->](.3,-.3)(0,0)\psline(0,0)(0,.3)
        \psline[ArrowInside=->](0,.3)(.3,.6)
        \rput(-.1,-.1){$c_1$}
        \psline[ArrowInside=->](.3,-.3)(.6,0)\psline(.6,0)(.6,.3)
        \psline[ArrowInside=->](.6,.3)(.3,.6)
        \rput(.75,-.1){$c_2$}
        \psline(.4,-.4)(.3,-.3)
        \psline[linestyle=dotted](.7,-.7)(.4,-.4)
        \rput(.6,-.4){$g_2$}\rput(.6,.7){$g_2$}
        \psline(.8,-.8)(.7,-.7)
        \psline(.8,-.8)(1.6,0)
        \psline[ArrowInside=->](1.3,-.3)(1.6,0)\psline(1.6,0)(1.6,.3)
        \psline[ArrowInside=->](1.6,.3)(1.3,.6)\psline(1.3,.6)(.8,1.1)
        \psline(.8,1.1)(.7,1)\psline[linestyle=dotted](.7,1)(.4,.7)\psline(.4,.7)(.3,.6)
        \rput(1.75,-.1){$c_n$}\rput(1.2,1.2){$z$}
        \psline[ArrowInside=->](.8,1.1)(1.1,1.4)
        \rput(-.3,.15){$\odot$}\rput(.3,.15){$\odot$}\rput(.9,.15){$\dots$}\rput(1.3,.15){$\odot$}
        \psline[linestyle=dashed](-.3,.25)(-.3,.35)(.95,1.6)(1.2,1.6)(1.75,1.05)(1.75,.15)(1.6,.15)
        \psline[linestyle=dashed](-.2,.15)(-.1,.15)
        \psline[linestyle=dashed](.1,.15)(.2,.15)\psline[linestyle=dashed](.4,.15)(.5,.15)
        \psline[linestyle=dashed](.65,.15)(.75,.15)
        \psline[linestyle=dashed](1.1,.15)(1.2,.15)
        \psline[linestyle=dashed](1.4,.15)(1.5,.15)
        \psline(.8,-.8)(-2.1,-3.7)\psline[ArrowInside=->](-.5,-2.1)(-.2,-1.8)\rput(-.2,-2){$z$}
                }
\end{pspicture}}.
\end{equation}

Cutting along the partition boundary (dashed lines), we have
\begin{equation}\label{eq:psicuttpsphere}
\begin{split}
\ket{\psi_{\text{cut}}}=\sum_{\substack{\vec{a},\vec{b}, \vec{c},\\ \vec{e}, \vec{f},\vec{g}
 }}& \frac{1}{\mathcal{D}^{l+m+n-5}\left(d_x d_y d_z\right)^{3/4}}
\begin{pspicture}[shift=-1.5](.3,-3.6)(5.9,.3)
\scriptsize
         \rput(2.9,-3.2){$(\AD)$}
        \psline(.8,-.8)(1.9,-1.9)
        \psline[ArrowInside=->](1.8,-1.8)(1.5,-1.5)\rput(1.7,-1.5){$x$}
        \psline[ArrowInside=->](2,-1.8)(2.3,-1.5)\rput(2.1,-1.5){$y$}
        \psline(1.9,-1.9)(3.5,-.3)
        \psline(2.85,-2.85)(5.4,-.3)
        \psline(2.85,-2.85)(1.9,-1.9)
        \psline[ArrowInside=->](2.8,-2.8)(2.5,-2.5)
        \psline[ArrowInside=->](2.9,-2.8)(3.2,-2.5)
        \rput(3,-2.5){$z$}
        \rput(2.35,-1.2){$\otimes$}
        \psellipse[border=1.5pt](2.5,-1.2)(.5,.25)\psline[ArrowInside=->](2.55,-1.45)(2.4,-1.44)\rput(2.65,-1.65){$\omega_0$}
        \rput(1.5,-1.2){$\otimes$}
        \psellipse[border=1.5pt](1.25,-1.2)(.5,.25)\psline[ArrowInside=->](1.25,-1.44)(1.1,-1.43)\rput(1.2,-1.65){$\omega_0$}
        \psline[border=1.5pt](.8,-.8)(1.2,-1.2)
        \psline[ArrowInside=->](.3,-.3)(0,0)\rput(0,.2){$a_1$}
        \psline[ArrowInside=->](.3,-.3)(.6,0)\rput(.6,.2){$a_2$}
        \psline(.4,-.4)(.3,-.3)
        \psline[linestyle=dotted](.7,-.7)(.4,-.4)
        \rput(.6,-.4){$e_2$}
        \psline(.8,-.8)(.7,-.7)
        \psline(.8,-.8)(1.6,0)
        \psline[ArrowInside=->](1.3,-.3)(1.6,0)\rput(1.6,.2){$a_l$}
        \psline[border=1.5pt](3,-.8)(2.5,-1.3)
        \psline[ArrowInside=->](2.5,-.3)(2.2,0)\rput(2.2,.2){$b_1$}
        \psline[ArrowInside=->](2.5,-.3)(2.8,0)\rput(2.8,.2){$b_2$}
        \psline(2.5,-.3)(2.6,-.4)
        \psline[linestyle=dotted](2.6,-.4)(2.9,-.7)
        \rput(2.8,-.4){$f_2$}
        \psline(2.9,-.7)(3,-.8)
        \psline[ArrowInside=->](3.5,-.3)(3.8,0)\rput(3.8,.2){$b_m$}
        \psline[border=1.5pt](4.9,-.8)(4.5,-1.2)
        \psline(2.85,-2.85)(4,-1.7)
        \psline[ArrowInside=->](4.4,-.3)(4.1,0)\rput(4.1,.2){$c_1$}
        \psline[ArrowInside=->](4.4,-.3)(4.7,0)\rput(4.7,.2){$c_2$}
        \psline(4.4,-.3)(4.5,-.4)
        \psline[linestyle=dotted](4.5,-.4)(4.7,-.6)
        \rput(4.7,-.4){$g_2$}
        \psline(4.8,-.7)(4.9,-.8)
        \psline[ArrowInside=->](5.4,-.3)(5.7,0)\rput(5.7,.2){$c_n$}
    \end{pspicture}
   \\ & \quad \quad
   \begin{pspicture}[shift=-1.5](.3,-1.6)(3,.3)
        \scriptsize
        \psline[ArrowInside=->](.3,-.3)(0,0)\rput(0,.2){$z$}
        \psline[ArrowInside=->](.9,-.3)(.6,0)\rput(.6,.2){$c_n$}
        \psline[ArrowInside=->](1.9,-.3)(1.6,0)\rput(1.6,.2){$c_2$}
        \psline[ArrowInside=->](1.9,-.3)(2.2,0)\rput(2.1,.2){$c_1$}
        \rput(1.6,-.35){$g_2$}
        \psline(1.9,-.3)(1.8,-.4)
        \psline[linestyle=dotted](1.8,-.4)(1.5,-.7)
        \psline(1.5,-.7)(1.1,-1.1)(.3,-.3)
        \psline[ArrowInside=->](1.1,-1.1)(1.4,-.8)
        \psline(.9,-.3)(1.4,-.8)
    \end{pspicture}
    \begin{pspicture}[shift=-1.5](.3,-1.6)(3,.3)
        \scriptsize
        \rput(1.1,-1.4){$(\bar{\mathbb{A}})$}
        \psline[ArrowInside=->](.3,-.3)(0,0)\rput(0,.2){$y$}
        \psline[ArrowInside=->](.9,-.3)(.6,0)\rput(.6,.2){$b_m$}
        \psline[ArrowInside=->](1.9,-.3)(1.6,0)\rput(1.6,.2){$b_2$}
        \psline[ArrowInside=->](1.9,-.3)(2.2,0)\rput(2.1,.2){$b_1$}
        \rput(1.6,-.35){$f_2$}
        \psline(1.9,-.3)(1.8,-.4)
        \psline[linestyle=dotted](1.8,-.4)(1.5,-.7)
        \psline(1.5,-.7)(1.1,-1.1)(.3,-.3)
        \psline[ArrowInside=->](1.1,-1.1)(1.4,-.8)
        \psline(.9,-.3)(1.4,-.8)
    \end{pspicture}
    \begin{pspicture}[shift=-1.5](.3,-1.6)(3,.3)
        \scriptsize
        \psline[ArrowInside=->](.3,-.3)(0,0)\rput(0,.2){$x$}
        \psline[ArrowInside=->](.9,-.3)(.6,0)\rput(.6,.2){$a_l$}
        \psline[ArrowInside=->](1.9,-.3)(1.6,0)\rput(1.6,.2){$a_2$}
        \psline[ArrowInside=->](1.9,-.3)(2.2,0)\rput(2.1,.2){$a_1$}
        \rput(1.6,-.35){$e_2$}
        \psline(1.9,-.3)(1.8,-.4)
        \psline[linestyle=dotted](1.8,-.4)(1.5,-.7)
        \psline(1.5,-.7)(1.1,-1.1)(.3,-.3)
        \psline[ArrowInside=->](1.1,-1.1)(1.4,-.8)
        \psline(.9,-.3)(1.4,-.8)
    \end{pspicture},
\end{split}
\end{equation}
where we have chosen to represent the region $\mathbb{A}$ in an analogous basis to that chosen in Eq.~(\ref{eq:psi-cut-annulus}) for $\ket{\psi_\text{cut}}$ of the annulus.
The diagram for region $\mathbb{A}$ embedded in three-dimensional space looks like

\begin{center}
\includegraphics[width=.9\linewidth]{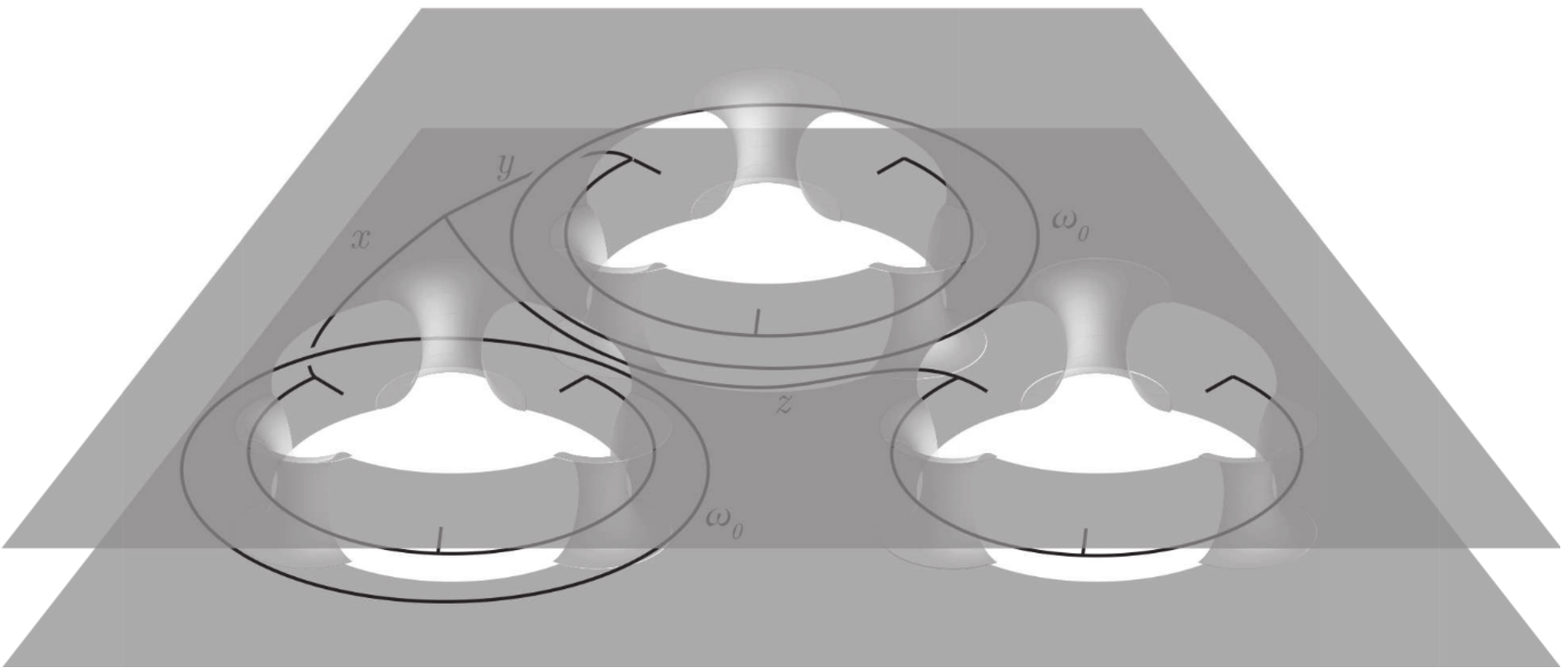}
\end{center}

Finally, we can trace over region $\bar{\mathbb{A}}$ to find the anyonic reduced density matrix for region $\mathbb{A}$ (restoring the vertex labels):
\begin{equation}\label{eq:rho-three-punct-sphere}
\rhoD=\sum_{\substack{\vec{a},\vec{b}, \vec{c},\\ \vec{e}, \vec{f},\vec{g},\\ \vec{\mu}, \vec{\nu},\vec{\lambda}
 }} \frac{\sqrt{d_{\vec{a}}d_{\vec{b}}d_{\vec{c}}}}{\mathcal{D}^{2(l+m+n-5)}d_x d_y d_z}
 \begin{pspicture}[shift=-3.25](.3,-6.5)(4,.3)
        \scriptsize
        \psline(.8,-.8)(1.9,-1.9)
        \psline[ArrowInside=->](1.8,-1.8)(1.5,-1.5)\rput(1.7,-1.5){$x$}
        \psline[ArrowInside=->](2,-1.8)(2.3,-1.5)\rput(2.1,-1.5){$y$}
        \psline(1.9,-1.9)(3.5,-.3)
        \psline(2.85,-2.85)(5.4,-.3)
        \psline(2.85,-2.85)(1.9,-1.9)
        \psline[ArrowInside=->](2.8,-2.8)(2.5,-2.5)
        \psline[ArrowInside=->](2.9,-2.8)(3.2,-2.5)
        \rput(3,-2.5){$z$}
        \rput(2.35,-1.2){$\otimes$}
        \psellipse[border=1.5pt](2.5,-1.2)(.5,.25)\psline[ArrowInside=->](2.55,-1.45)(2.4,-1.44)\rput(2.65,-1.65){$\omega_0$}
        \rput(1.5,-1.2){$\otimes$}
        \psellipse[border=1.5pt](1.25,-1.2)(.5,.25)\psline[ArrowInside=->](1.25,-1.44)(1.1,-1.43)\rput(1.2,-1.65){$\omega_0$}
        \psline[border=1.5pt](.8,-.8)(1.2,-1.2)
        \psline[ArrowInside=->](.3,-.3)(0,0)\rput(0,.2){$a_1$}
        \psline[ArrowInside=->](.3,-.3)(.6,0)\rput(.6,.2){$a_2$}
        \psline(.4,-.4)(.3,-.3)
        \psline[linestyle=dotted](.7,-.7)(.4,-.4)
        \rput(.1,-.45){$\mu_2$}\rput(.6,-.4){$e_2$}
        \psline(.8,-.8)(.7,-.7)
        \psline(.8,-.8)(1.6,0)
        \psline[ArrowInside=->](1.3,-.3)(1.6,0)\rput(1.6,.2){$a_l$}
        \rput(.6,-.85){$\mu_l$}
        \psline[border=1.5pt](3,-.8)(2.5,-1.3)
        \psline[ArrowInside=->](2.5,-.3)(2.2,0)\rput(2.2,.2){$b_1$}
        \psline[ArrowInside=->](2.5,-.3)(2.8,0)\rput(2.8,.2){$b_2$}
        \psline(2.5,-.3)(2.6,-.4)
        \psline[linestyle=dotted](2.6,-.4)(2.9,-.7)
        \rput(2.3,-.4){$\nu_2$}\rput(2.8,-.4){$f_2$}
        \rput(3.3,-.9){$\nu_m$}
        \psline(2.9,-.7)(3,-.8)
        \psline[ArrowInside=->](3.5,-.3)(3.8,0)\rput(3.8,.2){$b_m$}
        \psline[border=1.5pt](4.9,-.8)(4.5,-1.2)
        \psline(2.85,-2.85)(4,-1.7)
        \psline[ArrowInside=->](4.4,-.3)(4.1,0)\rput(4.1,.2){$c_1$}
        \psline[ArrowInside=->](4.4,-.3)(4.7,0)\rput(4.7,.2){$c_2$}
        \psline(4.4,-.3)(4.5,-.4)
        \psline[linestyle=dotted](4.5,-.4)(4.7,-.6)
        \rput(4.2,-.4){$\lambda_2$}\rput(4.7,-.4){$g_2$}
        \rput(5.2,-.9){$\lambda_n$}
        \psline(4.8,-.7)(4.9,-.8)
        \psline[ArrowInside=->](5.4,-.3)(5.7,0)\rput(5.7,.2){$c_n$}
        \rput(1.5,-4.5){$x$}\rput(2.3,-4.5){$y$}
        \rput(3,-3.8){$z$}
        \psline[ArrowInside=->](3.2,-3.7)(2.9,-3.4)
        \psline[ArrowInside=->](2.6,-3.7)(2.9,-3.4)
        \psline(2.9,-3.4)(.7,-5.6)
        \psline[linestyle=dotted](.4,-5.9)(.7,-5.6)
        \rput(.1,-5.9){$\mu_2$}\rput(.6,-5.9){$e_2$}
        \rput(.6,-5.45){$\mu_l$}
        \psline(.3,-6)(.4,-5.9)
        \psline[ArrowInside=->](0,-6.3)(.3,-6)\rput(0,-6.5){$a_1$}
        \psline(2.9,-3.4)(5.5,-6)
        \psline[ArrowInside=->](5.8,-6.3)(5.5,-6)\rput(5.7,-6.5){$c_n$}
        \psline[ArrowInside=->](.6,-6.3)(.3,-6)\rput(.6,-6.5){$a_2$}
        \psline[ArrowInside=->](1.6,-6.3)(1.3,-6)\rput(1.6,-6.5){$a_l$}
        \psline(1.3,-6)(.8,-5.5)
        \psline[ArrowInside=->](2.2,-6.3)(2.5,-6)\rput(2.2,-6.5){$b_1$}
        \psline[ArrowInside=->](2.8,-6.3)(2.5,-6)\rput(2.8,-6.5){$b_2$}
        \psline(2.5,-6)(2.6,-5.9)
        \psline[linestyle=dotted](2.6,-5.9)(2.8,-5.7)
        \rput(2.3,-5.9){$\nu_2$}\rput(2.8,-5.9){$f_2$}
        \rput(3.3,-5.5){$\nu_m$}
        \psline(2.9,-5.6)(3,-5.5)
        \psline[ArrowInside=->](3.8,-6.3)(3.5,-6)\rput(3.8,-6.5){$b_m$}
        \psline(3.5,-6)(3,-5.5)
        \psline(3,-5.5)(1.9,-4.4)
        \psline[ArrowInside=->](4.3,-6.3)(4.6,-6)\rput(4.3,-6.5){$c_1$}
        \psline[ArrowInside=->](4.9,-6.3)(4.6,-6)\rput(4.9,-6.5){$c_2$}
        \psline(4.6,-6)(4.7,-5.9)
        \psline[linestyle=dotted](4.6,-6)(4.9,-5.7)
        \rput(4.35,-5.9){$\lambda_2$}\rput(4.95,-5.95){$g_2$}
        \rput(5.3,-5.5){$\lambda_n$}
        \psline(4.95,-5.65)(5.05,-5.55)
        \psellipse[border=1.5pt](2.5,-4.9)(.5,.25)\psline[ArrowInside=->](2.55,-5.15)(2.4,-5.14)\rput(2.5,-5.35){$\omega_0$}
        \psellipse[border=1.5pt](1.25,-4.9)(.5,.25)\psline[ArrowInside=->](1.25,-5.14)(1.1,-5.13)\rput(1.35,-5.35){$\omega_0$}
        \rput(1.15,-4.9){$\otimes$}
        \rput(2.65,-4.9){$\otimes$}
        \psline[border=1.5pt](1.9,-4.4)(1.3,-5.)
        \psline[border=1.5pt](1.9,-4.4)(2.5,-5)
        \psline[ArrowInside=->](1.6,-4.7)(1.9,-4.4)
        \psline[ArrowInside=->](2.2,-4.7)(1.9,-4.4)
        \psline(1.8,-4.5)(2.1,-4.2)
    \end{pspicture}.
\end{equation}

Applying similar steps to those used in Eqs.~(\ref{eq:disk-ARE-start})-(\ref{eq:disk-ARE-end}) for the disk and in Eqs.~(\ref{eq:rhoannulus-squared})-(\ref{eq:annulus-ARE-end}) for the annulus, we find
that the anyonic R\'enyi entropy is
\begin{equation}
\aS^{(\alpha)}\left(\rhoD\right) = \frac{1}{1-\alpha} \log \left[ \sum_{\substack{\vec{a}, \vec{b},\vec{c}, \vec{e},\vec{f},\vec{g},\\ \vec{\mu},\vec{\nu},\vec{\lambda}}} \left( \frac{d_{\vec{a}}d_{\vec{b}}d_{\vec{c}}}{\mathcal{D}^{2(l+m+n-3)}d_x d_y d_z}\right)^\alpha\right].
\end{equation}
Taking the limit $\alpha\to 1$ yields the AEE for region $\mathbb{A}$
\begin{equation}
\aS(\rhoD ) = \lim_{\alpha\to 1} \aS^{(\alpha)}(\rhoD) = - (l+m+n) \sum_{a} \frac{d_{a}^2}{\mathcal{D}^2} \log \left(\frac{d_{a}}{\mathcal{D}^2}\right) - 6 \log \mathcal{D}  + \log \left(d_x d_y d_x\right).
\end{equation}
After taking into account the doubling of the surface, the topological contribution to the entanglement entropy for the original (un-doubled) region $A$, i.e. the 3-punctured sphere, is given by
\begin{eqnarray}
\aS_A &=& -\frac{l+m+n}{2} \sum_{a} \frac{d_{a}^2}{\mathcal{D}^2} \log \left( \frac{d_{a}}{\mathcal{D}^2} \right)  - 3 \log \mathcal{D} +\log d_{x} +\log d_{y} +\log d_{z}
\notag \\
&=& -\frac{l+m+n}{2} \aS(\arho_{\partial \mathbb{A}_j}) - 3 S_{\text{topo}} + \aS_x + \aS_y + \aS_z
.
\end{eqnarray}
We see the entanglement entropy of region $A$ is equal to the sum of the entanglement entropies of three disks with matching boundary charge values [see Eq.~\eqref{eq:AEEdisk}], as it should. Crucially, this implies that each separate boundary component of the region $A$ contributes a universal $\mathcal{O}(1)$ topological term $\log \left( d_c/\mathcal{D}\right)$ to the entanglement entropy, where $c$ is the total topological charge on the corresponding boundary component.

\subsection{General Result}
\label{sec:general}

Given the results of the prior examples, we can deduce the result for the general case of an arbitrary partitioning of a compact, orientable surface with genus $g$ and arbitrary number of punctures or quasiparticles that carry topological charge. For a partitioning of the surface into regions $A$ and $\bar{A}$, let us assume the joint boundary between $A$ and $\bar{A}$ (i.e. $\partial A \cap \partial \bar{A}$) has $N$ connected components, $\partial A^{(1)},\ldots, \partial A^{(N)}$. We denote the topological state of the system by $\tilde{\rho}$, which can be described using the anyonic formalism of fusion trees of topological charge lines of the punctures/quasiparticles and charge lines winding around non-contractible cycles. We denote the topological state of the (un-doubled) region $A$, including the boundaries, by $\tilde{\rho}_{A}$. We denote by $p_{c}^{(k)}$ the probability of the state $\tilde{\rho}_{A}$ being in a configuration wherein $\partial A^{(k)}$ carries topological charge $c$.

The topological contribution to the entanglement entropy associated with $\partial A^{(k)}$ is
\begin{eqnarray}
\label{eq:general}
\aS_{\partial A^{(k)}} &=& -\frac{n_k}{2} \sum_a \frac{d_a^2}{\mathcal{D}^2} \log \left( \frac{d_a}{\mathcal{D}^2} \right)- \log \mathcal{D} + \sum_{c} p_{c}^{(k)} \log d_{c}
\\
&=&
\frac{n_k}{2} \aS(\arho_{\partial \mathbb{A}_j}) +S_{\text{topo}} + \aS(\arho_{\partial A^{(k)}})
.
\end{eqnarray}
Here, $n_k \sim L_k / \ell$ is a non-universal quantity that is essentially the discretized length of the $k$th component of the partition boundary using some regularization.

The topological contribution to the entanglement entropy between regions $A$ and $\bar{A}$ is given by
\begin{equation}
\label{eq:more-general}
\aS_{A} = \sum_{k=1}^{N} \aS_{\partial A^{(k)}} + \aS(\tilde{\rho}_{A})
.
\end{equation}
That is, it is the sum of the contributions from each of the partition boundary components and the anyonic entropy of the reduced density matrix of region $A$ (including the boundary charges).  Eq.~(\ref{eq:more-general}) is consistent with previous studies on the entanglement entropy of orientable, higher genus surfaces supporting an SU(2)$_{k}$ Chern-Simons theory~\cite{Dong08, Wen16a}.

Generally, the superposition of charges on different partition boundary components cannot be described by independent probability distributions.  As an example, the three-punctured sphere considered in Section~\ref{sec:three-punctured-sphere} could be generalized to the case where the punctures have charges $x$, $y$, and $z$ with probability $p_{xyz}$.  The constant terms in the AEE would then depend on the probability distribution $\{p_{xyz}\}$ and it would not be possible to completely separate the terms associated with the disk containing charge $x$ from the terms associated with the disk containing charge $y$.  Therefore, we see that the entanglement entropy is highly state-dependent, even when we neglect the boundary-law term.  Nonetheless, the $\mathcal{O}(1)$ partition boundary terms show up in a universal way by contributing a term $\sum\limits_{c} p_{c}^{(k)} \log \left(\frac{d_{c} }{\mathcal{D}} \right)$ for the corresponding $k$th component of the partition boundary.

Thus, we have determined that the entanglement entropy for a topological phase on an arbitrary compact, orientable surface (possibly including genus, punctures, and quasiparticles) partitioned into regions $A$ and $\bar{A}$ will take the form
\begin{equation}
\label{eq:general_result2}
S_{A} = \sum_{k=1}^{N} \left( \alpha L_{k} - \log \mathcal{D} + \sum_{c} p_{c}^{(k)} \log d_{c} \right) + \aS(\tilde{\rho}_{A}) + \mathcal{O}(L_{k}^{-1})
,
\end{equation}
where $L_k$ is the length of the $k$th connected component of the partition boundary.

\section{Discussion}
\label{sec:discussion}

In this paper, we have investigated the rich entanglement structure of two-dimensional topological phases with anyons by applying the standard notions of entropy to the diagrammatic representation of the TQFT.  In Section~\ref{sec:anyonicEntropy}, we probed the correlations between subsystems of anyons using the anyonic entanglement entropy (AEE) and the entropy of anyonic charge entanglement.  We found that the fusion tensor category structure of the Hilbert space gives rise to entanglement associated with the topological charge line connecting two subsystems, a type of correlation not present in traditional quantum systems.
We further found, in Sections~\ref{sec:ATEEI} and \ref{sec:ATEE}, that the TEE is naturally explained from a decrease in the entropy (increase in order) evoked by a nonlocal (topological) constraint imposed on any region of the system by its topological order.  The total fusion channel of topological charges encoding local correlations across the partition boundary is fixed when the system is cut, resulting in a very specific reduction of the AEE. We now place our results in a broader context.  First, we discuss the relation of our results to the string-net formalism of Ref.~\cite{Levin05a}. Then, we explain how our analysis also applies to topological defects and generalizes straightforwardly to fermionic topological phases.  Finally, we discuss possible extensions of our methods to non-orientable surfaces and $(3+1)$-dimensional topologically ordered systems.

\subsection{Relation to String-Net Models}\label{sec:stringnets}

String-nets are exactly solvable models of topological phases~\cite{Levin05a} in which ``strings," labeled by the elements of a unitary fusion tensor category (UFTC) $\mathcal{F}$, lie on the links of a lattice.  A set of fusion rules constrains which strings may meet at a vertex.  In general, the string-net model built from $\mathcal{F}$ realizes a topological phase described by the Drinfeld center D$(\mathcal{F})$ of $\mathcal{F}$.  In the special case where $\mathcal{F}$ describes the fusion structure of a MTC $\mathcal{C}$, the Drinfeld center takes the form D$(\mathcal{F})=\mathcal{C}\times\overline{\mathcal{C}}$.

Ref.~\cite{Levin06a} found that the entanglement entropy of the (fixed point) string-net ground state of the plane partitioned into a disk region $A$ whose boundary is crossed by $n$ links of the lattice is
\begin{equation}
\label{eq:LevinEE}
S_A=-n\sum_{i\in \mathcal{F}} \frac{d_i^2}{D}\log \left(\frac{d_i}{D}\right)-\log D,
\end{equation}
where $i$ and $d_i$ are the labels and quantum dimensions, respectively, of the lattice strings.  The quantity
\begin{equation}
D = \mathcal{D}_{\mathcal{F}}^{2} = \sum_{i\in \mathcal{F}} d_i^2  = \sqrt{\sum_{a\in \text{D}(\mathcal{F})} d_{a}^2}=\mathcal{D}_{\text{D}(\mathcal{F})}
\end{equation}
is equal to the total quantum dimension $\mathcal{D}_{\text{D}(\mathcal{F})}$ of the emergent TQFT $\text{D}(\mathcal{F})$.

In this paper, we found the entanglement entropy for a topological phase described by a UMTC $\mathcal{C}$ by pairing the system with its time-reversal conjugate described by $\bar{\mathcal{C}}$, and inserting wormholes along the partition boundary to glue the two surfaces together. This process can be related to a string-net model based on the UFTC $\mathcal{F}$ describing the fusion structure of the UMTC $\mathcal{C}$.  More specifically, the graph of anyon charge lines representing the state of the system in the basis where all anyon charge lines are between the two (doubled) layers of the surface (hosting $\mathcal{C}$ and $\bar{\mathcal{C}}$) is instead interpreted as the underlying lattice of the string-net model hosting $\mathcal{F}$.  The lattice can be thought of as defining a surface (the original surface in the prior approach) and the wormholes are now thought of as passing through the (empty space at the) center of the plaquettes of the lattice.  The plaquette operator $B_p$ imposes trivial flux through the plaquettes, i.e. the $\omega_0$-loops circling the wormholes.
Consequently, our result in Eq.~\eqref{eq:S_A_doubled_disk}, the AEE obtained from doubling a disk region of the original system, is identical to Eq.~\eqref{eq:LevinEE}, the string-net result, when both $a_j$ and $i$ belong to $\mathcal{C}$, so that $D=\mathcal{D}_{\mathcal{C}}^2$.

Furthermore, when the UMTC $\mathcal{C}$ describing the topological phase can itself be written as $\mathcal{C}=\mathcal{E}\times\bar{\mathcal{E}}$ for some UMTC $\mathcal{E}$, then this phase can realized by the string-net model built out of the UFTC $\mathcal{E}$.~\footnote{In this case, the string-net lattice model provides a microscopic regularization of the theory.} In this case, Eq.~(\ref{eq:Sactual}), the topological contribution to the entanglement entropy for a topological phase described by $\mathcal{C}$, equals Eq.~(\ref{eq:LevinEE}) for the corresponding string-net model built from $\mathcal{E}$, where $a_j \in \mathcal{C}$ and $i \in \mathcal{E}$. While the TEE for a general UMTC $\mathcal{C}$ always agrees with the string-net computation, since $D=\mathcal{D}_{\text{D}(\mathcal{F})}$, it is interesting that the boundary length ($n$) dependent terms matches in this case where $\mathcal{C}=\mathcal{E}\times\overline{\mathcal{E}}$, that is
\begin{equation}
\frac{1}{2} \sum_{a \in \mathcal{C}} \frac{d_{a}^2}{\mathcal{D}_{\mathcal{C}}^2} \log \left( \frac{d_{a}}{\mathcal{D}_{\mathcal{C}}^2}\right)=\frac{1}{2} \sum_{\substack{a_L \in \mathcal{E} \\ a_R \in \overline{\mathcal{E}} }} \frac{d_{a_L}^2 d_{a_R}^2}{\mathcal{D}_{\mathcal{E}}^2 \mathcal{D}_{\overline{\mathcal{E}}}^2} \log \left( \frac{d_{a_L} d_{a_R}}{\mathcal{D}_{\mathcal{E}}^2 \mathcal{D}_{\overline{\mathcal{E}}}^2}\right)= \sum_{i\in \mathcal{E}} \frac{d_i^2}{D}\log \frac{ d_i }{D}.
\end{equation}
In the case where a UFTC $\mathcal{F}$ does \emph{not} describe the fusion structure of any UMTC, so that $\text{D}(\mathcal{F}) \neq \mathcal{E}\times\overline{\mathcal{E}}$ for any $\mathcal{E}$, it is not necessarily the case that there is equality between $\frac{1}{2} \sum\limits_{a\in \text{D}(\mathcal{F})} \frac{d_{a}^2}{ \mathcal{D}_{\text{D}(\mathcal{F})}^2 } \log \left( \frac{d_{a}} { \mathcal{D}_{\text{D}(\mathcal{F})}^2 } \right)$ and
$\sum\limits_{i\in \mathcal{F}} \frac{d_i^2}{D}\log \frac{ d_i }{D}$, so the linear terms (proportional to $n$) of Eq.~(\ref{eq:Sactual}) and Eq.~(\ref{eq:LevinEE}) do not generally agree.

Finally, we note that the string-net formalism gives an intuitive understanding for the form of $S_{\text{topo}}$.  Consider a string-net model built out of an Abelian UFTC $\mathcal{F}$. Then $d_i=1$ for $i\in \mathcal{F}$, and $D$ is simply the number of underlying string types ${D=|\mathcal{F}|=N}$.  From Eq.~\eqref{eq:LevinEE}, we see that the entanglement entropy is given by ${S_A = (n-1) \log N}$.  We can understand the form of $S_A$ in this case as follows.  The state space of each link in the lattice hosting the string-net has dimension $N$.  Without conservation of topological charge, the entanglement entropy would be the sum of each link lying across the partition boundary, i.e., $n\log N$.  The constraint on the total charge of the lattice strings on the boundary essentially fixes the state of the last link, reducing the entanglement entropy by ${\log N=\log D=-S_{\text{topo}}}$.  For a string-net built out of a UFTC $\mathcal{F}$ describing a non-Abelian theory, the probability of a link carrying a given string is weighted by the quantum dimension of that string type, which also enters the entanglement entropy when a boundary component carries a corresponding topological charge.

\subsection{Topological defects}

The analysis in this paper also applies to $(2+1)$-dimensional topologically ordered systems that contain topological defects whose universal properties can be described by ``$G$-crossed UMTCs.'' This includes on-site symmetry defects~\cite{Barkeshli14} and translational symmetry defects~\cite{Cheng15}. In such cases, the topological defects in the system have fusion and associativity properties that are precisely the same as that of quasiparticles, and they have a generalization of braiding that incorporates the symmetry action. In particular, this means the defects have quantum dimensions in the same sense as do quasiparticles. There is also a generalization of modular transformations in the presence of defects and defect branch lines, which allows one to apply the methods of our paper in a straightforward manner. Specifically, a wormhole with trivial topological flux threading it can be re-expressed in terms of the inside basis with topological charge lines circling the throat of the wormhole. In this case, if there is a ${\bf g}$-defect branch line around the location where the wormhole is inserted, the modular $\mathcal{S}$-transformation maps from the ${\bf 0}$-sector for the outside basis, where topological charge lines threading the throat of the wormhole correspond to quasiparticles, to the ${\bf g}$-sector for the inside basis, where topological charge lines circling the throat of the wormhole correspond to ${\bf g}$-defects. Since the charge line through the wormhole is trivial, the amplitudes of the defect charge lines of the inside basis are proportional to their quantum dimensions, i.e. $\mathcal{S}_{0 a_{\bf g}}^{({\bf 0},{\bf g})} = \frac{d_{a_{\bf g}}}{\mathcal{D}_{\bf 0}}$, where the total quantum dimension $\mathcal{D}_{\bf 0}$ is that of the quasiparticle sector of the $G$-crossed theory, i.e. the total quantum dimension of the UMTC that describes the topological order without defects (see Ref.~\cite{Barkeshli14} for more details). It follows that the results in the presence of topological defects are exactly the same as in Eqs.~(\ref{eq:general})-(\ref{eq:general_result2}), but the partition boundary components are now allowed to carry topological charges corresponding to quasiparticles or defects from the $G$-crossed MTC describing the system. This has been confirmed in the case of ``twist defects'' in the toric code model~\cite{Brown11}.

\subsection{Fermionic Topological Phases}

The analysis in this paper utilizes $(2+1)$-dimensional TQFTs, which describe bosonic topological phases of matter in two spatial dimensions.  However, the results are straightforwardly generalized to fermionic topological phases by utilizing $(2+1)$-dimensional fermionic TQFTs, also known as topological spin theories~\cite{Dijkgraaf90}.  A fermionic topological phase includes a physical fermion $\psi$, which has trivial braiding statistics with all quasiparticles in the theory, i.e. the physical fermion is transparent. The quasiparticles of the theory (including the physical fermion) are described by a super-modular tensor category (SMTC) $\mathcal{C}_{\bf 0}$, which is a unitary braided tensor category in which the fermion $\psi$ is transparent and the braiding is only two-fold degenerate, i.e. the degeneracy associated with the fermion. While charges in a bosonic topological phase are described by superselection sectors $a$ of the corresponding UMTC, for the fermionic case we must think in terms of supersectors, $\hat{a}=\{a,a\times \psi\}$, with associated quantum dimension $d_{\hat{a}}=d_a$. Forming supersectors, we find that the topological $S$-matrix takes the form $S = S_{\text{fermion}} \otimes \hat{S}$, where $S_{\text{fermion}}$ is the degenerate $2\times 2$ $S$-matrix of a trivial fermion theory (i.e. the only topological charges are the vacuum and the fermion) and $\hat{S}$ is the $S$-matrix of supersectors. The two-fold braiding degeneracy is equivalent to the condition that $\hat{S}$ is unitary. For modular transformations of the fermionic topological phase, we must specify the spin structure for every nontrivial cycle of the surface (i.e., we must fix periodic or antiperiodic boundary conditions of the $\psi$ Wilson loop for every nontrivial cycle), as this plays a crucial role in the structure of the fermionic modular transformations (see Ref.~\cite{Bonderson_FMT} for further details).

\begin{table}
\[
\begin{array}{c|l}
\hat{\mathcal{D}}^{2} & \text{SMTCs} \quad \mathcal{C}_0 \\
\hline \hline
1 & \mathbb{Z}_2^{(1)} \quad \text{(Trivial)} \\
\hline
2 & \mathbb{Z}_2^{(1)} \times \mathbb{Z}_2^{(1/2)} \\
\hline
3 & \mathbb{Z}_2^{(1)} \times \mathbb{Z}_3^{(p)}, ~p=1,2  \\
\hline
\phi+2 &  \mathbb{Z}_2^{(1)}\times \text{Fib}^{\pm 1}\\
\hline
4 &  \mathbb{Z}_2^{(1)}\times \mathcal{K}_{\nu}, ~\nu=0,1,\ldots,7 \\
\hline
5 & \mathbb{Z}_2^{( 1)}\times \mathbb{Z}_5^{(p)}, ~p=1,2 \\
\hline
6 &  \mathbb{Z}_2^{(1)}\times \mathbb{Z}_6^{(p)}, ~p=\frac{1}{2},\frac{5}{2} \\
\hline
4+2\sqrt{2} &  \text{SO}(3)_6\\
\hline
7  &  \mathbb{Z}_2^{( 1)}\times \mathbb{Z}_7^{(p)}, ~p=1,3 \\
\hline
\end{array}
\]
\caption{The quasiparticle sector of a fermionic TQFT in ($2+1$)D is described by a SMTC, which can be classified according to its value of the super total quantum dimension $\hat{\mathcal{D}}$. This table lists all distinct SMTCs with $\hat{\mathcal{D}}^{2} \leq 7$, as determined from Refs.~\cite{Bonderson07b,Lan16, BondersonWIP}. ($\phi = \frac{1+\sqrt{5}}{2} \approx 1.6$ is the Golden ratio.) For most values of $\hat{\mathcal{D}}$, there are very few possible SMTCs. Moreover, the SMTCs with a given value of $\hat{\mathcal{D}}$ are usually very closely related. Additional details may be found in \ref{sec:BTCs}.}
\label{Table:SMTC}
\end{table}

Given a fermionic TQFT, one can carry out the same steps and analogous calculations for fermionic topological phases as in the method presented in this paper for bosonic topological phases. The main differences in the analysis will be that each wormhole will carry a trivial supersector flux $\hat{0}=\{0,\psi\}$, the choice of spin structures on the surfaces must be specified, and fermionic modular transformations, which act on spin structures, are used. It turns out, however, that the choice of spin structure does not affect the TEE result. We find that the TEE associated with each distinct partition boundary component for a fermionic topological phase is
\begin{equation}
\hat{S}_{\text{topo}} = -\log\hat{\mathcal{D}}
,
\end{equation}
where we have defined the super total quantum dimension by
\begin{equation}
\hat{\mathcal{D}}=\sqrt{\sum_{\hat{a} \in \hat{\mathcal{C}}_{\bf 0} } d_{\hat{a}}^2} = \sqrt{ \frac{1}{2} \sum_{a \in \mathcal{C}_{\bf 0} } d_{a}^2} .
\end{equation}
This result has been confirmed for various fermionic fractional quantum Hall states~\cite{Haque07,Zozulya07,Rodriguez08,Zozulya08,Zaletel12,Zaletel13,Grushin15}.

Similar to the case of UMTCs, there are only a finite number of possible SMTCs for a particular value of $\hat{\mathcal{D}}$. In Table~\ref{Table:SMTC}, we list all SMTCs for $\hat{\mathcal{D}}^2 \leq 7$.

\subsection{Non-orientable surfaces}

An interesting future direction would be to generalize our analysis to study the entanglement entropy on non-orientable surfaces. We expect the construction of the reduced density matrix outlined in the beginning of  Section~\ref{sec:ATEE} will differ for non-orientable surfaces in step 3.  That is, the $\mathcal{S}$-transformation on a non-orientable surface will no longer necessarily result in an $\omega_0$-loop. Rather, the superposition of charges circling each wormhole will be a subset of all charges in the theory (see Ref.~\cite{Barkeshli16} for a discussion of state sums on non-orientable surfaces).  Nonetheless, we anticipate that the TEE will still originate from the conservation of topological charge.

\subsection{Three dimensional topological phases}

Finally, one could also extend our method of calculating the entanglement entropy to $(3+1)$-dimensional topological phases. Previous investigations of the TEE in $(3+1)$-dimensions have utilized a linear combination of spatial regions to isolate the boundary-independent contribution to the entanglement entropy, similarly to the $(2+1)$-dimensional Kitaev-Preskill method~\cite{Castelnovo08,Grover11,Kim15,Bullivant16}.  Dividing the partition boundary into smaller regions, as in our method for $(2+1)$-dimensions, could elucidate how the conservation of more general topological quantum numbers results in a reduction of the entanglement entropy in $(3+1)$-dimensions.  This analysis could be carried out for exactly solvable models~\cite{Castelnovo08,Zozulya08,vonKeyserlingk12,Walker12}, or more generally using TQFT methods.

\section*{Acknowledgments}
We are grateful to David Aasen, Matthew Hastings, Roger Mong, Zhenghang Wang, and Brayden Ware for helpful discussions.  P.B. acknowledges the Aspen Center for Physics, where part of this work was performed and which is supported by National Science Foundation grant PHY-1066293. C.K. acknowledges support from the NSF GRFP under Grant No. DGE $114085$.


\appendix

\section{Anyon Models on a Sphere}\label{sec:anyonmodelssphere}

In this appendix, we review the description of anyon models on a sphere~\cite{Bonderson07b,Bonderson07c}. Since punctures may be represented by anyons existing on their boundaries, this section also applies to spheres with punctures, e.g., a disk.

\subsection{Fusion Algebra}

Anyon models, or modular tensor categories (MTCs), consist of a finite set of objects, or \emph{anyons}, which obey a commutative, associative fusion algebra:
\begin{equation}
    a\times b=\sum_c N_{ab}^c c,
\end{equation}
where $N_{ab}^c$ is a non-negative integer that specifies the number of different ways anyons $a$ and $b$ can fuse to $c$. An anyon $a$ is \emph{non-Abelian} if $\sum_c N_{ab}^c>1$ for some $b$, and \emph{Abelian} otherwise.

The fusion algebra must obey certain conditions. There must exist a unique \emph{vacuum} anyon $0$ such that $N_{a0}^c=\delta_{ac}$, and each anyon $a$ must have a \emph{dual} anyon $\bar{a}$ such that $N_{ab}^0=\delta_{b\bar{a}}$. We also have the important relation
\begin{equation}\label{eq:Nabc}
    d_a d_b = \sum_c N_{ab}^c d_c,
\end{equation}
where $d_a$, the \emph{quantum dimension} of $a$, is the largest eigenvalue of the fusion matrix $N_a$, (whose elements are $[N_a]_{bc}=N_{ab}^c$.) For non-Abelian anyons, $d_a>1$, while for Abelian anyons, $d_a=1$.

The total quantum dimension of an anyon model $\mathcal{C}$ is
\begin{equation}
\mathcal{D}= \sqrt{\sum_{a\in \mathcal{C}}d_a^2}.
\end{equation}

\subsection{Anyonic Hilbert Space}

The \emph{anyonic Hilbert space} of topological system consists of all of its possible topologically distinct states. It can be constructed and expressed diagramatically as follows.

\subsubsection{Basis}

The building blocks of the anyonic Hilbert space for the sphere is the space $V_c^{ab}$ of two anyons $a$ and $b$ with definite total charge $c$, which is spanned by the vectors
\begin{equation}
    \ket{a,b;c,\mu}=\left(\frac{d_c}{d_a d_b}\right)^{1/4}
    \begin{pspicture}[shift=-0.4](0,0)(1.2,1)
        \scriptsize
        \psline[ArrowInside=->](0.5,0.5)(0,1)\rput(0.1,0.75){$a$}
        \psline[ArrowInside=->](0.5,0.5)(1,1)\rput(0.95,0.75){$b$}
        \psline[ArrowInside=->](0.5,0)(0.5,0.5)\rput(0.65,0.2){$c$}
        \rput(0.65,0.45){$\mu$}
    \end{pspicture},
\end{equation}
where $\mu=1,\dots,N_{ab}^c$. The dual space $V_{ab}^c$ is spanned by the covectors
\begin{equation}
    \bra{a,b;c,\mu}=\left(\frac{d_c}{d_a d_b}\right)^{1/4}
    \begin{pspicture}[shift=-0.4](0,0)(1.2,1)
        \scriptsize
        \psline[ArrowInside=->](0,0)(0.5,0.5)\rput(0.1,0.25){$a$}
        \psline[ArrowInside=->](1,0)(0.5,0.5)\rput(0.95,0.25){$b$}
        \psline[ArrowInside=->](0.5,0.5)(0.5,1)\rput(0.65,0.8){$c$}
        \rput(0.65,0.55){$\mu$}
    \end{pspicture}.
\end{equation}

Larger spaces are constructed by taking tensor products. For example, the space $V^{abc}_d$ of three anyons $a$, $b$, and $c$ with definite total charge $d$ can be constructed as
\begin{equation}
    V^{abc}_d\cong\bigoplus_e V^{ab}_e\otimes V^{ec}_d,
\end{equation}
which is spanned by
\begin{equation}
    \ket{a,b;e,\mu}\ket{e,c;d,\nu}=\left(\frac{d_d}{d_a d_b d_c}\right)^{1/4}
    \begin{pspicture}[shift=-0.7](0,0)(2,1.5)
        \scriptsize
        \psline[ArrowInside=->](0.5,1)(0,1.5)\rput(0.1,1.25){$a$}
        \psline[ArrowInside=->](0.5,1)(1,1.5)\rput(0.9,1.25){$b$}
        \psline[ArrowInside=->](1,0.5)(2,1.5)\rput(1.65,1){$c$}
        \psline[ArrowInside=->](1,0)(1,0.5)\rput(1.15,0.25){$d$}
        \psline[ArrowInside=->](1,0.5)(0.5,1)\rput(0.6,0.75){$e$}
        \rput(0.35,1){$\mu$}
        \rput(1.15,0.5){$\nu$}
    \end{pspicture},
\end{equation}
where $\mu=1,\dots,N_{ab}^e$, $\nu=1,\dots,N_{ec}^d$, and $e$ is any anyon such that $N_{ab}^e\ge1$ and $N_{ec}^d\ge1$. The space $V^{abc}_d$ can also be constructed as
\begin{equation}
    V^{abc}_d\cong\bigoplus_e V^{bc}_e\otimes V^{ae}_d,
\end{equation}
which is spanned by
\begin{equation}
    \ket{b,c;e,\mu}\ket{a,e;d,\nu}=\left(\frac{d_d}{d_a d_b d_c}\right)^{1/4}
    \begin{pspicture}[shift=-0.6](0.2,0)(2,1.5)
        \scriptsize
        \psline[ArrowInside=->](1,0.5)(0,1.5)\rput(0.35,1){$a$}
        \psline[ArrowInside=->](1.5,1)(1,1.5)\rput(1.1,1.25){$b$}
        \psline[ArrowInside=->](1.5,1)(2,1.5)\rput(1.9,1.25){$c$}
        \psline[ArrowInside=->](1,0)(1,0.5)\rput(1.15,0.25){$d$}
        \psline[ArrowInside=->](1,0.5)(1.5,1)\rput(1.4,0.75){$e$}
        \rput(1.65,1){$\mu$}
        \rput(1.15,0.5){$\nu$}
    \end{pspicture}.
\end{equation}
where $\mu=1,\dots,N_{bc}^e$, $\nu=1,\dots,N_{ae}^d$, and $e$ is any anyon such that $N_{bc}^e\ge1$ and $N_{ae}^d\ge1$.
These constructions are isomorphic, and their basis vectors are related by an \emph{$F$-move}:
\begin{equation}
    \begin{pspicture}[shift=-0.6](0,0)(2,1.5)
        \scriptsize
        \psline[ArrowInside=->](0.5,1)(0,1.5)\rput(0.1,1.25){$a$}
        \psline[ArrowInside=->](0.5,1)(1,1.5)\rput(0.9,1.25){$b$}
        \psline[ArrowInside=->](1,0.5)(2,1.5)\rput(1.65,1){$c$}
        \psline[ArrowInside=->](1,0)(1,0.5)\rput(1.15,0.25){$d$}
        \psline[ArrowInside=->](1,0.5)(0.5,1)\rput(0.6,0.75){$e$}
        \rput(0.35,1){$\mu$}
        \rput(1.15,0.5){$\nu$}
    \end{pspicture}
    =\sum_f \left[F^{abc}_d\right]_{(e,\mu,\nu)(f,\alpha,\beta)}
    \begin{pspicture}[shift=-0.6](0.2,0)(2,1.5)
        \scriptsize
        \psline[ArrowInside=->](1,0.5)(0,1.5)\rput(0.35,1){$a$}
        \psline[ArrowInside=->](1.5,1)(1,1.5)\rput(1.1,1.25){$b$}
        \psline[ArrowInside=->](1.5,1)(2,1.5)\rput(1.9,1.25){$c$}
        \psline[ArrowInside=->](1,0)(1,0.5)\rput(1.15,0.25){$d$}
        \psline[ArrowInside=->](1,0.5)(1.5,1)\rput(1.4,0.75){$f$}
        \rput(1.65,1){$\alpha$}
        \rput(1.15,0.5){$\beta$}
    \end{pspicture},
\end{equation}
where the \emph{$F$-symbols} $F^{abc}_d$ are unitary matrices that must satisfy the Pentagon consistency equations.

In general, the space $V^{a_1\dots a_n}_c$ of anyons $a_1$, \dots, $a_n$ with definite combined charge $c$ can be constructed as
\begin{equation}
    V^{a_1\dots a_n}_c\cong\bigoplus_{\vec{b}} V^{a_1 a_2}_{b_2} \otimes V^{b_2 a_3}_{b_3}\otimes \cdots \otimes V^{b_{n-1} a_n}_c,
\end{equation}
which is spanned by
\begin{align}
    \ket{\vec{a},\vec{b},\vec{\alpha};c}&=\ket{a_1,a_2;b_2,\alpha_2}\cdots\ket{b_{{n-1}},a_n;c,\alpha_n} \notag \\
    &=\left(\frac{d_{c}}{d_{a_1}\cdots d_{a_n}}\right)^{1/4}
    \begin{pspicture}[shift=-0.6](-0.2,0)(2.2,1.5)
        \scriptsize
        \psline[ArrowInside=->](0.5,1)(0,1.5)\rput(0.05,1.25){$a_1$}
        \psline[ArrowInside=->](0.5,1)(1,1.5)\rput(1,1.25){$a_2$}
        \psline[ArrowInside=->](1,0.5)(2,1.5)\rput(1.8,1){$a_n$}
        \psline[ArrowInside=->](1,0)(1,0.5)\rput(1.2,0.2){$c$}
        \psline[ArrowInside=->](0.7,0.8)(0.5,1)\rput(0.5,0.75){$b_2$}
        \psline[ArrowInside=->](1,0.5)(0.8,0.7)\rput(0.6,0.5){$b_{n-1}$}
        \rput(0.25,1){$\alpha_2$}
        \rput(1.3,0.5){$\alpha_n$}
        \rput(0.75,0.75){.}
        \rput(1.5,1.5){\dots}
    \end{pspicture}.
\end{align}
where $\vec{b}$ and $\vec{\alpha}$ take values that are allowed by fusion.

We can also write the $F$-move with two lower and two upper legs.  This basis change is given by
\begin{equation}
 \begin{pspicture}[shift=-1.05](1.1,-3.3)(2.5,-1.2)
        \scriptsize
        \psline[border=1.5pt](1.7,-2.1)(2.3,-2.4)
        \psline[ArrowInside=->](2.3,-2.4)(1.3,-1.9)
        \psline[ArrowInside=->](1.3,-1.8)(1.3,-1.5)
        \psline[ArrowInside=->](1.3,-3)(1.3,-2.5)
        \psline(1.3,-3)(1.3,-1.5)
        \psline[ArrowInside=->](2.3,-1.8)(2.3,-1.5)
        \psline[ArrowInside=->](2.3,-3)(2.3,-2.5)
        \psline(2.3,-3)(2.3,-1.5)
        \rput(2.3,-1.3){$b$}
        \rput(1.3,-1.3){$a$}
        \rput(2.3,-3.2){$d$}
        \rput(1.3,-3.2){$c$}
        \rput(2.15,-2.125){$e$}
 \end{pspicture}
 =\sum_{f,\mu,\nu} \left[F^{ab}_{cd} \right]_{(e,\alpha,\beta)(f,\mu,\nu)}
 \begin{pspicture}[shift=-1.05](1.1,-3.3)(2.8,-1.2)
        \scriptsize
        \psline(2.6,-1.2)(1.9,-1.9)
        \psline(1.2,-1.2)(1.9,-1.9)
        \psline[ArrowInside=->](1.7,-1.7)(1.4,-1.4)\rput(1.6,-1.4){$a$}
        \psline[ArrowInside=->](2.1,-1.7)(2.4,-1.4)\rput(2.2,-1.4){$b$}
        \psline(2.6,-3.2)(1.9,-2.5)
        \psline(1.2,-3.2)(1.9,-2.5)
        \psline[ArrowInside=->](1.4,-3)(1.7,-2.7)\rput(1.6,-3){$c$}
        \psline[ArrowInside=->](2.4,-3)(2.1,-2.7)\rput(2.15,-3){$d$}
        \psline[ArrowInside=->](1.9,-2.5)(1.9,-1.9)\rput(1.7,-2.2){$f$}
    \end{pspicture},
\end{equation}
where the $F$-symbol in the above equation is related to the regular $F$-symbol by
\begin{equation}
\left[ F^{ab}_{cd}\right]_{(e,\alpha,\beta)(f,\mu,\nu)} = \sqrt{\frac{d_e d_f}{d_a d_d}}\left[F^{ceb}_f \right]^*_{(a,\alpha,\mu)(d,\beta,\nu)}
\end{equation}
and is also a unitary transformation.

\subsubsection{Dimension}

The dimension of $V^{a_1\dots a_n}_c$ is given by
\begin{align}\label{eq:dim-V}
    \dim(V^{a_1\dots a_n}_c)&=\sum_{\vec{b}} N_{a_1 a_2}^{b_2} N_{b_2 a_3}^{b_3} \dots N_{b_{n-1} a_n}^c
    \equiv N_{a_1\dots a_n}^c.
\end{align}

The total dimension of the space of anyons $a_1, \dots, a_n$ is
\begin{equation}
    \sum_c\dim(V^{a_1\dots a_n}_c)=\sum_c N_{a_1\dots a_n}^c\equiv N_{a_1\dots a_n},
\end{equation}
In particular, if $a_1=\cdots=a_n=a$, then the dimension grows as $N_{a\dots a}\sim d_a^n$ for large $n$. Note that a collection of Abelian anyons can only produce 1-dimensional spaces, but non-Abelian anyons can give rise to higher dimensional spaces. When considered by itself, a single anyon does not possess a multi-dimensional Hilbert space, so, from the perspective of individual anyons, the meaning of the quantum dimension is not so clear.  We also define
\begin{equation}
    d_{\vec{a}}\equiv d_{a_1}\cdots d_{a_n}=\sum_c N_{a_1 \dots a_n}^c d_c.
\end{equation}
Note that $N_{a_1\dots a_n}=\text{Tr}(\mathbb{1}_{a_1\dots a_n})$ and $d_{\vec{a}}=\aTr(\mathbb{1}_{a_1\dots a_n})$, where $\text{Tr}$ and $\aTr$ are defined below, and that they both grow with the same scaling as $n\rightarrow\infty$.

\subsubsection{Inner Product}

Inner products can be evaluated by stacking diagrams, e.g. the fact that
\begin{equation}
    \langle a',b';c',\mu'|a,b;c,\mu\rangle=\delta_{a,a'}\delta_{b,b'}\delta_{c,c'}\delta_{\mu,\mu'} \mathbb{1}_{c}
\end{equation}
can be expressed as
\begin{equation}
    \left(\frac{d_c^2}{d_a d_b d_{a'} d_{b'}}\right)^{1/4}
    \begin{pspicture}[shift=-0.9](-0.2,0)(1.2,2)
        \scriptsize
        \psline[ArrowInside=->](0.5,0.5)(0,1)\rput(0.1,0.75){$a$}
        \psline[ArrowInside=->](0.5,0.5)(1,1)\rput(0.95,0.75){$b$}
        \psline[ArrowInside=->](0.5,0)(0.5,0.5)\rput(0.65,0.2){$c$}
        \rput(0.65,0.45){$\mu$}
        \rput(0,1){
        \psline[ArrowInside=->](0,0)(0.5,0.5)\rput(0.1,0.25){$a'$}
        \psline[ArrowInside=->](1,0)(0.5,0.5)\rput(1,0.25){$b'$}
        \psline[ArrowInside=->](0.5,0.5)(0.5,1)\rput(0.7,0.8){$c'$}
        \rput(0.7,0.55){$\mu'$}}
    \end{pspicture}
    =
    \delta_{a,a'}\delta_{b,b'}\delta_{c,c'}\delta_{\mu,\mu'}
    \begin{pspicture}[shift=-0.9](-0.2,0)(0.4,2)
        \scriptsize
        \psline[ArrowInside=->](0,0)(0,2)\rput(0.2,1){$c$}
    \end{pspicture}.
\end{equation}
Note that in the diagramatic notation, $\delta_{a,a'}$ and $\delta_{b,b'}$ ensure that the branches of the splitting vertex can be joined with those of the fusion vertex, while $\delta_{c,c'}$ enforces the conservation of anyonic charge. More complicated diagrams can be similarly evaluated.

\subsubsection{Operators}

The space $V_{a_1\dots a_n}^{a_1'\dots a_n'}$ of operators acting on anyons $a_1$, \dots, $a_n$ can be constructed as
\begin{equation}
    V_{a_1\dots a_n}^{a_1'\dots a_n'}=\bigoplus_c V^c_{a_1\dots a_n}\otimes V^{a_1'\dots a_n'}_c,
\end{equation}
which is spanned by
\begin{align}
    \ket{\vec{a}',\vec{b}',\vec{\alpha}';c}&\bra{\vec{a},\vec{b},\vec{\alpha};c} =\left(\frac{d_c^2}{d_{\vec{a}}d_{\vec{a}'}}\right)^{1/4}
    \begin{pspicture}[shift=-1.5](-0.2,-1.5)(2.2,1.5)
        \scriptsize
        \psline[ArrowInside=->](0.5,1)(0,1.5)\rput(0,1.25){$a'_1$}
        \psline[ArrowInside=->](0.5,1)(1,1.5)\rput(1,1.25){$a'_2$}
        \psline[ArrowInside=->](1,0.5)(2,1.5)\rput(1.8,1){$a'_n$}
        \psline[ArrowInside=->](1,0)(1,0.5)\rput(1.15,0.25){$c$}
        \psline[ArrowInside=->](0.7,0.8)(0.5,1)\rput(0.5,0.75){$b'_2$}
        \psline[ArrowInside=->](1,0.5)(0.8,0.7)\rput(0.6,0.5){$b'_{n-1}$}
        \rput(0.2,1){$\alpha'_2$}
        \rput(1.3,0.5){$\alpha'_n$}
        \rput(0.75,0.75){.}
        \rput(1.5,1.5){\dots}
        \rput(0,0.5){
        \psline[ArrowInside=->](0,-1.5)(0.5,-1)\rput(0.05,-1.25){$a_1$}
        \psline[ArrowInside=->](1,-1.5)(0.5,-1)\rput(1,-1.25){$a_2$}
        \psline[ArrowInside=->](2,-1.5)(1,-0.5)\rput(1.8,-1){$a_n$}
        \psline[ArrowInside=->](0.5,-1)(0.7,-0.8)\rput(0.5,-0.75){$b_2$}
        \psline[ArrowInside=->](0.8,-0.7)(1,-0.5)\rput(0.6,-0.45){$b_{n-1}$}
        \rput(0.25,-1){$\alpha_2$}
        \rput(1.3,-0.5){$\alpha_n$}
        \rput(0.75,-0.75){.}
        \rput(1.5,-1.5){\dots}}
    \end{pspicture},
\end{align}
where $\vec{b}$, $\vec{\alpha}$, $\vec{b}'$, and $\vec{\alpha}'$ take values that are allowed by fusion.

For example, the identity operator for a pair of anyons $a$ and $b$ is
\begin{equation}
    \mathbb{1}_{ab}=\sum_{c,\mu} \ket{a,b;c,\mu}\bra{a,b;c,\mu},
\end{equation}
or, diagramatically,
\begin{equation}
\label{eq:identity-op}
    \begin{pspicture}[shift=-0.6](-0.2,0)(1.4,1.5)
        \scriptsize
        \psline[ArrowInside=->](0,0)(0,1.5)\rput(0.2,0.75){$a$}
        \psline[ArrowInside=->](1,0)(1,1.5)\rput(1.2,0.75){$b$}
    \end{pspicture}
    =\sum_{c,\mu}[F_{ab}^{ab}]_{0,(c,\mu,\nu)}
    \begin{pspicture}[shift=-0.6](-0.2,0)(1.2,1.5)
        \scriptsize
        \psline[ArrowInside=->](0,0)(0.5,0.5)\rput(0,0.25){$a$}
        \psline[ArrowInside=->](1,0)(0.5,0.5)\rput(1,0.25){$b$}
        \rput(0.7,0.5){$\nu$}
        \psline[ArrowInside=->](0.5,0.5)(0.5,1)\rput(0.65,0.75){$c$}
        \rput(0.7,1){$\mu$}
        \psline[ArrowInside=->](0.5,1)(0,1.5)\rput(0,1.25){$a$}
        \psline[ArrowInside=->](0.5,1)(1,1.5)\rput(1,1.25){$b$}
    \end{pspicture}
    =\sum_{c,\mu} \sqrt{\frac{d_c}{d_a d_b}}
    \begin{pspicture}[shift=-0.6](-0.2,0)(1.2,1.5)
        \scriptsize
        \psline[ArrowInside=->](0,0)(0.5,0.5)\rput(0,0.25){$a$}
        \psline[ArrowInside=->](1,0)(0.5,0.5)\rput(1,0.25){$b$}
        \rput(0.7,0.5){$\mu$}
        \psline[ArrowInside=->](0.5,0.5)(0.5,1)\rput(0.65,0.75){$c$}
        \rput(0.7,1){$\mu$}
        \psline[ArrowInside=->](0.5,1)(0,1.5)\rput(0,1.25){$a$}
        \psline[ArrowInside=->](0.5,1)(1,1.5)\rput(1,1.25){$b$}
    \end{pspicture},
\end{equation}
and the braiding operator for the pair is
\begin{equation}
R^{ab}=\sum_{c,\mu} [R^{ab}_c]_{\mu\nu} \ket{a,b;c,\mu}\bra{b,a;c,\nu},
\end{equation}
or, diagramatically,
\begin{equation}
    \begin{pspicture}[shift=-0.6](-0.2,0)(1.2,1.5)
        \scriptsize
        \psline(1,0)(0,1.5)
        \psline[border=1pt](0,0)(1,1.5)
        \psline[ArrowInside=->](1,0)(0.66,0.5)\rput(1,0.3){$a$}
        \psline[ArrowInside=->](0,0)(0.33,0.5)\rput(0,0.3){$b$}
    \end{pspicture}
    =\sum_{c,\mu,\nu} \sqrt{\frac{d_c}{d_a d_b}} [R^{ab}_c]_{\mu\nu}
    \begin{pspicture}[shift=-0.6](-0.2,0)(1.2,1.5)
        \scriptsize
        \psline[ArrowInside=->](0,0)(0.5,0.5)\rput(0,0.25){$b$}
        \psline[ArrowInside=->](1,0)(0.5,0.5)\rput(1,0.25){$a$}
        \rput(0.7,0.5){$\nu$}
        \psline[ArrowInside=->](0.5,0.5)(0.5,1)\rput(0.65,0.75){$c$}
        \rput(0.7,1){$\mu$}
        \psline[ArrowInside=->](0.5,1)(0,1.5)\rput(0,1.25){$a$}
        \psline[ArrowInside=->](0.5,1)(1,1.5)\rput(1,1.25){$b$}
    \end{pspicture},
\end{equation}
where the \emph{$R$ symbols} $R^{ab}_c$ are unitary matrices that must satisfy the Hexagon consistency equations.

\subsubsection{$\mathcal{S}$-matrix}\label{sec:S-matrix}

The topological $S$-matrix is defined by
\begin{equation}
S_{ab}=\frac{1}{\mathcal{D}}\aTr\left(R^{b\bar{a}}R^{\bar{a}b}\right).
\end{equation}
The quantum dimension is related to the $S$-matrix by
\begin{equation}
d_a = \frac{S_{0a}}{S_{00}}.
\end{equation}

For a modular tensor category (MTC), the $S$-matrix is unitary and provides a unitary projective representation of the modular $\mathcal{S}$-transformations. In this case, the fusion coefficients can be expressed in terms of the $\mathcal{S}$-matrix by the Verlinde formula
\begin{equation}
N_{a b}^c=\sum_x \frac{\mathcal{S}_{a x} \mathcal{S}_{b x} \mathcal{S}_{c x}^{\ast} }{ \mathcal{S}_{0 x} }.
\end{equation}
It follows that the dimension of $V_{c}^{a_1\dots a_n}$, given in Eq.~(\ref{eq:dim-V}), can also be expressed in terms of the $\mathcal{S}$-matrix as
\begin{equation}
N_{a_1\dots a_n}^c=\sum_x  \mathcal{S}_{0 x}^{1-n} \mathcal{S}_{a_1 x} \cdots \mathcal{S}_{a_n x} \mathcal{S}_{c x}^*.
\end{equation}

\subsubsection{$\omega_a$-loops}

The $\omega_a$-loop is defined by
\begin{equation}
\omega_a
\begin{pspicture}[shift=-.3](-.6,-.4)(.5,.3)
  \scriptsize
  \psellipse[linecolor=black,border=0](0,0)(.5,.3)
  \psline[ArrowInside=-<](-.1,-.279)(.075,-.2825)
  \end{pspicture}
  =\sum_x \mathcal{S}_{0a}\mathcal{S}_{ax}^*
\begin{pspicture}[shift=-.3](-.6,-.4)(.6,.4)
\scriptsize
  \psellipse[linecolor=black,border=0](0,0)(.5,.3)  \rput(0,-.5){$x$}
  \psline[ArrowInside=-<](-.1,-.279)(.075,-.2825)
  \end{pspicture},
\end{equation}
and acts a projector on all charges threading the loop,
\begin{equation}
\begin{pspicture}[shift=-.4](-.9,-.5)(.6,.5)
  \psellipse[linecolor=black,border=0](0,-.1)(.5,.2)
  \psline[ArrowInside=-<](-.05,-.284)(.15,-.277)
  \psline(0,-.5)(0,-.35)
  \psline[border=1.5pt](0,-.22)(0,.5)
  \psline[ArrowInside=->](0,-.1)(0,.1)
  \rput(.2,.25){$b$}
  \rput(-.75,0){$\omega_a$}
  \end{pspicture}
  = \delta_{ab}
  \begin{pspicture}[shift=-.4](-.2,-.5)(.4,.5)
  \psline[ArrowInside=->](0,-.5)(0,.5)
  \rput(.2,.25){$b$}
  \end{pspicture}.
\end{equation}

\subsubsection{Trace}\label{sec:trace}

The \emph{trace} of an operator is defined, as usual, to be the sum of its diagonal elements, e.g.
\begin{equation}\label{eq:egtrace}
    \text{Tr}(\ket{a',b';c,\mu'}\bra{a,b;c,\mu})=\delta_{a,a'}\delta_{b,b'}\delta_{\mu,\mu'}
\end{equation}
Its diagramatic equivalent is the \emph{quantum trace} $\aTr$, (also called the \emph{anyonic trace},) which is obtained by joining the outgoing anyon lines of the operator's diagram back onto the corresponding incoming lines, e.g.
\begin{eqnarray}
\aTr\Bigg(\left(\frac{d_c^2}{d_a d_b d_{a'} d_{b'}}\right)^{1/4}
    \begin{pspicture}[shift=-0.6](-0.2,0)(1.2,1.5)
        \scriptsize
        \psline[ArrowInside=->](0,0)(0.5,0.5)\rput(0,0.25){$a$}
        \psline[ArrowInside=->](1,0)(0.5,0.5)\rput(1,0.25){$b$}
        \rput(0.7,0.5){$\mu$}
        \psline[ArrowInside=->](0.5,0.5)(0.5,1)\rput(0.65,0.75){$c$}
        \rput(0.7,1){$\mu'$}
        \psline[ArrowInside=->](0.5,1)(0,1.5)\rput(0,1.25){$a'$}
        \psline[ArrowInside=->](0.5,1)(1,1.5)\rput(1,1.25){$b'$}
    \end{pspicture}
    \Bigg)
   &=& \left(\frac{d_c^2}{d_a d_b d_{a'} d_{b'}}\right)^{1/4}
    \begin{pspicture}[shift=-1.2](-0.2,-0.6)(1.8,1.7)
        \scriptsize
        \psline[ArrowInside=->](0,0)(0.5,0.5)\rput(0,0.25){$a$}
        \psline[ArrowInside=->](1,0)(0.5,0.5)\rput(1,0.25){$b$}
        \rput(0.7,0.5){$\mu$}
        \psline[ArrowInside=->](0.5,0.5)(0.5,1)\rput(0.65,0.75){$c$}
        \rput(0.7,1){$\mu'$}
        \psline[ArrowInside=->](0.5,1)(0,1.5)(0.75,2.25)(1.75,1.25)(1.75,0.25)(0.75,-0.75)(0,0)\rput(0,1.25){$a'$}
        \psline[ArrowInside=->](0.5,1)(1,1.5)(1.5,1)(1.5,0.5)(1,0)\rput(1,1.25){$b'$}
    \end{pspicture}
\notag \\
&=& d_c \delta_{a,a'}\delta_{b,b'}\delta_{\mu,\mu'},
\end{eqnarray}
which agrees with Eq.~(\ref{eq:egtrace}) except for the factor of $d_c$. In general, the anyonic trace of an operator $X\in V^{a_1 \dots a_n}_{a'_1 \dots a'_n}$ is related to its ordinary trace by
\begin{align}
    \aTr(X) =\sum_c d_c\text{Tr}([X]_c),\\
    \text{Tr}(X) =\sum_c \frac{1}{d_c}\aTr([X]_c)
\end{align}
where $[X]_c = \Pi_c X \Pi_c\in V^{a_1 \dots a_n}_c\otimes V^c_{a'_1 \dots a'_n}$ is the projection of $X$ onto definite total charge $c$, with $X=\sum_c [X]_c$.

The \emph{partial anyonic trace} is obtained by joining only the outgoing and incoming lines of the anyons being traced over, e.g.
\begin{align}
    \aTr_b\Bigg(\left(\frac{d_c^2}{d_a d_b d_{a'} d_{b'}}\right)^{1/4}
    \begin{pspicture}[shift=-0.6](-0.2,0)(1.2,1.5)
        \scriptsize
        \psline[ArrowInside=->](0,0)(0.5,0.5)\rput(0,0.25){$a$}
        \psline[ArrowInside=->](1,0)(0.5,0.5)\rput(1,0.25){$b$}
        \rput(0.7,0.5){$\mu$}
        \psline[ArrowInside=->](0.5,0.5)(0.5,1)\rput(0.65,0.75){$c$}
        \rput(0.7,1){$\mu'$}
        \psline[ArrowInside=->](0.5,1)(0,1.5)\rput(0,1.25){$a'$}
        \psline[ArrowInside=->](0.5,1)(1,1.5)\rput(1,1.25){$b'$}
    \end{pspicture}
    \Bigg)
    &=\left(\frac{d_c^2}{d_a d_b d_{a'} d_{b'}}\right)^{1/4}
    \begin{pspicture}[shift=-0.6](-0.2,0)(1.7,1.5)
        \scriptsize
        \psline[ArrowInside=->](0,0)(0.5,0.5)\rput(0,0.25){$a$}
        \psline[ArrowInside=->](1,0)(0.5,0.5)\rput(1,0.25){$b$}
        \rput(0.7,0.5){$\mu$}
        \psline[ArrowInside=->](0.5,0.5)(0.5,1)\rput(0.65,0.75){$c$}
        \rput(0.7,1){$\mu'$}
        \psline[ArrowInside=->](0.5,1)(0,1.5)\rput(0,1.25){$a'$}
        \psline[ArrowInside=->](0.5,1)(1,1.5)(1.5,1)(1.5,0.5)(1,0)\rput(1,1.25){$b'$}
    \end{pspicture} \notag \\
    &=\frac{d_c}{d_a} \delta_{a,a'}\delta_{b,b'}\delta_{\mu,\mu'}
    \begin{pspicture}[shift=-0.6](-0.2,0)(0.4,1.5)
        \scriptsize
        \psline[ArrowInside=->](0,0)(0,1.5)\rput(0.2,0.75){$a$}
    \end{pspicture}.
\end{align}
Before computing the partial trace, all the anyons being traced over must moved to the edge of the diagram by braiding them past the other anyons, a process which is not necessarily unique. In general, the partial anyonic trace of $X\in V^{a_1 \dots a_n b_1 \dots b_m}_{a'_1 \dots a'_n b'_1 \dots b'_m}$ over the anyons $b_1$, \dots, $b_m$ is related to its ordinary partial trace by
\begin{align}\label{eq:partial-trace}
    \aTr_{b_1\dots b_m}(X)&=\sum_{c,a}\frac{d_c}{d_a}[\text{Tr}_{b_1\dots b_m}([X]_{c})]_{a},\\
    \text{Tr}_{b_1\dots b_m}(X)&=\sum_{c,a}\frac{d_a}{d_c}[\aTr_{b_1\dots b_m}([X]_{c})]_{a}.
\end{align}

\subsection{Anyonic Density Matrix}

An \emph{anyonic density matrix} $\arho$  is an anyonic operator normalized by the quantum trace $\aTr \arho=1$, that describes the topological state of the system. The anyonic density matrix $\arho$ determines the expectation value of anyonic operators acting on the system, ${\langle X \rangle = \aTr(\arho X)}$. For example, the density matrix describing a pair of anyons $a$ and $b$ with definite total charge $c$ is
\begin{equation}
    \arho_{ab}=\frac{1}{d_c}\ket{a,b;c,\mu}\bra{a,b;c,\mu}=\frac{1}{\sqrt{d_a d_b d_c}}
    \begin{pspicture}[shift=-0.6](-0.2,0)(1.2,1.5)
        \scriptsize
        \psline[ArrowInside=->](0,0)(0.5,0.5)\rput(0,0.25){$a$}
        \psline[ArrowInside=->](1,0)(0.5,0.5)\rput(1,0.25){$b$}
        \rput(0.7,0.5){$\mu$}
        \psline[ArrowInside=->](0.5,0.5)(0.5,1)\rput(0.65,0.75){$c$}
        \rput(0.7,1){$\mu$}
        \psline[ArrowInside=->](0.5,1)(0,1.5)\rput(0,1.25){$a$}
        \psline[ArrowInside=->,](0.5,1)(1,1.5)\rput(1,1.25){$b$}
    \end{pspicture},
\end{equation}
which is normalized such that $\aTr(\arho^{ab})=1$, while the most general state for the pair is given by
\begin{align}
    \arho_{ab}&=\sum_{\substack{a, b, \mu \\ c \\ a', b', \mu'}} \frac{\rho_{(a,b; c,\mu)(a',b';c,\mu')}}{d_c}\ket{a,b;c,\mu}\bra{a',b';c,\mu'}
\notag \\
&=\sum_{\substack{a, b, \mu \\ c \\ a', b', \mu'}} \frac{\rho_{(a,b;c,\mu)(a',b';c,\mu')}}{(d_a d_b d_{a'} d_{b'} d_c^2)^{1/4}}
    \begin{pspicture}[shift=-0.6](-0.2,0)(1.1,1.5)
        \scriptsize
        \psline[ArrowInside=->](0,0)(0.5,0.5)\rput(0,0.25){$a$}
        \psline[ArrowInside=->](1,0)(0.5,0.5)\rput(1,0.25){$b$}
        \rput(0.7,0.5){$\mu$}
        \psline[ArrowInside=->](0.5,0.5)(0.5,1)\rput(0.65,0.75){$c$}
        \rput(0.7,1){$\mu'$}
        \psline[ArrowInside=->](0.5,1)(0,1.5)\rput(0,1.25){$a'$}
        \psline[ArrowInside=->](0.5,1)(1,1.5)\rput(1,1.25){$b'$}
    \end{pspicture},
\end{align}
where the coefficients are normalized such that $\sum_{a,b,\mu,c}\rho_{(a,b;c,\mu)(a,b;c,\mu)}=1$.

For a collection of anyons $a_1$, \dots, $a_n$, $b_1$, \dots, $b_n$, the reduced anyonic density matrix
\begin{equation}
    \arho_{a_1\dots a_n}=\aTr_{b_1\dots b_n}(\arho_{a_1\dots a_n b_1\dots b_n})
\end{equation}
describes the topological state of the anyons $a_1$, \dots, $a_n$, i.e. for any operator $X\in V^{a_1 \dots a_n}_{a'_1 \dots a'_n}$,
\begin{equation}
    \langle X \rangle = \aTr(\arho_{a_1\dots a_n b_1\dots b_n} X)=\aTr(\arho_{a_1\dots a_n} X).
\end{equation}

\section{Examples of Braided Tensor Categories}
\label{sec:BTCs}

In this Appendix, we provide additional details of the braided tensor categories (BTCs) mentioned in this paper. In particular, we list the fusion rules (which are commutative), quantum dimensions, and topological twist factors. (The $F$-symbols and $R$-symbols for these theories are uniquely determined, up to gauge freedom, by this data, and can be found in the literature, such as Ref.~\cite{Bonderson07b}.)

\subsection{$\mathbb{Z}_N^{(p)}$}
\label{sec:znmodel}

The $\mathbb{Z}_N^{(p)}$ BTC for $N$ a positive integer can have $p\in\mathbb{Z}$ for all $N$ and $p\in\mathbb{Z}+\frac{1}{2}$ for $N$ even. The total quantum dimension is $\mathcal{D}^{2} = N$. This BTC has $N$ topological charges labeled by $\{0,1,\dots, N-1 \}$, for which the fusion rules, quantum dimensions, and twist factors are
\begin{eqnarray}
a\times b &=& [a+b]_N, \\
d_{a} &=& 1, \\
\theta_a &=& e^{i \frac{2 \pi p}{N} a^2 },
\end{eqnarray}
where $[a]_{N} = a (\text{mod } N)$.

For odd $N$, $\mathbb{Z}_N^{(p)}$ is modular when $[p]_{N} \neq 0$ and $\gcd(N , [p]_{N}) =1$. For even $N$, $\mathbb{Z}_N^{(p)}$ is modular when $p\in\mathbb{Z}+\frac{1}{2}$ and $\gcd(N , 2[p]_{N}) =1$.
Notice that $p$ is periodic in $N$, so we can restrict our attention to $0 \leq p < N$. In some cases, there is a redundancy where distinct values of $p$ describe the same BTC when the topological charge values are relabeled (i.e. $a \mapsto a' = [na]_{N}$ for some integer $n$). For example, in the case of $\mathbb{Z}_5^{(p)}$, $p=1$ and $4$ are the same BTC, and $p=2$ and $3$ are the same BTC; in the case of $\mathbb{Z}_7^{(p)}$, $p=1$, $2$, and $4$ are the same BTC, and $p=3$, $5$, and $6$ are the same BTC.

The trivial fermion SMTC is described by $\mathbb{Z}_2^{(1)}$.

\subsection{Fib$^{\pm 1}$}
\label{sec:Fib}

The Fibonacci (Fib$^{\pm 1}$) MTCs has two topological charges $\{ 0,1 \}$, for which the fusion rules are given by
\begin{equation}
0 \times a = a, \quad 1 \times 1 = 0+1
.
\end{equation}
The quantum dimensions are given by
\begin{equation}
d_0 = 1, \quad d_1 = \phi
,
\end{equation}
where $\phi = \frac{1 + \sqrt{5}}{2}$ is the Golden ratio, so $\mathcal{D}^{2}= \phi +2$. The twist factors are
\begin{equation}
\theta_0 = 1, \quad \theta_1 = e^{\pm i \frac{4 \pi}{5}}
.
\end{equation}

\subsection{$\mathcal{K}_{\nu}$}
\label{sec:Kitaev_16}

We use the notation $\mathcal{K}_{\nu}$ with $\nu = 0,1,\ldots,15$ to denote Kitaev's 16-fold way of MTCs~\cite{Kitaev06a}, which have chiral central charge $c_{-} (\text{mod})8 = \nu$ and total quantum dimension $\mathcal{D}^2 = 4$.

For $\nu$ odd, there are three topological charge values, which we denote $\{ I , \sigma, \psi\}$, where the vacuum charge here is denoted $I$. The fusion rules are given by
\begin{equation}
I \times a = a, \quad  \psi\times \psi =I, \quad \psi \times \sigma = \sigma, \quad  \sigma \times \sigma = I+ \psi
.
\end{equation}
The quantum dimensions and twist factors are given by
\begin{equation}
\begin{array}{lllll}
d_I = 1, & & d_\sigma = \sqrt{2}, & & d_{\psi} = 1, \\
\theta_I = 1, & & \theta_\sigma=e^{i\frac{\pi }{8}\nu}, & & \theta_\psi=-1 .
\end{array}
\end{equation}
$\nu=1$ corresponds to the Ising TQFT, $\nu=3$ corresponds to SU$(2)_2$, and $\nu \geq 5$ can be realized by SO$(\nu)_1$ Chern-Simons field theory.

For $\nu$ even, there are four topological charge values, all of which have quantum dimension $d_{a}=1$. It is useful to further split them into two categories, as follows.

For $\nu = 0$, $4$, $8$, and $12$, the fusion rules are $\mathbb{Z}_{2} \times \mathbb{Z}_{2}$. The twist factors are
\begin{equation}
\theta_{(0,0)} = 1, \quad \theta_{(0,1)} = \theta_{(1,0)} = e^{i\frac{\pi }{8}\nu}, \quad \theta_{(1,1)} = -1
.
\end{equation}
$\nu = 0$ corresponds to the toric code D$(\mathbb{Z}_2)$, $\nu = 8$ corresponds to the three fermion theory SO$(8)_1$, and $\nu = 4$ and $12$ correspond to $\mathbb{Z}_{2}^{(\pm 1/2)} \times \mathbb{Z}_{2}^{(\pm 1/2)}$, respectively.

For $\nu = 2$, $6$, $10$, and $14$, the fusion rules are $\mathbb{Z}_{4}$. The twist values are
\begin{equation}
\theta_{0} = 1, \quad \theta_{1} = \theta_{3} = e^{i\frac{\pi }{8}\nu}, \quad \theta_{2} = -1
.
\end{equation}
Thus, these correspond to the $\mathbb{Z}_{4}^{(\nu/4)}$ MTCs.

\subsection{SO$(3)_{6}$}
\label{sec:SO(3)_6}

The SO$(3)_{6}$ SMTC can be obtained as the restriction of the SU$(2)_{6}$ MTC to its integer spin topological charge values. It has four topological charge values $\{0,1,2,3\}$, which have the fusion rules
\begin{equation}
0 \times a = a, \quad 3 \times a = 3-a, \quad  1 \times 2 = 1 + 2 + 3 , \quad 1 \times 1 = 2 \times 2 = 0 + 1 + 2
.
\end{equation}
The quantum dimensions and twist factors are given by
\begin{equation}
\begin{array}{lllllll}
d_0 = 1, & & d_1 = 1+ \sqrt{2}, & & d_2 = 1+ \sqrt{2}, & & d_3 = 1, \\
\theta_0 = 1, & & \theta_1=i, & & \theta_2= -i, & & \theta_3=-1 .
\end{array}
\end{equation}

\section{Proofs} \label{app:proofs}

We now prove various properties of anyonic entropy $\aS$, following Ref.~\cite{Nielsen11} and adapting the proofs appropriately.  We make use of the following definitions: the anyonic relative entropy is
\begin{equation}
    \aS(\arho \| \tilde{\sigma}) \equiv \aTr(\arho \log \arho -\arho \log \tilde{\sigma}).
\end{equation}
and the anyonic mutual information between the two subsystems is
\begin{equation}
    \tilde{I}(A:B) \equiv \aS(\arho_A) + \aS(\arho^B) - \aS(\arho^{AB}).
\end{equation}

\subsubsection{Anyonic Entropy is non-negative}

\noindent {\bf Statement}: $\aS\left( \arho\right) \geq 0$ with equality iff $\arho$ is pure.

\noindent {\bf Proof}: Positivity follows from the definition. To see this, it may be helpful to write the anyonic density matrices in diagonalized form
\begin{equation}
\arho = \sum_{c,\alpha_c} \frac{p_{\alpha_c}}{d_c} \left| \alpha_c \right\rangle \left\langle \alpha_c \right|
\end{equation}
where $\left| \alpha_c \right\rangle$ are orthonormal states with total charge $c$. This gives
\begin{eqnarray}
\aS\left( \arho\right) &=& - \sum_{c,\alpha_c} p_{\alpha_c} \log \left( \frac{p_{\alpha_c}}{d_c} \right) \\
&=& H \left( \{ p_{\alpha_c} \} \right) + \sum_{c,\alpha_c} p_{\alpha_c} \log d_c ,
\end{eqnarray}
which is positive, since $d_c  \geq 1$ (and $d_c = 1$ iff $c$ is Abelian).

\subsubsection{Relative Anyonic Entropy is non-negative}

\noindent {\bf Statement}: $\aS\left( \arho \| \tilde{\sigma} \right) \geq 0$ with equality iff $\arho = \tilde{\sigma}$.

\noindent {\bf Proof}: Start by diagonalizing the anyonic density matrices
\begin{eqnarray}
\arho &=& \sum_{c,\alpha_{c}} \frac{p_{\alpha_{c}}}{d_{c}} \left| \alpha_{c} \right\rangle \left\langle \alpha_{c} \right|, \\
\tilde{\sigma} &=& \sum_{c,\beta_{c}} \frac{q_{\beta_{c}}}{d_{c}} \left| \beta_{c} \right\rangle \left\langle \beta_{c} \right|,
\end{eqnarray}
where $\left| \alpha_{c} \right\rangle$ and $\left| \beta_{c} \right\rangle$ are possibly different orthonormal bases for the space of states with total charge $c$. Now we can write
\begin{eqnarray}
&&\aS\left( \arho \| \tilde{\sigma }\right) = \sum_{c,\alpha_{c}} \left[ p_{\alpha_{c}} \log \left( \frac{ p_{\alpha_{c}} }{ d_{c} } \right) - d_{c} \left\langle \alpha_{c} \right| \arho \log \tilde{\sigma} \left| \alpha_{c} \right\rangle \right] \notag \\
&& = \sum_{c,\alpha_{c}} p_{\alpha_{c}} \left[  \log \left( \frac{ p_{\alpha_{c}} }{ d_{c} } \right) - \sum_{\beta_{c}} P_{\alpha_{c},\beta_{c} } \log \left( \frac{ q_{\beta_{c}} }{ d_{c} } \right) \right] \notag \\
&& = \sum_{c,\alpha_{c}} p_{\alpha_{c}} \left[  \log p_{\alpha_{c}} - \sum_{\beta_{c}} P_{\alpha_{c},\beta_{c} } \log q_{\beta_{c}}  \right]
,
\end{eqnarray}
where we used
\begin{equation}
P_{\alpha_{c},\beta_{c} } \equiv \left\langle \alpha_{c} \left| \beta_{c} \right. \right\rangle \left\langle \beta_{c} \left| \alpha_{c} \right. \right\rangle \geq 0 ,
\end{equation}
and the fact that it satisfies
\begin{equation}
\sum_{\alpha_{c}} P_{\alpha_{c},\beta_{c} } =\sum_{\beta_{c}} P_{\alpha_{c},\beta_{c} } = 1
\end{equation}
because the basis states are orthonormal. Now the rest of the proof from Ref.~\cite{Nielsen11} applies.

\subsubsection{Maximum of Anyonic Entropy}

\noindent {\bf Statement}: The entropy for a state $\arho$ of anyons with topological charges $a_1,\dots,a_n$ satisfies the bound
\begin{equation}
\aS\left( \arho\right) \leq \log \left( \prod_{i=1}^n d_{a_i}\right) = \sum_{j} \log d_{a_j},
\end{equation}
with equality obtained iff
\begin{equation}
\arho = \frac{\mathbb{1}_{a_1 \dots a_n}}{\prod_{i=1}^n d_{a_i}}=\arho_{a_1}\otimes\arho_{a_2}\otimes \dots \otimes \arho_{a_n} .
\end{equation}

\noindent {\bf Proof}: Using the relative entropy with $\tilde{\sigma} = \frac{\mathbb{1}_{a_1 \dots a_n}}{\prod_{i=1}^n d_{a_i}}$, we see
\begin{equation}
0 \leq \aS\left( \arho \|  \tilde{\sigma} \right) = - \aS\left( \arho \right) + \log \left( \prod_{i=1}^n d_{a_i}\right)
\end{equation}

\subsubsection{Anyonic Entanglement Entropy of Pure States}

\noindent {\bf Statement}:  The entanglement entropy of a composite system in a pure state $\arho_{AB}=\ket{\psi_c}\bra{\psi_c}$ has $\aS\left( \arho_A\right) = \aS\left( \arho_B \right)$.

\emph{Corollary}: For a pure state $\arho_{AB}$, $I\left( A:B \right) = 2 \aS\left( \arho_A\right)$.

\noindent {\bf Proof}: Begin by Schmidt decomposing the state
\begin{equation}
\left| \psi_{c} \right\rangle = \sum_{a, \alpha_{a}} \sqrt{ p_{\alpha_{a} } } \left| \alpha_{a} \right\rangle_{A} \left| \alpha_{b} \right\rangle_{B}
,
\end{equation}
where $b = \bar{a} \times c$ is uniquely determined by $a$ and has $d_{b} = d_{a}$, since $c$ is Abelian. Now we have
\begin{eqnarray}
\arho_{A} &=& \sum_{a, \alpha_{a}} \frac{p_{\alpha_{a}}}{d_{a}} \left| \alpha_{a} \right\rangle \left\langle \alpha_{a} \right| \\
\arho_{B} &=& \sum_{a, \alpha_{a}} \frac{p_{\alpha_{a}}}{d_{a}} \left| \alpha_{\bar{a} \times c} \right\rangle \left\langle \alpha_{\bar{a} \times c} \right|
\end{eqnarray}
which clearly gives
\begin{equation}
\aS\left( \arho_A\right) = \aS\left( \arho_B\right) = - \sum_{a, \alpha_{a}} p_{\alpha_{a}} \log \left( \frac{p_{\alpha_{a}}}{d_{a}} \right)
.
\end{equation}

\subsubsection{Entropy of Tensor Product of States}

\noindent {\bf Statement}:  The entropy of the tensor product $\arho_{AB} = \arho_{A} \otimes \arho_{B}$ of two states is $\aS\left( \arho_{AB} \right) = \aS\left( \arho_{A} \right) + \aS\left( \arho_{B} \right)$.

\emph{Corollary}: If $\arho_{AB} = \arho_{A} \otimes \arho_{B}$, then $\tilde{I}\left( A:B \right) = 0$

\noindent {\bf Proof}: Same as proof in Ref.~\cite{Nielsen11}.

\subsubsection{Entropy of Distribution of Orthogonal States}

\noindent {\bf Statement}: For a probability distribution $p_{i}$ of states $\arho_{i}$ with orthogonal support ($ \arho_{i} \arho_{j} =0 $ for $i \neq j$), the entropy is
\begin{eqnarray}
\aS\left( \sum_{i} p_{i} \arho_{i} \right)
&=& H\left(\{ p_{i}\} \right) + \sum_{i} p_{i} \aS\left( \arho_{i} \right)
.
\end{eqnarray}

\noindent {\bf Proof}:
Begin by decomposing the density matrix $\arho_i$ as
\begin{equation}
\arho_i = \sum_{c,\alpha_c^{(i)}} \frac{q_{\alpha_c}^{(i)}}{d_c} \ket{\alpha_c^{(i)}}\bra{\alpha_c^{(i)}}.
\end{equation}
It follows that
\begin{equation}
\begin{split}
\aS &\left( \sum_i p_i \arho_i\right) = -\sum_{i,c,\alpha_c^{(i)}} p_i q_{\alpha_c}^{(i)} \log \left( \frac{p_i q_{\alpha_c^{(i)}}}{d_c}\right)
\\ &= -\sum_i p_i \log p_i -\sum_i p_i \left( \sum_{c,\alpha_c^{(i)}} q_{\alpha_c^{(i)}} \log \left( \frac{q_{\alpha_c^{(i)}}}{d_c}\right)\right)
\\ &= H(\{p_i\}) +\sum_i p_i \aS \left( \arho_i\right).
\end{split}
\end{equation}

\subsubsection{Joint Entropy}

\noindent {\bf Statement}:  For a set of states $\arho_{i}$ and an orthogonal set of pure states $\left| i \right\rangle \left\langle i \right|$, then
\begin{equation}
\aS\left( \sum_{i} p_{j} \left| i \right\rangle \left\langle i \right| \otimes \arho_{j} \right) = H\left( \{p_{i}\} \right) + \sum_{i} p_{i} \aS\left( \arho_{i} \right)
.
\end{equation}

\noindent {\bf Proof}: This follows from the previous result. If necessary, we could introduce a set of unpure orthogonal states $\left| i \right\rangle \left\langle i \right|$ with non-Abelian collective charge, which will require modification of this equation.

\subsubsection{Decoherence Due to Projective Measurement Increases Anyonic Entropy}

\noindent {\bf Statement}:  Consider a projective measurement given by the complete, orthogonal set of projectors $\Pi_{i}$. The decoherence of a state $\arho$ due to this measurement is given by the transformation $ \arho^{\prime} = \sum_{i} \Pi_i \arho \Pi_i$. Then $\aS\left( \arho^{\prime} \right) \geq \aS\left( \arho \right)$, with equality iff $\arho = \arho^{\prime}$.

\noindent {\bf Proof}: We use the fact that
\begin{eqnarray}
\aTr \left[ \arho \log \arho^{\prime} \right] &=& \aTr \left[ \arho \log \left( \sum_{i} \Pi_i \arho \Pi_i \right) \right] \notag \\
&=& \aTr \left[ \sum_{j} \Pi_{j} \arho \log \left( \sum_{i} \Pi_i \arho \Pi_i \right) \Pi_{j} \right] \notag \\
&=& \aTr \left[ \sum_{j} \Pi_{j} \arho \Pi_{j} \log \left( \sum_{i} \Pi_i \arho \Pi_i \right) \right] \notag \\
&=& \aTr \left[ \arho^{\prime} \log \arho^{\prime} \right]
\end{eqnarray}
and the previous results to get
\begin{eqnarray}
0 \leq \aS\left( \arho \| \arho^{\prime} \right) &=& -\aS\left( \arho \right) - \aTr \left[ \arho \log \arho^{\prime} \right] \notag \\
&=& -\aS\left( \arho \right) + \aS\left( \arho^{\prime} \right)
.
\end{eqnarray}

\subsubsection{Subadditivity}

\noindent {\bf Statement}:  For a composite state $\arho_{AB}$, we have
\begin{equation}
\aS\left(\arho_{AB}\right) \leq \aS\left( \arho_A \right) + \aS\left( \arho_B\right),
 \end{equation}
with equality iff $\arho_{AB} = \arho_{A} \otimes \arho_{B}$.

\noindent {\bf Proof}: Let $\arho = \arho_{AB}$ and $\tilde{\sigma} =\arho_{A} \otimes \arho_{B}$. Then we have
\begin{eqnarray}
0 \leq \aS\left( \arho \| \tilde{\sigma} \right) &=& -\aS\left( \arho \right) - \aTr \left[ \arho_{AB} \log \tilde{\sigma} \right] \notag \\
&=& -\aS\left( \arho_{AB} \right) + \aS\left( \arho_{A} \right) + \aS\left( \arho_{B} \right)
.
\end{eqnarray}

\subsubsection{Triangle Inequality}

\noindent {\bf Statement}:  For a composite state $\arho_{AB}$, we have $\aS\left( \arho_{AB}\right) \geq \left| \aS\left( \arho_A \right) - \aS\left( \arho_B \right) \right|$, with equality iff $\arho_A$ is already maximally entangled with the environment by its existing correlations with $\arho_B$.

\noindent {\bf Proof}: Let $R$ be a system which purifies systems $A$ and $B$.  Then $\aS(\arho_{AR})=\aS(\arho_B)$ and $\aS(\arho_R)=\aS(\arho_{AB})$ because $\arho_{ABR}$ is a pure state.  If we consider the composite state of $\arho_{AR}$, then from subadditivity we have
\begin{equation}
\begin{split}
\aS(\arho_{AR})&\leq \aS (\arho_A)+\aS(\arho_R)
\\ \aS(\arho_B) &\leq \aS( \arho_A)+\aS(\arho_{AB})
\\ \aS(\arho_{AB}) &\geq \aS ( \arho_B) -\aS(\arho_A).
\end{split}
\end{equation}
Similarly,
\begin{equation}
\begin{split}
\aS(\arho_{BR}) &\leq \aS (\arho_B) +\aS(\arho_R)
\\ \aS(\arho_A) &\leq \aS (\arho_B) +\aS(\arho_{AB})
\\ \aS(\arho_{AB}) &\geq \aS(\arho_A)-\aS(\arho_B).
\end{split}
\end{equation}
Taken together, the above equations imply
\begin{equation}
\aS(\arho_{AB}) \geq |\aS(\arho_A)-\aS(\arho_B)|.
\end{equation}

From subadditivity we know that ${\tilde{S}(\arho_{AR})=\aS(\arho_A)+\aS(\arho_R)}$ iff $\arho_{AR}=\arho_A\otimes \arho_R$.

\subsubsection{Concavity}

\noindent {\bf Statement}:  ${\displaystyle{\aS(\sum_j p_j \arho_j)\geq \sum_j p_j \aS(\arho_j)}}$, with equality iff all the $\arho_j$ are the same.

\noindent {\bf Proof}: Let the sum on $j$ run from $1$ to $n$.  We introduce an auxillary system $B$ whose state space has an orthonormal basis $\{\ket{\psi_k}\}$, such that at least $n$ basis states have Abelian total charge.  We enlarge the set $\{p_j\}$ by setting $p_j=0$ for $j>n$.  One choice of auxillary system is for a particular basis state $\ket{\psi_k}$ to correspond to $k$ copies of $\bar{c}$ and $c$ fusing to vacuum for some nontrivial charge $c$ in the anyon model describing the system.  The proof from here follows that in Ref.~\cite{Nielsen11}.


\begin{thebibliography}{72}
\expandafter\ifx\csname natexlab\endcsname\relax\def\natexlab#1{#1}\fi
\expandafter\ifx\csname bibnamefont\endcsname\relax
  \def\bibnamefont#1{#1}\fi
\expandafter\ifx\csname bibfnamefont\endcsname\relax
  \def\bibfnamefont#1{#1}\fi
\expandafter\ifx\csname citenamefont\endcsname\relax
  \def\citenamefont#1{#1}\fi
\expandafter\ifx\csname url\endcsname\relax
  \def\url#1{\texttt{#1}}\fi
\expandafter\ifx\csname urlprefix\endcsname\relax\def\urlprefix{URL }\fi
\providecommand{\bibinfo}[2]{#2}
\providecommand{\eprint}[2][]{\url{#2}}

\bibitem[{\citenamefont{Schr\"{o}dinger}(1935)}]{Schroedinger35}
\bibinfo{author}{\bibfnamefont{E.}~\bibnamefont{Schr\"{o}dinger}},
  \bibinfo{journal}{Proceedings of the Cambridge Philosophical Society}
  \textbf{\bibinfo{volume}{31}}, \bibinfo{pages}{555} (\bibinfo{year}{1935}).

\bibitem[{\citenamefont{{Calabrese} and {Cardy}}(2004)}]{Calabrese04}
\bibinfo{author}{\bibfnamefont{P.}~\bibnamefont{{Calabrese}}} \bibnamefont{and}
  \bibinfo{author}{\bibfnamefont{J.}~\bibnamefont{{Cardy}}},
  \bibinfo{journal}{Journal of Statistical Mechanics: Theory and Experiment}
  \textbf{\bibinfo{volume}{6}}, \bibinfo{pages}{06002} (\bibinfo{year}{2004}),
  \eprint{hep-th/0405152}.

\bibitem[{\citenamefont{{Calabrese} and {Cardy}}(2009)}]{Calabrese09}
\bibinfo{author}{\bibfnamefont{P.}~\bibnamefont{{Calabrese}}} \bibnamefont{and}
  \bibinfo{author}{\bibfnamefont{J.}~\bibnamefont{{Cardy}}},
  \bibinfo{journal}{Journal of Physics A Mathematical General}
  \textbf{\bibinfo{volume}{42}}, \bibinfo{eid}{504005} (\bibinfo{year}{2009}),
  \eprint{arXiv:0905.4013}.

\bibitem[{\citenamefont{{Savary} and {Balents}}(2017)}]{Savary16}
\bibinfo{author}{\bibfnamefont{L.}~\bibnamefont{{Savary}}} \bibnamefont{and}
  \bibinfo{author}{\bibfnamefont{L.}~\bibnamefont{{Balents}}},
  \bibinfo{journal}{Reports on Progress in Physics}
  \textbf{\bibinfo{volume}{80}}, \bibinfo{eid}{016502} (\bibinfo{year}{2017}),
  \eprint{arXiv:1601.03742}.

\bibitem[{\citenamefont{Wen}(1990)}]{Wen90}
\bibinfo{author}{\bibfnamefont{X.-G.} \bibnamefont{Wen}},
  \bibinfo{journal}{International Journal of Modern Physics B}
  \textbf{\bibinfo{volume}{4}}, \bibinfo{pages}{239} (\bibinfo{year}{1990}).

\bibitem[{\citenamefont{{Nayak} et~al.}(2008)\citenamefont{{Nayak}, {Simon},
  {Stern}, {Freedman}, and {Das Sarma}}}]{Nayak08}
\bibinfo{author}{\bibfnamefont{C.}~\bibnamefont{{Nayak}}},
  \bibinfo{author}{\bibfnamefont{S.~H.} \bibnamefont{{Simon}}},
  \bibinfo{author}{\bibfnamefont{A.}~\bibnamefont{{Stern}}},
  \bibinfo{author}{\bibfnamefont{M.}~\bibnamefont{{Freedman}}},
  \bibnamefont{and} \bibinfo{author}{\bibfnamefont{S.}~\bibnamefont{{Das
  Sarma}}}, \bibinfo{journal}{Reviews of Modern Physics}
  \textbf{\bibinfo{volume}{80}}, \bibinfo{pages}{1083} (\bibinfo{year}{2008}),
  \eprint{arXiv:0707.1889}.

\bibitem[{\citenamefont{Leinaas and Myrheim}(1977)}]{Leinaas77}
\bibinfo{author}{\bibfnamefont{J.~M.} \bibnamefont{Leinaas}} \bibnamefont{and}
  \bibinfo{author}{\bibfnamefont{J.}~\bibnamefont{Myrheim}},
  \bibinfo{journal}{Nuovo Cimento B} \textbf{\bibinfo{volume}{37B}},
  \bibinfo{pages}{1} (\bibinfo{year}{1977}).

\bibitem[{\citenamefont{Wilczek}(1982)}]{Wilczek82b}
\bibinfo{author}{\bibfnamefont{F.}~\bibnamefont{Wilczek}},
  \bibinfo{journal}{Phys. Rev. Lett.} \textbf{\bibinfo{volume}{49}},
  \bibinfo{pages}{957} (\bibinfo{year}{1982}).

\bibitem[{\citenamefont{Goldin et~al.}(1985)\citenamefont{Goldin, Menikoff, and
  Sharp}}]{Goldin85}
\bibinfo{author}{\bibfnamefont{G.~A.} \bibnamefont{Goldin}},
  \bibinfo{author}{\bibfnamefont{R.}~\bibnamefont{Menikoff}}, \bibnamefont{and}
  \bibinfo{author}{\bibfnamefont{D.~H.} \bibnamefont{Sharp}},
  \bibinfo{journal}{Phys. Rev. Lett.} \textbf{\bibinfo{volume}{54}},
  \bibinfo{pages}{603} (\bibinfo{year}{1985}).

\bibitem[{\citenamefont{Fredenhagen et~al.}(1989)\citenamefont{Fredenhagen,
  Rehren, and Schroer}}]{Fredenhagen89}
\bibinfo{author}{\bibfnamefont{K.}~\bibnamefont{Fredenhagen}},
  \bibinfo{author}{\bibfnamefont{K.~H.} \bibnamefont{Rehren}},
  \bibnamefont{and} \bibinfo{author}{\bibfnamefont{B.}~\bibnamefont{Schroer}},
  \bibinfo{journal}{Commun. Math. Phys.} \textbf{\bibinfo{volume}{125}},
  \bibinfo{pages}{201} (\bibinfo{year}{1989}).

\bibitem[{\citenamefont{Fr\"{o}hlich and Gabbiani}(1990)}]{Froehlich90}
\bibinfo{author}{\bibfnamefont{J.}~\bibnamefont{Fr\"{o}hlich}}
  \bibnamefont{and} \bibinfo{author}{\bibfnamefont{F.}~\bibnamefont{Gabbiani}},
  \bibinfo{journal}{Rev. Math. Phys.} \textbf{\bibinfo{volume}{2}},
  \bibinfo{pages}{251} (\bibinfo{year}{1990}).

\bibitem[{\citenamefont{Kitaev}(2003)}]{Kitaev03}
\bibinfo{author}{\bibfnamefont{A.~Y.} \bibnamefont{Kitaev}},
  \bibinfo{journal}{Ann. Phys.} \textbf{\bibinfo{volume}{303}},
  \bibinfo{pages}{2} (\bibinfo{year}{2003}), \eprint{quant-ph/9707021}.

\bibitem[{\citenamefont{Freedman}(1998)}]{Freedman98}
\bibinfo{author}{\bibfnamefont{M.~H.} \bibnamefont{Freedman}},
  \bibinfo{journal}{Proc. Natl. Acad. Sci. USA} \textbf{\bibinfo{volume}{95}},
  \bibinfo{pages}{98} (\bibinfo{year}{1998}).

\bibitem[{\citenamefont{Kitaev and Preskill}(2006)}]{Kitaev06b}
\bibinfo{author}{\bibfnamefont{A.}~\bibnamefont{Kitaev}} \bibnamefont{and}
  \bibinfo{author}{\bibfnamefont{J.}~\bibnamefont{Preskill}},
  \bibinfo{journal}{Phys. Rev. Lett.} \textbf{\bibinfo{volume}{96}},
  \bibinfo{eid}{110404} (\bibinfo{year}{2006}), \eprint{hep-th/0510092}.

\bibitem[{\citenamefont{Levin and Wen}(2006)}]{Levin06a}
\bibinfo{author}{\bibfnamefont{M.}~\bibnamefont{Levin}} \bibnamefont{and}
  \bibinfo{author}{\bibfnamefont{X.-G.} \bibnamefont{Wen}},
  \bibinfo{journal}{Phys. Rev. Lett.} \textbf{\bibinfo{volume}{96}},
  \bibinfo{eid}{110405} (\bibinfo{year}{2006}), \eprint{cond-mat/0510613}.

\bibitem[{\citenamefont{{Hamma} et~al.}(2005)\citenamefont{{Hamma},
  {Ionicioiu}, and {Zanardi}}}]{Hamma05}
\bibinfo{author}{\bibfnamefont{A.}~\bibnamefont{{Hamma}}},
  \bibinfo{author}{\bibfnamefont{R.}~\bibnamefont{{Ionicioiu}}},
  \bibnamefont{and}
  \bibinfo{author}{\bibfnamefont{P.}~\bibnamefont{{Zanardi}}},
  \bibinfo{journal}{Physics Letters A} \textbf{\bibinfo{volume}{337}},
  \bibinfo{pages}{22} (\bibinfo{year}{2005}), \eprint{quant-ph/0406202}.

\bibitem[{\citenamefont{Dong et~al.}(2008)\citenamefont{Dong, Fradkin, Leigh,
  and Nowling}}]{Dong08}
\bibinfo{author}{\bibfnamefont{S.}~\bibnamefont{Dong}},
  \bibinfo{author}{\bibfnamefont{E.}~\bibnamefont{Fradkin}},
  \bibinfo{author}{\bibfnamefont{R.~G.} \bibnamefont{Leigh}}, \bibnamefont{and}
  \bibinfo{author}{\bibfnamefont{S.}~\bibnamefont{Nowling}},
  \bibinfo{journal}{JHEP} \textbf{\bibinfo{volume}{05}}, \bibinfo{pages}{016}
  (\bibinfo{year}{2008}), \eprint{arXiv:0802.3231}.

\bibitem[{\citenamefont{{Castelnovo} and {Chamon}}(2008)}]{Castelnovo08}
\bibinfo{author}{\bibfnamefont{C.}~\bibnamefont{{Castelnovo}}}
  \bibnamefont{and} \bibinfo{author}{\bibfnamefont{C.}~\bibnamefont{{Chamon}}},
  \bibinfo{journal}{Phys. Rev. B} \textbf{\bibinfo{volume}{78}},
  \bibinfo{eid}{155120} (\bibinfo{year}{2008}), \eprint{arXiv:0804.3591}.

\bibitem[{\citenamefont{{Rodr{\'{\i}}guez} and {Sierra}}(2009)}]{Rodriguez08}
\bibinfo{author}{\bibfnamefont{I.~D.} \bibnamefont{{Rodr{\'{\i}}guez}}}
  \bibnamefont{and} \bibinfo{author}{\bibfnamefont{G.}~\bibnamefont{{Sierra}}},
  \bibinfo{journal}{Phys. Rev. B} \textbf{\bibinfo{volume}{80}},
  \bibinfo{eid}{153303} (\bibinfo{year}{2009}), \eprint{arXiv:0811.2188}.

\bibitem[{\citenamefont{{Grover} et~al.}(2011)\citenamefont{{Grover}, {Turner},
  and {Vishwanath}}}]{Grover11}
\bibinfo{author}{\bibfnamefont{T.}~\bibnamefont{{Grover}}},
  \bibinfo{author}{\bibfnamefont{A.~M.} \bibnamefont{{Turner}}},
  \bibnamefont{and}
  \bibinfo{author}{\bibfnamefont{A.}~\bibnamefont{{Vishwanath}}},
  \bibinfo{journal}{Phys. Rev. B} \textbf{\bibinfo{volume}{84}},
  \bibinfo{eid}{195120} (\bibinfo{year}{2011}), \eprint{arXiv:1108.4038}.

\bibitem[{\citenamefont{Brown et~al.}(2013)\citenamefont{Brown, Bartlett,
  Doherty, and Barrett}}]{Brown11}
\bibinfo{author}{\bibfnamefont{B.~J.} \bibnamefont{Brown}},
  \bibinfo{author}{\bibfnamefont{S.~D.} \bibnamefont{Bartlett}},
  \bibinfo{author}{\bibfnamefont{A.~C.} \bibnamefont{Doherty}},
  \bibnamefont{and} \bibinfo{author}{\bibfnamefont{S.~D.}
  \bibnamefont{Barrett}}, \bibinfo{journal}{Phys. Rev. Lett.}
  \textbf{\bibinfo{volume}{111}}, \bibinfo{pages}{220402}
  (\bibinfo{year}{2013}), \eprint{arXiv:1303.4455}.

\bibitem[{\citenamefont{Kim}(2013)}]{Kim13}
\bibinfo{author}{\bibfnamefont{I.~H.} \bibnamefont{Kim}},
  \bibinfo{journal}{Phys. Rev. Lett.} \textbf{\bibinfo{volume}{111}},
  \bibinfo{pages}{080503} (\bibinfo{year}{2013}), \eprint{arXiv:1304.3925}.

\bibitem[{\citenamefont{Kim and Brown}(2015)}]{Kim15}
\bibinfo{author}{\bibfnamefont{I.~H.} \bibnamefont{Kim}} \bibnamefont{and}
  \bibinfo{author}{\bibfnamefont{B.~J.} \bibnamefont{Brown}},
  \bibinfo{journal}{Phys. Rev. B} \textbf{\bibinfo{volume}{92}},
  \bibinfo{pages}{115139} (\bibinfo{year}{2015}), \eprint{arXiv:1410.7411}.

\bibitem[{\citenamefont{{Wen} et~al.}(2016)\citenamefont{{Wen}, {Matsuura}, and
  {Ryu}}}]{Wen16a}
\bibinfo{author}{\bibfnamefont{X.}~\bibnamefont{{Wen}}},
  \bibinfo{author}{\bibfnamefont{S.}~\bibnamefont{{Matsuura}}},
  \bibnamefont{and} \bibinfo{author}{\bibfnamefont{S.}~\bibnamefont{{Ryu}}},
  \bibinfo{journal}{Phys. Rev. B} \textbf{\bibinfo{volume}{93}},
  \bibinfo{eid}{245140} (\bibinfo{year}{2016}), \eprint{arXiv:1603.08534}.

\bibitem[{\citenamefont{{Bullivant} and {Pachos}}(2016)}]{Bullivant16}
\bibinfo{author}{\bibfnamefont{A.}~\bibnamefont{{Bullivant}}} \bibnamefont{and}
  \bibinfo{author}{\bibfnamefont{J.~K.} \bibnamefont{{Pachos}}},
  \bibinfo{journal}{Phys. Rev. B} \textbf{\bibinfo{volume}{93}},
  \bibinfo{eid}{125111} (\bibinfo{year}{2016}), \eprint{arXiv:1504.02868}.

\bibitem[{\citenamefont{Furukawa and Misguich}(2007)}]{Furukawa07}
\bibinfo{author}{\bibfnamefont{S.}~\bibnamefont{Furukawa}} \bibnamefont{and}
  \bibinfo{author}{\bibfnamefont{G.}~\bibnamefont{Misguich}},
  \bibinfo{journal}{Phys. Rev. B} \textbf{\bibinfo{volume}{75}},
  \bibinfo{pages}{214407} (\bibinfo{year}{2007}), \eprint{cond-mat/0612227}.

\bibitem[{\citenamefont{Haque et~al.}(2007)\citenamefont{Haque, Zozulya, and
  Schoutens}}]{Haque07}
\bibinfo{author}{\bibfnamefont{M.}~\bibnamefont{Haque}},
  \bibinfo{author}{\bibfnamefont{O.}~\bibnamefont{Zozulya}}, \bibnamefont{and}
  \bibinfo{author}{\bibfnamefont{K.}~\bibnamefont{Schoutens}},
  \bibinfo{journal}{Phys. Rev. Lett.} \textbf{\bibinfo{volume}{98}},
  \bibinfo{pages}{060401} (\bibinfo{year}{2007}), \eprint{cond-mat/0609263}.

\bibitem[{\citenamefont{Zozulya et~al.}(2007)\citenamefont{Zozulya, Haque,
  Schoutens, and Rezayi}}]{Zozulya07}
\bibinfo{author}{\bibfnamefont{O.~S.} \bibnamefont{Zozulya}},
  \bibinfo{author}{\bibfnamefont{M.}~\bibnamefont{Haque}},
  \bibinfo{author}{\bibfnamefont{K.}~\bibnamefont{Schoutens}},
  \bibnamefont{and} \bibinfo{author}{\bibfnamefont{E.~H.}
  \bibnamefont{Rezayi}}, \bibinfo{journal}{Phys. Rev. B}
  \textbf{\bibinfo{volume}{76}}, \bibinfo{pages}{125310}
  (\bibinfo{year}{2007}), \eprint{arXiv:0705.4176}.

\bibitem[{\citenamefont{{Zozulya} et~al.}(2009)\citenamefont{{Zozulya},
  {Haque}, and {Regnault}}}]{Zozulya08}
\bibinfo{author}{\bibfnamefont{O.~S.} \bibnamefont{{Zozulya}}},
  \bibinfo{author}{\bibfnamefont{M.}~\bibnamefont{{Haque}}}, \bibnamefont{and}
  \bibinfo{author}{\bibfnamefont{N.}~\bibnamefont{{Regnault}}},
  \bibinfo{journal}{Phys. Rev. B} \textbf{\bibinfo{volume}{79}},
  \bibinfo{eid}{045409} (\bibinfo{year}{2009}), \eprint{arXiv:0809.1589}.

\bibitem[{\citenamefont{Yao and Qi}(2010)}]{Yao10}
\bibinfo{author}{\bibfnamefont{H.}~\bibnamefont{Yao}} \bibnamefont{and}
  \bibinfo{author}{\bibfnamefont{X.-L.} \bibnamefont{Qi}},
  \bibinfo{journal}{Phys. Rev. Lett.} \textbf{\bibinfo{volume}{105}},
  \bibinfo{pages}{080501} (\bibinfo{year}{2010}), \eprint{arXiv:1001.1165}.

\bibitem[{\citenamefont{{Isakov} et~al.}(2011)\citenamefont{{Isakov},
  {Hastings}, and {Melko}}}]{Isakov11}
\bibinfo{author}{\bibfnamefont{S.~V.} \bibnamefont{{Isakov}}},
  \bibinfo{author}{\bibfnamefont{M.~B.} \bibnamefont{{Hastings}}},
  \bibnamefont{and} \bibinfo{author}{\bibfnamefont{R.~G.}
  \bibnamefont{{Melko}}}, \bibinfo{journal}{Nature Physics}
  \textbf{\bibinfo{volume}{7}}, \bibinfo{pages}{772} (\bibinfo{year}{2011}),
  \eprint{arXiv:1102.1721}.

\bibitem[{\citenamefont{{Sterdyniak} et~al.}(2012)\citenamefont{{Sterdyniak},
  {Chandran}, {Regnault}, {Bernevig}, and {Bonderson}}}]{Sterdyniak11}
\bibinfo{author}{\bibfnamefont{A.}~\bibnamefont{{Sterdyniak}}},
  \bibinfo{author}{\bibfnamefont{A.}~\bibnamefont{{Chandran}}},
  \bibinfo{author}{\bibfnamefont{N.}~\bibnamefont{{Regnault}}},
  \bibinfo{author}{\bibfnamefont{B.~A.} \bibnamefont{{Bernevig}}},
  \bibnamefont{and}
  \bibinfo{author}{\bibfnamefont{P.}~\bibnamefont{{Bonderson}}},
  \bibinfo{journal}{Phys. Rev. B} \textbf{\bibinfo{volume}{85}},
  \bibinfo{eid}{125308} (\bibinfo{year}{2012}), \eprint{arXiv:1111.2810}.

\bibitem[{\citenamefont{Zhang et~al.}(2012)\citenamefont{Zhang, Grover, Turner,
  Oshikawa, and Vishwanath}}]{Zhang12}
\bibinfo{author}{\bibfnamefont{Y.}~\bibnamefont{Zhang}},
  \bibinfo{author}{\bibfnamefont{T.}~\bibnamefont{Grover}},
  \bibinfo{author}{\bibfnamefont{A.}~\bibnamefont{Turner}},
  \bibinfo{author}{\bibfnamefont{M.}~\bibnamefont{Oshikawa}}, \bibnamefont{and}
  \bibinfo{author}{\bibfnamefont{A.}~\bibnamefont{Vishwanath}},
  \bibinfo{journal}{Phys. Rev. B} \textbf{\bibinfo{volume}{85}},
  \bibinfo{pages}{235151} (\bibinfo{year}{2012}), \eprint{arXiv:1111.2342}.

\bibitem[{\citenamefont{{Jiang} et~al.}(2012)\citenamefont{{Jiang}, {Wang}, and
  {Balents}}}]{Jiang12}
\bibinfo{author}{\bibfnamefont{H.-C.} \bibnamefont{{Jiang}}},
  \bibinfo{author}{\bibfnamefont{Z.}~\bibnamefont{{Wang}}}, \bibnamefont{and}
  \bibinfo{author}{\bibfnamefont{L.}~\bibnamefont{{Balents}}},
  \bibinfo{journal}{Nature Physics} \textbf{\bibinfo{volume}{8}},
  \bibinfo{pages}{902} (\bibinfo{year}{2012}), \eprint{arXiv:1205.4289}.

\bibitem[{\citenamefont{Zaletel and Mong}(2012)}]{Zaletel12}
\bibinfo{author}{\bibfnamefont{M.~P.} \bibnamefont{Zaletel}} \bibnamefont{and}
  \bibinfo{author}{\bibfnamefont{R.~S.~K.} \bibnamefont{Mong}},
  \bibinfo{journal}{Phys. Rev. B} \textbf{\bibinfo{volume}{86}},
  \bibinfo{pages}{245305} (\bibinfo{year}{2012}), \eprint{arXiv:1208.4862}.

\bibitem[{\citenamefont{Zaletel et~al.}(2013)\citenamefont{Zaletel, Mong, and
  Pollmann}}]{Zaletel13}
\bibinfo{author}{\bibfnamefont{M.~P.} \bibnamefont{Zaletel}},
  \bibinfo{author}{\bibfnamefont{R.~S.~K.} \bibnamefont{Mong}},
  \bibnamefont{and} \bibinfo{author}{\bibfnamefont{F.}~\bibnamefont{Pollmann}},
  \bibinfo{journal}{Phys. Rev. Lett.} \textbf{\bibinfo{volume}{110}},
  \bibinfo{pages}{236801} (\bibinfo{year}{2013}), \eprint{arXiv:1211.3733}.

\bibitem[{\citenamefont{Cincio and Vidal}(2013)}]{Cincio13}
\bibinfo{author}{\bibfnamefont{L.}~\bibnamefont{Cincio}} \bibnamefont{and}
  \bibinfo{author}{\bibfnamefont{G.}~\bibnamefont{Vidal}},
  \bibinfo{journal}{Phys. Rev. Lett.} \textbf{\bibinfo{volume}{110}},
  \bibinfo{pages}{067208} (\bibinfo{year}{2013}), \eprint{arXiv:1208.2623}.

\bibitem[{\citenamefont{Estienne et~al.}(2015)\citenamefont{Estienne, Regnault,
  and Bernevig}}]{Estienne14}
\bibinfo{author}{\bibfnamefont{B.}~\bibnamefont{Estienne}},
  \bibinfo{author}{\bibfnamefont{N.}~\bibnamefont{Regnault}}, \bibnamefont{and}
  \bibinfo{author}{\bibfnamefont{B.}~\bibnamefont{Bernevig}},
  \bibinfo{journal}{Phys. Rev. Lett.} \textbf{\bibinfo{volume}{114}},
  \bibinfo{pages}{186801} (\bibinfo{year}{2015}), \eprint{arXiv:1406.6262}.

\bibitem[{\citenamefont{Grushin et~al.}(2015)\citenamefont{Grushin, Motruk,
  Zaletel, and Pollmann}}]{Grushin15}
\bibinfo{author}{\bibfnamefont{A.~G.} \bibnamefont{Grushin}},
  \bibinfo{author}{\bibfnamefont{J.}~\bibnamefont{Motruk}},
  \bibinfo{author}{\bibfnamefont{M.~P.} \bibnamefont{Zaletel}},
  \bibnamefont{and} \bibinfo{author}{\bibfnamefont{F.}~\bibnamefont{Pollmann}},
  \bibinfo{journal}{Phys. Rev. B} \textbf{\bibinfo{volume}{91}},
  \bibinfo{pages}{035136} (\bibinfo{year}{2015}), \eprint{arXiv:1407.6985}.

\bibitem[{\citenamefont{Moore and Seiberg}(1989)}]{Moore89b}
\bibinfo{author}{\bibfnamefont{G.}~\bibnamefont{Moore}} \bibnamefont{and}
  \bibinfo{author}{\bibfnamefont{N.}~\bibnamefont{Seiberg}},
  \bibinfo{journal}{Commun. Math. Phys.} \textbf{\bibinfo{volume}{123}},
  \bibinfo{pages}{177} (\bibinfo{year}{1989}).

\bibitem[{\citenamefont{Turaev}(1994)}]{Turaev94}
\bibinfo{author}{\bibfnamefont{V.~G.} \bibnamefont{Turaev}},
  \emph{\bibinfo{title}{Quantum Invariants of Knots and 3-Manifolds}}
  (\bibinfo{publisher}{Walter de Gruyter}, \bibinfo{address}{Berlin, New York},
  \bibinfo{year}{1994}).

\bibitem[{\citenamefont{Bakalov and Kirillov}(2001)}]{Bakalov01}
\bibinfo{author}{\bibfnamefont{B.}~\bibnamefont{Bakalov}} \bibnamefont{and}
  \bibinfo{author}{\bibfnamefont{A.}~\bibnamefont{Kirillov}},
  \emph{\bibinfo{title}{Lectures on Tensor Categories and Modular Functors}},
  vol.~\bibinfo{volume}{21} of \emph{\bibinfo{series}{University Lecture
  Series}} (\bibinfo{publisher}{American Mathematical Society},
  \bibinfo{year}{2001}).

\bibitem[{\citenamefont{Preskill}()}]{Preskill-lectures}
\bibinfo{author}{\bibfnamefont{J.}~\bibnamefont{Preskill}},
  \emph{\bibinfo{title}{lecture notes}},
  \\
  \bibinfo{howpublished}{\url{http://www.theory.caltech.edu/~preskill/ph219/topological.pdf}}.

\bibitem[{\citenamefont{Kitaev}(2006)}]{Kitaev06a}
\bibinfo{author}{\bibfnamefont{A.}~\bibnamefont{Kitaev}},
  \bibinfo{journal}{Annals Phys.} \textbf{\bibinfo{volume}{321}},
  \bibinfo{pages}{2} (\bibinfo{year}{2006}), \eprint{cond-mat/0506438}.

\bibitem[{\citenamefont{{Bonderson}}(2007)}]{Bonderson07b}
\bibinfo{author}{\bibfnamefont{P.~H.} \bibnamefont{{Bonderson}}}, Ph.D. thesis,
  \bibinfo{school}{California Institute of Technology} (\bibinfo{year}{2007}).

\bibitem[{\citenamefont{Bonderson
  et~al.}(2008{\natexlab{a}})\citenamefont{Bonderson, Shtengel, and
  Slingerland}}]{Bonderson07c}
\bibinfo{author}{\bibfnamefont{P.}~\bibnamefont{Bonderson}},
  \bibinfo{author}{\bibfnamefont{K.}~\bibnamefont{Shtengel}}, \bibnamefont{and}
  \bibinfo{author}{\bibfnamefont{J.~K.} \bibnamefont{Slingerland}},
  \bibinfo{journal}{Annals of Physics} \textbf{\bibinfo{volume}{323}},
  \bibinfo{pages}{2709} (\bibinfo{year}{2008}{\natexlab{a}}),
  \eprint{arXiv:0707.4206}.

\bibitem[{\citenamefont{Nielsen and Chuang}(2011)}]{Nielsen11}
\bibinfo{author}{\bibfnamefont{M.~A.} \bibnamefont{Nielsen}} \bibnamefont{and}
  \bibinfo{author}{\bibfnamefont{I.~L.} \bibnamefont{Chuang}},
  \emph{\bibinfo{title}{Quantum Computation and Quantum Information: 10th
  Anniversary Edition}} (\bibinfo{publisher}{Cambridge University Press},
  \bibinfo{address}{New York, NY, USA}, \bibinfo{year}{2011}),
  \bibinfo{edition}{10th} ed., ISBN \bibinfo{isbn}{1107002176, 9781107002173}.

\bibitem[{\citenamefont{Turaev and Viro}(1992)}]{Turaev92}
\bibinfo{author}{\bibfnamefont{V.~G.} \bibnamefont{Turaev}} \bibnamefont{and}
  \bibinfo{author}{\bibfnamefont{O.~Y.} \bibnamefont{Viro}},
  \bibinfo{journal}{Topology} \textbf{\bibinfo{volume}{31}},
  \bibinfo{pages}{865} (\bibinfo{year}{1992}).

\bibitem[{\citenamefont{Levin and Wen}(2005)}]{Levin05a}
\bibinfo{author}{\bibfnamefont{M.~A.} \bibnamefont{Levin}} \bibnamefont{and}
  \bibinfo{author}{\bibfnamefont{X.-G.} \bibnamefont{Wen}},
  \bibinfo{journal}{Phys. Rev. B} \textbf{\bibinfo{volume}{71}},
  \bibinfo{eid}{045110} (\bibinfo{year}{2005}), \eprint{cond-mat/0404617}.

\bibitem[{\citenamefont{Li and Haldane}(2008)}]{Li08}
\bibinfo{author}{\bibfnamefont{H.}~\bibnamefont{Li}} \bibnamefont{and}
  \bibinfo{author}{\bibfnamefont{F.~D.~M.} \bibnamefont{Haldane}},
  \bibinfo{journal}{Phys. Rev. Lett.} \textbf{\bibinfo{volume}{101}},
  \bibinfo{pages}{010504} (\bibinfo{year}{2008}), \eprint{arXiv:0805.0332}.

\bibitem[{\citenamefont{{Bruillard} et~al.}(2016)\citenamefont{{Bruillard},
  {Ng}, {Rowell}, and {Wang}}}]{Bruillard13}
\bibinfo{author}{\bibfnamefont{P.}~\bibnamefont{{Bruillard}}},
  \bibinfo{author}{\bibfnamefont{S.-H.} \bibnamefont{{Ng}}},
  \bibinfo{author}{\bibfnamefont{E.~C.} \bibnamefont{{Rowell}}},
  \bibnamefont{and} \bibinfo{author}{\bibfnamefont{Z.}~\bibnamefont{{Wang}}},
  \bibinfo{journal}{J. Amer. Math. Soc.} \textbf{\bibinfo{volume}{29}}
  (\bibinfo{year}{2016}), \eprint{arXiv:1310.7050}.

\bibitem[{\citenamefont{Rowell et~al.}(2009)\citenamefont{Rowell, Stong, and
  Wang}}]{Rowell07}
\bibinfo{author}{\bibfnamefont{E.}~\bibnamefont{Rowell}},
  \bibinfo{author}{\bibfnamefont{R.}~\bibnamefont{Stong}}, \bibnamefont{and}
  \bibinfo{author}{\bibfnamefont{Z.}~\bibnamefont{Wang}},
  \bibinfo{journal}{Comm. Math. Phys.} \textbf{\bibinfo{volume}{292}},
  \bibinfo{pages}{343} (\bibinfo{year}{2009}), \eprint{arXiv:0712.1377}.

\bibitem[{\citenamefont{Bonderson and Slingerland}()}]{BondersonWIP}
\bibinfo{author}{\bibfnamefont{P.}~\bibnamefont{Bonderson}} \bibnamefont{and}
  \bibinfo{author}{\bibfnamefont{J.~K.} \bibnamefont{Slingerland}},
  \bibinfo{note}{in preparation}.

\bibitem[{\citenamefont{{Bonderson} et~al.}(2009)\citenamefont{{Bonderson},
  {Freedman}, and {Nayak}}}]{Bonderson09}
\bibinfo{author}{\bibfnamefont{P.}~\bibnamefont{{Bonderson}}},
  \bibinfo{author}{\bibfnamefont{M.}~\bibnamefont{{Freedman}}},
  \bibnamefont{and} \bibinfo{author}{\bibfnamefont{C.}~\bibnamefont{{Nayak}}},
  \bibinfo{journal}{Annals of Physics} \textbf{\bibinfo{volume}{324}},
  \bibinfo{pages}{787} (\bibinfo{year}{2009}), \eprint{arXiv:0808.1933}.

\bibitem[{\citenamefont{Bonderson et~al.}(2009)\citenamefont{Bonderson,
  Freedman, and Nayak}}]{Bonderson08b}
\bibinfo{author}{\bibfnamefont{P.}~\bibnamefont{Bonderson}},
  \bibinfo{author}{\bibfnamefont{M.}~\bibnamefont{Freedman}}, \bibnamefont{and}
  \bibinfo{author}{\bibfnamefont{C.}~\bibnamefont{Nayak}},
  \bibinfo{journal}{Annals Phys.} \textbf{\bibinfo{volume}{324}},
  \bibinfo{pages}{787} (\bibinfo{year}{2009}), \eprint{arXiv:0808.1933}.

\bibitem[{\citenamefont{Bonderson
  et~al.}(2008{\natexlab{b}})\citenamefont{Bonderson, Freedman, and
  Nayak}}]{Bonderson08a}
\bibinfo{author}{\bibfnamefont{P.}~\bibnamefont{Bonderson}},
  \bibinfo{author}{\bibfnamefont{M.}~\bibnamefont{Freedman}}, \bibnamefont{and}
  \bibinfo{author}{\bibfnamefont{C.}~\bibnamefont{Nayak}},
  \bibinfo{journal}{Phys. Rev. Lett.} \textbf{\bibinfo{volume}{101}},
  \bibinfo{pages}{010501} (\bibinfo{year}{2008}{\natexlab{b}}),
  \eprint{arXiv:0802.0279}.

\bibitem[{\citenamefont{{Hikami}}(2008)}]{Hikami08}
\bibinfo{author}{\bibfnamefont{K.}~\bibnamefont{{Hikami}}},
  \bibinfo{journal}{Annals of Physics} \textbf{\bibinfo{volume}{323}},
  \bibinfo{pages}{1729} (\bibinfo{year}{2008}), \eprint{arXiv:0709.2409}.

\bibitem[{\citenamefont{{Kato} et~al.}(2014)\citenamefont{{Kato}, {Furrer}, and
  {Murao}}}]{Kato14}
\bibinfo{author}{\bibfnamefont{K.}~\bibnamefont{{Kato}}},
  \bibinfo{author}{\bibfnamefont{F.}~\bibnamefont{{Furrer}}}, \bibnamefont{and}
  \bibinfo{author}{\bibfnamefont{M.}~\bibnamefont{{Murao}}},
  \bibinfo{journal}{Phys. Rev. A} \textbf{\bibinfo{volume}{90}},
  \bibinfo{eid}{062325} (\bibinfo{year}{2014}), \eprint{arXiv:1310.4140}.

\bibitem[{\citenamefont{{Pfeifer}}(2014)}]{Pfeifer14}
\bibinfo{author}{\bibfnamefont{R.~N.~C.} \bibnamefont{{Pfeifer}}},
  \bibinfo{journal}{Phys. Rev. B} \textbf{\bibinfo{volume}{89}},
  \bibinfo{eid}{035105} (\bibinfo{year}{2014}), \eprint{arXiv:1310.0373}.

\bibitem[{\citenamefont{{Cardy}}(2006)}]{Cardy04}
\bibinfo{author}{\bibfnamefont{J.}~\bibnamefont{{Cardy}}},
  \bibinfo{journal}{Encyclopedia of Mathematical Physics}
  (\bibinfo{year}{2006}), \eprint{hep-th/0411189}.

\bibitem[{\citenamefont{{Pfeifer} et~al.}(2012)\citenamefont{{Pfeifer},
  {Buerschaper}, {Trebst}, {Ludwig}, {Troyer}, and {Vidal}}}]{Pfeifer10}
\bibinfo{author}{\bibfnamefont{R.~N.~C.} \bibnamefont{{Pfeifer}}},
  \bibinfo{author}{\bibfnamefont{O.}~\bibnamefont{{Buerschaper}}},
  \bibinfo{author}{\bibfnamefont{S.}~\bibnamefont{{Trebst}}},
  \bibinfo{author}{\bibfnamefont{A.~W.~W.} \bibnamefont{{Ludwig}}},
  \bibinfo{author}{\bibfnamefont{M.}~\bibnamefont{{Troyer}}}, \bibnamefont{and}
  \bibinfo{author}{\bibfnamefont{G.}~\bibnamefont{{Vidal}}},
  \bibinfo{journal}{Phys. Rev. B} \textbf{\bibinfo{volume}{86}},
  \bibinfo{eid}{155111} (\bibinfo{year}{2012}), \eprint{arXiv:1005.5486}.

\bibitem[{\citenamefont{Witten}(1989)}]{Witten89}
\bibinfo{author}{\bibfnamefont{E.}~\bibnamefont{Witten}},
  \bibinfo{journal}{Comm. Math. Phys.} \textbf{\bibinfo{volume}{121}},
  \bibinfo{pages}{351} (\bibinfo{year}{1989}).

\bibitem[{\citenamefont{Stevens}(2006)}]{Stevens06}
\bibinfo{author}{\bibfnamefont{T.}~\bibnamefont{Stevens}},
  \emph{\bibinfo{title}{Speech Opposing Network Neutrality}}
  (\bibinfo{publisher}{United States Senate}, \bibinfo{address}{Washington DC,
  USA}, \bibinfo{year}{2006}).

\bibitem[{\citenamefont{{Flammia} et~al.}(2009)\citenamefont{{Flammia},
  {Hamma}, {Hughes}, and {Wen}}}]{Flammia09}
\bibinfo{author}{\bibfnamefont{S.~T.} \bibnamefont{{Flammia}}},
  \bibinfo{author}{\bibfnamefont{A.}~\bibnamefont{{Hamma}}},
  \bibinfo{author}{\bibfnamefont{T.~L.} \bibnamefont{{Hughes}}},
  \bibnamefont{and} \bibinfo{author}{\bibfnamefont{X.-G.} \bibnamefont{{Wen}}},
  \bibinfo{journal}{Phys. Rev. Lett.} \textbf{\bibinfo{volume}{103}},
  \bibinfo{eid}{261601} (\bibinfo{year}{2009}), \eprint{arXiv:0909.3305}.

\bibitem[{\citenamefont{{Barkeshli} et~al.}(2014)\citenamefont{{Barkeshli},
  {Bonderson}, {Cheng}, and {Wang}}}]{Barkeshli14}
\bibinfo{author}{\bibfnamefont{M.}~\bibnamefont{{Barkeshli}}},
  \bibinfo{author}{\bibfnamefont{P.}~\bibnamefont{{Bonderson}}},
  \bibinfo{author}{\bibfnamefont{M.}~\bibnamefont{{Cheng}}}, \bibnamefont{and}
  \bibinfo{author}{\bibfnamefont{Z.}~\bibnamefont{{Wang}}}
  (\bibinfo{year}{2014}), \eprint{arXiv:1410.4540}.

\bibitem[{\citenamefont{{Cheng} et~al.}(2016)\citenamefont{{Cheng}, {Zaletel},
  {Barkeshli}, {Vishwanath}, and {Bonderson}}}]{Cheng15}
\bibinfo{author}{\bibfnamefont{M.}~\bibnamefont{{Cheng}}},
  \bibinfo{author}{\bibfnamefont{M.}~\bibnamefont{{Zaletel}}},
  \bibinfo{author}{\bibfnamefont{M.}~\bibnamefont{{Barkeshli}}},
  \bibinfo{author}{\bibfnamefont{A.}~\bibnamefont{{Vishwanath}}},
  \bibnamefont{and}
  \bibinfo{author}{\bibfnamefont{P.}~\bibnamefont{{Bonderson}}},
  \bibinfo{journal}{Phys. Rev. X} \textbf{\bibinfo{volume}{6}},
  \bibinfo{eid}{041068} (\bibinfo{year}{2016}), \eprint{arXiv:1511.02263}.

\bibitem[{\citenamefont{Dijkgraaf and Witten}(1990)}]{Dijkgraaf90}
\bibinfo{author}{\bibfnamefont{R.}~\bibnamefont{Dijkgraaf}} \bibnamefont{and}
  \bibinfo{author}{\bibfnamefont{E.}~\bibnamefont{Witten}},
  \bibinfo{journal}{Commun. Math. Phys.} \textbf{\bibinfo{volume}{129}},
  \bibinfo{pages}{393} (\bibinfo{year}{1990}).

\bibitem[{\citenamefont{Bonderson et~al.}()\citenamefont{Bonderson, Cheng,
  Mong, and Tran}}]{Bonderson_FMT}
\bibinfo{author}{\bibfnamefont{P.}~\bibnamefont{Bonderson}},
  \bibinfo{author}{\bibfnamefont{M.}~\bibnamefont{Cheng}},
  \bibinfo{author}{\bibfnamefont{R.}~\bibnamefont{Mong}}, \bibnamefont{and}
  \bibinfo{author}{\bibfnamefont{A.}~\bibnamefont{Tran}}, \bibinfo{note}{in
  preparation}.

\bibitem[{\citenamefont{{Lan} et~al.}(2016)\citenamefont{{Lan}, {Kong}, and
  {Wen}}}]{Lan16}
\bibinfo{author}{\bibfnamefont{T.}~\bibnamefont{{Lan}}},
  \bibinfo{author}{\bibfnamefont{L.}~\bibnamefont{{Kong}}}, \bibnamefont{and}
  \bibinfo{author}{\bibfnamefont{X.-G.} \bibnamefont{{Wen}}},
  \bibinfo{journal}{Phys. Rev. B} \textbf{\bibinfo{volume}{94}},
  \bibinfo{eid}{155113} (\bibinfo{year}{2016}), \eprint{arXiv:1507.04673}.

\bibitem[{\citenamefont{{Barkeshli} et~al.}(2016)\citenamefont{{Barkeshli},
  {Bonderson}, {Jian}, {Cheng}, and {Walker}}}]{Barkeshli16}
\bibinfo{author}{\bibfnamefont{M.}~\bibnamefont{{Barkeshli}}},
  \bibinfo{author}{\bibfnamefont{P.}~\bibnamefont{{Bonderson}}},
  \bibinfo{author}{\bibfnamefont{C.-M.} \bibnamefont{{Jian}}},
  \bibinfo{author}{\bibfnamefont{M.}~\bibnamefont{{Cheng}}}, \bibnamefont{and}
  \bibinfo{author}{\bibfnamefont{K.}~\bibnamefont{{Walker}}}
  (\bibinfo{year}{2016}), \eprint{arXiv:1612.07792}.

\bibitem[{\citenamefont{{von Keyserlingk} et~al.}(2013)\citenamefont{{von
  Keyserlingk}, {Burnell}, and {Simon}}}]{vonKeyserlingk12}
\bibinfo{author}{\bibfnamefont{C.~W.} \bibnamefont{{von Keyserlingk}}},
  \bibinfo{author}{\bibfnamefont{F.~J.} \bibnamefont{{Burnell}}},
  \bibnamefont{and} \bibinfo{author}{\bibfnamefont{S.~H.}
  \bibnamefont{{Simon}}}, \bibinfo{journal}{Phys. Rev. B}
  \textbf{\bibinfo{volume}{87}}, \bibinfo{eid}{045107} (\bibinfo{year}{2013}),
  \eprint{arXiv:1208.5128}.

\bibitem[{\citenamefont{{Walker} and {Wang}}(2012)}]{Walker12}
\bibinfo{author}{\bibfnamefont{K.}~\bibnamefont{{Walker}}} \bibnamefont{and}
  \bibinfo{author}{\bibfnamefont{Z.}~\bibnamefont{{Wang}}},
  \bibinfo{journal}{Frontiers of Physics} \textbf{\bibinfo{volume}{7}},
  \bibinfo{pages}{150} (\bibinfo{year}{2012}), \eprint{arXiv:1104.2632}.

\end{thebibliography}
\end{document}